%% file: thesis.tex
\documentclass[a4paper,12pt,times,numbered,print,index,openany]{Classes/PhDThesisPSnPDF}
\input{Preamble/preamble}
\ifdefineAbstract
\fi
\begin{document}
%%%%%%%%%%%%%%%%%%%%%%%%%%%%%%%%%
%%%% Define Titlepage Format
%%%%%%%%%%%%%%%%%%%%%%%%%%%%%%%%%
\newcommand{\address}[1]{\newcommand{\@address}{#1}}
\newcommand{\institute}[1]{\newcommand{\@institute}{#1}}
\newcommand{\rf}[1]{\textcolor{black}{#1}}
\renewcommand{\maketitle}{
	\begin{titlepage}
		{\flushright
			\includegraphics[width=0.3\textwidth]{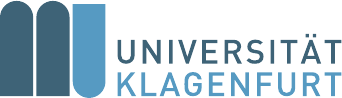}\par\vspace{1.5cm}
		}
		\centering
		{\large\noindent\ignorespaces\@ Reza Farahani\par}\vskip 1em 
		
		\noindent\rule{\textwidth}{1pt}%		
		\vspace{0.8\baselineskip}
		{%
			\LARGE%
			\@Network-Assisted Delivery of Adaptive Video Streaming Services through CDN, SDN, and MEC
		}
		\vspace{0.5\baselineskip}
		\noindent\rule{\textwidth}{1pt}
		
		\vskip .5em
		
		{\LARGE\noindent\ignorespaces \textsc{Dissertation}\par}\vskip 2.5em
		{\large\noindent\ignorespaces \small {zur Erlangung des akademischen Grades\\Doktor der Technischen Wissenschaften\par}}
		
		\vskip 1em
		%{\Large\textsc{Studium}\par}\vskip 0.1em
		{\large{\vspace{0.5\baselineskip} \small{Doktoratsstudium der Technischen Wissenschaften\\\raggedbottom im Dissertationsgebiet Angewandte Informatik}\par}}
		\vskip 1em
		
		{\large\noindent\ignorespaces \small {Alpen-Adria-Universität Klagenfurt\\Fakultät für Technische Wissenschaften\par}}
		
		\vskip 2em
		
		\begin{minipage}{0.55\textwidth}
			\flushleft\large
			\textsc{Erstbetreuer}\\
			\footnotesize
			Univ.-Prof. DI Dr. \textbf{Hermann Hellwagner}\\
			Institut für Informationstechnologie\\
			Alpen-Adria-Universität Klagenfurt
		\end{minipage}
		\begin{minipage}{0.44\textwidth}
			\flushright\large
			\textsc{Zweitbetreuer}\\
			\footnotesize
			Univ.-Prof. DI Dr. \textbf{Christian Timmerer}\\
			Institut für Informationstechnologie\\
			Alpen-Adria-Universität Klagenfurt
		\end{minipage}
%%%%%
\vskip 1.5em
		\begin{minipage}{0.55\textwidth}
			\flushleft\large
			\textsc{Erstgutachter}\\
			\footnotesize
			Prof. Dr. Ir. \textbf{Filip De Turck}\\
			Department of Information Technology (Intec)\\
			Ghent University
		\end{minipage}
		\begin{minipage}{0.44\textwidth}
			\flushright\large
			\textsc{Zweitgutachter}\\
			\footnotesize
			Prof. Dr. \textbf{Tobias Hoßfeld}\\
			Institut für Informatik\\
			University of Würzburg
		\end{minipage}  
  \vskip 4em
		
		{\normalsize\noindent\ignorespaces\@\small Klagenfurt am Wörthersee, \@July 2023}
	\end{titlepage}
}
\makeatother
\maketitle
\pagenumbering{roman}
\include{Chapters/Dedication/dedication.tex}

\include{Chapters/Declaration/declaration.tex}
\include{Chapters/Acknowledgement/acknowledgement.tex}
\include{Chapters/Abstract/abstract.tex}
\include{Chapters/Abstract/Kurzfassung}
% *********************** Adding TOC and List of Figures ***********************
\tableofcontents
%%%
\setcounter{page}{5}
%%%%
\listoffigures
\listoftables
\input{Chapters/Abbreviations.tex}

% ******************************** Main Matter *********************************
\mainmatter
%%%
\input{Chapters/Chapter1/1-1-Intro.tex}
\input{Chapters/Chapter1/1-2-Motivation.tex}

\input{Chapters/Chapter1/1-3-RQs.tex}

\input{Chapters/Chapter1/1-4-Methodology.tex}

\input{Chapters/Chapter1/1-5-Contribution.tex}

\input{Chapters/Chapter1/1-6-Publications}
\input{Chapters/Chapter1/1-7-Organization}

%%%
\input{Chapters/Chapter2/2-0-Intro}

\input{Chapters/Chapter2/2-1-0-Background}

\input{Chapters/Chapter2/2-1-1-HAS}

\input{Chapters/Chapter2/2-1-2-CDN}
\input{Chapters/Chapter2/2-1-3-P2P}
\input{Chapters/Chapter2/2-1-4-SDN}
\input{Chapters/Chapter2/2-1-5-NFV}
\input{Chapters/Chapter2/2-1-6-SFC}
\input{Chapters/Chapter2/2-1-7-MEC}

\input{Chapters/Chapter2/2-2-SOTA}

%%%
\input{Chapters/Chapter3/3-1-Intro.tex}

\input{Chapters/Chapter3/3-2-ESHAS.tex}

\input{Chapters/Chapter3/3-3-CSDN}

%%%
\input{Chapters/Chapter4/4-2-SARENA}
%%%
\input{Chapters/Chapter5/5-1-Intro.tex}
\input{Chapters/Chapter5/5-2-LEADER}
\input{Chapters/Chapter5/5-3-ARARAT}
%%%
\input{Chapters/Chapter6/6-1-Intro.tex}
\input{Chapters/Chapter6/6-2-RICHTER}

\input{Chapters/Chapter6/6-3-ALIVE}

%%%
\input{Chapters/Chapter7/7-1-Intro.tex}

\input{Chapters/Chapter7/7-2-Summary}

\input{Chapters/Chapter7/7-3-outlook}
%%%
\begin{spacing}{0.8}
\bibliographystyle{plainnat} % use this to have URLs listed in References
\cleardoublepage
% \bibliography{thesis} % Path to your References.bib file
%%%%%%%%%%%%%%%%%

\end{spacing}
\printthesisindex 
\end{document}

%% file: Chapters/Dedication/dedication.tex
\begin{dedication} 
I would like to dedicate this thesis to my loving parents and my wife, \textit{Sepideh}, who have always motivated me to pursue and accomplish my dreams~\dots
\end{dedication}

%% file: Chapters/Declaration/declaration.tex
% ******************************* Thesis Declaration ***************************
\chapter*{Eidesstattliche Erklärung}
%\vskip 2em
{
	\singlespacing
	Ich versichere an Eides statt, dass ich	
	\begin{itemize}[noitemsep]
		\item[--] die eingereichte wissenschaftliche Arbeit selbstständig verfasst und keine anderen als die ange­gebenen Hilfsmittel benutzt habe,
		\item[--] die während des Arbeitsvorganges von dritter Seite erfahrene Unterstützung, ein­schließ­lich signifikanter Betreuungshinweise, vollständig offengelegt habe,
		\item[--] die Inhalte, die ich aus Werken Dritter oder eigenen Werken wortwörtlich oder sinn­gemäß übernommen habe, in geeigneter Form gekennzeichnet und den Ursprung der                Infor­mation durch möglichst exakte Quellenangaben (z.B. in Fußnoten) ersichtlich gemacht habe,
		\item[--] die eingereichte wissenschaftliche Arbeit bisher weder im Inland noch im Ausland einer Prüfungsbehörde vorgelegt habe und
		\item[--] bei der Weitergabe jedes Exemplars (z.B. in gebundener, gedruckter oder digitaler Form) der wissenschaftlichen Arbeit sicherstelle, dass diese mit der eingereichten digitalen Version übereinstimmt.
	\end{itemize}
	\noindent Ich bin mir bewusst, dass eine tatsachenwidrige Erklärung rechtliche Folgen haben wird.}
\vskip 6em

\begin{minipage}{\columnwidth}
\begin{minipage}{0.49\columnwidth}
	Reza Farahani\\Klagenfurt am Wörthersee, July 2023
\end{minipage} 
\end{minipage}

\chapter*{Affidavit}
{	
	\singlespacing
	I hereby declare in lieu of an oath that
	\begin{itemize}
		\item[--] the submitted academic paper is entirely my own work and that no auxiliary materials have been used other than those indicated,
		\item[--] I have fully disclosed all assistance received from third parties during the process of writing the thesis, including any significant advice from supervisors,
		\item[--] any contents taken from the works of third parties or my own works that have been included either literally or in spirit have been appropriately marked and the respective source of the information has been clearly identified with precise bibliographical references (e.g. in footnotes),
		\item[--] to date, I have not submitted this paper to an examining authority either in Austria or abroad and that
		\item[--] when passing on copies of the academic thesis (e.g. in bound, printed or digital form), I will ensure that each copy is fully consistent with the submitted digital version.
	\end{itemize}
	I am aware that a declaration contrary to the facts will have legal consequences.	
}
\vskip 6em
\begin{minipage}{\columnwidth}
	\begin{minipage}{0.49\columnwidth}
	Reza Farahani\\Klagenfurt am Wörthersee, July 2023
	\end{minipage} 
	\end{minipage}

%% file: Chapters/Acknowledgement/acknowledgement.tex
\begin{acknowledgements} 
Three and a half incredible years of research work are enclosed in the pages of this thesis. During this time, I had the honor of encountering wonderful moments, making significant discoveries, overcoming substantial obstacles, and meeting outstanding individuals. I owe my success to the numerous kind-hearted and professional people who supported and assisted me throughout my Ph.D. journey.  I would like to take this opportunity to give them credit and thank them for their help and support.

First and foremost, I would like to thank my excellent supervisors, Professor \textbf{Hermann Hellwagner} and Professor \textbf{Christian Timmerer}. 
It has been a great pleasure to be able to work under their supervision, who guided my first steps as a researcher. I will be forever grateful to them for their patience, precious time, tremendous guidance, and endless support. 

I want to thank my ex-professor \textbf{Mohammad Ghanbari} at the University of Essex. I had the opportunity to continue collaborating with him and receive his valuable comments and feedback to improve my work. 

I also would like to express my appreciation to Professor \textbf{Radu Prodan}, with whom I had the opportunity to collaborate on this dissertation's last contribution. I thank him for the insightful discussions and fruitful communications that gave me valuable experience to prepare for future projects.

I also like to thank my Ph.D. visiting mentor and collaborator at the University of Surrey, Dr. \textbf{Mohammad Shojafar}. I am thankful for all the online discussions and for hosting my three-month research stay at the 5GIC/6GIC, University of Surrey. I had the opportunity to work closely with Dr. \textbf{Abdelhak Bentaleb}, a postdoctoral researcher at the National University of Singapore at the time and currently an assistant professor at Concordia University. 

I want to thank all of my \textbf{co-authors}, also my colleagues at the ATHENA Christian Doppler (CD) Laboratory, plus my colleagues in the Department of Information Technology (ITEC) of the University of Klagenfurt, especially \textbf{Nina} and \textbf{Martina}, for their constant support. \rf{I would also like to acknowledge Professor \textbf{Filip De Turck} and Professor \textbf{Tobias Hoßfeld} for their time and insightful feedback.}

I would also like to mention the \textbf{CloudLab} team, who have constantly supported me in running my experiments, no matter how unreasonable my demands were. 
The research leading to this dissertation has received funding from the Austrian Federal Ministry for Digital and Economic Affairs, the National Foundation for Research, Technology, and Development, and the Christian Doppler Research Association; their financial support is gratefully acknowledged. \textbf{Christian Doppler Laboratory ATHENA}: \url{https://athena.itec.aau.at/}.
%%%%%
\begin{figure}[!h]
	\centering
	\includegraphics[scale=.3]{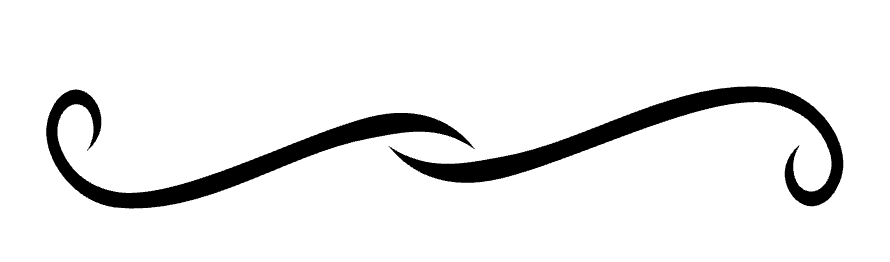}
\end{figure}
%%%%%%

\textit{Last but not least, I also want to thank my family, especially my parents and my wife: \textbf{Majid}, \textbf{Masoumeh}, and \textbf{Sepideh}. None of this would have been possible without their continued support and endless love; I will forever be indebted to them for making me who I am now.}
\\
\\
Reza Farahani,\\
Klagenfurt, July 2023
\end{acknowledgements}

%% file: Chapters/Abstract/abstract.tex
\begin{abstract}
\vspace{-1cm}
Multimedia applications, mainly video streaming services, are currently the dominant source of network load worldwide. 
In recent Video-on-Demand (VoD) and live video streaming services, traditional streaming delivery techniques have been replaced by adaptive solutions based on the HTTP protocol. Current trends toward high-resolution (\eg 8K) and/or low-latency VoD and live video streaming pose new challenges to end-to-end (E2E) bandwidth demand and have stringent delay requirements. To do this, video providers typically rely on \textit{Content Delivery Networks} (CDNs) to ensure that they provide scalable video streaming services. To support future streaming scenarios involving millions of users, it is necessary to increase the CDNs' efficiency. It is widely agreed that these requirements may be satisfied by adopting emerging networking techniques to present \textit{Network-Assisted Video Streaming} (NAVS) methods. Motivated by this, this thesis goes one step beyond traditional pure client-based HAS algorithms by incorporating (an) in-network component(s) with a broader view of the network to present \rf{completely transparent} NAVS solutions for HAS clients.

Our first contribution concentrates on leveraging the capabilities of the \textit{Software Defined Networking} (SDN), \textit{Network Function Virtualization} (NFV), and \textit{Multi-Access Edge Computing} (MEC) paradigms to introduce \texttt{ES-HAS} and \texttt{CSDN} as edge- and SDN-assisted frameworks, \rf{mainly for VoD and live streaming, respectively}. \texttt{ES-HAS} and \texttt{CSDN} introduce \textit{Virtual Network Functions} (VNFs) named \textit{Virtual Reverse Proxy} (VRP) servers at the edge of \rf{an SDN-enabled network} to collect HAS clients' requests and retrieve networking information. \rf{The SDN controller in these systems manages a single domain network.} VRP servers perform optimization models as server/segment selection policies to serve clients' requests with the shortest fetching time by selecting the most appropriate cache server/video segment quality or by reconstructing the requested quality through transcoding at the edge. Deployment of \texttt{ES-HAS} and \texttt{CSDN} on the cloud-based testbeds and \rf{estimation of users' Quality of Experience (QoE) using objective metrics} demonstrates how clients' requests can be served with higher QoE (by 40\%) and lower bandwidth usage (by 63\%) compared to state-of-the-art approaches.

Our second contribution designs an architecture that simultaneously supports various types of video streaming \rf{(live and VoD)}, considering their versatile QoE and latency requirements. To this end, the SDN, NFV, and MEC paradigms are leveraged, and three VNFs, \ie \textit{Virtual Proxy Function} (VPF), \textit{Virtual Cache Function} (VCF), and \textit{Virtual Transcoding Function} (VTF), are designed. We build a series of these function chains through the \textit{Service Function Chaining} (SFC) paradigm, utilize all CDN and edge server resources, and present \texttt{SARENA}, an SFC-enabled architecture for adaptive video streaming applications. We equip \texttt{SARENA}'s SDN controller with a \textit{lightweight request scheduler} and \textit{edge configurator} to make it deployable in practical environments and to dynamically scale edge servers based on service requirements, respectively. Experimental results show that \texttt{SARENA} outperforms baseline schemes in terms of \rf{higher users' QoE figures} by 39.6\%, \rf{lower E2E} latency by 29.3\%, and \rf{lower backhaul traffic usage by 30\%} for live and VoD services.

Our third contribution aims to use the idle resources of edge servers and employ the capabilities of the SDN controller to establish a collaboration between edge servers in addition to collaboration between edge servers and the SDN controller. We introduce two collaborative edge-assisted frameworks \rf{working for HAS-based live or VoD scenarios} named \texttt{LEADER} and \texttt{ARARAT}. \texttt{LEADER} utilizes sets of actions (\eg transcode the requested quality in the local edge server or a neighboring edge server with the highest available resources), presented in a so-called \textit{Action Tree}, formulates the problem as a central optimization model \rf{to enhance the HAS clients' serving time, subject to the network's and edge servers' resource constraints}, and proposes a lightweight heuristic algorithm \rf{to solve the model}. \texttt{ARARAT} extends \texttt{LEADER}'s \textit{Action Tree}, considers network cost in the optimization, devises multiple heuristic algorithms, and runs extensive scenarios. Evaluation results show that \texttt{LEADER} and \texttt{ARARAT} improve users' QoE by 22\%, decrease the streaming cost by 47\%, and enhance network utilization by 13\%, as compared to their competitors.

Our final contribution focuses on incorporating the capabilities of both peer-to-peer (P2P) networks and CDNs, utilizing NFV and edge computing techniques, and then presenting \texttt{RICHTER} and \texttt{ALIVE} as \textit{hybrid P2P-CDN} frameworks for live streaming scenarios. \texttt{RICHTER} and \texttt{ALIVE} particularly use HAS clients' (\ie peers') potential idle computational resources besides their available bandwidth to provide distributed video processing services, \eg video transcoding and video super-resolution. Both frameworks introduce multi-layer architectures and design \textit{Action Trees} that consider all feasible resources (\ie storage, computation, and bandwidth) provided by peers, edge, and CDN servers for serving clients' requests with acceptable latency and quality. Moreover, \texttt{RICHTER} proposes an online learning method and \texttt{ALIVE} utilizes a lightweight algorithm distributed over in-network virtualized components, which are designed to play decision-maker roles in large-scale practical scenarios. Evaluation results show that \texttt{RICHTER} and \texttt{ALIVE} improve the users’ QoE by 22\%, decrease cost incurred for the streaming service provider by 34\%, shorten clients’ serving latency by 39\%, enhance edge server energy consumption by 31\%, and reduce backhaul bandwidth usage by 24\% compared to the baseline approaches.
\end{abstract}

%% file: Chapters/Abbreviations.tex
\chapter*{List of Acronyms and Abbreviations}
\addcontentsline{toc}{chapter}{List of Acronyms and Abbreviations}\label{chap:abbreviations}

\LARGE{\color{gray}{\textbf{0-9}}}\normalsize\\
\textbf{3GPP}~3rd Generation Partnership Project\\
\\
%%%%%%%%%%%%%%%%%%%%%%%
\LARGE{\color{gray}{\textbf{A}}}\normalsize\\
\textbf{ABR}~Adaptive Bitrate \\
\textbf{ALTO}~Application Layer Traffic Optimization  \\
\textbf{AP}~Access Point  \\
\textbf{AR}~Augmented Reality \\
\textbf{AVC}~Advanced Video Coding \\
\\
%%%%%%%%%%%%%%%%%%%%%%%
\LARGE{\color{gray}{\textbf{C}}}\normalsize\\
\textbf{CAPEX}~Capital Expenditure  \\
\textbf{CDN}~Content Delivery Network \\
\textbf{CMAF}~Common Media Application Format \\
\textbf{CMCD}~Common Media Client Data \\
\textbf{CMSD}~Common Media Server Data \\
\textbf{CNN}~Convolutional Neural Network  \\
\textbf{CTA}~Consumer Technology Association \\
\\
%%%%%%%%%%%%%%%%%%%%%%%
\LARGE{\color{gray}{\textbf{D}}}\normalsize\\
\textbf{DANE}~DASH-Aware Network Element  \\
\textbf{DASH}~Dynamic Adaptive Streaming over HTTP \\
\textbf{DNN}~Deep Neural Network  \\
\textbf{DNS}~Domain Name System  \\
\textbf{DRL}~Deep Reinforcement Learning  \\
\textbf{DRM}~Digital Rights Management  \\
\\
\\
%%%%%%%%%%%%%%%%%%%%%%%
\LARGE{\color{gray}{\textbf{E}}}\normalsize\\
\textbf{E2E}~End-to-End\\
\textbf{EC}~Edge Computing  \\
\textbf{ETR}~Edge Transcoding  \\
\textbf{ETSI}~European Telecommunications Standards Institute  \\
\\
%%%%%%%%%%%%%%%%%%%%%%%
% \LARGE{\color{gray}{\textbf{F}}}\normalsize\\
% \textbf{FC}~Fog Computing  \\
% \\
%%%%%%%%%%%%%%%%%%%%%%%
\LARGE{\color{gray}{\textbf{H}}}\normalsize\\
\textbf{HAS}~HTTP Adaptive Streaming \\
\textbf{HEVC}~High Efficiency Video Coding  \\
\textbf{HLS}~HTTP Live Streaming  \\
\textbf{HTTP}~Hypertext Transfer Protocol\\
\\
%%%%%%%%%%%%%%%%%%%%%%%
\LARGE{\color{gray}{\textbf{I}}}\normalsize\\
\textbf{IEC}~International Electrotechnical Commission \\
\textbf{IETF}~Internet Engineering Task Force  \\
\textbf{IoT}~Internet of Things \\
\textbf{IRTF}~Internet Research Task Force  \\
\textbf{ISO}~International Organization for Standardization \\ 
\textbf{ISP}~Internet Service Provider \\
\textbf{ITU-T}~ITU Telecommunication Standardization \\
\\
%%%%%%%%%%%%%%%%%%%%%%%
\LARGE{\color{gray}{\textbf{K}}}\normalsize\\
\textbf{KPI}~Key Performance Indicator \\
\\
%%%%%%%%%%%%%%%%%%%%%%%
\LARGE{\color{gray}{\textbf{L}}}\normalsize\\
\textbf{LES}~Local Edge Server\\
\textbf{LP}~Linear Programming\\
\\
%%%%%%%%%%%%%%%%%%%%%%%
\LARGE{\color{gray}{\textbf{M}}}\normalsize\\
\textbf{MEC}~Multi-Access Edge Computing \\ 
\textbf{MILP}~Mixed Integer Linear Programming \\ 
\textbf{MINLP}~Mixed Integer Non Linear Programming \\ 
\textbf{ML}~Machine Learning   \\
\textbf{MNO}~Mobile Network Operator \\
\textbf{MOS}~Mean Opinion Score  \\
\textbf{MPD}~Media Presentation Description  \\
\textbf{MPEG}~Moving Picture Experts Group \\
\textbf{MPTCP}~Multi-Path TCP  \\
\textbf{MS}~Multimedia Service \\
\textbf{MSS}~Microsoft Smooth Streaming \\
\\
%%%%%%%%%%%%%%%%%%%%%%%
\LARGE{\color{gray}{\textbf{N}}}\normalsize\\
\textbf{NAVS}~Network Assisted Video Streaming   \\
\textbf{NES}~Neighboring Edge Server   \\
\textbf{NDN}~Named Data Networking   \\
\textbf{NFV}~Network Function Virtualization  \\
\textbf{NGN}~Next Generation Networks   \\
\textbf{NN}~Neural Network  \\
\\
%%%%%%%%%%%%%%%%%%%%%%%
\LARGE{\color{gray}{\textbf{O}}}\normalsize\\
\textbf{OF}~OpenFlow  \\
\textbf{OL}~Online Learning  \\
\textbf{ONF}~Open Networking Foundation  \\
\textbf{OPEX}~Operational Expenditure  \\
\textbf{OTT}~Over-the-Top \\
\\
%%%%%%%%%%%%%%%%%%%%%%%
\LARGE{\color{gray}{\textbf{P}}}\normalsize\\
\textbf{P2P}~Peer to Peer Network \\
\textbf{PED}~Parameters Enhancing Delivery  \\ 
\textbf{PER}~Parameters Enhancing Reception  \\ 
\textbf{PSNR}~Peak Signal to Noise Ratio  \\
\textbf{PSR}~Peer Super-Resolution  \\
\textbf{PTR}~Peer Transcoding  \\
\\
%%%%%%%%%%%%%%%%%%%%%%%
\LARGE{\color{gray}{\textbf{Q}}}\normalsize\\
\textbf{QoE}~Quality of Experience \\ 
\textbf{QoS}~Quality of Service \\
\textbf{QUIC}~Quick UDP Internet Connections\\ 
\\
%%%%%%%%%%%%%%%%%%%%%%%
\LARGE{\color{gray}{\textbf{R}}}\normalsize\\
\textbf{RA}~Resource Allocation \\
\textbf{RAN}~Radio Access Network\\ 
\textbf{RL}~Reinforcement Learning\\ 
\textbf{RTCP}~Real-Time Control Protocol\\
\textbf{RTMP}~Real-Time Messaging Protocol \\
\textbf{RTP}~Real-Time Transport Protocol \\ 
\textbf{RTSP}~Real-Time Streaming Protocol \\ 
\\
%%%%%%%%%%%%%%%%%%%%%%%
\LARGE{\color{gray}{\textbf{S}}}\normalsize\\
\textbf{SAND}~Server and Network Assisted DASH \\
\textbf{SDN}~Software-Defined Networking  \\ 
\textbf{SFC}~Service Function Chaining \\
\textbf{SLA}~Service Level Agreement \\
\textbf{SR}~Super-Resolution \\
\\
%%%%%%%%%%%%%%%%%%%%%%%
\LARGE{\color{gray}{\textbf{T}}}\normalsize\\
\textbf{TCP}~Transmission Control Protocol \\ 
\textbf{TR}~Transcoding \\ 
\\
%%%%%%%%%%%%%%%%%%%%%%%
\LARGE{\color{gray}{\textbf{U}}}\normalsize\\
\textbf{UDP}~User Datagram Protocol \\ 
\textbf{UE}~User Equipment \\ 
\\
%%%%%%%%%%%%%%%%%%%%%%%
\LARGE{\color{gray}{\textbf{V}}}\normalsize\\
\textbf{VCF}~Video Cache Function \\ 
\textbf{VMAF}~Video Multi-Method Assessment Fusion \\  
\textbf{VNF}~Virtual Network Function \\
\textbf{VPF}~Virtual Proxy Function \\
\textbf{VQA}~Video Quality Assessment \\
\textbf{VR}~Virtual Reality \\
\textbf{VRP}~Virtual Reverse Proxy \\
\textbf{VTF}~Virtual Trascoding Function \\
\textbf{VoD}~Video-on-Demand \\
\\
%%%%%%%%%%%%%%%%%%%%%%%
\LARGE{\color{gray}{\textbf{W}}}\normalsize\\
\textbf{WebRTC}~Web Real-Time Communications \\ 
%%%%%%%%%%%%%%%%%%%%%%%

%% file: Chapters/Chapter1/1-1-Intro.tex
%************************************************
\chapter{General Introduction}\label{chap:Introduction}
%************************************************
\doublespacing
In the first section of this chapter (Section~\ref{chap:Introduction:Motivation}), we describe the primary motivation of the thesis and state the main problem. Section~\ref{chap:Introduction:RQs} outlines four fundamental research questions (RQ1--RQ4) that must be taken into account to solve the expressed problem. Section~\ref{chap:Introduction:methodology} explains the research methodology utilized to address the RQs. Sections~\ref{chap:Introduction:Contributions} and~\ref{chap:Introduction:Publications} discuss the principal contributions and show outcomes of the thesis, which are published in or submitted to partially highly prestigious journals or conferences/workshops, respectively. Finally, Section~\ref{chap:Introduction:Organizations} outlines the organization of the thesis.

%% file: Chapters/Chapter1/1-2-Motivation.tex
\section{Motivation and Problem Statement}\label{chap:Introduction:Motivation}
\begin{figure}[!t]
    \centering
    \includegraphics[width=1\textwidth]{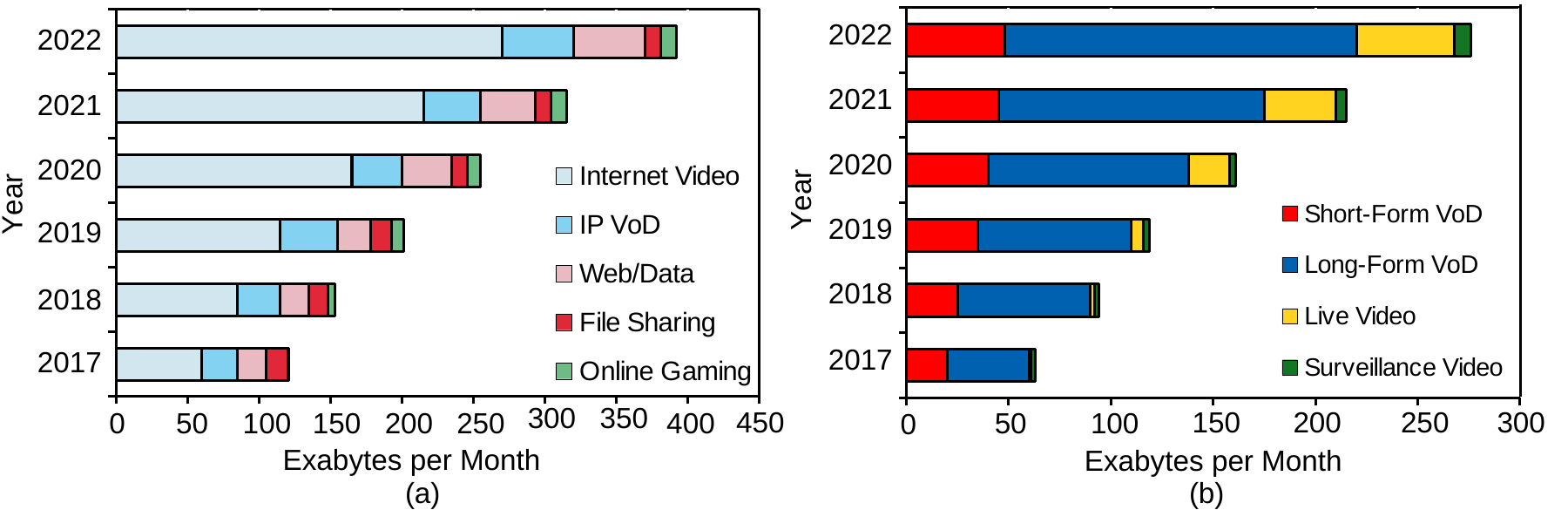}
    \caption{\small Global (a) IP traffic by application category and (b) Internet video by subsegment~\cite{cisco2018cisco}.}
    \hspace{.5cm}
    \label{cisco-rep12}   
\end{figure}

\doublespacing
The emergence of modern networks (\eg 5G and 6G), the development of Over-The-Top (OTT) content services (\eg YouTube, Netflix, Facebook, Amazon Prime), and steadily growing numbers of users who prefer to stream video content online rather than using classical TV, have made video the dominant traffic on the Internet~\cite{cisco2018cisco}. According to the Cisco Annual Internet Report, video traffic will make up more than 60\% of the entire IP network traffic by 2023. As shown in Fig.~\ref{cisco-rep12} (a), in 2022 consumer Internet video traffic exceeded 300 Exabytes per month, which is equivalent to over 60 billion DVDs per month or 82 million DVDs per hour. This rise in traffic is closely linked to the continuous development of video services. Nowadays, video content is transitioning from Full HD (FHD) to 8K resolution. There are also growing demands for Virtual (VR) and Augmented Reality (AR) equipment with high-bandwidth requirements.
%%%%%%%%%%
%%%%%%%%%%
Concerning video traffic, video streaming applications, such as video-on-demand (VoD) or live video streaming, represent about 66\% of the total traffic over the Internet~\cite{Sandvine}. Indeed, VoD and live video have gained tremendous popularity in the last few years, where over 270 Exabytes per month of both types were streamed by 2022 (see Fig.~\ref{cisco-rep12} (b)). These trends are expected to continue; for instance, live video streaming will account for more than 17\% of total video traffic by 2023~\cite{Sandvine}. 
\raggedbottom

% These increases, combined with the progressively growing network demands of new applications, pose considerable challenges to both OTT companies and Internet Service Providers (ISPs). Ensuring a satisfactory quality perceived by the end user, so-called Quality of Experience (QoE), is a vital business incentive for these stakeholders in the Internet ecosystem as it helps mitigate the constant threat of customer attrition~\cite{chu2007toward}. 
Ensuring a satisfactory degree of satisfaction or a decreasing annoyance quality perceived by the end user of an application or service, the so-called Quality of Experience (QoE)~\cite{brunnstrom2013qualinet}, is a vital business incentive for these stakeholders in the Internet ecosystem, as it helps mitigate the constant threat of customer attrition. Various video delivery technologies have emerged in recent years in order to improve the performance of video streaming services. The scale of current streaming platforms has prompted a transition from protocols relying on the Real-Time Transport Protocol (RTP)~\cite{RTP} to those based on \textit{HTTP Adaptive Video Streaming} (HAS). Among all HAS-based solutions, the Moving Picture Experts Group (MPEG) standard \textit{Dynamic Adaptive Streaming over HTTP} (DASH)~\cite{sodagar2011mpeg} and Apple's \textit{HTTP Live Streaming} (HLS)~\cite{HLS} are two predominant HAS delivery systems that OTT providers use for video streaming applications. 
% OTT providers to mainly use them for video streaming applications. 

In HAS-enabled systems, the video sequences are encoded into various bitrates or resolutions, known as representations, and then divided into small and fixed segments, each consisting of a few seconds of the video. A manifest file (\eg Media Presentation Description (MPD) in DASH systems) stored on the server side specifies the structure of the video in terms of available representations and segments. At the start of the streaming session, the client downloads this manifest file from the server and then requests the video segments in a chronological sequence. Once one or more video segments are downloaded and saved in the client's buffer, the video playback commences. In addition, HAS enables switching the video's representations dynamically during the streaming session. The switching strategy, known as the Adaptive Bitrate Algorithm (ABR), is entirely placed on the client side, which makes this approach highly scalable. The objective of the adaptation algorithm is determining the most appropriate video representation according to the network bandwidth or/and client buffer, with a key emphasis on providing uninterrupted playback while maximizing the streamed quality~\cite{bentaleb2018survey, barakabitze2019qoe}. 

Although HAS-enabled delivery systems certainly offer benefits over non-HAS streaming systems, most of the existing HAS-based methods are pure client-based solutions, where the players' ABR algorithms often make bitrate adaptation decisions by only considering local \rf{device} parameters (\rf{\eg a smartphone's buffer status and estimated available bandwidth)} without any global network consideration. Given the large-scale distributed nature of the Internet, individual nodes only observe part of the video streaming system, which results in imperfect adaptations and poor long-tail performance~\cite{bentaleb2018survey}. Moreover, various studies indicate that the adaptation mechanisms in HAS players suffer performance problems in competitive situations when numerous players compete for a shared bottleneck network link~\cite{akhshabi2012happens,huang2012confused}. 
 
Collectively, the aforementioned problems and the requirements of assuring the users' QoE have motivated OTT companies and ISPs to collaborate, upgrade their video delivery systems, and introduce an emerging direction called \textit{Network-Assisted Video Streaming} (NAVS). NAVS systems refer to techniques used to optimize video delivery over the Internet to devices that support HAS-based streaming. In such systems, one or multiple in-network components with a broader view of the network can be employed to \rf{provide fair network resources to video players and} assist \rf{them} in making precise adaptation decisions, hence meeting video usage with extremely high-bandwidth demands under high-quality and low-latency constraints. Therefore, to provide modern NAVS services, ISPs must overhaul their traditional network structures, which are based on IP architecture principles. 

This thesis aims to advance streaming services and content delivery performance by introducing modern NAVS systems. Our NAVS systems employ promising network models, \eg virtualized and softwarized networks and edge computing, and incorporate overlay networks such as Content Delivery Networks (CDNs) and/or Peer-to-Peer (P2P) networks to provide support and assistance for VoD and live HAS clients. \rf{One of the important services offered by such NAVS systems is video transcoding services. Video transcoding solutions, such as ``AWS Elemental MediaConvert''~\cite{AET}, are typically used to adjust video quality based on network conditions, device capabilities, and user preferences. Indeed, converting a video file/segment from one format to another, from a higher bitrate/resolution to the requested one, makes it suitable for different devices with different bitrate/resolution or codec support. The usage of such services is expected to grow in the future due to the increasing demand for
high-quality and high-resolution (\eg 4K or 8K) video sequences, newer video codecs (\eg  HEVC and VVC), and bandwidth-intensive immersive streaming.}
\rf{The NAVS frameworks presented in this thesis offer entirely transparent services to HAS clients to ensure that HAS clients' adaptation decisions are free from any kind of confusion or risk.} As a result, clients can obtain acceptable quality video sequences with enhanced latencies. \rf{While some of the developed techniques may be relevant to other types of video transport, such as conversational or immersive video sequences (tile-based 360-degree VR), these use cases are not within the scope of our work}. The thesis studies video in terms of its transport and delivery, and thus video encoding aspects are also considered out of our scope. Given the aforementioned factors, this thesis aims to identify and tackle the following challenges:
%%%
\begin{enumerate}[label=\textit{Challenge}\arabic*,noitemsep]
\item \textbf{Improving users' QoE during streaming sessions.} The users' QoE in HAS-enabled systems is affected by different factors, particularly stalling and quality switches, as defined in the following. Since the HAS clients are not aware of the real network conditions, if they are not assisted in improving the delivered QoE, they encounter unnecessary quality switches and video playout interruptions. Consequently, improving the users' QoE should be considered as the primary goal of designing NAVS systems.
\begin{itemize}[noitemsep]
\item \textbf{Stalls: } Stalls represent video playback interruptions during the streaming session, making users experience unexpected waiting times. The video typically stalls due to a rebuffering event, \ie when the player buffer is emptied. Therefore, the video playback can only be resumed once the buffer is refilled and the next segment is available. There are two key performance indicators (KPIs) associated with stalling that impact the QoE: the \textit{stalling duration} and the overall \textit{number of stalls} happening throughout the video playback. Studies have indicated that the users' QoE is more degraded by frequent short interruptions than by less frequent longer interruptions~\cite{hossfeld2013pippi}.
\item \textbf{Quality switches: }The dynamic nature of HAS resulting in varying video quality during playback can lead to viewer distraction. Two crucial KPIs related to quality switches are the overall \textit{number of quality switches} and the \textit{switching amplitudes}. In order to ensure a satisfactory QoE, a balance must be struck between maximizing the delivered video quality and minimizing the frequency and amplitudes of quality switches~\cite{moldovan2017keep}.
\end{itemize}
%%%%%%
\item \textbf{Establishing collaboration between different stakeholders.}  Collaboration among different business stakeholders, \eg ISPs, CDN providers, OTT companies, and P2P networks, is required for implementing NAVS systems. To this aim, edge nodes are enabled to host third-party services, \eg \textit{proxy} servers of OTT providers, of mobile network operators (\eg MobiledgeX), and of CDN providers (\eg Cloudflare and Limelight).
%%%%%%
\item \textbf{Presenting NAVS solutions compatible with HAS systems using minimum modifications on the client side.} Some of the NAVS systems impose modifications on HAS clients or require specific software to be installed on the client devices to play out the requested video sequences. Thus, designing NAVS systems for HAS systems that can only leverage existing HAS protocols and standards with minimum modification or extra software/plugin installment on the HAS clients is challenging.
%%%%%%
\item \textbf{Enhancing networking resource utilization.} Video streaming often involves dynamic and unpredictable network conditions, such as fluctuations in available bandwidth. On the other hand, different video streaming applications have varying requirements for network resources. Therefore enhancing resource usage while achieving a desirable performance goal, \eg ensuring high-quality video delivery in designing NAVS systems, adds to the complexity of the problem.
%%%%%%
\item \textbf{Reducing streaming costs.} Video streaming systems often require significant amounts of CDN/origin servers' storage and/or CDN/edge servers' computations for running computationally intensive tasks like video transcoding. Therefore, devising NAVS systems that can efficiently distribute video content with acceptable quality and latency across many users while avoiding excessive use of storage and computation requirements is a complex task.
%%%%%%
% a...
%%%%%%
\end{enumerate}

%% file: Chapters/Chapter1/1-3-RQs.tex
\section{Research Questions}\label{chap:Introduction:RQs}
\doublespacing
As mentioned in the previous section, the aim of this thesis is to propose innovative NAVS systems for HAS clients to overcome the drawbacks of pure client-based HAS systems. To achieve this goal, we take into account the perspectives of different stakeholders (\ie OTTs, CDNs, and ISPs) within the HAS ecosystem and then elaborate on the following research questions (\textit{RQs}): 
%%%%%%%%
\begin{figure}[!t]
	\centering
	\includegraphics[width=1\linewidth]{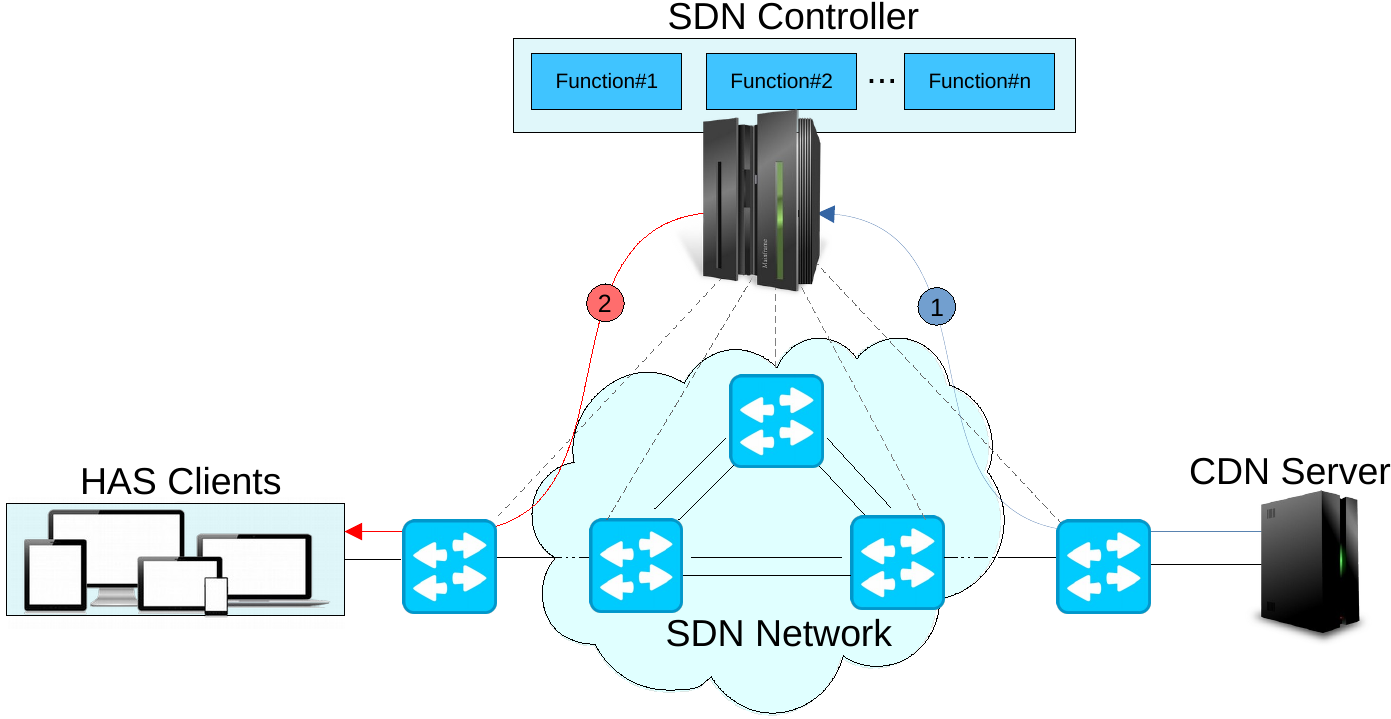}
	\caption{\small An example of the first research question (\textit{RQ1}).}
        \vspace{.5cm}
	\label{RQ1}
\end{figure}
%%%%%%%%
\begin{enumerate}[label=\textit{RQ}\arabic*,noitemsep]
%%%%
    \item \textbf{How can SDNs/CDNs provide assistance for HAS clients in order to improve media delivery services?}
    
    As illustrated in Fig.~\ref{RQ1}, the SDN controller with a broader network view could host multiple Virtualized Network Functions (VNFs) with particular responsibilities to collect information (\eg network map, path information, cache occupancy, throughput measurements) from different network components such as CDN servers or network switches (step 1 in Fig.~\ref{RQ1}).  
    The SDN controller \rf{controlling a single domain network (\ie Autonomous System)} can use this network-wide information to improve delivery services. For instance, collecting, processing, and distributing information about cache occupancy by a central controller (\ie the SDN controller) can help clients satisfy their content requests from an appropriate cache server (step 2).  
    % \item \textbf{Will assistance by HAS clients for the SDNs/CDNs (and client-network collaboration) work, which assistance, how?}    
    
    \item \textbf{How can resources (\ie computation, storage, bandwidth) provided by the HAS clients be used to improve media delivery services?} 

    % This research question is widely unexplored. 
    \rf{Although several works introduced P2P or hybrid P2P-CDN video delivery approaches (as discussed in Section \ref{chap:SOTA:CDN}), there is still a need to further investigate this research question.} An example is shown in Fig.~\ref{RQ2}, where the SDN controller could employ an intermediate server holding multiple functionalities to receive information provided by HAS clients (step 1 in Fig.~\ref{RQ2}), such as available resources, user behavior, content popularity, content prefetching, and/or representation selection hints. This information can then be used by the SDN controller (step 2) for several purposes, \eg task offloading into the P2P network, caching, or dynamic routing policies to improve media delivery and network utilization. Moreover, HAS clients can be used within a hybrid P2P-CDN NAVS system to provide low-latency video streaming, improve network bandwidth usage (\ie backhaul and fronthaul) as well as enhance CDNs' performance. 
%%%%%%%%%%
%%%%%%%%%%
\begin{figure}[!t]
	\centering
	\includegraphics[width=1\linewidth]{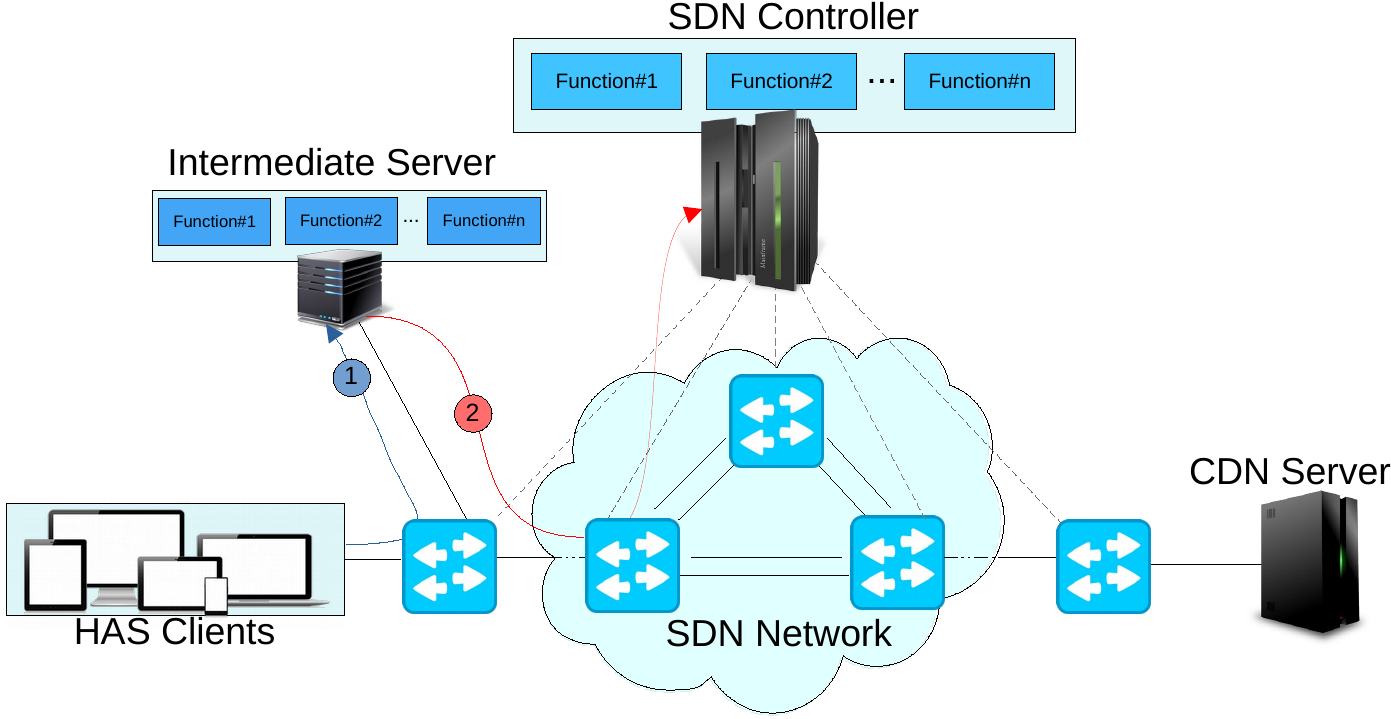}
	\caption{\small An example of the second research question (\textit{RQ2}).}
        \vspace{.5cm}
	\label{RQ2}
\end{figure}
%%%%%%%%%%
    \item \textbf{What is the utility of the proposed assistance and collaboration service?} 

% To avoid such problems, objective quality assessment through objective metrics offers scalability,

    \rf{Although subjective quality assessment methods, \eg using metrics like Mean Opinion Score (MOS)~\cite{streijl2016mean}, are still among the prevalent methods for evaluating the impact of NAVS systems on the users' perception, they typically involve a large number of participants to rate the quality. Therefore, such studies can be time-consuming and costly, particularly when dealing with large-scale scenarios. To avoid such problems, \rf{objective quality assessment methods via objective metrics estimate users' QoE and  
    % objective quality assessment. 
    offer} scalability, reproducibility, real-time monitoring, interpretability, systematic analysis and cost-effectiveness as advantages.} Thus, improving (1) \rf{users' objective QoE parameters, \ie application QoS parameters} (\eg average quality bitrate, the number of quality switches, delay duration, number of delays, standardized perceived quality metrics such as the one obtained from P.1203~\cite{p1203}), (2) CDN utilization (\eg server load, cache hit rates), (3) \rf{network} QoS parameters (\eg delay, throughput), (4) network utilization (backhaul/fronthaul bandwidth, computation usage), (5) SDN controller's load (messages to/from the controller, load), (6) content providers' costs (video bitrate ladder), (7) edge server costs (\eg  computational or storage), \etc~are considered as the principal utilities of the proposed NAVS frameworks.    
    %%%%
    \item \textbf{How can the utility of the proposed NAVS frameworks be thoroughly evaluated, both theoretically and practically?} 
    
     Large-scale evaluations employing realistic settings, assumptions, networking topologies, video datasets, and network traces to conduct testbed experiments are considered to evaluate the utility of the proposed approaches. \rf{Given the wide scope of what can and needs to be investigated, conducting subjective quality studies in such large-scale evaluations is impossible. However, estimating the objective QoE through objective quality assessment methods, such as application QoS parameters via P.1203 metric, is feasible.}
    \end{enumerate}

%% file: Chapters/Chapter1/1-4-Methodology.tex
\section{Research Methodology}\label{chap:Introduction:methodology}
\doublespacing
The \textit{design} approach introduced by the Association for Computing Machinery (ACM)~\cite{comer1989computing} will be used to address the RQs given in Section~\ref{chap:Introduction:RQs} with the following steps:

\begin{enumerate}[label=\textit{Step}\arabic*,noitemsep]
    \item \textbf{State requirements:} First, we investigate which type of information provided by CDNs, SDNs, or HAS clients could be feasible and valuable to establish a collaboration between them. Network maps, cache server occupancy, application analytics information to an SDN/CDN, user behavior, content popularity, content prefetching, representation selection hints or/and predictions to a CDN might be considered as important information. Surveying state-of-the-art works, standard specifications, and software/platforms' capabilities are taken into account in this step.
    
    \item \textbf{State specifications:} After finding appropriate information, we must define specifications by designing innovative system architectures considering HAS capabilities, novel networking paradigms, and communication protocol requirements. We also require formulating the problem in mathematical terms with a set of constraints.
    
    \item \textbf{Design and implement the system:} According to the system model, we will design possible solutions that satisfy the constraints and meet the objectives through optimization or heuristic approaches. Afterward, those solutions will be implemented carefully in SDN/CDN environments.
    
    \item \textbf{Test the system:} Extensive experiments using large-scale testbeds under various context settings will be conducted. The results of the proposed NAVS solutions will be compared with other state-of-the-art and baseline approaches in terms of metrics mentioned in \textit{RQ3} to show their performance.
\end{enumerate}

%% file: Chapters/Chapter1/1-5-Contribution.tex
\section{Thesis Contributions}\label{chap:Introduction:Contributions}
\doublespacing
The scientific contributions of this thesis are summarized in Fig.~\ref{Thesis-contribution}. In this figure, the author's scientific publications are categorized according to the employed techniques, \ie \textit{(i)} Emerging Networking Paradigms, \textit{(ii)} Delivery Networks Techniques, \textit{(iii)} Learning-based Solutions, and \textit{(iv)} Emerging Communication Protocols to propose NAVS solutions assisting HAS clients. 
%The dotted line in the ``Emerging Networking Paradigms'' box denotes the \texttt{SARENA}~\cite{farahani2023sarena} framework that utilizes other networking paradigms in addition to SFC. 
%On the other hand, the line in the ``Delivery Networks Techniques'' box shows that the \texttt{TQPM}~\cite{TQPM}, \texttt{VQA-TIF}~\cite{menon2022etps}, and \texttt{Towards LLL}~\cite{kalan2022towards} papers do not consider any caching strategy. 
Moreover, the color markers denote the research questions (RQs) addressed by the scientific publications. The main contributions of this thesis are summarized as follows:
%%%%%%%%
\begin{figure}[!t]
	\centering
	\includegraphics[width=1\linewidth]{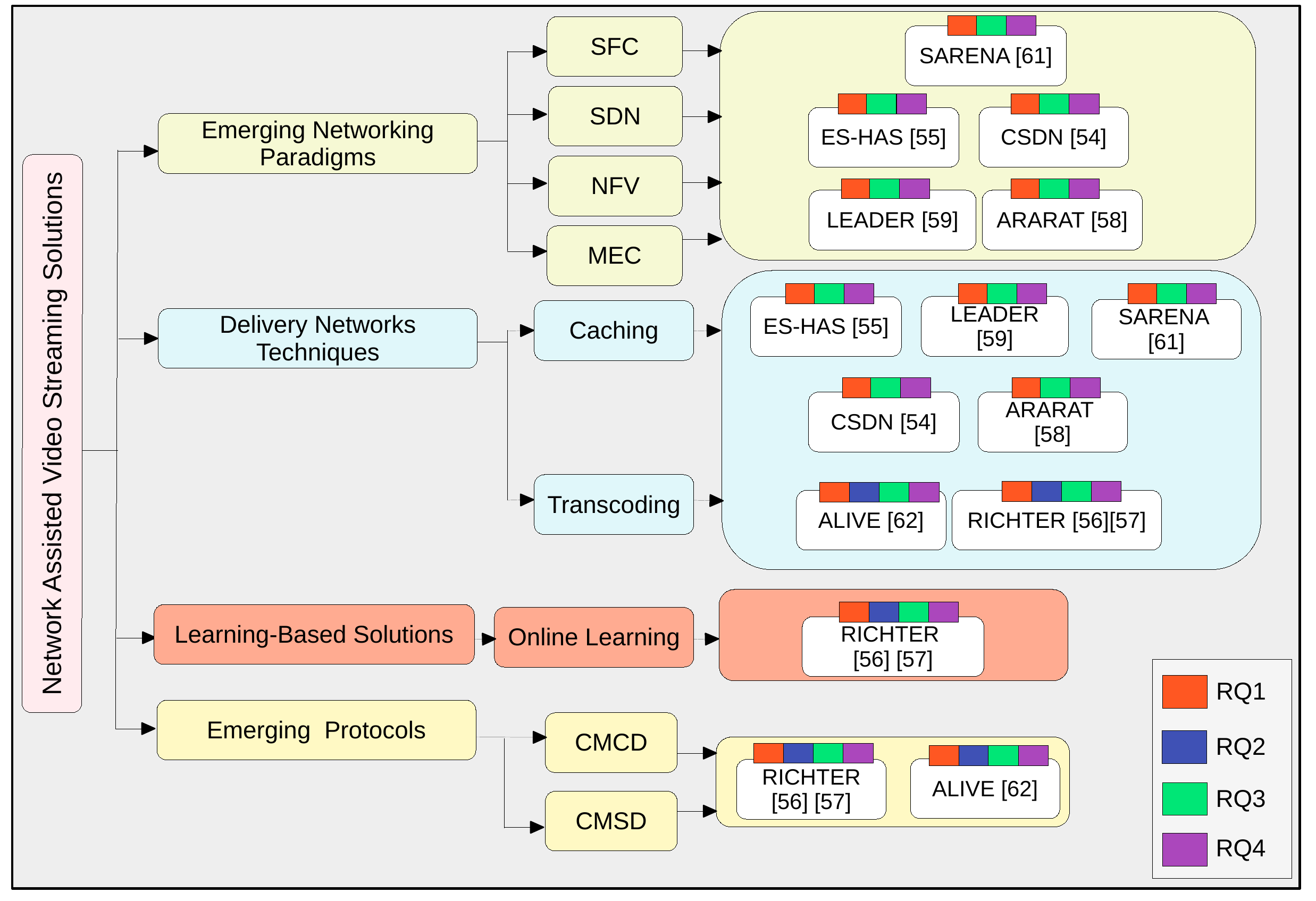}
	\caption{\small The contributions of this thesis are  classified along the scientific publications conducted by the author. The small rectangles indicate which research questions (RQs) are addressed by each publication.}
        \vspace{.5cm}
	\label{Thesis-contribution}\end{figure}
%%%%%%%%
%%%
\begin{enumerate}[noitemsep]
%%%
\item \textbf{Edge and SDN-assisted NAVS frameworks (\textit{Chapter~\ref{chap:EdgeSDN}}).} The first contribution of the dissertation is focused on the utilization of the SDN, NFV, and MEC paradigms to introduce two frameworks called \texttt{ES-HAS}~\cite{Farahani2021eshas} and \texttt{CSDN}~\cite{Farahani2021csdn}. The aim of these frameworks is to assist VoD HAS clients by introducing Virtual Network Functions (VNFs) at the edge of a single domain SDN-based network. These VNFs are named Virtual Reverse Proxy (VRP) servers, and their primary responsibilities are to \textit{(i)} gather requests from HAS clients, \textit{(ii)} retrieve networking information in a time-slot-based manner, and then \textit{(iii)} utilize optimization models as server/segment selection policies to serve clients' requests with the shortest fetching time. Indeed, this is achieved by selecting the most appropriate cache server/video segment quality or by reconstructing the requested quality through transcoding at the edge. To demonstrate the effectiveness of the \texttt{ES-HAS} and \texttt{CSDN} frameworks, they were deployed on cloud-based testbeds. The results showed that these frameworks serve clients' requests with higher QoE (by at least 40\%) and lower bandwidth usage (by at least 63\%) compared to state-of-the-art approaches. 
%%%
\item \textbf{SFC-enabled architecture for adaptive video streaming applications (\textit{Chapter~\ref{chap:SFCEnabled}}).} Our second contribution is focused on designing an architecture for supporting different types of video streaming services (live and VoD) with versatile QoE and latency requirements. To this end, we use the SDN, NFV, and MEC paradigms and propose the \texttt{SARENA} architecture~\cite{farahani2023sarena}. \texttt{SARENA} introduces three multimedia VNFs, \ie Virtual Proxy Function (VPF), Virtual Cache Function (VCF), and Virtual Transcoding Function (VTF), and builds various chains of them using the SFC paradigm. The SDN controller of \texttt{SARENA} is equipped with a lightweight request scheduler and an edge configurator to make it practical in real \rf{large-scale} scenarios and to enable dynamic edge server configuration (scaling up/down) based on service requirements. The experimental results demonstrate that \texttt{SARENA} outperforms baseline schemes in terms of QoE by at least 39.6\%, latency by 29.3\%, and network utilization by 30\% for two live and VoD multimedia services. \rf{It must be noted that other aspects of designing SFC-enabled architectures, such as stability and robustness, are not considered in the SARENA design and need further investigation.}  
%%%
\item \textbf{Collaborative edge-assisted NAVS frameworks (\textit{Chapter~\ref{chap:CollaborativeEdge}}).} In the third contribution, we utilize the potential idle resources of edge servers and the SDN controller's capabilities to establish collaboration between the SDN controller and edge servers and between edge servers themselves. We introduce two collaborative edge-assisted frameworks for live HAS clients called \texttt{LEADER}~\cite{farahani2022leader} and \texttt{ARARAT}~\cite{farahani2022ararat}. \texttt{LEADER} introduces sets of actions (\eg fetch the requested quality from the local edge server or a neighboring edge server that holds that and has the highest available bandwidth; transcode the requested quality in the local edge server or a neighboring edge server that has the highest available computational resources) called \textit{Action Tree} for serving client requests that consider all feasible resources (\ie storage, computation, and bandwidth) provided by edge, CDN and origin servers for serving clients' requests with acceptable latency and quality. It also formulates the problem as a central optimization model and proposes a lightweight heuristic algorithm to solve the proposed optimization model. \texttt{ARARAT} extends \texttt{LEADER}'s \textit{Action Tree}, considers network cost in the optimization model and proposes extra coarse- and fine-grained heuristic algorithms to solve the proposed optimization model. The evaluation results show that \texttt{LEADER} and \texttt{ARARAT} enhance users' QoE by at least 22\%, decrease the streaming cost, including bandwidth and computational costs, by at least 47\%, and enhance network utilization by at least 13\% compared to state-of-the-art approaches.
%%%
\item \textbf{Hybrid P2P-CDN NAVS frameworks for live streaming (\textit{Chapter~\ref{chap:CollaborativeEdge}}).} Our final contribution aims to combine the strengths of both P2P networks and CDNs and leverage the use of MEC and NFV techniques to develop \texttt{RICHTER}~\cite{farahani2022hybrid, farahani2022RICHTER} and \texttt{ALIVE}~\cite{farahani2023alive} as hybrid P2P-CDN frameworks for live streaming scenarios. These frameworks utilize idle computational resources and available bandwidth of HAS clients (\ie peers) to offer distributed video processing services, such as video transcoding and video super-resolution. \texttt{RICHTER} and \texttt{ALIVE} formulate the problems as optimization models, considering all feasible resources (\ie storage, computation, and bandwidth), provided by peers, edge, and CDN servers to serve requests with the most appropriate action of the proposed \textit{Action Trees}. They also propose heuristic methods (based on online learning or lightweight algorithms) that are designed to play decision-maker roles in large-scale practical scenarios. The evaluation results demonstrate that \texttt{RICHTER} and \texttt{ALIVE} improve users' QoE by at least 22\%, decrease streaming service provider costs by at least 34\%, shorten clients' serving latency by at least 39\%, reduce edge server energy consumption by at least 31\%, and reduce backhaul bandwidth usage by at least 24\% compared to the baseline approaches.
%%%
\end{enumerate}

%% file: Chapters/Chapter1/1-6-Publications.tex
\section{Publications}\label{chap:Introduction:Publications}
\begin{figure}[!t]
	\centering
	\includegraphics[width=.9\linewidth]{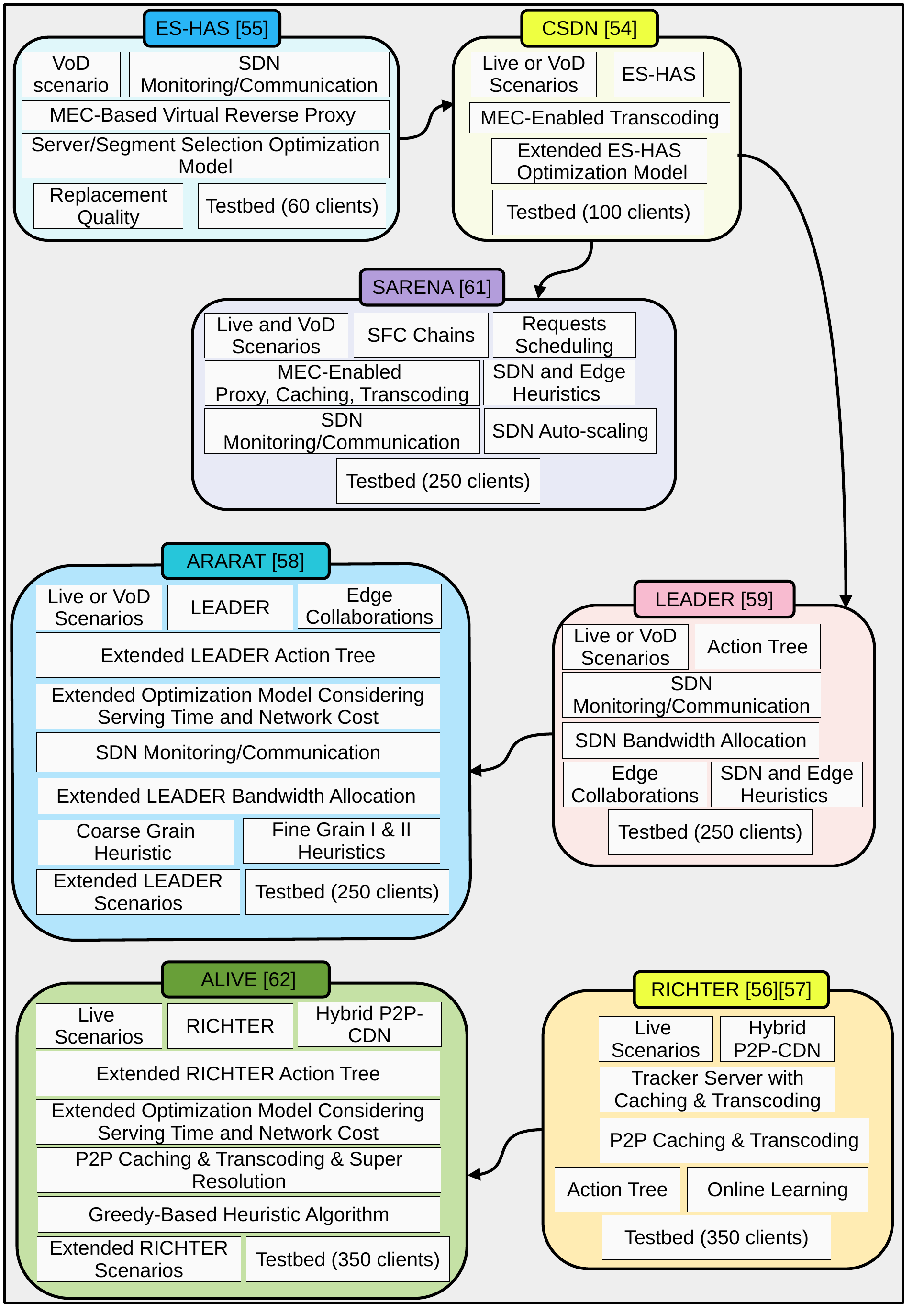}
	\caption{\small \rf{Differences and relations between frameworks introduced in the dissertation.}}
        \vspace{.5cm}
	\label{Thesis-publication}\end{figure}

\doublespacing
The scientific papers that have been published are an essential metric to measure progress and to highlight the accomplishments achieved so far. The results of this dissertation have been published in/submitted to partially highly prestigious journals and presented at flagship conferences/workshops. The following list provides an overview of the publications during this Ph.D. research. The seven works below marked with \textcolor{cyan}{$\bigstar$} are included in this doctoral dissertation; the other publications are not discussed further in this dissertation. \rf{The overview in Fig.~\ref{Thesis-publication} highlights the differences and relations between the frameworks introduced in this dissertation~\cite{Farahani2021csdn,Farahani2021eshas,farahani2022RICHTER,farahani2022hybrid,farahani2022ararat,farahani2022leader,farahani2023sarena,farahani2023alive}.}
\clearpage
%%%%%%%%
%%%%%%%%
\subsection{First Author Publications}
\singlespacing
\begin{enumerate}
%1
\item \textcolor{cyan}{$\bigstar$}\textbf{R. Farahani}, E. \c{C}etinkaya, C. Timmerer, M. Shojafar, M. Ghanbari, H. Hellwagner. ALIVE: A Latency- and Cost-Aware Hybrid P2P-CDN Framework for Live Video Streaming. \textit{Submitted to IEEE Transactions on Network and Service Management (TNSM)}, 2023.
%2
\item \textcolor{cyan}{$\bigstar$}\textbf{R. Farahani}, M. Shojafar, C. Timmerer, F. Tashtarian, M. Ghanbari, H. Hellwagner. ARARAT: A Collaborative Edge-Assisted Framework for HTTP Adaptive Video Streaming. \textit{IEEE Transactions on Network and Service Management (TNSM)}, 2022.
%3
\item \textcolor{cyan}{$\bigstar$}\textbf{R. Farahani}, A. Bentaleb, C. Timmerer, M. Shojafar, R. Prodan, H. Hellwagner. SARENA: SFC-Enabled Architecture for Adaptive Video Streaming Applications. \textit{IEEE International Conference on Communications (ICC)}, 2023.
%4
\item \textcolor{cyan}{$\bigstar$}\textbf{R. Farahani}, A. Bentaleb, E. \c{C}etinkaya , C. Timmerer, R. Zimmermann, and H. Hellwagner. Hybrid P2P-CDN Architecture for Live Video Streaming: An Online Learning Approach. \textit{IEEE Global Communications Conference (GLOBECOM)}, 2022.
%5
\item \textcolor{cyan}{$\bigstar$}\textbf{R. Farahani}, F. Tashtarian, C. Timmerer, M. Ghanbari, and H. Hellwagner. LEADER: A Collaborative Edge- and SDN-Assisted Framework for HTTP Adaptive Video. \textit{IEEE International Conference on Communications (ICC)}, 2022.
%6
\item \textcolor{cyan}{$\bigstar$}\textbf{R. Farahani}, F. Tashtarian, H. Amirpour, C. Timmerer, M. Ghanbari, and H. Hellwagner. CSDN: CDN-Aware QoE Optimization in SDN-Assisted HTTP Adaptive Video Streaming. \textit{IEEE 46th Conference on Local Computer Networks (LCN)}, 2021.
%7
\item \textcolor{cyan}{$\bigstar$} \textbf{R. Farahani}, F. Tashtarian, A. Erfanian, C. Timmerer, M. Ghanbari, and H. Hellwagner. ES-HAS: An Edge- and SDN-Assisted Framework for HTTP Adaptive Video Streaming. \textit{31st ACM Workshop on Network and Operating Systems Support for Digital Audio and Video (NOSSDAV)}, 2021.
%8
\item \textbf{R. Farahani}. CDN and SDN Support and Player Interaction for HTTP Adaptive Video Streaming. \textit{12th ACM Multimedia Systems Conference (MMSys)}, 2021.
%9
\item \textbf{R. Farahani}, A. Bentaleb, M. Shojafar, H. Hellwagner. CP-Steering: CDN- and Protocol-Aware Content Steering Solution for HTTP Adaptive Video Streaming. \textit{ACM Mile High Video (MHV)}, 2023.
%10
\item \textbf{R. Farahani}, H. Amirpour, F. Tashtarian, A. Bentaleb, C. Timmerer, H. Hellwagner, and R. Zimmermann. RICHTER: Hybrid P2P-CDN Architecture for Low Latency Live
Video Streaming. \textit{ACM Mile-High Video (MHV)}, 2022.
\end{enumerate}
%%%%%%%%%%%%%%%%%%%%%%%%%%%%%%%%%%%%%%%%%%%%%%%%%%%%%%%%%%%%%%%%%%%%%%%%%%%%%%%
\subsection{Co-authored Publications}
\begin{enumerate}\setcounter{enumi}{10}
%12
\item V. V Menon, P. T Rajendran, \textbf{R. Farahani}, K. Schöffmann, C. Timmerer. Video Quality Assessment with Texture Information Fusion for Streaming Applications. \textit{Submitted to the IEEE International Conference on Visual Communications and Image Processing (VCIP)}, 2023.
%13
\item V. V Menon, \textbf{R. Farahani}, P. T Rajendran, H. Hellwagner, M. Ghanbari, C. Timmerer. Reduced Reference Transcoding Quality Prediction for Video Streaming Applications. \textit{ACM Mile High Video (MHV)}, 2023.
%14
\item S. Chellappa, \textbf{R. Farahani}, R. Bartos, H. Hellwagner. Context-Aware HTTP Adaptive Video Streaming Utilizing QUIC’s Stream Priority. \textit{ACM Mile High Video (MHV)}, 2023.
%15
\item A. Bentaleb, \textbf{R. Farahani}, F. Tashtarian, H. Hellwagner, R. Zimmermann. Which CDN to Download From? A Client and Server Strategies. \textit{ACM Mile High Video (MHV)}, 2023.
%16
\item R. Shokri Kalan, \textbf{R. Farahani}, E. Karsli, C. Timmerer, and H. Hellwagner. Towards Low Latency Live Streaming: Challenges in Real-World Deployment. \textit{13th ACM Multimedia Systems Conference (MMSys)}, 2022.
%17
\item F. Tashtarian, A. Bentaleb, \textbf{R. Farahani}, M. Nguyen, C. Timmerer, H. Hellwagner, and R. Zimmermann. A Distributed Delivery Architecture for User Generated Content Live Streaming over HTTP. \textit{IEEE 46th Conference on Local Computer Networks (LCN)}, 2021.
%18
\item  A. Erfanian, F. Tashtarian, \textbf{R. Farahani}, C. Timmerer, and H. Hellwagner. On Optimizing Resource Utilization in AVC-based Real-time Video Streaming. \textit{6th IEEE Conference on Network Softwarization (NetSoft)}, 2020.
\end{enumerate}
%%%%%%%%%%%%%%%%%%%%%%%%%%%%%%%%%%%%%%%%%%%%%%%%%%%%%%%%%%%%%%%%%%%%%%%%%%%%%%%

%% file: Chapters/Chapter1/1-7-Organization.tex
\section{Thesis Organization}\label{chap:Introduction:Organizations}
\doublespacing
% The structure of this thesis is shown in Fig.~\ref{Thesis-structure2}. 
In Chapter~\ref{chap:BackgroundandRelated}, we provide the general technical background of modern networking paradigms utilized in designing NAVS frameworks. Furthermore, this chapter surveys state-of-the-art approaches in the areas of the described research contributions. 

The focus of Chapter~\ref{chap:EdgeSDN} is utilizing edge and SDN capabilities to present NAVS solutions for HAS clients. This chapter describes the details of the \texttt{ES-HAS}~\cite{Farahani2021eshas} and \texttt{CSDN}~\cite{Farahani2021csdn} frameworks. More specifically, in each section, we first express our primary motivation through examples and define problems.  We then describe the details of each system's architecture and formulate each problem as a mathematical optimization model. At the end of each section, we discuss the details of the testbed design and obtain and analyze results. Finally, we summarize the chapter's achievements and learned lessons.

In Chapter~\ref{chap:SFCEnabled}, we introduce an SFC-enabled architecture for adaptive video streaming applications called \texttt{SARENA}~\cite{farahani2023sarena}. We first elaborate on the \texttt{SARENA} multi-layer architecture and formulate the problem as an optimization model. We then propose a lightweight heuristic model before discussing the details of the designed testbed and achieved results. We finally conclude the chapter by highlighting achievements and the findings. 

Chapter~\ref{chap:CollaborativeEdge} proposes two collaborative edge-assisted frameworks for HAS clients, \texttt{LEADER}\\~\cite{farahani2022leader} and \texttt{ARARAT}~\cite{farahani2022ararat}, in two sections. In each section, we first describe the system design perspective, including the problem statement, architecture, and system model. Afterward, we propose the heuristic solution(s), which are designed to remedy the high time complexity of the proposed optimization models. We finally describe evaluation setups and methods and perform in-depth analyses of these collaborative systems before emphasizing the achievements and learned lessons of the chapter.

Chapter~\ref{chap:Hybrid-P2PCDN} explains the details of two hybrid P2P-CDN architectures for HAS-based systems, called \texttt{RICHTER}~\cite{farahani2022RICHTER} and \texttt{ALIVE}~\cite{farahani2023alive}, in two sections. Each section first discusses the problem statement, particular architecture, and problem formulation. It then discusses the proposed heuristic scheme(s) before providing a comprehensive testbed evaluation and model analysis. At the end of this chapter, we summarize the outcomes and valuable insights that we have gained.

This thesis is concluded in Chapter~\ref{chap:Conclusion} by briefly summarizing the conducted studies, the obtained results, and the derived insights.
Furthermore, four promising future research directions are discussed at the end of this chapter.

%% file: Chapters/Chapter2/2-0-Intro.tex
% \singlespacing
\chapter{Technical Background and Related Work}\label{chap:BackgroundandRelated}
%************************************************
\doublespacing
Improving video quality for HAS users relying solely on embedded players' ABR algorithms is a complex and challenging task. This is because pure client-based HAS solutions use limited awareness of streaming and network conditions. To address this challenge, Network-Assisted Video Streaming (NAVS) solutions have emerged. In NAVS systems, an in-network component (\eg edge server or SDN controller) with a broader view of the network is employed to assist video players in making precise adaptation decisions, hence improving users’ QoE and network QoS KPIs. In this chapter, we first discuss the end-to-end (E2E) workflow of one of the popular HAS-based video streaming applications, \ie live streaming. We also explain the concept of fundamental and emerging networking paradigms in six different categories, which are leveraged in designing our proposed NAVS solutions. Finally, we review prior work in the network-assisted video streaming domain in two categories, highlight several standardization activities, \rf{and describe research gaps in the literature for designing NAVS systems.}

%% file: Chapters/Chapter2/2-1-0-Background.tex
\section{Technical Background}\label{chap:Background:Intro}
This section explains the details of the \textit{HTTP Adaptive Streaming} (HAS) technology, plus the concept of six key networking paradigms. These paradigms are utilized in the subsequent technical chapters (Chapters~\ref{chap:EdgeSDN}--\ref{chap:Hybrid-P2PCDN}) to introduce NAVS streaming frameworks for HAS clients.

%% file: Chapters/Chapter2/2-1-1-HAS.tex
\subsection{HTTP Adaptive Streaming (HAS)}\label{chap:Background:HAS}
Although traditional video delivery technologies such as Real-time Transport Protocol (RTP)~\cite{RTP}, Real-Time Messaging Protocol (RTMP)~\cite{RTMP}, and Real Time Streaming Protocol (RTSP)~\cite{RTSP} have their own merits, many works have revealed their limitations to achieve scalability and vendor independency, and their high maintenance costs~\cite{bentaleb2018survey}. Many technologies have emerged in recent years to improve video streaming services. Among them, \textit{HTTP Adaptive Streaming} (HAS)-based solutions, such as \DASH (DASH)~\cite{sodagar2011mpeg} or Apple's \textit{HTTP Live Streaming} (HLS)~\cite{HLS}, with delivering more than 50\% of video streams, have become the most prevalent technologies utilized by Over-the-Top (OTT) providers for VoD and live video streaming applications~\cite{Sandvine}. In HAS-enabled systems, video sequences are split into short- and fixed-duration segments (\ie between one and ten seconds durations), and then each segment is encoded into multiple versions called \textit{representations} (\ie resolutions or bitrate levels). Information on the segments, including their locations on media servers, \eg CDN servers or origin server, is stored within a manifest file such as DASH's \textit{Media Presentation Description} (MPD). HAS players then process MPDs and consider the current network status and/or capabilities of the user equipment (\eg playout buffer) to adaptively download appropriate representations from media servers using \textit{Adaptive Bitrate} (ABR) algorithms, consequently improving users' QoE~\cite{dao2022contemporary}. 
%%%%%%%%
\begin{figure}[!t]
	\centering
	\includegraphics[width=1\linewidth]{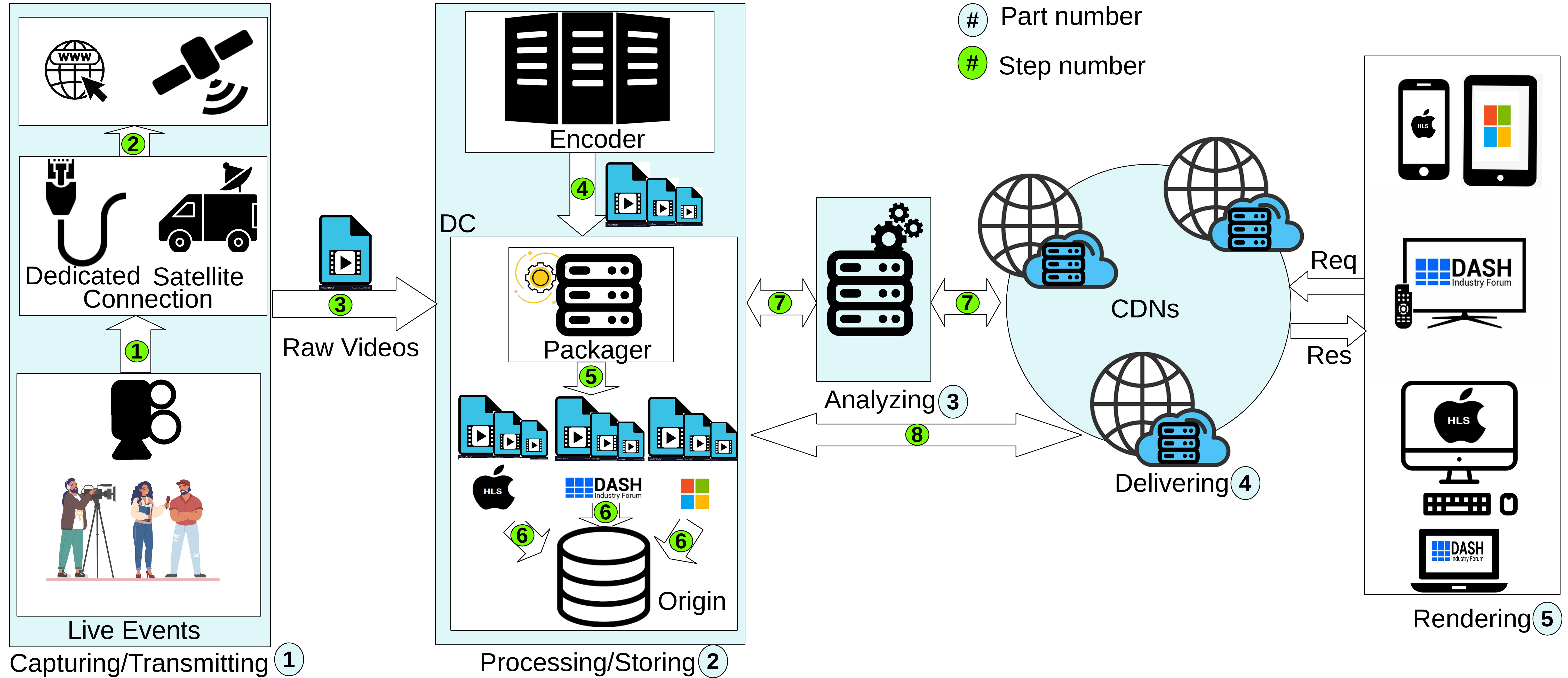}
	\caption{\small Example of an E2E HTTP adaptive live video streaming system~\cite{kalan2022towards}.}
        \vspace{.5cm}
	\label{bk-HAS}
\end{figure}
%%%%%%%%%%

The workflow of a real-world E2E HAS-based system for live video streaming is shown in Fig.~\ref{bk-HAS}, which includes five core parts as follows (part and step numbering as in the figure):
\begin{enumerate}[noitemsep]
\item \textbf{Capturing/Transmitting. } A live event is captured (step \tiny{\circled[text=black,fill=green!78!white,draw=black]{\scriptsize{1}}}\normalsize) and then transmitted via satellite or dedicated network connection to the OTT headquarters (steps \tiny{\circled[text=black,fill=green!78!white,draw=black]{\scriptsize{2}}}\normalsize~ and \tiny{\circled[text=black,fill=green!78!white,draw=black]{\scriptsize{3}}}\normalsize). 
\item \textbf{Processing/Storing. }The raw input video is encoded through encoders to produce multiple representations of a single video. After encoding, the produced representations are sent to the data center (DC) (step \tiny{\circled[text=black,fill=green!78!white,draw=black]{\scriptsize{4}}}\normalsize). The DC can be employed for both packaging (\ie into a specific HAS format like DASH) and caching functions (step \tiny{\circled[text=black,fill=green!78!white,draw=black]{\scriptsize{5}}}\normalsize) to improve the origin server utilization by \textit{(i)} avoiding to package all bitrates and \textit{(ii)} storing a minimum number of bitrates.
When a client’s request for segment $s$ and bitrate $b$ comes to the origin server, it does on-the-fly packaging and returns the requested ($s,b$) segment to the client. To prevent running computation-intensive tasks, \ie packaging ($s,b$) for another client request, ($s,b$) is cached on the origin server (step \tiny{\circled[text=black,fill=green!78!white,draw=black]{\scriptsize{6}}}\normalsize). 
\item \textbf{Analyzing. } Monitoring and analyzing tools can be used, focusing on network conditions and client quality parameters. To do that, monitoring software can be used to monitor both the origin and the CDN servers (step \tiny{\circled[text=black,fill=green!78!white,draw=black]{\scriptsize{7}}}\normalsize). The provided information could be used for load balancing or the selection of the best origin/CDN server. Moreover, monitoring software can be employed as an online video analyzer tool to investigate client access patterns for some purposes like producing popular representations.
\item \textbf{Delivering. } In order to provide fast delivery and a high level of reliability, CDN provider services can be used as well. When ($s,b$) is unavailable in the CDN, it will be fetched from the origin server (step \tiny{\circled[text=black,fill=green!78!white,draw=black]{\scriptsize{8}}}\normalsize). 
\item \textbf{Rendering. }Different types of clients, \eg Web browsers, mobile devices, and TVs, can be connected to and served by such an envirenment, where the number of connected clients and requested bitrates changes dynamically. 
\end{enumerate}

This thesis uses existing and standard HAS technology to design NAVS systems for VoD and live streaming.

%% file: Chapters/Chapter2/2-1-2-CDN.tex
\subsection{Content Delivery Networks (CDNs)}\label{chap:Background:CDN}
\begin{figure}[!t]
	\centering
	\includegraphics[width=.8\linewidth]{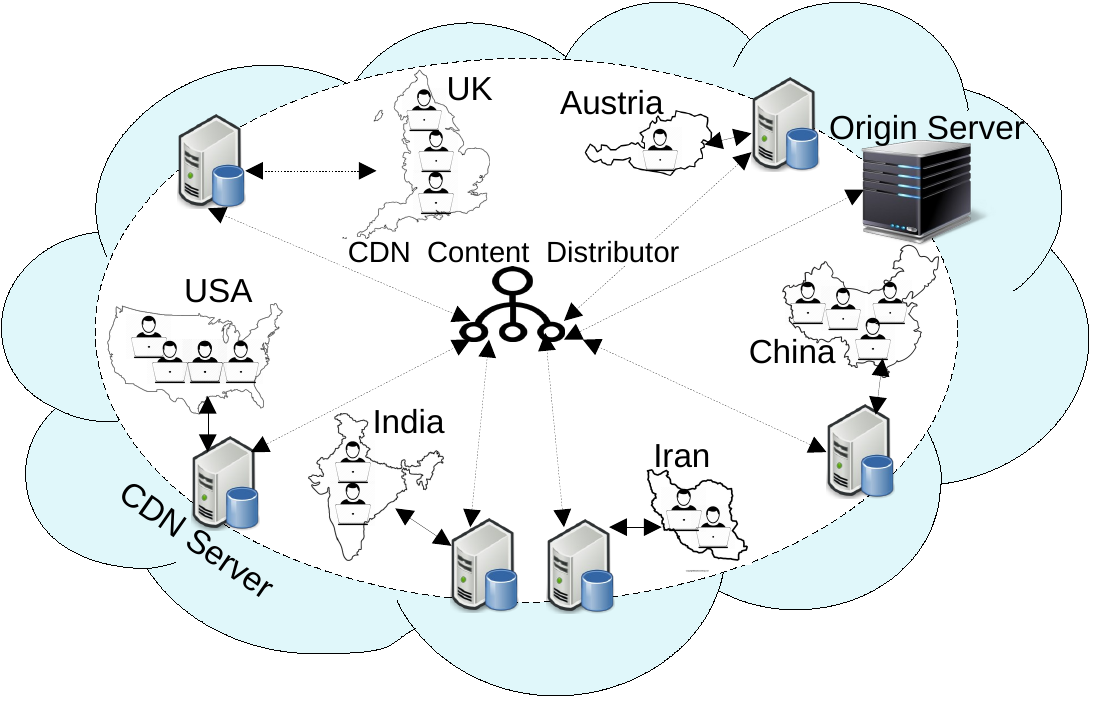}
	\caption{\small Illustration of a CDN architecture.}
        \vspace{.5cm}
	\label{bk-CDN}
\end{figure}

A \textit{Content Delivery Network} (CDN) is a distributed network of servers that work together to deliver content to users. The content could be any type of digital assets, such as data, web pages, or multimedia content. A conceptual architecture of CDNs, including \textit{three} key components, is illustrated in Fig.~\ref{bk-CDN}~\cite{pathan2007taxonomy}. These components are:
\begin{enumerate}[noitemsep]
\item \textbf{CDN Servers. }These servers (also so-called surrogate, cache, or edge servers) are located in different geographical locations close to the end-users to cache and deliver content quickly and efficiently. Indeed, when a user requests content, the request is first directed to the nearest CDN server. Afterward, the CDN server checks the availability of the content in its cache. If the CDN server holds the content, it serves the request directly, much faster than retrieving it from the origin server. Otherwise, the content will be fetched and served from the origin server, at rates slower than serving from the CDN server.
\item \textbf{Origin Server. }This server is responsible for generating the content, storing the original content, and ensuring it is updated. When a CDN server does not cache the requested content, the origin server serves the request and then feeds the CDN servers for future demands based on the CDN content distributor policy.
\item \textbf{CDN Content Distributor. }This component manages content distribution across the CDN and defines important policies for improving the CDN network performance. For example, it monitors the CDN network to detect frequently accessed content and to have it cached on CDN servers, reducing the load on the origin server and improving the overall performance of content delivery. Furthermore, it monitors user requests to distribute them across multiple CDN servers to ensure that no single server is overloaded. These algorithms take into account the CDN's KPIs to ensure that content is delivered to end-users quickly and efficiently.
\end{enumerate}
%%%%%%%%
%%%%%%%%%%

These components work together to build a CDN infrastructure that offers scalable services and enhances website and application performance~\cite{pathan2007taxonomy}. Due to the ability of CDNs in delivering large volumes of data and the increasing demand for low-latency and high-quality video streaming services, CDNs have become the primary method for video delivery~\cite{cisco2018cisco}. In the following chapter, we utilize CDN caching services to introduce NAVS solutions for HAS clients.

%% file: Chapters/Chapter2/2-1-3-P2P.tex
\subsection{Peer-to-Peer (P2P) Networks}\label{chap:Background:P2P}
\begin{figure}[!t]
	\centering
	\includegraphics[width=.85\linewidth]{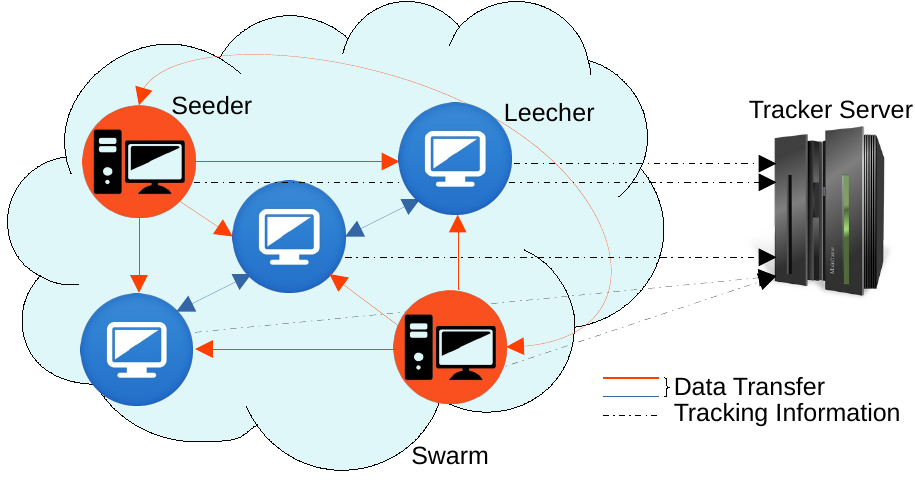}
	\caption{\small Illustration of a P2P network.}
        \vspace{.5cm}
	\label{bk-P2P}
\end{figure}

A \textit{Peer-to-Peer} (P2P) network is a decentralized content delivery network where each device, a so-called \textit{peer}, communicates and shares content directly with others rather than relying on a centralized server. In fact, each peer in a P2P network acts as both client and server, which enables content to be downloaded and uploaded simultaneously~\cite{siano2019survey}. It is worth noting that content refers to data, digital media (\ie audio, video), and software in this context. An example architecture of a P2P network is shown in Fig.~\ref{bk-P2P}, where the following components play critical roles:
\begin{enumerate}[noitemsep]
\item \textbf{Tracker Server. }This server is responsible for keeping track of the peers and coordinating their content exchange. When a peer requests content, it contacts the tracker server to fetch a list of other peers which hold the content for downloading. This list enables the peer to connect to others and download the requested content directly without relying on a central server. However, it is worth noting that in other P2P structures, peers can discover other peers through other methods, \eg \textit{distributed hash tables} (DHTs), instead of using a tracker server~\cite{anjum2017survey}.
\item \textbf{Seeders. }Seeders are one type of peers which have completely downloaded content and are willing to share it with other peers. Once a peer downloads the content from a seeder, it may become a seeder and continues sharing it with others. Holding high resources, \eg upload/download bandwidth and battery, is another characteristic of seeders, which allows them to remain connected to the P2P network for extended periods; consequently, other peers fetch their desired content from these stable peers.
\item \textbf{Leechers. }In contrast to seeders, peers which are still in the process of downloading the content and have not yet fetched the complete content are referred to as leechers. The P2P network would not operate effectively without seeders since leechers could not complete their downloads without them.
\end{enumerate}
%%%%%%%%
%%%%%%%%%%

A \textit{swarm} in Fig.~\ref{bk-P2P} refers to a group of peers which concurrently download and share content at a particular time. As more peers join the swarm and commence downloading and sharing the content, the number of sources from which fetches can occur increases, resulting in more efficient and rapid downloads. 
Considering features provided by the P2P networks, many companies, \eg Peer5\footnote{\url{https://peer5.com/}; last access: 2023-03-16.}, Strivecast\footnote{\url{https://strivecast.com/}; last access: 2023-03-16.}, and Livepeer\footnote{\url{https://livepeer.org/}; last access: 2023-03-16.} have been utilizing peer-assisted networks with emerging communication protocols such as \textit{Web Real-Time Communications} (WebRTC)~\cite{roach2016webrtc,jansen2018performance} to assist CDNs and provide low-latency and high-quality video streaming services.
In Chapter~\ref{chap:Hybrid-P2PCDN}, we leverage the decentralized property of a P2P network, show how computational tasks, \eg video transcoding and video super-resolution, can be offloaded to peers, and introduce novel hybrid P2P-CDN frameworks for HAS clients.

%% file: Chapters/Chapter2/2-1-4-SDN.tex
\subsection{Software Defined Networking (SDN)}\label{chap:Background:SDN}
\begin{figure}[!t]
	\centering
	\includegraphics[width=1\linewidth]{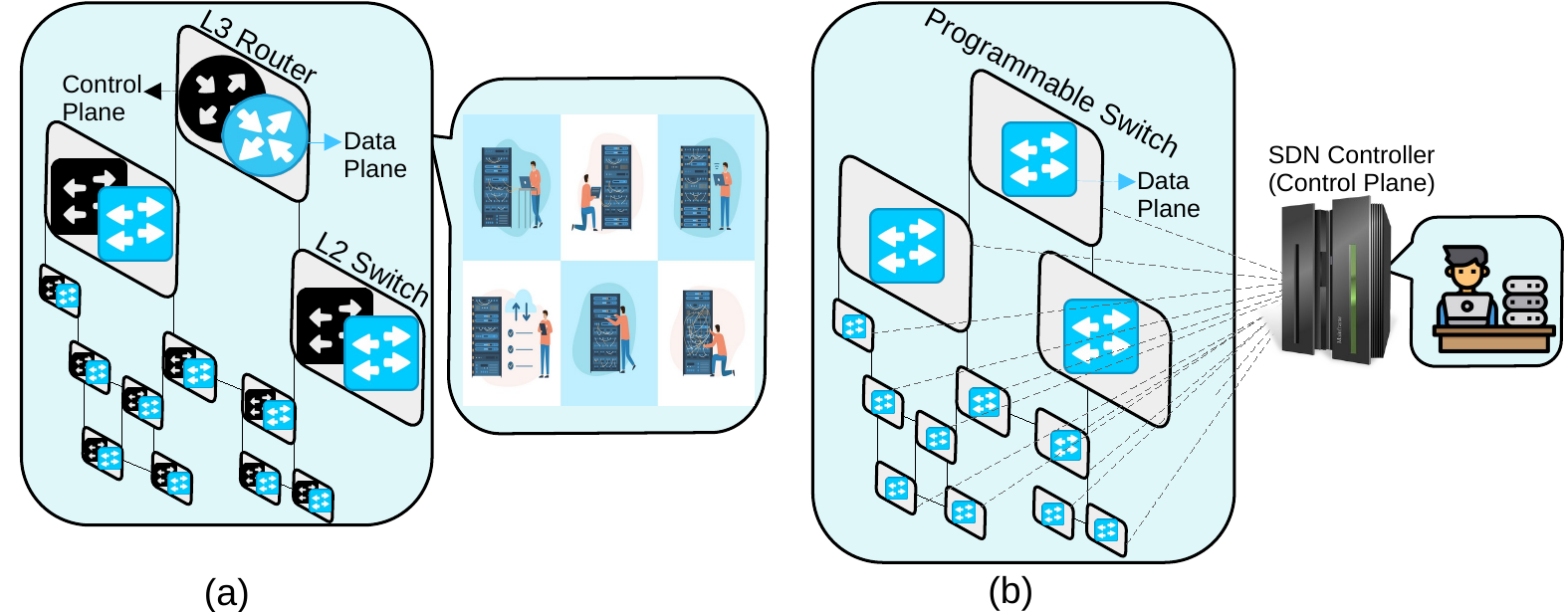}
	\caption{(a) Traditional and (b) SDN-based network architectures.}
         \vspace{.5cm}
	\label{bk-SDN}
\end{figure}

In traditional networks, all networking devices, such as layer 3 routers and layer 2 switches, combine the control plane and data plane into a single unit. Indeed, both functions are tightly integrated, and each proprietary networking device performs both functions within the same hardware (Fig.~\ref{bk-SDN}(a)).
The control plane executes decision-making tasks such as routing, switching, or security policy enforcement, while data plane functions involve forwarding network traffic. As a result, managing such networks poses considerable challenges, such as:
\begin{enumerate}[noitemsep]
\item \textbf{Complexity. }Network administrators must configure individual network devices separately, which can lead to human error and misconfiguration, resulting in complex network management. 
\item \textbf{Lack of flexibility. }Since each network device can only provide information about its immediate neighbors, achieving a broader view of the network topology is impossible. Consequently, when there is a problem in the network, it can be complicated to troubleshoot the fault.
\item \textbf{Lack of automation. }Network administrators must manually execute all management and configuration tasks without any automation procedure, which can be time-consuming.
\end{enumerate}
\raggedbottom
%%%%%%%%
%%%%%%%%%

The aforementioned issues led to the developing \textit{network softwarization} paradigm, which replaces traditional networks with \textit{Software Defined Networking} (SDN)-enabled networks~\cite{kreutz2014software}. As depicted in Fig.~\ref{bk-SDN} (b), the control plane and data plane are decoupled in the SDN approach, and the control plane is placed in a central entity called the \textit{SDN controller}. This decoupling allows the SDN controller to communicate with programmable network nodes, manage traffic flows more efficiently, simplify network management, and then apply network policies to satisfy network services' requirements (\eg video streaming), overall addressing the above-mentioned challenges.
One considerable difference between traditional and SDN-based networks is that SDN presents a communication interface, such as standardized OpenFlow (OF)~\cite{mckeown2008openflow}, that connects network devices to the SDN controllers. This feature allows network developers, administrators, and managers to define network-aware policies for any type of application, such as video streaming, that can dynamically adjust to changing network conditions and take advantage of network resources more effectively. In Chapters~\ref{chap:EdgeSDN}--\ref{chap:CollaborativeEdge}, we show how the SDN controller characteristics can be utilized to develop network-assisted video streaming frameworks for HAS clients.
 

%% file: Chapters/Chapter2/2-1-5-NFV.tex
\subsection{Network Function Virtualization (NFV)}\label{chap:Background:NFV}
As previously stated, the use of dedicated network devices in traditional networks gives rise to numerous issues, such as high deployment costs, limited adaptability, and difficult-to-manage services. 
%%%%%%%%
\begin{figure}[!t]
	\centering
	\includegraphics[width=1\linewidth]{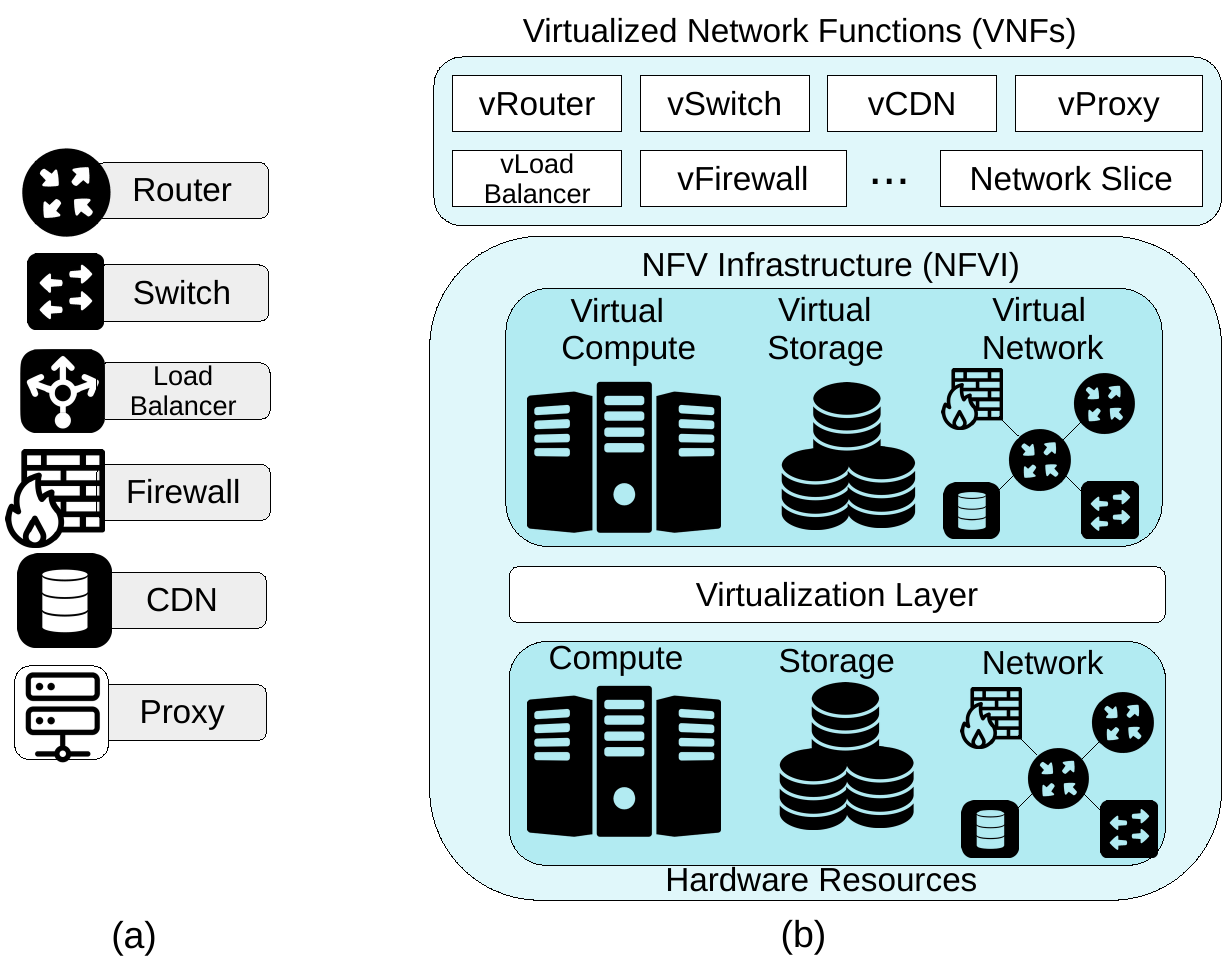}
	\caption{(a) Traditional and (b) NFV-based network components.}
        \vspace{.5cm}
	\label{bk-NFV}
\end{figure}
%%%%%%%%%%
To cope with these issues, a group of leading service providers, including AT\&T, Verizon, and Deutsche Telekom, established the \textit{Network Functions Virtualization} (NFV) Industry Specification Group within the European Telecommunications Standards Institute (ETSI) in 2012~\cite{etsi2014network}. The NFV group aimed to add the idea of virtualization to the network, which enables network functions to be implemented as \textit{Virtualized Network Functions} (VNFs) on any network appliance. As Fig.~\ref{bk-NFV} illustrates, traditional network functions such as routers, switches, or firewalls are virtualized and run as software applications, \eg vRouter, vSwitch, or vFirewall on general-purpose appliances, rather than being dependent on proprietary network devices. Moreover, the NFV paradigm provides network layer virtualization to present a slice of the \textit{NFV Infrastructure} as an NFVI service~\cite{chiosi2012network}.

Considering the aforementioned benefits, the combination of SDN and NFV paradigms leads to an agile, scalable, and innovative network. In such networks, the SDN controller orchestrates the network resources and VNFs to manage the network functions; thus, operators can provide E2E network services that are highly automated, programmable, and adaptable to changing business needs. For instance, the SDN controller has the potential to partition the network into multiple virtual networks (\ie slices) for different purposes (\eg video delivery), thus enabling operators to provide customized network services to various users or applications~\cite{viola2023survey}. In the next technical chapters, we elaborate on how multimedia functions can be presented as VNFs and how the SDN controller can orchestrate the network and VNFs to deliver multimedia services.

%% file: Chapters/Chapter2/2-1-6-SFC.tex
\subsection{Service Function Chaining (SFC)}\label{chap:Background:SFC}
Providing a range of network services in an NFV-enabled infrastructure using only one type of VNF is impossible. This is because network services require their traffic to be steered through a set of VNFs in a specific order. To address this issue, \textit{Service Function Chaining} (SFC), relying on the NFV paradigm, has been introduced by the Internet Engineering Task Force (IETF) to direct network traffic through a sequence or chain of VNFs~\cite{halpern2015service}. 
%%%%%%%%
\begin{figure}[!t]
	\centering
	\includegraphics[width=1\linewidth]{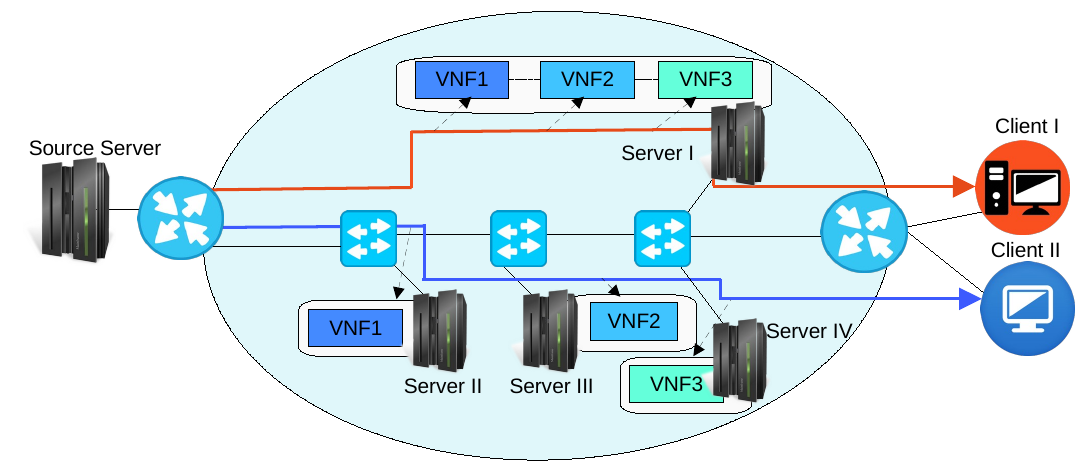}
	\caption{Illustration of an SFC-enabled system.}
        \vspace{.5cm}
	\label{bk-SFC}
\end{figure}
%%%%%%%%%%

\rf{An example of an SFC-enabled system is shown in Fig.~\ref{bk-SFC},} where two different network services from the source server are requested by client I (red) and client II (blue). Suppose that both network services must be steered through VNF1$\rightarrow$VNF2$\rightarrow$VNF3, and each intermediate server (\ie server I-IV) holds a single or a set of these VNFs. Moreover, assume the priority of the first service (issued by client I) is higher than the second service. Based on the network service requirements and VNF availability, chain VNF1$\rightarrow$VNF2$\rightarrow$VNF3 of the server I can be chosen to serve client I. Furthermore, VNF1 of server II, VNF2 of server III, and VNF3 of server IV, are selected to form the mentioned chain and serve client II. 
Note that defining appropriate service VNFs, and suitable \textit{VNF placement, service scheduling, and resource allocation} algorithms involve many vital considerations and decisions. 
%network services requests for different network services, and \textit{allocating} networking resources to various VNFs can alter the sample decision 
Chapter~\ref{chap:SFCEnabled} shows how the SDN and SFC paradigms can be leveraged to present an architecture for serving various types of multimedia services.

%% file: Chapters/Chapter2/2-1-7-MEC.tex
\subsection{Multi-Access Edge Computing (MEC)}\label{chap:Background:MEC}
Cloud computing has become the de-facto technology in computing, enabling users to quickly invoke a range of servers and deploy a variety of user-customized infrastructure services (\eg storage, resource management, scaling, monitoring) on them. The edge computing paradigm, \eg \textit{Multi-Access Edge Computing} (MEC)~\cite{taleb2017multi},
has been proposed as a complementary technology to cloud computing that provides storage and compute resources close to end-users at the network’s edge, thus offering lower data processing latencies, better user experience, and lower cost and bandwidth consumption than cloud-based computing. In contrast to the cloud, edge servers often include limited resources.
%%%%%%%%
\begin{figure}[!t]
	\centering
	\includegraphics[width=1\linewidth]{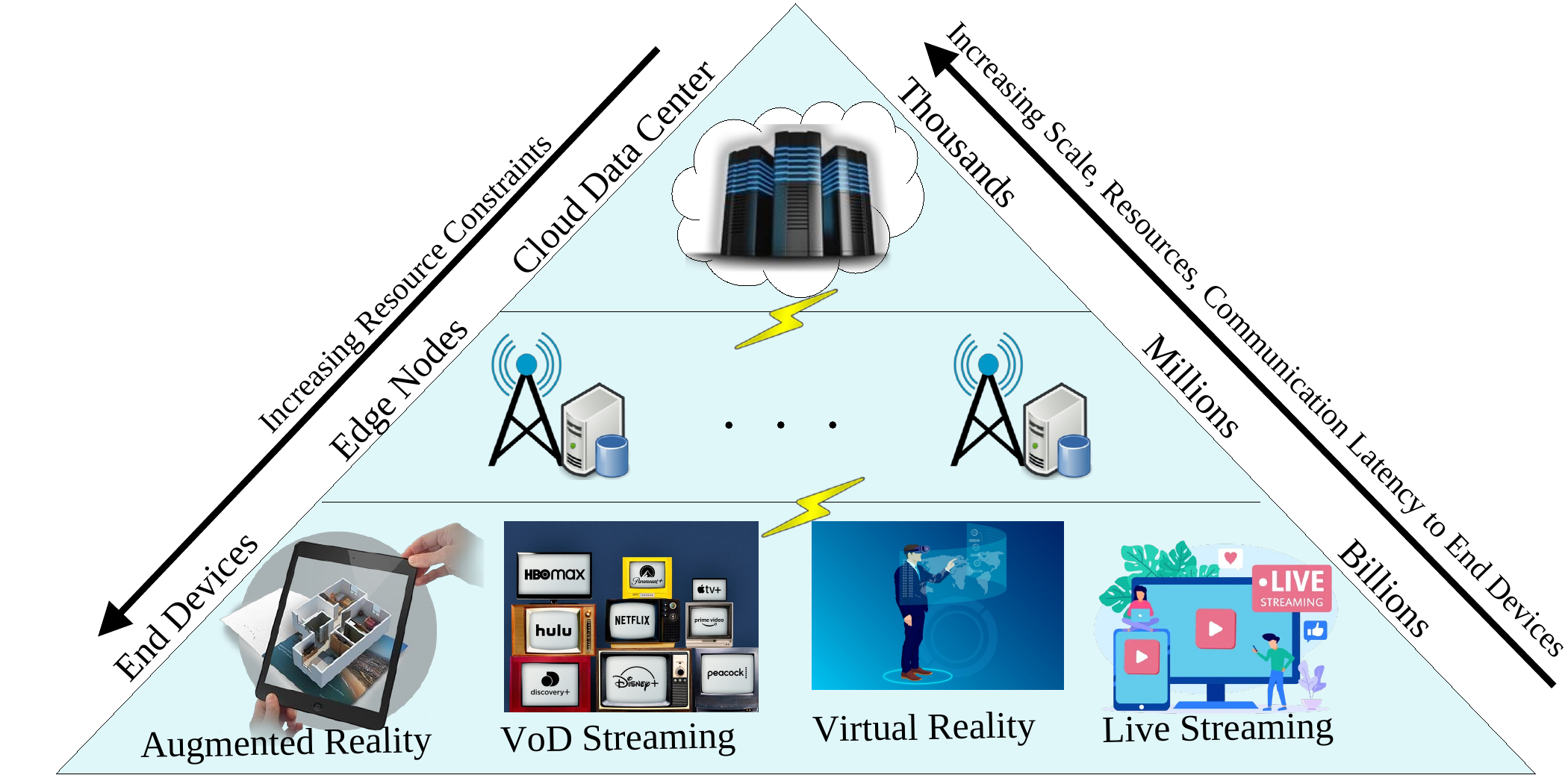}
	\caption{Illustration of cloud and edge computing.}
         \vspace{.5cm}
	\label{bk-Edge}
\end{figure}
%%%%%%%%%%
Due to the nature of edge computing (\eg improved latency, network congestion, scalability, reliability), it is one of the most suitable techniques to support video streaming applications~\cite{khan2022survey} (see Fig.~\ref{bk-Edge} for cloud and edge comparison). It can offer several merits in various types of video streaming applications, including:
\begin{enumerate}[noitemsep]
\item \textbf{VoD Streaming. }Applying techniques such as most popular content caching or content prefetching in advance at the edge enhances the scalability and reliability of VoD services while reducing the load on cloud servers (\eg origin server), enabling faster content delivery, and boosting user QoE. 
\item \textbf{Live Streaming. }Employing caching, transcoding, and processing content closer to the end user, where data is generated and consumed, reduces latency and improves live video streaming quality. 
\item \textbf{Immersive Video Streaming. }Processing and rendering content closer to the end user reduces latency and improves performance for immersive streaming applications. This is because such applications, \eg Augmented Reality (AR) or Virtual Reality (VR), require real-time processing and rendering of video content.
\end{enumerate}
In all our contributions, we leverage MEC capabilities to improve user QoE and network utilization for the two first applications, \ie VoD and live streaming.

%% file: Chapters/Chapter2/2-2-SOTA.tex
\section{Related Work}\label{chap:SOTA}
This section surveys state-of-the-art solutions and highlights standardization activities in two categories. At the end of this section, we summarize related work properties in Table~\ref{tab:Related}.
%%%%%%%%%%%%%%%%%
\subsection{SDN- and/or Edge-Based NAVS Solutions}
\label{chap:SOTA:SDN}
Innovations in network architecture, specifically in core and edge infrastructures, have enabled large CDNs to manage their caching resources better and enhance the quality of user experience. The SDN model is a noteworthy advancement in this area, as it utilizes software-based principles to enable CDNs to have more flexible and real-time traffic management, which can be easily administered. Mukerjee~\etal~investigated the use of the control plane for domain name system (DNS) load-balancing to improve the QoE of live video delivery~\cite{mukerjee2015practical}. 
The work by Kleinrouweler~\etal~proposed an SDN-assisted architecture for DASH clients utilizing a modified version of OpenFlow (OF) switches to support QoS differentiation~\cite{kleinrouweler2016delivering}. They did not use the edge and NFV paradigms, and QoS differentiation was their only objective. Wamser~\etal~proposed a dynamic re-routing strategy to avoid link congestion during streaming sessions~\cite{wamser2015modelling}. Since congested links primarily occur in the network edge, and traffic re-routing can only be set within the core network, this method is only partially effective in optimizing the behavior of HAS clients. 

Bentaleb~\etal~introduced an NAVS solution called SDNDASH~\cite{bentaleb2016sdndash}, which uses SDN capabilities to dynamically allocate network resources based on QoS criteria to increase the users' QoE. They employed a client-side probe to estimate the available network resources and allocate bandwidth resources dynamically to each client. Using this method, they prevent quality instability, unfair distribution of bandwidth, and inefficient use of network resources among multiple DASH clients that share bottleneck network links.
However, Bhat~\cite{bhat2017network} showed that such a probe-based solution as proposed by SDNDASH leads to an overhead both for the network and control plane. 
As an extension to this work, Bentaleb~\etal~proposed another NAVS system called SDNHAS~\cite{bentaleb2017sdnhas} to support a cluster-based estimation for bandwidth requests and address the scalability issue of SDNDASH.
SDNHAS employed a reinforcement learning (RL)-based optimizer component that utilizes the Structural Similarity Index Plus (SSIMplus)~\cite{rehman2015display} perceptual quality metric to create a logical network topology and group the HAS players into virtual clusters. In such a way, clients belonging to the same cluster receive similar recommendations (\ie bitrate levels and their corresponding qualities) and bandwidth slice allocations.

Bhat~\etal~proposed network-assisted content distribution for a QoE-driven ABR video streaming system, entitled SABR~\cite{bhat2018sabr}. In an SABR-enabled system, DASH clients are enabled to receive support from an SDN controller by requesting information on available video sources, such as a list of CDN servers, a list of video segments/qualities, and network conditions, \eg bandwidth history, for making well-informed decisions on content requests. The authors also proposed a network-assisted caching mechanism that helps to reduce the amount of data that needs to be transferred over the network, resulting in improved video playback quality. However, the client must be modified to be compatible with the SABR strategy. Bentaleb~\etal~utilized an online reinforcement learning (ORL) model to introduce a QoE optimization framework for SDN-enabled HAS systems, named ORL-SDN~\cite{bentaleb2018orl}. ORL-SDN groups the HAS players into disjoint clusters based on a perceptual quality index and then formulates the bitrate selection problem as a partially observable Markov decision process. It finally executes an online Q-learning algorithm to solve the QoE optimization problem and determine the optimal bitrate decision for each cluster. Nevertheless, ORL-SDN imposes considerable client-side modifications, making it incompatible with the most common HAS players.

Kalan~\etal~presented an SDN-assisted HAS system for assisting clients in selecting the video codec, bitrate, and routing the video packet over the appropriate network paths~\cite{kalan2021sdn}. However, layered coding used in their work is out of this thesis's scope.  
Ozcelik and Ersoy presented a chunk-size aware SDN-assisted DASH mechanism for streaming called CSASDN~\cite{ozcelik2019chunk}. CSASDN employs various segment durations over the standard HTTP 1.1 GET and POST requests to provide fair QoE for the DASH players while avoiding network underutilization. In the proposed system, each player shares the device characteristics and the content details of the requested video with the SDN controller, and the SDN controller supports them in finding the optimal maximum bitrate. Ge~\etal~devised a novel framework named MVP (mobile edge virtualization with adaptive prefetching) that assures 4K quality for VoD scenarios in 5G networks. This method employs a context-aware adaptive video prefetching scheme deployed at the mobile edge to achieve an improved QoE~\cite{ge2017toward}. The authors in~\cite{ge2018qoe} introduced an edge-based transient holding strategy for live segments called ETHELE. ETHELE uses QoE information from the users' devices and RAN (Radio Access Network) conditions provided by RNIS (Radio Network Information Service) to conduct transient holding of a minimum number of video segments at the edge servers, achieving a seamless 4K live streaming experience by eliminating buffering and reducing initial startup delay and live stream latency. In fact, utilizing the holding method assists the clients in selecting a suitable initial video segment (IVS) that best matches their network throughput to start the playback. However, all of these approaches modified the client side to have a compatible system with any ABR algorithm. In addition, they do not propose any solution for making an efficient collaboration between SDN controller and edge devices as well as among edge devices. 

As discussed in Section~\ref{chap:Background:MEC}, edge computing is a technology for providing computation and caching services for video streaming applications to satisfy the QoE requirements. Bayhan~\etal~proposed the EdgeDASH~\cite{bayhan2020edgedash} framework, which caches popular contents at a WiFi Access Point (AP) and redirects clients' requests to this edge node instead of serving them from the origin server. EdgeDASH's AP also assists DASH clients in making adaptation decisions by offering a few quality levels higher or lower than the requested ones if the alternative quality provides a better QoE. The authors showed that EdgeDASH improved cache hits, decreased buffer stalls, and increased video bitrates.  
Mehrabi~\etal~in~\cite{mehrabi2018edge}~introduced an edge-enabled rate adaptation system through a greedy client/server mapping strategy to jointly maximize the QoE and fairness of competing mobile video streaming clients. They demonstrated that the proposed solution outperforms client-based rate adaptation heuristics. However, due to increasing content volume and the limited cache size of the edge nodes (\eg EdgeDASH's WiFi AP), it is impossible to cache all the contents locally through such approaches. Considering the resource limitations of edge servers (\ie storage, computation, and bandwidth) and the heterogeneity of users' requests, some works like~\cite{tran2018adaptive, baccour2020collaborative} proposed collaborative caching and processing schemes, mainly when several clients' requests exceed the available edge servers' resources. However, they did not utilize the SDN controller capabilities as a network orchestrator.

Since Initial Video Segment (IVS) parameters affect QoE performance, Wang~\etal~designed Rldish~\cite{wang2020rldish}, a reinforcement learning (RL)-based strategy, which is deployed at the edge server to dynamically select a suitable IVS for new live viewers. In fact, Rldish monitors users' QoE at the edge servers and uses QoE values as a reward function of the RL algorithm. The QoE metrics in this system are defined as a weighted combination of a maximum start-up delay and buffering time. However, training RL models in real-time and achieving convergence in dynamic environments can be challenging.
Zhang~\etal~introduced a video super-resolution (SR) and caching system named VISCA~\cite{zhang2021video}, which utilizes a video super-resolution technique for upscaling the requested video from a low-resolution video at the edge of the network and designs an edge-based ABR algorithm to improve users' QoE. However, since the output of the super-resolution module is raw-input video, recompression is needed to transmit the reconstructed video segment. Thus, extra delay and distortion will be imposed by the system. Therefore, the best place for applying the super-resolution technique can be the peers, where the decoded segments will be immediately played out after reconstruction. Moreover, edge servers in the aforementioned systems do not have comprehensive information about the network topology; thus, they individually could not be the most suitable candidates to assist HAS users. Therefore, employing a network entity with a broader view of the network, \eg an SDN controller, besides edge servers, could further improve users' QoE and network utilization. 

Although edge computing pushes the services to the edge closer to the end user with lower latency and higher throughput, this approach comes with several challenges, like limited computational resources and handling heterogeneous users' requests. Hence, combining the SDN and MEC paradigms and integrating them with the existing network infrastructure brings advantages, \eg improved interoperability between different network components through the coordination of the SDN controller~\cite{baktir2017can,jiang2021survey}.
Liotou~\etal~devised a video segment selection and caching strategy using SDN and edge computing called QoE-SDN~\cite{liotou2018qoe}.
Their approach allows feedback from the Mobile Network Operator (MNOs) to the Video Service Provider (VSPs) toward a better network-aware video segment selection. However, scalability needs to be investigated further in this solution. Erfanian~\etal~leveraged the SDN and NFV concepts to 
form a multicast tree(s) for live streaming and then introduce a cost-aware real-time video streaming approach~\cite{erfanian2020optimizing}.
Their solution uses a set of edge virtualized components to gather clients' information, cooperate with the SDN controller and transmit them through a hybrid multicast/unicast approach. 
However, the edge components are limited to functioning independently, as the SDN controller did not include any techniques to enable edge servers to facilitate the use of idle resources of other edge servers or network components. Thus, collaboration among edge servers can result in improving the utilization of available resources. 

% Nacakli~\etal~combined the SDN and edge computing paradigms to present a novel hybrid P2P-CDN live video streaming service~\cite{nacakli2020controlling}. In their proposed architecture, the edge data centers fully control the service with the support of an SDN controller to reduce CDN server loads, overcome the QoE fluctuations in each flow, and improve fairness between heterogeneous clients that request different video qualities/resolutions.
%%%%%%%%%%%%%%
\subsection{CDN- and/or P2P-Based NAVS Solutions}
\label{chap:SOTA:CDN}
With emerging HAS-based technologies, OTT companies often utilize distributed CDN servers in different geographical locations to get more reliable and cost-effective content delivery. 
Intense research in a variety of aspects relevant to content delivery performance and costs, like user request routing~\cite{chen2015end,zhu2022best}, content placement and replication~\cite{passarella2012survey,li2019flexible}, server offloading~\cite{fang2022cooperative}, and transport costs~\cite{rizk2017model}, as well as content popularity and content awareness~\cite{koch2018category,yao2019mobile} has been pursued in recent years. Many works such as~\cite{pedersen2015enhancing,ma2017understanding,choi2018wireless,zhang2022optimizing} propose new caching strategies to improve CDN hit rates and eventually enhance video QoE. The detailed ``interplay of quality adaptation mechanisms, network bandwidth information, as well as caching strategies''~\cite{bhat2017network} is barely addressed in works on global control for CDN-based video delivery, \eg \cite{ganjam2015c3}. 

Some works like~\cite{akhshabi2013server} use a server-assisted HAS adaptation approach where an element, \eg CDN server, can assist HAS players in deciding the next segment's representation. The proposed method in~\cite{akhshabi2013server} aims to detect quality fluctuations by analyzing client requests. Once such oscillations are identified, a corresponding server-side algorithm is triggered, restricting the maximum segment's quality level the client can access. This restriction is implemented to stabilize the stream quality level and mitigate further oscillations. Although such solutions improve the HAS adaptation performance compared to the pure client-based approaches with only local parameters' consideration, they encounter significant problems in large-scale deployment. A network-assistance approach can be taken into account as an alternative solution, where both in-network elements and servers support the bitrate adaptation. For example, the authors in~\cite{el2017improving} introduced an enhanced server and client cooperation called ESTC. ESTC facilitates rapid convergence among different clients’ bandwidth levels towards the estimated bandwidth through enabling close collaboration between the server and client sides. It also appropriately assigns the allocated bitrate to clients, where the clients have the responsibility for making the right bitrate adaptation decisions for efficiency and service stability insurance. In addition, the server leverages information such as the number of connected clients, current download bitrates, and bottleneck link bandwidth to ensure fairness among competing clients.

The Server and Network-Assisted Dynamic Adaptive Streaming over HTTP (SAND) standard~\cite{thomas2016applications} was introduced by MPEG. SAND tries to mitigate DASH performance problems by enabling protocol messages to be exchanged among network components, \eg CDN servers. However, SAND does not specify how to handle such messages efficiently. SAND defines four message types, \ie status messages, metrics messages like buffer occupancy, packets enhancing reception, and packets enhancing delivery. If a network component can process all mentioned messages or a subset of them, it is called a DASH-Aware Network Element (DANE). A novel delivery scheme that combined both SAND and Content Centric Networks (CCN) was introduced in~\cite{jmal2017network}. The authors demonstrated how the SAND architecture could solve the bandwidth oscillation problems of DASH-based systems over traditional delivery architectures by the existence of cache proxies in the CCN delivery network. However, the proposed approach primarily targets CCN-based architectures, which are distinct from the more widely used CDN-based architectures. Although SAND-based systems can improve client-side adaptation decisions, they are not straightforward to implement, and only a few papers like~\cite{kalan2019optimal,thomas2018application, kalan2021improving} have pursued this approach yet~\cite{CDNSDNSupport2021}. 

A specification by the Consumer Technology Association (CTA) in the Web Application Video Ecosystem (WAVE) project has introduced solutions termed \textit{Common Media Client Data} (CMCD) and \textit{Common Media Server Data} (CMSD)~\cite{CTA,CTA-wave}. The primary motivation behind designing the CMCD and CMSD was to address questions as follows~\cite{lim2022benefits}: \textit{(i)} How could a streaming client relay media-related information (\eg segment type/duration/format, content/session IDs) to the CDN in a way that allows the CDN to associate individual GET requests with playback sessions? \textit{(ii)} How could a streaming client relay playback-related information (\eg current buffer length and latency) to the CDN in a way that enables accurate generation of dashboard metrics, such as delivery performance, player software issues, and viewer experience? \textit{(iii)} How could the CDN react to the time constraints implicit in media segment requests (\eg prioritize delivery for urgent requests)? 

The CMCD protocol outlines data collected by a client and sent to the CDNs, which facilitates the correlation of CDN data with client data. This correlation enables content providers to enhance CDN performance while ensuring that viewers receive high-quality content. For instance, by utilizing the CMCD information, CDNs can improve their performance via \textit{(i)} intelligently responding to the implicit time constraints of each request and \textit{(ii)} identifying performance issues that may be linked to particular devices or player software versions. The CMSD specification enables the CDNs to send useful information back to the clients~\cite{begen2021road}. Bentaleb~\etal~\cite{bentaleb2021common} built an initial proof-of-concept system that adheres to the CMCD specification on both the client and server ends. Their proposed system deployed a buffer-aware bandwidth allocation mechanism to enhance the clients' QoE for clients that share network bandwidth. Begen~\etal~in~\cite{begen2021road} presented an overview of the CMCD and CMSD specifications and demonstrated various applications of CMCD to improve HAS-based streaming. However, this work did not provide any implementations.

Several prior works like~\cite{xie2008p4p} proposed solutions for establishing a collaboration between network service providers and content providers leading to an Internet Engineering Task Force (IETF) standard called \textit{Application-Layer Traffic Optimization} (ALTO)~\cite{alto}. ALTO enables P2P applications to obtain abstract maps of network information, including from CDNs, to optimize network resource consumption and traffic delivery while maintaining application performance. However, some studies reveal that the ALTO protocol still has drawbacks, and its usage is limited~\cite {ellouze2013bidirectional}. As another well-known peer-assisted product, Akamai introduced the NetSession interface~\cite{zhao2013peer} that provides peer-assisted delivery and supports peer-CDN cooperation. However, it is required to install the software on the client's device. Our works differ from the above-mentioned approaches because we do not modify the client side. As a consequence, the proposed NAVS frameworks can be compatible with any ABR algorithm. 
In the last decade, both academia and industry have considered combining P2P and CDN delivery networks to improve performance and reduce the cost of OTT video delivery services.
A P2P-CDN NAVS system is a hybrid architecture that aims to use the advantages of both P2P and CDN systems. In such systems, the most challenging research question is: ``\textit{How should the traditional CDN and P2P networks be integrated to enable OTT providers to efficiently employ users' resources and benefit from both CDN and P2P networks' advantages?}''

Xuening~\etal~proposed a commercial hybrid CDN-P2P system for live video streaming applications called LiveSky~\cite{yin2009design}. 
LiveSky used the CDN servers to deliver the video sequences from the source to the edge of the network, while the P2P network was employed to deliver the video sequences from the edge to end users. Moreover, the authors propose an adaptive scaling mechanism to provide fairness in video delivery by dynamically adjusting the contribution of each peer in the P2P network based on its available bandwidth and buffer status. Ha~\etal~presented a dedicated hybrid P2P-CDN system for live video streaming to reduce CDN usage significantly~\cite{ha2017design}. In their proposed system, the video segments are portioned further into small chunks of equal size. The HAS player modules were customized and equipped with an additional P2P module that prefetches the video chunks ahead of time. However, Yousef~\etal~\cite{yousef2020enabling} showed that the prefetching process in such a system might be incomplete due to the peers' heterogeneity. 

Khan~\etal~\cite{khan2022code} proposed a Computation Offloading in the D2D-Edge framework called CODE. They proposed Integer Linear Programming (ILP) models, \ie maximal offloading with delay constraint (MOD), minimum delay offloading (MDO), and a lightweight heuristic to form cooperation between edge servers and peers in MEC networks to efficiently utilize the edge computational resources during large-events live streaming. Ma~\etal~proposed learning-based methods for hybrid P2P-CDN live video streaming systems that enable their trackers to perform peer selection~\cite{ma2021locality}. They first analyzed a large-scale hybrid CDN-P2P live streaming system and collected a dataset of popular channels in Europe and North Africa, and then trained a neural network (NN)-based model. 
All the systems mentioned above use only the storage and bandwidth resources of peers, not their computational resources, for reconstructing the requested video from available higher quality (\eg by transcoding) or from lower resolution (\eg by super-resolution), besides CDN servers to enhance users' QoE and network utilization. To fill this gap, we will use all feasible peers' computational resources and propose two hybrid P2P-CDN NAVS frameworks in Chapter~\ref{chap:Hybrid-P2PCDN}.

Nacakli~\etal~combined the SDN and edge computing paradigms to present a novel hybrid P2P-CDN live video streaming service~\cite{nacakli2020controlling}. In their proposed architecture, the edge data centers fully control the service with the support of an SDN controller. In fact, their proposed edge servers employ a P2P grouping method and a P2P scheduling algorithm considering CDN locations to reduce CDN server loads, overcome the QoE fluctuations in each flow, and improve fairness between heterogeneous clients that request different video qualities/resolutions. The authors in~\cite{kumar2022software} leveraged the SDN concept and presented the concept of ``CDN at the edge'' where end hosts can collaborate for video streaming in a pure P2P manner to provide better QoE. They used the capability of the SDN controller to provide timely information about leaving or joining peers. Additionally, the proposed method involves the SDN controller calculating the E2E bandwidth and latency between potential candidate peers and the requesting peer. The calculated values are subsequently communicated back to the requesting peers, enabling them to select the optimal peer to retrieve the video segment. However, the proposed systems only rely on overlay networks' bandwidth and storage capacities to serve clients' requests. In other words, the computational resources that can be leveraged to perform various video processing functionalities at edge servers or P2P networks (\eg video transcoding) are not taken into account in such systems. 

\begin{figure}
	\centering
	\includegraphics[width=1.5\linewidth,{angle=90}]{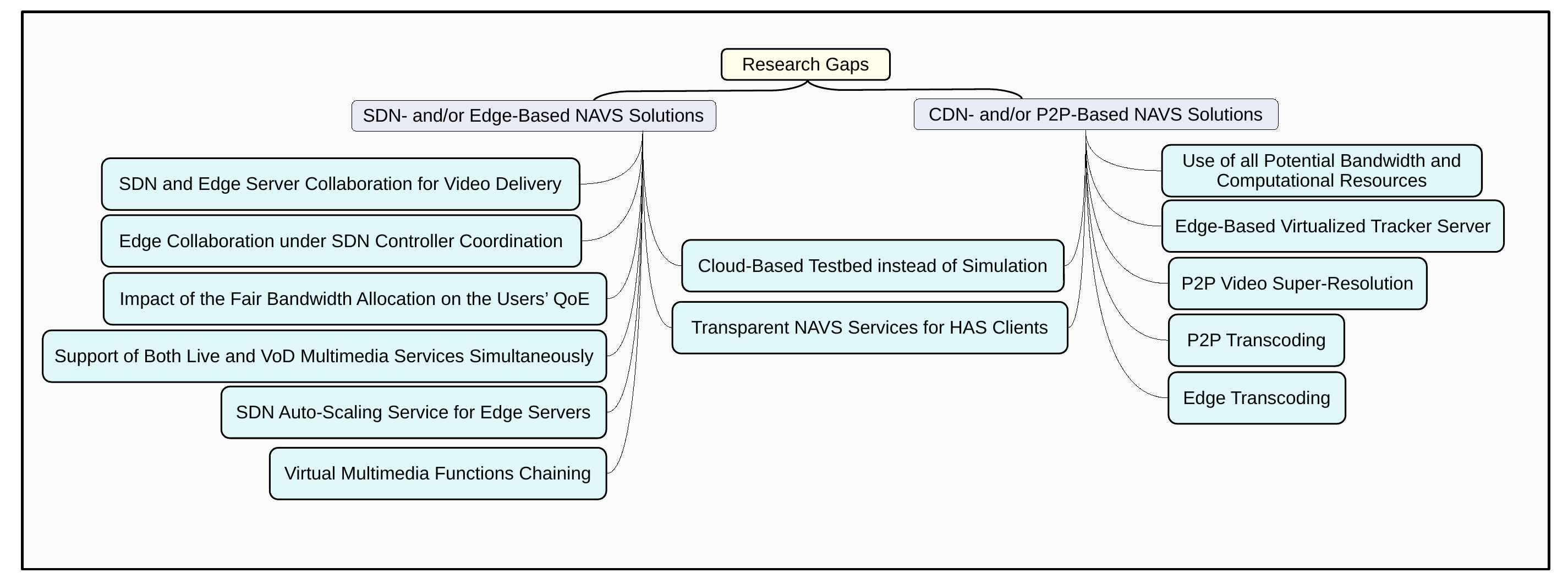}
	\caption{\rf{Research gaps in the literature.}}
         \vspace{.5cm}
	\label{research-gap}
\end{figure}

\rf{As a summary, Fig.~\ref{research-gap} depicts the research gaps in the literature that the technical chapters of this dissertation address. Moreover,} in Table~\ref{tab:Related}, we outline an analysis of the important state-of-the-art works, with a focus on their streaming type (\ie VoD or live), networking paradigms, deployed functions on edge, P2P, and/or CDN nodes, and the environment in which their evaluations took place. We also show that they are either formulated via an optimization model and/or solved by heuristic or learning-based methods. In this table, $\times$ means that a particular related paper does not include any information about the specific aspect.
%%%%%%%%%%%%%
%%%%%%%%%%%%%%%
\begin{landscape}
\begin{table}[!t]
\caption{\small Overview of related approaches; HE := Heuristic; ML := Machine Learning; OL := Online Learning; RL := Reinforcement Learning; TR := Transcoding; SR := Super-resolution.}
\label{tab:Related}
\fontsize{8.5pt}{11pt}\selectfont
\begin{tabular}{@{}lllllllll@{}}\toprule
\multicolumn{1}{c}{Paper(s)}&\multicolumn{1}{c}{Streaming Type}&\multicolumn{1}{c}{Networking Paradigm(s)}& \multicolumn{1}{c}{CDN Function(s)}& \multicolumn{1}{c}{MEC Function(s)}& \multicolumn{1}{c}{Peer Function(s)}& \multicolumn{1}{c}{Optimization}& \multicolumn{1}{c}{Heuristic}& \multicolumn{1}{c}{Evaluation}\\ \midrule
%%%%%%%%%%
\multicolumn{1}{c}{\cite{wang2020rldish}}& \multicolumn{1}{c}{Live} & \multicolumn{1}{c}{CDN,NFV} & \multicolumn{1}{c}{Caching, RL Agent} & \multicolumn{1}{c}{$\times$} & \multicolumn{1}{c}{$\times$} & \multicolumn{1}{c}{$\times$} & \multicolumn{1}{c}{RL} & \multicolumn{1}{c}{Real Testbed} \\ 
\multicolumn{1}{c}{\cite{baccour2020collaborative}}& \multicolumn{1}{c}{$\times$} & \multicolumn{1}{c}{MEC,P2P} & \multicolumn{1}{c}{$\times$} & \multicolumn{1}{c}{Caching, TR} & \multicolumn{1}{c}{Caching} & \multicolumn{1}{c}{$\checkmark$} & \multicolumn{1}{c}{HE} & \multicolumn{1}{c}{Python} \\ 
\multicolumn{1}{c}{\cite{bayhan2020edgedash}}& \multicolumn{1}{c}{$\times$} & \multicolumn{1}{c}{MEC} & \multicolumn{1}{c}{$\times$} & \multicolumn{1}{c}{Caching} & \multicolumn{1}{c}{$\times$} & \multicolumn{1}{c}{$\checkmark$}  & \multicolumn{1}{c}{HE} & \multicolumn{1}{c}{Python} \\ 
\multicolumn{1}{c}{\cite{bentaleb2016sdndash}}& \multicolumn{1}{c}{VoD} & \multicolumn{1}{c}{SDN} & \multicolumn{1}{c}{$\times$} & \multicolumn{1}{c}{$\times$} & \multicolumn{1}{c}{$\times$} & \multicolumn{1}{c}{$\checkmark$} & \multicolumn{1}{c}{HE} & \multicolumn{1}{c}{50 Nodes}\\ 
\multicolumn{1}{c}{\cite{bhat2017network}}& \multicolumn{1}{c}{$\times$} & \multicolumn{1}{c}{SDN,CDN} & \multicolumn{1}{c}{Caching} & \multicolumn{1}{c}{$\times$} & \multicolumn{1}{c}{$\times$} & \multicolumn{1}{c}{$\times$}   & \multicolumn{1}{c}{$\times$} & \multicolumn{1}{c}{Real Testbed}\\ 
\multicolumn{1}{c}{\cite{bhat2018sabr}}& \multicolumn{1}{c}{$\times$} & \multicolumn{1}{c}{SDN,CDN} & \multicolumn{1}{c}{Caching} & \multicolumn{1}{c}{$\times$} & \multicolumn{1}{c}{$\times$} & \multicolumn{1}{c}{$\times$}  & \multicolumn{1}{c}{HE} & \multicolumn{1}{c}{Real Testbed} \\ 
\multicolumn{1}{c}{\cite{borcoci2010novel}}& \multicolumn{1}{c}{$\times$} & \multicolumn{1}{c}{NFV,CDN} & \multicolumn{1}{c}{Caching} & \multicolumn{1}{c}{$\times$} & \multicolumn{1}{c}{$\times$} & \multicolumn{1}{c}{$\times$} & \multicolumn{1}{c}{$\times$} & \multicolumn{1}{c}{$\times$}  \\ 
\multicolumn{1}{c}{\cite{el2017improving}}& \multicolumn{1}{c}{$\times$} & \multicolumn{1}{c}{CDN} & \multicolumn{1}{c}{Caching} & \multicolumn{1}{c}{$\times$} & \multicolumn{1}{c}{$\times$} & \multicolumn{1}{c}{$\times$} & \multicolumn{1}{c}{HE} & \multicolumn{1}{c}{NS-3} \\ 
\multicolumn{1}{c}{\cite{erfanian2020optimizing}}& \multicolumn{1}{c}{Live} & \multicolumn{1}{c}{SDN,NFV,MEC} & \multicolumn{1}{c}{Caching} & \multicolumn{1}{c}{$\times$} & \multicolumn{1}{c}{$\times$} & \multicolumn{1}{c}{$\checkmark$} & \multicolumn{1}{c}{$\times$} & \multicolumn{1}{c}{Mininet-WiFi}  \\ 
\multicolumn{1}{c}{\cite{ge2017toward}}& \multicolumn{1}{c}{VoD} & \multicolumn{1}{c}{NFV,MEC} & \multicolumn{1}{c}{$\times$} & \multicolumn{1}{c}{Caching} & \multicolumn{1}{c}{$\times$} & \multicolumn{1}{c}{$\times$}  & \multicolumn{1}{c}{HE} & \multicolumn{1}{c}{Real Testbed} \\ 
\multicolumn{1}{c}{\cite{ge2018qoe}}& \multicolumn{1}{c}{Live} & \multicolumn{1}{c}{NFV,MEC} & \multicolumn{1}{c}{$\times$} & \multicolumn{1}{c}{Caching} & \multicolumn{1}{c}{$\times$} & \multicolumn{1}{c}{$\times$} & \multicolumn{1}{c}{HE} & \multicolumn{1}{c}{Real Testbed}  \\ 
\multicolumn{1}{c}{\cite{ha2017design}}& \multicolumn{1}{c}{Live} & \multicolumn{1}{c}{P2P,CDN} & \multicolumn{1}{c}{Caching} & \multicolumn{1}{c}{$\times$} & \multicolumn{1}{c}{Caching} & \multicolumn{1}{c}{$\times$} & \multicolumn{1}{c}{HE} & \multicolumn{1}{c}{10 Nodes}  \\ 
\multicolumn{1}{c}{\cite{jmal2017network}}& \multicolumn{1}{c}{$\times$} & \multicolumn{1}{c}{CCN} & \multicolumn{1}{c}{$\times$} & \multicolumn{1}{c}{$\times$} & \multicolumn{1}{c}{$\times$} & \multicolumn{1}{c}{$\times$} & \multicolumn{1}{c}{HE} & \multicolumn{1}{c}{ndnSIM}  \\ 
\multicolumn{1}{c}{\cite{kalan2021improving,kalan2021sdn,kalan2019optimal}}& \multicolumn{1}{c}{$\times$} & \multicolumn{1}{c}{SDN,CDN} & \multicolumn{1}{c}{Caching} & \multicolumn{1}{c}{$\times$} & \multicolumn{1}{c}{$\times$} & \multicolumn{1}{c}{$\times$} & \multicolumn{1}{c}{HE} & \multicolumn{1}{c}{Mininet}  \\ 
\multicolumn{1}{c}{\cite{kleinrouweler2016delivering}}& \multicolumn{1}{c}{$\times$} & \multicolumn{1}{c}{SDN} & \multicolumn{1}{c}{$\times$} & \multicolumn{1}{c}{$\times$} & \multicolumn{1}{c}{$\times$} & \multicolumn{1}{c}{$\times$}  & \multicolumn{1}{c}{HE} & \multicolumn{1}{c}{4 Nodes}  \\ 
\multicolumn{1}{c}{\cite{kumar2022software}}& \multicolumn{1}{c}{$\times$} & \multicolumn{1}{c}{SDN,P2P,CDN} & \multicolumn{1}{c}{Caching} & \multicolumn{1}{c}{$\times$} & \multicolumn{1}{c}{Caching} & \multicolumn{1}{c}{$\times$} & \multicolumn{1}{c}{HE} & \multicolumn{1}{c}{Mininet}  \\
%%%%%%%%%%%%%
\multicolumn{1}{c}{\cite{liotou2018qoe}}& \multicolumn{1}{c}{$\times$} & \multicolumn{1}{c}{SDN,MEC} & \multicolumn{1}{c}{$\times$} & \multicolumn{1}{c}{$\times$} & \multicolumn{1}{c}{$\times$} & \multicolumn{1}{c}{$\checkmark$}  & \multicolumn{1}{c}{HE} & \multicolumn{1}{c}{ViennaSim} \\
\multicolumn{1}{c}{\cite{ma2021locality}}& \multicolumn{1}{c}{Live} & \multicolumn{1}{c}{P2P,CDN} & \multicolumn{1}{c}{Caching} & \multicolumn{1}{c}{$\times$} & \multicolumn{1}{c}{Caching} & \multicolumn{1}{c}{$\times$} & \multicolumn{1}{c}{ML} & \multicolumn{1}{c}{Trace-Driven}  \\
\multicolumn{1}{c}{\cite{mehrabi2018edge}}& \multicolumn{1}{c}{$\times$} & \multicolumn{1}{c}{MEC} & \multicolumn{1}{c}{$\times$} & \multicolumn{1}{c}{$\times$} & \multicolumn{1}{c}{Caching} & \multicolumn{1}{c}{$\checkmark$} & \multicolumn{1}{c}{HE} & \multicolumn{1}{c}{SimuLTE}  \\
\multicolumn{1}{c}{\cite{mukerjee2015practical}}& \multicolumn{1}{c}{Live} & \multicolumn{1}{c}{CDN,SDN} & \multicolumn{1}{c}{Caching} & \multicolumn{1}{c}{$\times$} & \multicolumn{1}{c}{$\times$} & \multicolumn{1}{c}{$\times$} & \multicolumn{1}{c}{HE} & \multicolumn{1}{c}{10 Nodes}  \\
\multicolumn{1}{c}{\cite{nacakli2020controlling}}& \multicolumn{1}{c}{Live} & \multicolumn{1}{c}{SDN,MEC,P2P,CDN} & \multicolumn{1}{c}{Caching} & \multicolumn{1}{c}{$\times$} & \multicolumn{1}{c}{Caching} & \multicolumn{1}{c}{$\times$} & \multicolumn{1}{c}{HE} & \multicolumn{1}{c}{Mininet}  \\
%%%%%%%%%%%
\multicolumn{1}{c}{\cite{ozcelik2019chunk}}& \multicolumn{1}{c}{VoD} & \multicolumn{1}{c}{SDN} & \multicolumn{1}{c}{$\times$} & \multicolumn{1}{c}{$\times$} & \multicolumn{1}{c}{$\times$} & \multicolumn{1}{c}{$\times$} & \multicolumn{1}{c}{HE} & \multicolumn{1}{c}{Mininet}  \\
\multicolumn{1}{c}{\cite{tran2018adaptive}}& \multicolumn{1}{c}{$\times$} & \multicolumn{1}{c}{MEC} & \multicolumn{1}{c}{$\times$} & \multicolumn{1}{c}{Caching, TR} & \multicolumn{1}{c}{$\times$} & \multicolumn{1}{c}{$\checkmark$}  & \multicolumn{1}{c}{HE} & \multicolumn{1}{c}{Numerical} \\
%%%%%%%%%%%%%
%%
\multicolumn{1}{c}{\cite{xie2008p4p}}& \multicolumn{1}{c}{$\times$} & \multicolumn{1}{c}{P2P} & \multicolumn{1}{c}{$\times$} & \multicolumn{1}{c}{$\times$} & \multicolumn{1}{c}{Caching} & \multicolumn{1}{c}{$\checkmark$}  & \multicolumn{1}{c}{HE} & \multicolumn{1}{c}{PlanetLab} \\
\multicolumn{1}{c}{\cite{yin2009design}}& \multicolumn{1}{c}{Live} & \multicolumn{1}{c}{P2P,CDN} & \multicolumn{1}{c}{Caching} & \multicolumn{1}{c}{$\times$} & \multicolumn{1}{c}{Caching} & \multicolumn{1}{c}{$\times$} & \multicolumn{1}{c}{HE} & \multicolumn{1}{c}{Real Testbed}  \\
\multicolumn{1}{c}{\cite{yousef2020enabling}}& \multicolumn{1}{c}{$\times$} & \multicolumn{1}{c}{P2N,CDN} & \multicolumn{1}{c}{Caching} & \multicolumn{1}{c}{$\times$} & \multicolumn{1}{c}{Caching, Prefetching} & \multicolumn{1}{c}{$\times$}  & \multicolumn{1}{c}{HE} & \multicolumn{1}{c}{Matlab} \\
\multicolumn{1}{c}{\cite{zhang2021video}}& \multicolumn{1}{c}{$\times$} & \multicolumn{1}{c}{MEC} & \multicolumn{1}{c}{$\times$} & \multicolumn{1}{c}{Caching, SR} & \multicolumn{1}{c}{$\times$} & \multicolumn{1}{c}{$\checkmark$} & \multicolumn{1}{c}{HE} & \multicolumn{1}{c}{3 Nodes}  \\
%%%%%%%%%%%%%Contributions%%%%%%%%%%%
\multicolumn{1}{c}{\rf{\texttt{ES-HAS}}~\cite{Farahani2021eshas}}& \multicolumn{1}{c}{VoD} & \multicolumn{1}{c}{SDN, NFV, MEC, CDN} & \multicolumn{1}{c}{Caching} & \multicolumn{1}{c}{Caching} & \multicolumn{1}{c}{$\times$} & \multicolumn{1}{c}{$\checkmark$} & \multicolumn{1}{c}{$\times$} & \multicolumn{1}{c}{Real Testbed}  \\
\multicolumn{1}{c}{\rf{\texttt{CSDN}}~\cite{Farahani2021csdn}}& \multicolumn{1}{c}{Live, VoD} & \multicolumn{1}{c}{SDN, NFV, MEC, CDN} & \multicolumn{1}{c}{Caching} & \multicolumn{1}{c}{Caching, TR} & \multicolumn{1}{c}{$\times$} & \multicolumn{1}{c}{$\checkmark$} & \multicolumn{1}{c}{$\times$} & \multicolumn{1}{c}{Real Testbed}  \\
\multicolumn{1}{c}{\rf{\texttt{SARENA}}~\cite{farahani2023sarena}}& \multicolumn{1}{c}{Live, VoD} & \multicolumn{1}{c}{SDN, NFV, SFC, MEC, CDN} & \multicolumn{1}{c}{Caching} & \multicolumn{1}{c}{Caching, TR} & \multicolumn{1}{c}{$\times$} & \multicolumn{1}{c}{$\checkmark$} & \multicolumn{1}{c}{HE} & \multicolumn{1}{c}{Real Testbed}  \\
\multicolumn{1}{c}{\rf{\texttt{LEADER}}~\cite{farahani2022leader}}& \multicolumn{1}{c}{Live, VoD} & \multicolumn{1}{c}{SDN, NFV, MEC, CDN} & \multicolumn{1}{c}{Caching} & \multicolumn{1}{c}{Caching, TR} & \multicolumn{1}{c}{$\times$} & \multicolumn{1}{c}{$\checkmark$} & \multicolumn{1}{c}{HE} & \multicolumn{1}{c}{Real Testbed}  \\
\multicolumn{1}{c}{\rf{\texttt{ARARAT}}~\cite{farahani2022ararat}}& \multicolumn{1}{c}{Live, VoD} & \multicolumn{1}{c}{SDN, NFV, MEC, CDN} & \multicolumn{1}{c}{Caching} & \multicolumn{1}{c}{Caching, TR} & \multicolumn{1}{c}{$\times$} & \multicolumn{1}{c}{$\checkmark$} & \multicolumn{1}{c}{HE} & \multicolumn{1}{c}{Real Testbed}  \\
\multicolumn{1}{c}{\rf{\texttt{RICHTER}}~\cite{farahani2022RICHTER}}& \multicolumn{1}{c}{Live} & \multicolumn{1}{c}{MEC, NFV, P2P, CDN} & \multicolumn{1}{c}{Caching} & \multicolumn{1}{c}{Caching, TR} & \multicolumn{1}{c}{Caching, TR} & \multicolumn{1}{c}{$\checkmark$} & \multicolumn{1}{c}{OL} & \multicolumn{1}{c}{Real Testbed}  \\
\multicolumn{1}{c}{\rf{\texttt{ALIVE}}~\cite{farahani2023alive}}& \multicolumn{1}{c}{Live} & \multicolumn{1}{c}{MEC, NFV, P2P, CDN} & \multicolumn{1}{c}{Caching} & \multicolumn{1}{c}{Caching, TR} & \multicolumn{1}{c}{Caching, TR, SR} & \multicolumn{1}{c}{$\checkmark$} & \multicolumn{1}{c}{HE} & \multicolumn{1}{c}{Real Testbed}  \\

\bottomrule
\end{tabular}
\end{table}
\end{landscape}
%%%%%%%%%%%%%%

%% file: Chapters/Chapter3/3-1-Intro.tex
% \singlespacing
\chapter{Edge- and SDN-Assisted Frameworks for HAS}\label{chap:EdgeSDN}
%************************************************
\doublespacing
\vspace{-1cm}
This chapter leverages the SDN and NFV paradigms to propose two edge- and  SDN-assisted frameworks for HTTP adaptive video streaming, named \texttt{ES-HAS} and \texttt{CSDN} (\rf{\texttt{ES-HAS} extended by the transcoding capability), designed particularly for VoD and live streaming, respectively}. We introduce VNF components called \textit{Virtual Reverse Proxies} (VRPs) at the edge of \rf{single domain SDN-based networks}. These systems (via the SDN controller) provide VRPs with information on network conditions and available video sources (cache servers). In addition, VRPs collect clients' requests for video segments and aggregate them in time slots. Using this information, VRPs run optimization models as server/segment selection policies to help clients get their requested segment quality levels from cache servers with the shortest fetching time or receive from cache servers better replacement quality levels with minimum deviation from the original requested quality levels or, in \texttt{CSDN}, use transcoding functions to transcode qualities from a higher available quality level. We implement the proposed frameworks and their modules on a cloud-based large-scale testbed and conduct various experiments in different scenarios. We also evaluate the MILP models' behavior and compare the results with other state-of-the-art approaches. Experimental results demonstrate that, on average, \texttt{ES-HAS} and \texttt{CSDN} outperform their competitors in terms of common QoE and network KPIs.\\
\noindent
The first contribution, \ie ``\texttt{ES-HAS}: An Edge- and SDN-Assisted Framework for HTTP Adaptive Video Streaming'' was presented at the 31st Workshop on Network and Operating Systems Support for Digital Audio and Video (NOSSDAV)~\cite{Farahani2021eshas}. The second one, \ie ``\texttt{CSDN}: CDN-Aware QoE Optimization in SDN-Assisted HTTP Adaptive Video Streaming'', was presented at the 46th Conference on Local Computer Networks (LCN)~\cite{Farahani2021csdn}.

%% file: Chapters/Chapter3/3-2-ESHAS.tex
\doublespacing
\section{ES-HAS Framework}
\label{chap:EdgeSDN:ES-HAS}
This section presents the details of the \texttt{ES-HAS} framework~\cite{Farahani2021eshas}, which adopts and extends the core idea of SABR~\cite{bhat2018sabr} to introduce an edge- and SDN-assisted solution for HAS clients. Section~\ref{sec:EdgeSDN:ES-HAS:ESHAS-Motivation} motivates the \texttt{ES-HAS} method through an example before elaborating on the details of the proposed architecture and optimization model in Section~\ref{sec:EdgeSDN:ES-HAS:ESHAS-SystemModel}. The \texttt{ES-HAS} evaluation setup and results are discussed in Section~\ref{sec:EdgeSDN:ES-HAS:ES-HAS-PerformanceEvaluation}.
%%%%%%%%%%%%%%%%%%%%%%%%%%%%%%%%%%%%%%%%%%%%%%%%%%%%%%%%%%%%%%%%%%%%%%%%%%%%%%%%%%%%%%%%%%%%%%%%%%%%%%%%%%%%%%%%%%%%%%%%%%%%%%%%%%%%%%%%%%%%%%%%%%%%%%%%%%%%%%%%%%%%%%%%%%%
\subsection{ES-HAS Motivating Example and Problem Statement}
\label{sec:EdgeSDN:ES-HAS:ESHAS-Motivation}
\begin{figure}[!t]
	\centering
	\includegraphics[width=1\linewidth]{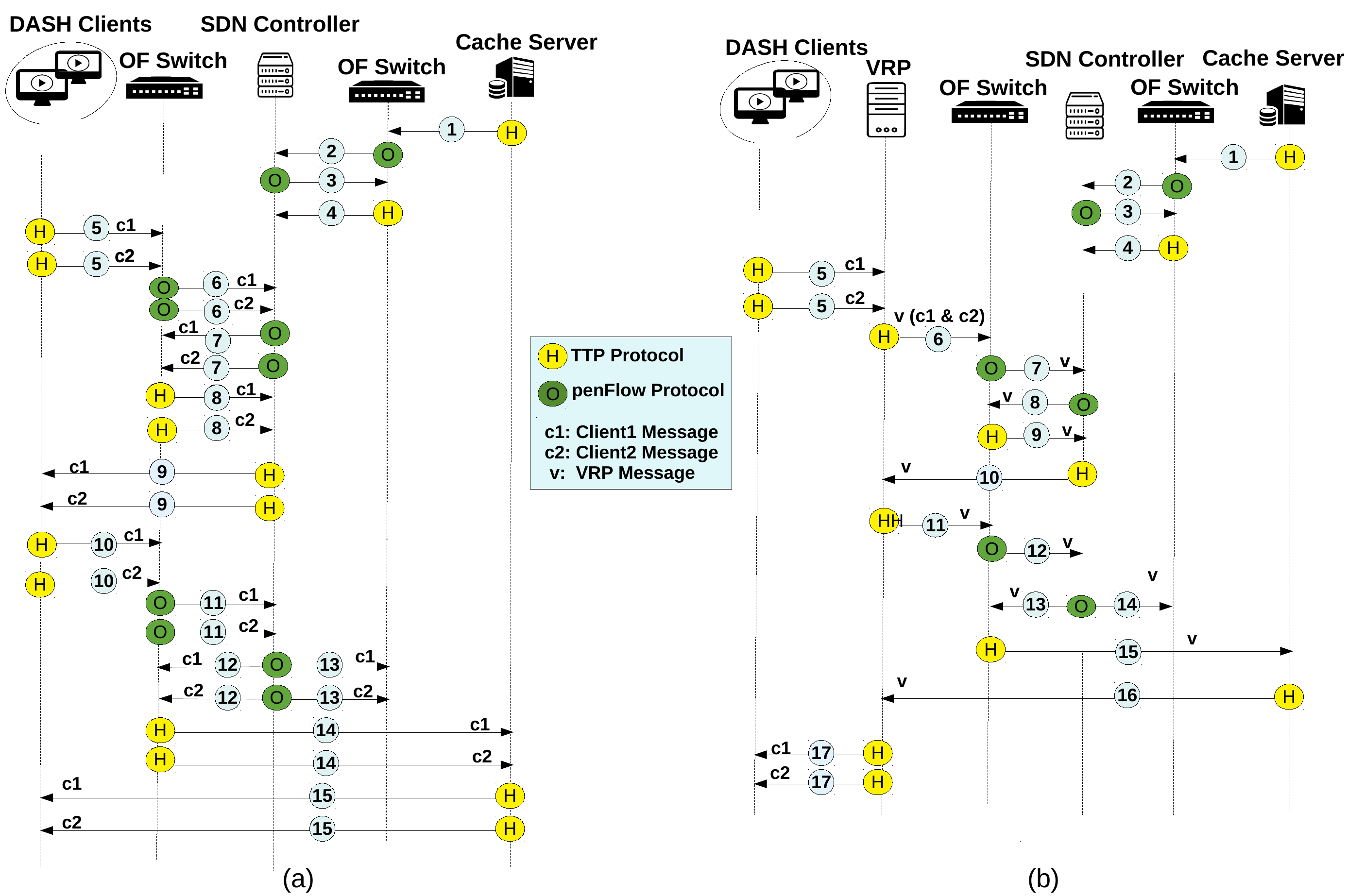}
	\caption{\small Exchanged messages in (a) SABR and (b) ES-HAS.}
 \vspace{.5cm}
	\label{es-has-Motivation}
\end{figure}

\begin{figure}[!t]
	\centering
	\includegraphics[width=1\linewidth]{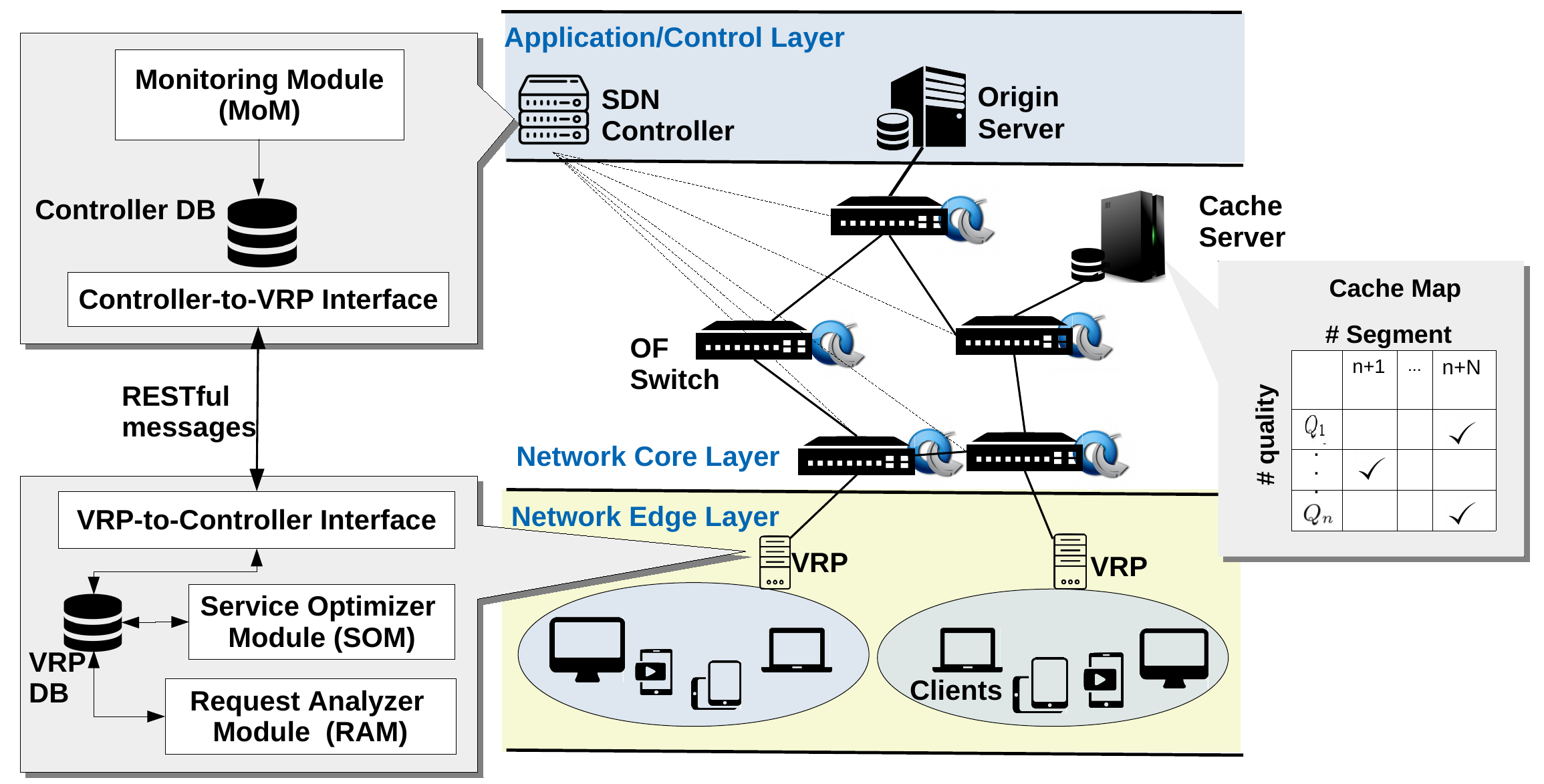}
	\caption{\small Proposed ES-HAS architecture.}
 \vspace{.5cm}
	\label{ESHAS-Arch}
\end{figure}

We present our main motivation by means of the following example. Concerning the components involved in the example, we refer to the \texttt{ES-HAS} architecture depicted in Fig.~\ref{ESHAS-Arch} and the fact that \texttt{ES-HAS} adopts and extends the core idea of the SABR framework~\cite{bhat2018sabr}. We consider a simplistic scenario with only one cache server, an SDN controller, several OpenFlow (OF) switches, one virtual reverse proxy (VRP) server, and two clients. We assume clients request their next video segments in an SABR-enabled system~\cite{bhat2018sabr} (Fig.~\ref{es-has-Motivation}(a)) and an \texttt{ES-HAS} system (Fig.~\ref{es-has-Motivation}(b)). As illustrated, the SDN controller should initially receive \textit{cache map} messages from cache servers (step $1$). Each cache map indicates the presence of representations (quality levels) of requested video segments. The structure of a cache map is illustrated in Fig.~\ref{ESHAS-Arch}. Whenever an OF CDN-side switch cannot find a matching rule for ($1$), it sends a \textit{packet-in} message ($2$) through the OF protocol to the SDN controller. The SDN controller replies to the packet-in ($3$) and installs a path for the cache server. Finally, the cache map will be received by the SDN controller ($4$).  
%%%%%%%%
%%%%%%%%%%

In an SABR-based system (Fig.~\ref{es-has-Motivation}(a)), the clients send requests ($5$) to the SDN controller for acquiring cache map and network status information. An OF client-side switch uses OF packet-in messages ($6$) for these non-matching HTTP requests. After obtaining the replies ($7$), the OF switch forwards these requests to the SDN controller ($8$). Assuming that the clients support SABR, after receiving the requested information from the SDN controller ($9$), they determine a joint cache map, including all cache servers for the requested segments. The desired segments are requested by the clients ($10-11$). Then, the SDN controller installs the corresponding routes between the clients and the determined cache servers ($12-13$). Consequently, the requested segments ($14$) are transferred to the clients ($15$). It is clear that when the number of DASH clients increases, the number of exchanged messages to/from the SDN controller (via OF and HTTP) will increase proportionally. Hence, system efficiency will decrease gradually.

Our proposed framework employs a VRP at the edge of the network to overcome the aforementioned problem. A VRP works in time slots as follows. As shown in Fig.~\ref{es-has-Motivation}(b), the clients send requests ($5$) to the VRP for the desired segments' qualities, and the VRP collects these received requests in each time slot. The VRP plays the role of a gateway for the client to the network and vice versa. Therefore, the VRP requests the cache map and network status information from the SDN controller for all collected requests ($6-10$). Using aggregate messages in these steps decreases the number of exchanged messages to/from the SDN controller (via OF and HTTP) as compared to SABR. After receiving the demanded information from the SDN controller ($10$), the VRP runs an optimization program to determine the optimal cache servers for the gathered clients' requests. For the sake of simplicity, let us assume that the requested segment quality levels are available on the cache servers. Then they are requested from the cache servers and transmitted to the VRP ($11-16$). A second reduction of exchanged messages (as compared to SABR) may occur in this stage since the VRP does not forward identical requests by clients redundantly to the SDN controller and cache servers. Finally, the fetched segments are transferred to the clients ($17$). The details of the \texttt{ES-HAS} framework will be discussed in the next sections. 

The number of communicated messages to/from the SDN controller for both systems ({Fig.~\ref{es-has-Motivation}}) is shown in Table~\ref{tab:table-mot}. In real scenarios with a large video dataset, this amount of exchanged messages could overload the SDN controller in an SABR-enabled system. 
In our proposed structure, we introduce VRPs at the edge of the system for collecting client-side requests, which increases costs and imposes additional delay to the system. However, the number of messages to/from the SDN controller, the load on the SDN controller, and the network bandwidth consumption are significantly reduced. With an increased number of clients, we can enlarge the VRP's resources or, alternatively, add another VRP to manage more clients efficiently.
%%%%%%%%
\begin{table}[!t]
    \centering
    \caption {\small Number of messages to/from SDN controller.}
    \label{tab:table-mot}
    \begin{tabular}[c]{ |c| c|c|c|} 
    \hline
         \textbf{Arch.}& \textbf{Number of $\ldots$ OF msgs.}& \textbf{$\ldots$ HTTP msgs.} \\\hline 
         \textbf{SABR} &  $2N_{cache}+5N_{client}$ & $N_{cache}+2N_{client}$ \\\hline
         \textbf{ES-HAS} &$2N_{cache}+2N_{VRP}+3N_{cache}N_{VRP}$ &$N_{cache}+2N_{VRP}$ \\
    \hline
    \end{tabular}
    \vspace{.5cm}
    \end{table}
%%%%%%%%
%%%%%%%%
%%%%%%%%%%%%%%%%%%%%%%%%%%%%%%%%%%%%%%%%%%%%%%%%%%%%%%%%%%%%%%%%%%%%%%%%%%%%%%%%%%%%%%%%%%%%%%%%%%%%%%%%%%%%%%%%%%%%%%%%%%%%%%%%%%%%%%%%%%%%%%%%%%%%%%%%%%%%%%%%%%%%%%%%%%%%%%%%%%%%%%%%%%%
\subsection{ES-HAS System Design}
\label{sec:EdgeSDN:ES-HAS:ESHAS-SystemModel} 
In this section, we first present the \texttt{ES-HAS} architecture and its main components and then explain the \texttt{ES-HAS} problem formulation.
\subsubsection{ES-HAS Architecture}
\label{sec:ESHAS-SystemModel:Arch} 
\texttt{ES-HAS} includes three main layers: \textit{(i)} Application/Control, \textit{(ii)} Network Core, and \textit{(iii)} Network Edge (Fig.~\ref{ESHAS-Arch}). On the application/control layer, the SDN controller periodically monitors available bandwidth and the cache servers' occupancy information (cache maps) and stores them in its database. Thus, we define a \textit{Monitoring Module} (MoM) as the controller's main application module to collect the aforementioned information from the OpenFlow switches. Therefore, the database in the controller has accurate information about cache servers and paths' available bandwidths and serves the VRPs' requests through the \textit{Controller-to-VRP Interface}.
The network core layer consists of the OpenFlow switches connected to the SDN controller, CDN components, including an origin server, and multiple cache servers.
In the network edge layer, we employ several VRPs, which are equipped with three main modules as follows: \textit{(i) Request Analyzer Module} (RAM), \textit{(ii) Service Optimizer Module} (SOM), and \textit{(iii) VRP-to-Controller Interface}.

\begin{figure}[!t]
	\centering
	\includegraphics[width=.8\linewidth]{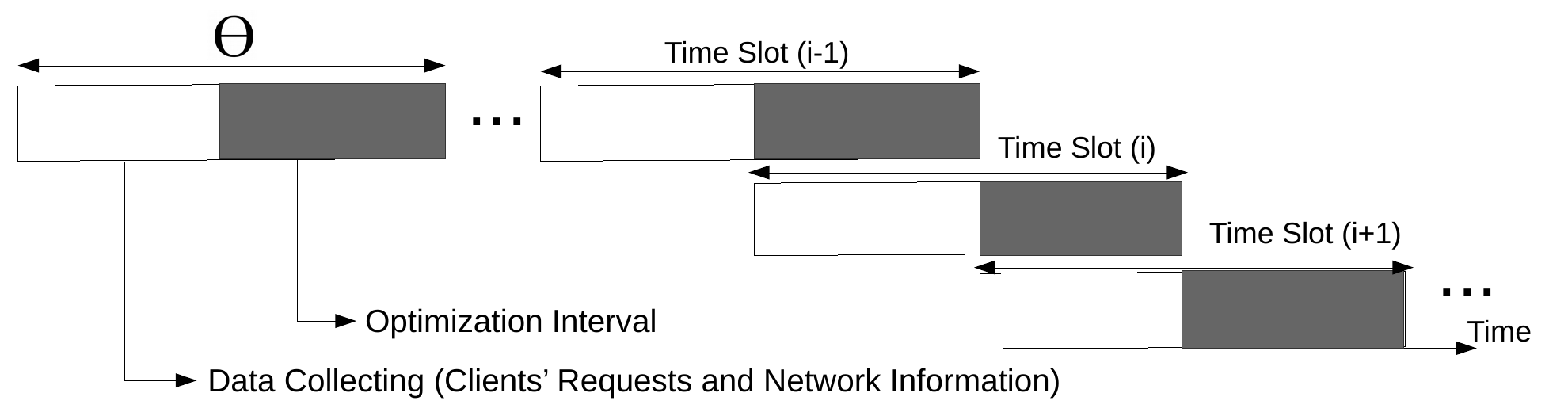}
	\caption{\small Proposed time slot structure.}
        \vspace{.5cm}
	\label{ESHAS-TS}
\end{figure}

Before describing the modules, it is noted that VRPs operate in time slots with an equal duration of~$\theta$. As shown in Fig.~\ref{ESHAS-TS}, each time slot consists of two intervals: \textit{(i)} Data Collecting and \textit{(ii)} Optimization interval. In the first interval, users' requests are gathered and aggregated by a VRP's RAM. To prevent sending identical requests (issued by multiple clients in a given time slot), RAM identifies them and considers only one request per segment. Moreover, using RESTful messages, the VRP periodically retrieves the required information (cache maps plus available bandwidths between each cache server and the VRP) from the SDN controller and stores them in its DB. Then, during the first interval, the VRP can fetch them from its database.
The SOM is executed by the VRP in the second interval to serve clients' requests optimally. 
%%%%%
%%%%%
Note that, according to the SOM's results, the data transmission will start in the first interval of the next time slot. Actually, for each request, using the optimization model presented later in this section, the SOM selects an appropriate cache server that hosts the requested quality. However, when the requested quality is not available in a cache server, the SOM either determines an optimal replacement quality level on a cache or fetches the originally requested quality from the origin server.

In this work, we employ the concept that a \textit{replacement quality} may be delivered to the client rather than the requested quality level of a segment, based on Consumer Technology Association's \textit{Common Media Client Data} (CMCD) standard~\cite{cta-wave-cmcd}. This standard defines the information that a media player can communicate to (within media object requests) and have processed and considered by CDNs. One piece of information is the ``requested maximum throughput that the client considers sufficient for delivery of a content asset''. In our system, we interpret this as conveying the information that a client will accept potential replacement (but only better) quality levels in lieu of the actually requested quality of a segment, up to a certain limit set by the client. 
For the sake of simplification, we assume that each DASH client can request just one segment during each time slot. However, it is possible to consider multiple requests from each client in each time slot without fundamental changes in the proposed model. Note that we will discuss how to determine the time slot duration in Section~\ref{sec:EdgeSDN:ES-HAS:ESHAS-Res1}.
\raggedbottom
%%%%%%
\begin{table}[!b]
\begin{center}
\caption {\small ES-HAS Notation.}
\label{tab:tablen}
{
\begin{tabular}{llllll}
\cline{1-2}
\multicolumn{2}{|c|}{\textbf{Input Parameters}}                                                       
&  &  &  &  \\ \cline{1-2}                                                                                      
\multicolumn{1}{|l|}{\begin{tabular}[c]{@{}l@{}}
$\mathcal{S}$\\ 
$\mathcal{C}$\\ 
$\mathcal{A}$\\ 
$a^{c,s}_q$\\ 
$v^{c,s}_q$\\  
$\mathcal{R}$\\ 
$R_s$\\ 
$i_c$\\ 
$m$\\ 
$\mathcal{K}_c$\\ 
\\$\delta^c_q$\\ 
$\pi^c_q$\\ 
$\theta$\\
\end{tabular}} & 
\multicolumn{1}{l|}{\begin{tabular}[c]{@{}l@{}}
Set of cache servers and origin server, $s\in \mathcal{S}$\\ 
Set of clients, $c\in \mathcal{C}$\\ 
Set of available qualities in $\mathcal{S}$\\
$a^{c,s}_q=1$ if quality $q$ requested by client $c$ is available in server $s$, $a^{c,s}_q=0$ otherwise\\
$v^{c,s}_q=1$ if the requested quality $q$ is available in any cache server, $v^{c,s}_q=0$ otherwise \\ 
Set of available bandwidth values\\
Available bandwidth between VRP and server $s$ \\ 
Quality level requested by client $c$\\ 
Integer number to limit the range of potential replacement quality levels for $i_c$\\ 
Set of eligible quality levels for a quality  requested by client $c$, where \\$\mathcal{K}_c=\{i_c,{i_c}+1,..., min[{i_c}+m,q_{max}]\}$\\ 
Size of the segment of quality level $q$ delivered to client $c$\\ 
Bitrate of the quality level $q\in \mathcal{K}_c$  \\ 
Time slot duration
\end{tabular}} 
&  &  &  &  \\ \cline{1-2}
\multicolumn{2}{|c|}{\textbf{Variables}}                                                                                                                                 &  &  &  &  \\ \cline{1-2}                   \multicolumn{1}{|l|}{\begin{tabular}[c]{@{}l@{}}$B^{c,s}_q$\\ $T^{c,s}_q$\\ $F_c$\\ $Q_c$\end{tabular}}                                                                & \multicolumn{1}{l|}{\begin{tabular}[c]{@{}l@{}}$B^{c,s}_q$=1 if quality $q$ requested by $c$ is served from server $s$\\ Required time to fetch quality $q$ for client $c$ from server $s$\\ Deviation of quality level to serve client $c$ w.r.t. $i_c$\\ Selected quality bitrate to serve client $c$\end{tabular}}                                                    &  &  &  &  \\ \cline{1-2}                                                                                                                                     &                                                                                &  &  &  &                                              
\end{tabular}}
\end{center}
\end{table}
%%%%%
\clearpage
\subsubsection{ES-HAS Optimization Problem Formulation}
\label{sec:EdgeSDN:ES-HAS:ESHAS-SystemModel:Opt} 
Let set $\mathcal{A}$ denote the cache map (\ie availability of segments/bitrates) that a VRP receives from the SDN controller, where $a^{c,s}_q=1$ means that the quality level $q$ requested by client $c \in \mathcal{C}$ is available on the server $s\in \mathcal{S}$ (see Table~\ref{tab:tablen} for notations). 
Moreover, let a VRP's database host set $\mathcal{R}$ containing available bandwidth values between the cache servers and the VRP. In the SOM module, we introduce a mixed-integer linear programming model (MILP) which tries to minimize the segment fetching time if the requested segment quality is available in at least one cache server; otherwise, for each non-cached segment quality, it determines whether to serve the client's request by a replacement quality from a cache server or by the original requested quality from the origin server. The proposed MILP model finds the optimal solution by minimizing the fetching time and the segments' quality level deviations while maximizing the selected quality bitrates. We derive the following constraints that must be satisfied to achieve an optimal solution.  

Let us define $i_c$ as the quality level requested by client $c\in\mathcal{C}$. We also define $\mathcal{K}_c=\{i_c,{i_c}+1,...,min[{i_c}+m,q^c_{max}]\}$ as the set of eligible (potentially, replacement) quality levels for the segment requested by client $c$, where $m$ and $q^c_{max}$ denote the maximum deviation from $i_c$ and the maximum quality level of the segment requested by $c$, respectively. In each optimization interval, we select only one server to serve client $c$. For this purpose, we introduce binary variable $B^{c,s}_q$, where $B^{c,s}_q=1$ indicates that quality level $q$ must be served to client $c$ by cache server $s$. As mentioned earlier, if $i_c$ is available in any cache server, we should force the model to select one cache server to serve the client's request in the original quality by setting the following constraint:
\begin{flalign}
&\sum_{s\in \mathcal{S}\setminus Origin} B^{c,s}_q~.~{a}^{c,s}_q =v^{c,s}_q,&&\forall c \in \mathcal{C}, q=i_c \label{ESHAS:eq:0}
\end{flalign}
where $v^{c,s}_q=1$ if the requested quality $q=i_c$ is available in any cache server; otherwise $v^{c,s}_q=0$.
In the case of a cache miss, \ie when $i_c$ is unavailable in any cache server, the VRP can fetch it from the origin server or use other quality levels available in cache servers. Thus, the following constraint must be satisfied:
%constraint2
\begin{flalign}
&\sum_{s\in \mathcal{S}}\sum_{q\in \mathcal{K}_c} B^{c,s}_q~.~{a}^{c,s}_q=1, &&\hspace{.4cm} \forall c \in \mathcal{C}\label{ESHAS:eq:1}
\end{flalign}
The required time to fetch the quality $q$ for client $c$ from server $s$, denoted by $T^{c,s}_q$, is determined by the following constraint: 
%constraint3
\begin{flalign}
&\delta^c_q~.~B^{c,s}_q\leq T^{c,s}_q~.~R_s,&&\forall c\in \mathcal{C},q\in \mathcal{K}_c,s\in \mathcal{S}
\label{ESHAS:eq:2}
\end{flalign}
where $\delta^c_q$ is the size of the quality $q$ requested by client $c$ and $R_s$ is the available bandwidth between the VRP and cache server $s$. 
We define $Q_c$ as the selected quality bitrate to serve client $c\in\mathcal{C}$ according to the following constraint:
%constraint4
\begin{flalign}
&\sum_{s\in \mathcal{S}}\sum_{q\in \mathcal{K}_c} B^{c,s}_q~.~\pi^c_q \geq Q_c&& \forall c\in \mathcal{C}\label{ESHAS:eq:3}
\end{flalign}
where $\pi^c_q$ is the bitrate of the selected quality level $q$ for serving client $c$. To determine the quality deviation when the requested quality $i_c$ is not available in $\mathcal{K}_c$ and a replacement quality has to be found, we introduce the following constraint:
%constraint5
\begin{flalign} 
&\sum_{s\in \mathcal{S}}\sum_{q\in \mathcal\mathcal{K}_c} | i_c-(B^{c,s}_q~.~q) |\leq F_c&& \forall c\in \mathcal{C}\label{ESHAS:eq:4}\end{flalign}

Finally, the proposed model is formulated as follows:
%objective
\begin{flalign}
\textit{Minimize}&\hspace{.3cm}\alpha_1\sum_{c\in \mathcal{C}}\sum_{s\in \mathcal{S}}\sum_{q\in \mathcal{K}_c}\frac{T^{c,s}_q}{T^*}+\sum_{c\in\mathcal{C}}( \alpha_2\frac{F_c}{F^*} - \alpha_3\frac{Q_c}{Q^*})\label{ESHAS:eq:5}\\
   s.t.&\hspace{.5cm}\text{Constraints}\hspace{.5cm} (\ref{ESHAS:eq:0})-(\ref{ESHAS:eq:4})&&\nonumber\\
   vars.&\hspace{.5cm} T^{c,s}_q,F_c,Q_c \geq 0, B^{c,s}_q\in\{0,1\}\nonumber
\end{flalign}
where $T^*$, $F^*$, and $Q^*$ are the maximum values for the fetching time, the quality level deviation, and the quality bitrate for client $c$, respectively. The SOM runs the above model for all clients' requested qualities in a time-slotted manner to minimize the fetching time and quality level deviations as well as to maximize the qualities delivered to the clients. Furthermore, in the objective function~(\ref{ESHAS:eq:5}), we set priorities for $T^{c,s}_q\text{, }F_c\text{, and } Q_c$ by adjusting the wightings $\alpha_1$, $\alpha_2$, and $\alpha_3$, respectively. The weighting values are tuned empirically by considering the network conditions or application policies. \rf{Section~\ref{sec:EdgeSDN:ES-HAS:ESHAS-Res1} discusses the impact of these weighting coefficients on the model's objectives. }
%%%%%%%%%%%%%%%%%%%%%%%%%%%%%%%%%%%%%%%%%%%%%%%%%%%%%%%%%%%%%%%%%%%%%%%%%%%%%%%%%%%%%%%%%%%%%%%%%%%%%%%%%%%%%%%%%%%%%%%%%%%%%%%%%%%%%%%%%%%%%%%%%%%%%%%%%%%%%%%%%%%%%%%%%%%%%%%%%%%%%%%%%%%%%%%%
\subsection{ES-HAS Performance Evaluation}
\label{sec:EdgeSDN:ES-HAS:ES-HAS-PerformanceEvaluation}
In this section, we evaluate the performance of \texttt{ES-HAS} compared to SABR~\cite{bhat2017network} and pure client-based approaches. 
\subsubsection{Evaluation Setup}
\label{sec:Evaluation Setup}
\begin{figure}[!t]
	\centering
	\includegraphics[width=.9\linewidth]{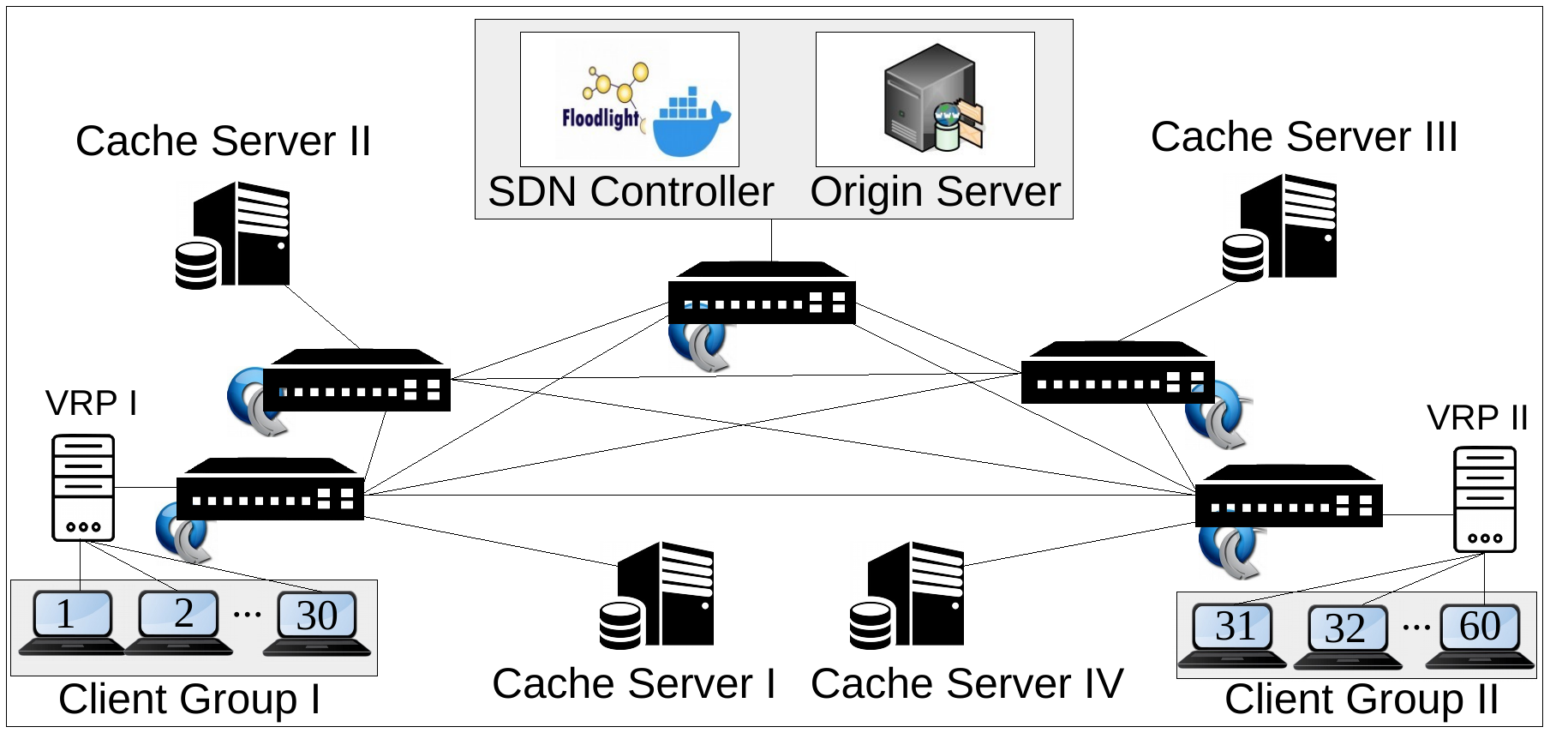}
	\caption{\small ES-HAS evaluation testbed.}
        \vspace{.5cm}
	\label{ESHAS-Testbed}
\end{figure}

Our testbed consists of 72 nodes running Ubuntu 18.04 LTS inside Xen virtual machines. The proposed network topology is built in the CloudLab~\cite{ricci2014introducing} environment. As shown in Fig.~\ref{ESHAS-Testbed}, it includes five OpenFlow (OF) switches, 60 AStream DASH players~\cite{juluri2015sara,AStream}, two VRP servers with the modules described in Section~\ref{sec:EdgeSDN:ES-HAS:ESHAS-SystemModel}. Moreover, we employ four cache servers and one extra server that jointly hosts an origin server and a dockerized SDN controller. The bandwidth values in different paths between each VRP and the cache servers are set to 100, 80, 60, and 40 Mbps, respectively, which explicitly gives higher priority to download segments from the local cache servers. Cache servers I and IV are local cache servers for client groups I and II, respectively, with 100 Mbps bandwidth. The bandwidth values to the origin server are set to 20 Mbps for both VRP servers. Apache~\cite{Apache} and MongoDB~\cite{Mongodb} with supporting RESTful APIs for cache map exchange are installed on all cache servers. 

Moreover, the Least Recently Used (LRU) is considered in all cache servers as the cache replacement policy. The policy on a cache miss is that the requested quality will be fetched from the origin server only to the related local cache server, \ie cache server I for client group I and cache server IV for client group II. Floodlight~\cite{Floodlight} is used as an SDN controller. Among other tasks, it monitors the network to find paths’ available bandwidth (in one-second intervals) and thus assigns paths with the highest available bandwidth between a VRP server and each cache server. For the sake of simplicity, we assume that all clients have already joined the network. Ten test video sequences~\cite{lederer2012dynamic} with 300 seconds durations are used in our experiments. These video sequences comprise two-second segments in five representations ({89k, 0.262M, 0.791M, 2.4M, 4.2M}bps). 60\% of the video sequences’ segments are stored in each cache server randomly. All clients run simultaneously in all scenarios. Each client requests one video where $video_1$ is streamed to clients ({1,11,21,31,41,51}), $video_2$ is streamed to clients ({2,12,22,32,42,52}), and so on.
The time slot duration is set to 52 milliseconds in all experiments. (We will discuss how to obtain the time slot duration in the next subsection.) Two different ABR algorithms, BOLA~\cite{spiteri2016bola} and SQUAD~\cite{wang2016squad}, representing buffer-based and hybrid approaches, are used in all experiments. Python and the PuLP library~\cite{PuLP} are employed to implement and solve the proposed MILP model.
%%%%%%%%%%%%%%%%%%%%%%%%%%%%%%%%%%%%%%%%%%%%%%%%%%%%%%%%%%%%%%%%%%%%%%%%%%%%%%%%%%%%%%%%%%%%%%%%%%%%%%%%%%%%%%%%%%%%%%%%%%%%%%%%%%%%%%%%%%%%%%%%%%%%%%%%%%%%%%%%%%%%%%%%%%%%%%%%%%%%%%%%%%%%%%%%
\subsubsection{Evaluation Results}
\label{sec:EdgeSDN:ES-HAS:ESHAS-Res1}
One of the critical parameters for analyzing the performance of \texttt{ES-HAS} is the time slot duration. In the adaptation process, the client's ABR algorithm estimates the network's bandwidth by measuring the time between sending the request to download a segment and having received the segment's last packet. An additional delay in the request-response interval (due to an overly long time slot duration in the VRP) may convey an incorrect network situation to the client; consequently, the client may request a lower quality level for the next segment. Hence, in the first experiment, we investigate various time slot durations for various segment durations in order to find suitable values of $\theta$. For each segment duration of 2, 4, and 6 seconds, we start the experiment with the initial value of $\theta = 10$ ms and gradually increase it in each run. The results show that for time slot durations longer than 52, 95, and 200 ms for 2, 4, and 6 sec.~segments, respectively, clients start to request lower quality levels. We conclude that these values are the maximum time slot durations without any negative effect on the clients' adaptation behavior.   

In the second experiment, we numerically investigate the scalability of the proposed MILP model execution. As illustrated in Fig.~\ref{ESHAS-TS}, each time slot includes data collecting and optimization intervals. In the worst case, requests arriving at the beginning of the optimization interval should wait to be processed in the next optimization interval. In other words, a request could possibly wait for two optimization intervals plus a collection interval. To avoid optimization interval overlapping, in this study, we assume the optimization interval should be less than or equal to $\frac{\theta}{2}$. Therefore, based on the first experiment, this value should be less than 26, 47, and 100 milliseconds for 2, 4, and 6 sec.~segments, respectively.
%%%%%%
\begin{figure}[!t]
	\centering
    \includegraphics[width=.9\linewidth]{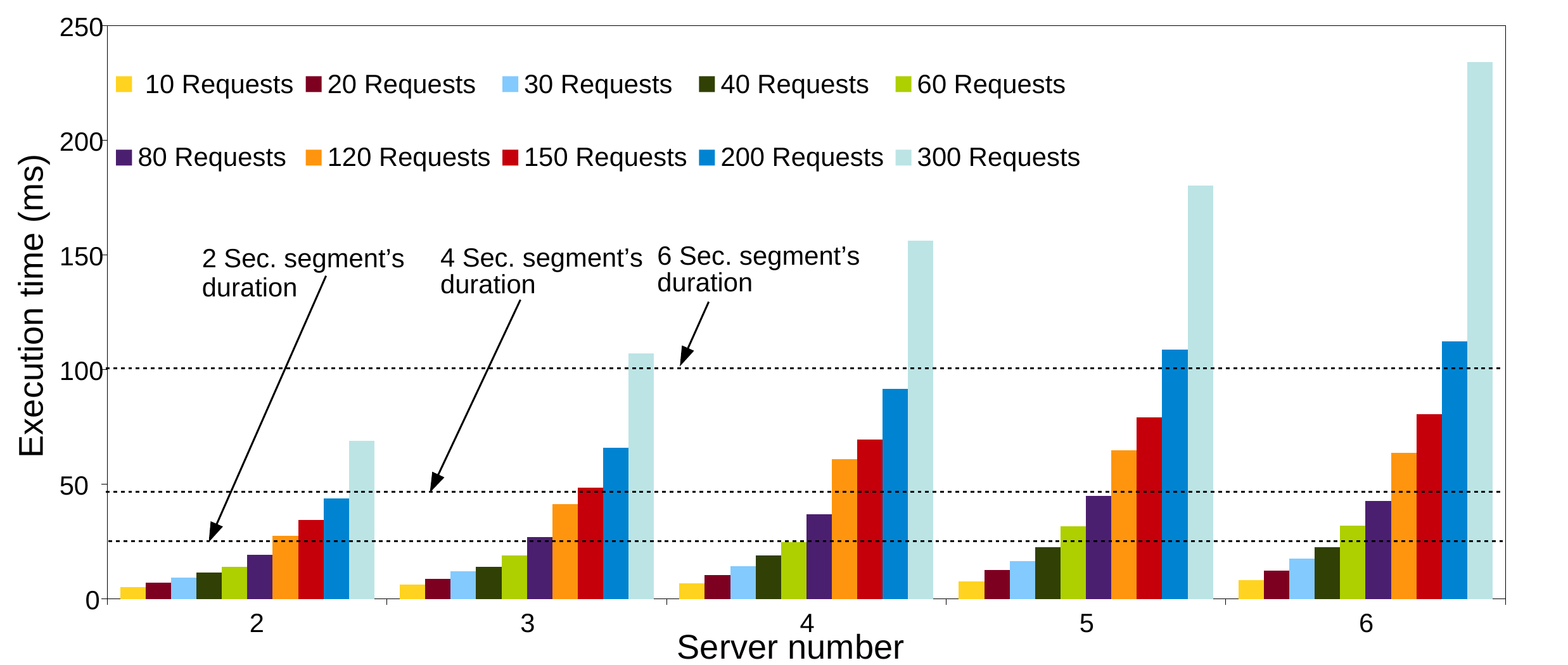}
	\caption{\small ES-HAS MILP model execution time for different numbers of segment requests and cache servers.}
        \vspace{.5cm}
	\label{OM-Beh}
\end{figure}
%%%%%%

We measure the MILP execution time for various numbers of clients' requests and cache servers (Fig.~\ref{OM-Beh}). Considering the optimization interval duration, the proposed MILP model can handle about 60, 100, and 210  different requests for 2, 4, and 6 sec.~segments, respectively, in a topology with four cache servers (see Fig.~\ref{OM-Beh}). The number of requests decreases when increasing the number of cache servers. 
We now calculate the maximum number of clients that each VRP can handle in a  topology with four cache servers. As we discussed earlier, in our experiments, each client sends 150 requests to download 150 segments with 2 seconds duration. We showed that the maximum time slot duration should be less than or equal to 52 milliseconds;
that means we have about 5700 ($\frac{300}{0.052}$) time slots in each experiment. Assuming a uniform distribution, each client sends a request in a given time slot with a probability of 0.026 ($\frac{150}{5700}$). For instance, in the case of 60 requests as the maximum number of requests that can be handled by the proposed MILP model in each time slot, one VRP can serve up to a maximum of 2300 clients and send 60 distinct requests per time slot ($2300\leq\frac{60}{0.026}$).

To investigate the behavior of the  framework regarding different values of $\alpha_1$, $\alpha_2$, and $\alpha_3$, we define three metrics:
\begin{enumerate}[noitemsep]
\item \textbf{ACS}: the usage percentage of cache servers with the shortest fetching time (since the first term of the objective function (Eq.~\ref{ESHAS:eq:5}) forces the model to select a cache server with the shortest fetching time). 
\item \textbf{AMD}: the average  (for different $m$) of the maximum deviation between requested quality and forwarded quality (second term of the objective function (Eq.~\ref{ESHAS:eq:5}). 
\item \textbf{AQB}: the average of the video quality bitrate for all received segments in Mbps (third term of the objective function (Eq.~\ref{ESHAS:eq:5}).
\end{enumerate}
%%%%%
\begin{figure}[!t]
\centering
	\includegraphics[width=1\linewidth]{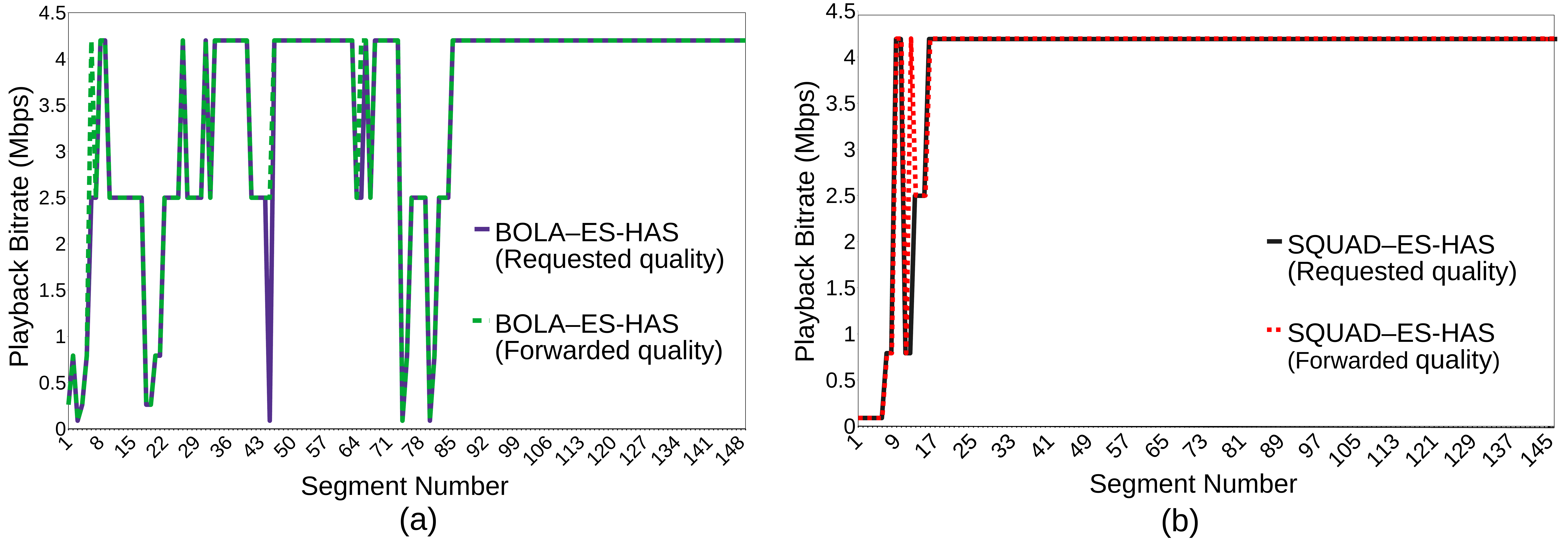}
\caption{\small Requested quality levels by the (a) BOLA and (b) SQUAD algorithms vs. forwarded quality levels for client 1.}	
\vspace{.5cm}
	\label{Client-BH-replacement}
\end{figure}
%%%%%
In the next experiment, we investigate the reactions of clients' ABR algorithms when their decisions are overwritten by replacement qualities. \rf{For this purpose, we prioritize the quality coefficient ($\alpha_3 = 0.8$) and select the identical small values for the fetching time and deviation coefficients ($\alpha_1=0.1,\alpha_2=0.1$) to force the model to choose appropriate replacement qualities in cache miss situations. Moreover, we set $m=3$ to give the model a wider range of options for selecting the potential replacement quality levels.} All requested qualities and forwarded qualities are logged by the VRP. As shown in  Fig.~\ref{Client-BH-replacement} (client I),
the requested quality levels by the BOLA and SQUAD algorithms are overwritten multiple times by better replacement qualities (\eg segments 46 and 15 requested by BOLA~(Fig.~\ref{Client-BH-replacement}(a)) and SQUAD~(Fig.~\ref{Client-BH-replacement}(b)) respectively); however, that does not have a negative impact on the overall playback quality of subsequent segments. Other clients have a similar reaction to the replacement quality. 

We extend our experiments by various values of $\alpha$ and $m$ and investigate their impact on the aforementioned metrics. Table~\ref{tab:aplpha-values} shows the results for group I of clients. As expected, most of the requested quality levels are delivered from cache server I (denoted $c_1$) for a high $\alpha_1$ value, which forces the system to emphasize low fetching time ($\alpha_1=0.8$, $\alpha_2=0.1$, $\alpha_3=0.1$).
On the other hand, setting $\alpha_2=0.8$ forces the model to deliver the quality levels as requested by the clients and keeps AMD to zero ($\alpha_1=0.1$, $\alpha_2=0.8$, $\alpha_3=0.1$). By setting $\alpha_3$ to a high value, the model can serve higher quality levels to clients ($\alpha_1=0.1$, $\alpha_2=0.1$, $\alpha_3=0.8$). Although increasing $m$ and $\alpha_3$ increases the AMD, the model performs negligible improvement in AQB since it has a wider range to select replacement quality levels ($\alpha_1=0.1$, $\alpha_2=0.1$, $\alpha_3=0.8$, and $m=3$). 
%%%%
\begin{table}[!t]
\centering
\caption {\small Impact of $m$, {$\alpha_1$}, $\alpha_2$, and $\alpha_3$ on MILP model behavior.}
\label{tab:aplpha-values}
\begin{tabular}{ll|l|l|l|}
\\\cline{1-5}
\multicolumn{1}{|l|}{\textbf{m}}  & {\textbf{Metric}} & {\textbf{($\alpha_1$,$\alpha_2$,$\alpha_3$): (.8,.1,.1)}} & \textbf{($\alpha_1$,$\alpha_2$,$\alpha_3$): (.1,.8,.1)} & 
\textbf{($\alpha_1$,$\alpha_2$,$\alpha_3$): (.1,.1,.8)}\\ \hline
%%%%%%%%%%%%%%%%%%%%
\multicolumn{1}{|l|}{}& \multicolumn{1}{c|}{\textbf{ACS}}&   \multicolumn{1}{c|}{$c_1:56\%$, $c_2:25\%$}&  
\multicolumn{1}{c|}{$c_1:51\%$, $c_2:31\%$}& \multicolumn{1}{c|}{$c_1:46\%$, $c_2:36\%$}      
\\ \cline{2-5} 
\multicolumn{1}{|l|}{\textbf{1}}& \multicolumn{1}{c|}{\textbf{AMD}}&     \multicolumn{1}{c|}{0}&\multicolumn{1}{c|}{0}&   \multicolumn{1}{c|}{0}
\\ \cline{2-5} 
\multicolumn{1}{|l|}{}  & \multicolumn{1}{c|}{\textbf{AQB}}& \multicolumn{1}{c|}{3.73}&   \multicolumn{1}{c|}{3.73}& \multicolumn{1}{c|}{3.73}                  
\\ \hline
%%%%%%%%%%%%%%%%%%
\multicolumn{1}{|l|}{}& \multicolumn{1}{c|}{\textbf{ACS}}&   \multicolumn{1}{c|}{$c_1:55\%$, $c_2:26\%$}&  
\multicolumn{1}{c|}{$c_1:53\%$, $c_2:29\%$}& \multicolumn{1}{c|}{$c_1:54\%$, $c_2:29\%$}      
\\ \cline{2-5} 
\multicolumn{1}{|l|}{\textbf{2}}& \multicolumn{1}{c|}{\textbf{AMD}}&     \multicolumn{1}{c|}{1}&\multicolumn{1}{c|}{0}&   \multicolumn{1}{c|}{1}
\\ \cline{2-5} 
\multicolumn{1}{|l|}{}  & \multicolumn{1}{c|}{\textbf{AQB}}& \multicolumn{1}{c|}{3.75}&   \multicolumn{1}{c|}{3.73}& \multicolumn{1}{c|}{3.75}                  
\\ \hline
%%%%%%%%%%%%%%%%%%
\multicolumn{1}{|l|}{}& \multicolumn{1}{c|}{\textbf{ACS}}&   \multicolumn{1}{c|}{$c_1:56\%$, $c_2:36\%$}&  
\multicolumn{1}{c|}{$c_1:52\%$, $c_2:31\%$}& \multicolumn{1}{c|}{$c_1:52\%$, $c_2:32\%$}      
\\ \cline{2-5} 
\multicolumn{1}{|l|}{\textbf{3}}& \multicolumn{1}{c|}{\textbf{AMD}}&     \multicolumn{1}{c|}{1}&\multicolumn{1}{c|}{0}&   \multicolumn{1}{c|}{.437}
\\ \cline{2-5} 
\multicolumn{1}{|l|}{}  & \multicolumn{1}{c|}{\textbf{AQB}}& \multicolumn{1}{c|}{3.75}&   \multicolumn{1}{c|}{3.75}& \multicolumn{1}{c|}{3.8}                  
\\ \hline
\end{tabular}
\vspace{.5cm}
\end{table}
%%%%%%%%%%%%
%%%%%%
\begin{figure}[!t]
    \centering
	\includegraphics[width=1\linewidth]{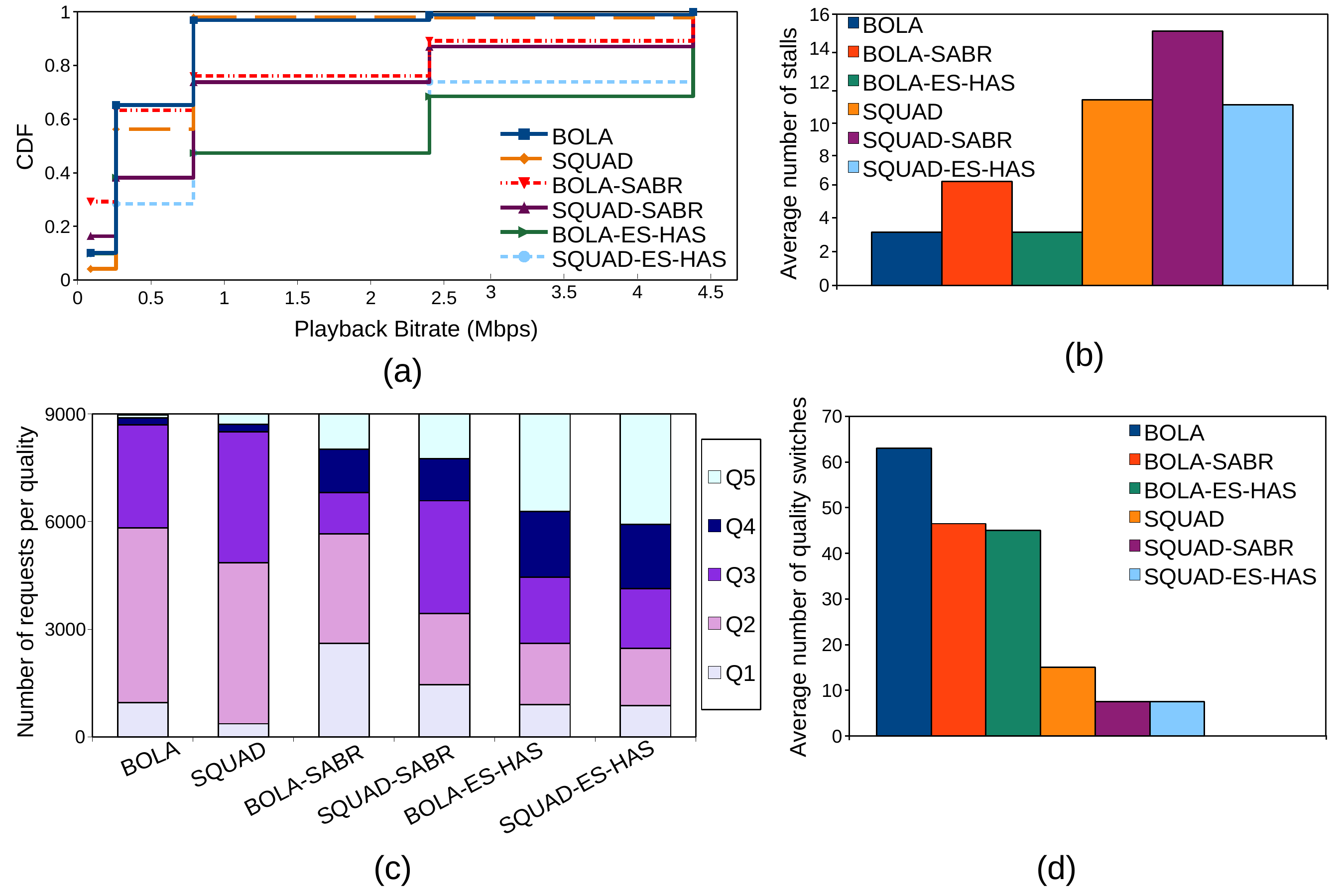}
	\caption{\small (a) CDF of average playback bitrate, (b) average number of stalls, (c) number of requests for each quality level, and (d) average number of quality switches in pure client-based ABR, SABR, and ES-HAS systems for 60 clients.}
 \vspace{.5cm}
\label{ESHAS-Results}
\end{figure}
%%%%%%%%

As another experiment, we compare the performance of \texttt{ES-HAS} with the pure client-based (video players in client groups I and II directly send their requests to CDN edge servers I and IV, respectively) and the SABR~\cite{bhat2017network} approaches. Because~\cite{bhat2017network} used different components than we have in our testbed, we implemented the SABR approach and reproduced its results without using its available source code. The $\alpha_1$, $\alpha_2$, $\alpha_3$, and $m$ values are set to 1, 0, 0, and 0, respectively, to have fair comparisons among the schemes; \ie the VRP only transmits the original requested quality from the optimal cache server. As illustrated in Fig.~\ref{ESHAS-Results} (a) and Fig.~\ref{ESHAS-Results}(d), the  \texttt{ES-HAS} framework outperforms the pure client-based method in terms of playback bitrate and the number of quality switches since it fetches segments from the cache server with the shortest fetching time. Although SABR and \texttt{ES-HAS} show (almost) identical results in terms of the number of quality switches for both ABR methods (Fig.~\ref{ESHAS-Results}(d)), \texttt{ES-HAS} results in better performance in terms of playback bitrate (Fig.~\ref{ESHAS-Results}(a)), the number of requests for the highest quality level (Fig.~\ref{ESHAS-Results}(c)), and the number of stalls compared to SABR (Fig.~\ref{ESHAS-Results}(b)).
Recognizing similar requests (video/segment/quality) and sending only one request instead of several requests to the selected cache server, plus employing an optimization-based approach for the cache/segment selection policy, are the main reasons for \texttt{ES-HAS} performance improvements over SABR.

%% file: Chapters/Chapter3/3-3-CSDN.tex
\doublespacing
\section{CSDN Framework}
\label{chap:EdgeSDN:CSDN}
This section introduces the \texttt{CSDN} framework~\cite{Farahani2021csdn} as a CDN-aware QoE optimization solution in SDN-assisted HTTP Adaptive Video Streaming. \rf{\texttt{CSDN} actually extends the \texttt{ES-HAS} framework by equipping its edge servers with transcoding capabilities, making it suitable for both VoD and live streaming scenarios.} We motivate this work through an example in Section~\ref{sec:EdgeSDN:CSDN:Motivating Example} and elaborate on its architecture and optimization model in Section~\ref{sec:EdgeSDN:CSDN:System Model}. The \texttt{CSDN} evaluation setup and obtained results are described in Section~\ref{sec:EdgeSDN:CSDN:PerformanceEvaluation}. 
%%%%%%%%%%%%%%%%%%%%%%%%%%%%%%%%%%%%%%%%%%%%%%%%%%%%%%%%%%%%%%%%%%%%%%%%%%%%%%%%%%%%%%%%%%%%%%%%%%%%%%%%%%%%%%%%%%%%%%%%%%%%%%%%%%%%%%%%%%%%%%%%%%%%%%%%%%%%%%%%%%%%%%%%%%%%%%%%%%%%%%%%%%%
\subsection{CSDN Motivating Example and Problem Description}
\label{sec:EdgeSDN:CSDN:Motivating Example}
\begin{figure}[!t]
\centering
\includegraphics[width=1\linewidth]{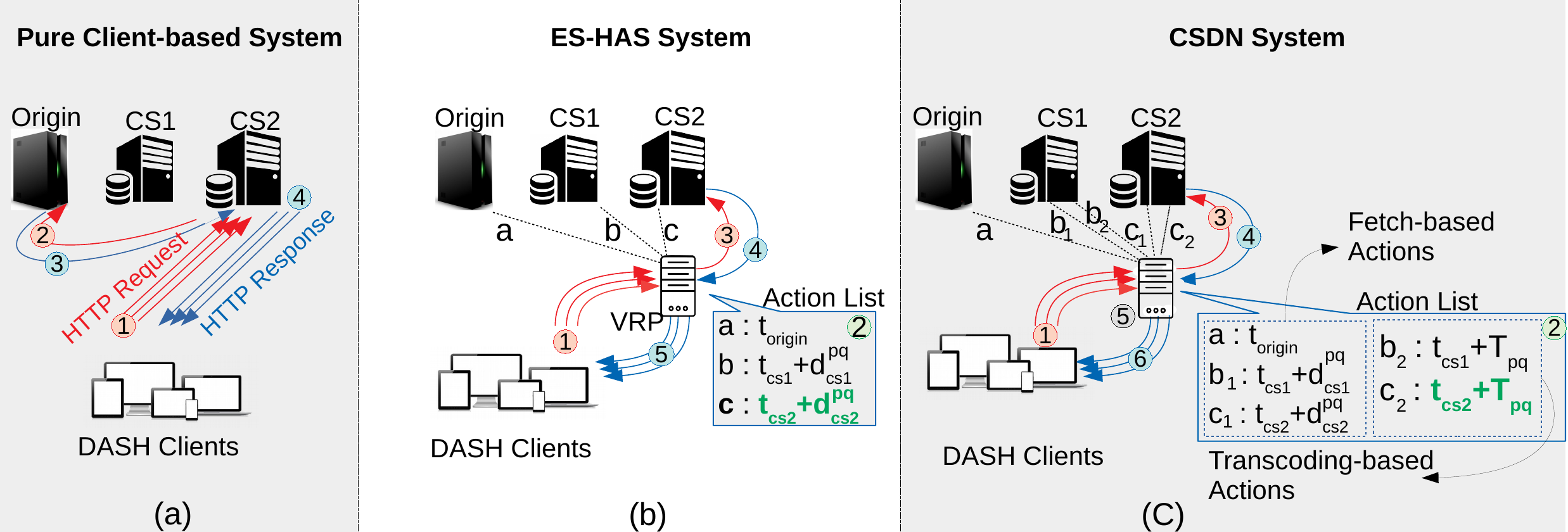}
\caption{\small (a) Pure client-based, (b) ES-HAS, and (c) CSDN approaches.} 
\vspace{.5cm}
\label{csdn-Motivation}	
\end{figure}

We present our primary motivation by an example shown in Figure~\ref{csdn-Motivation}. In this example, we consider a simplistic scenario with only two cache servers, an origin server, one VRP server, and a group of DASH clients. We assume DASH clients request the segments of video $v$ in \textit{(i)} a pure client-based approach (Fig.~\ref{csdn-Motivation}(a)), \textit{(ii)} the \texttt{ES-HAS} framework described in section~\ref{chap:EdgeSDN:ES-HAS} (Fig.~\ref{csdn-Motivation}(b)), and \textit{(iii)} a \texttt{CSDN}-enabled system (Fig.~\ref{csdn-Motivation}(c)). In the pure client-based approach, clients have to send their requests to only a preset cache server, \eg CS2 (the nearest cache server in terms of geographical location) (step $1$). In case of a cache hit, CS2 directly replies to the requests ($4$). However, the cache server must hold the requests and fetch the requested segments from the origin server for cache miss events ($2-3$). 

As illustrated in Figure~\ref{csdn-Motivation}(b) and explained previously, \texttt{ES-HAS} utilizes an edge component (called VRP) that aggregates clients' requests (\ie it forwards only one request for a set of identical requests) in a time-slotted fashion. After collecting requests for segments of a video $v$ ($1$) and associated network information (\ie the presence of representations of video segments and available bandwidth from cache servers), the VRP runs an optimization model with the depicted action list ($2$). In fact, \texttt{ES-HAS} improves users' QoE by \textit{(i)} selecting an optimal cache server that has the most available bandwidth or the least fetching time than other servers (denoted by $t_{origin},t_{cs1},t_{cs2}$ for origin server and cache servers, respectively) or \textit{(ii)} using a replacement quality level (denoted by $q$) to serve the originally requested quality (denoted by $p$) with the minimum deviation (only from cache servers denoted by $d_{cs1}^{pq},$ $d_{cs2}^{pq}$). Although \texttt{ES-HAS} solves the aforementioned problems of a pure client-based system, it still suffers from some constraints. Imagine a scenario that the demanded quality levels are available only on one cache server (\eg CS2); \texttt{ES-HAS} forces the VRP to fetch the original segments from CS2 ($2-4$). Finally, the fetched segments are transferred to clients ($5$). In this case, if another server (\ie CS1) with more available bandwidth could serve the requested segments with higher quality levels, \texttt{ES-HAS} ignores it. In another scenario, imagine the requested segments are unavailable in all cache servers, and the quality deviation is unacceptable for the clients. Hence, the desired quality levels must be fetched from the origin server to the VRP and cache servers, leading to a considerable backhaul bandwidth consumption.

\texttt{CSDN} framework equips the \texttt{ES-HAS} VRP with the transcoding capability to remedy the aforementioned problems. As shown in Figure~\ref{csdn-Motivation}(c), the clients send requests ($1$) to the VRP for the desired segments' qualities, and the VRP collects these requests in each time slot with the aggregation technique. After receiving the mentioned demanded network information, plus user preferences, the VRP runs an optimization program by taking into account two different groups of actions, \ie fetching- and transcoding-based actions ($2$). The first group, like \texttt{ES-HAS}, considers the fetching time and quality deviations, while the latter group is based on the serving time (\ie fetching time and transcoding time ($T_{pq}$) values). 
%%%%%%%%
%%%%%%%%%
For simplicity, let us proceed by turning to the considered scenarios. In both scenarios, \texttt{CSDN} uses a hybrid approach \ie fetches the quality level $q$ as a replacement quality from CS1 (\ie quality with the minimum fetching time, and cache server with a higher bandwidth value) ($3-4$), and transcodes it to the quality level $p$ at the edge ($5$). Next, it forwards the desired segments back to the clients ($6$). The details of \texttt{CSDN} will be discussed in the next section. In this work, we introduce the \texttt{CSDN} VRP at the edge of the network as an \texttt{ES-HAS} VRP extended with transcoding capabilities, which increases the computational costs of the system. However, the backhaul bandwidth consumption and users' QoE (based on their preferences) are significantly improved by the VRP possibly performing additional actions. According to the presented motivating example, we can define the following problem: ``\textit{How should VRPs serve many clients' requests to minimize the quality deviations and serving times (\ie fetching and transcoding time) based on the stated action list and different categories of clients?}''
%%%%%%%%%%%%%%%%%%%%%%%%%%%%%%%%%%%%%%%%%%%%%%%%%%%%%%%%%%%%%%%%%%%%%%%%%%%%%%%%%%%%%%%%%%%%%%%%%%%%%%%%%%%%%%%%%%%%%%%%%%%%%%%%%%%%%%%%%%%%%%%%%%%%%%%%%%%%%%%%%%%%%%%%%%%%%%%%%%%%%%%%%%%%%%%%
\subsection{CSDN System Design}
\label{sec:EdgeSDN:CSDN:System Model} 
This section introduces the \texttt{CSDN} architecture and its main components before elaborating on the \texttt{CSDN}'s problem formulation.
\subsubsection{CSDN Architecture}
\label{sec:EdgeSDN:CSDN:Architecture} 
\begin{figure}[!t]
	\centering
	\includegraphics[width=.9\linewidth]{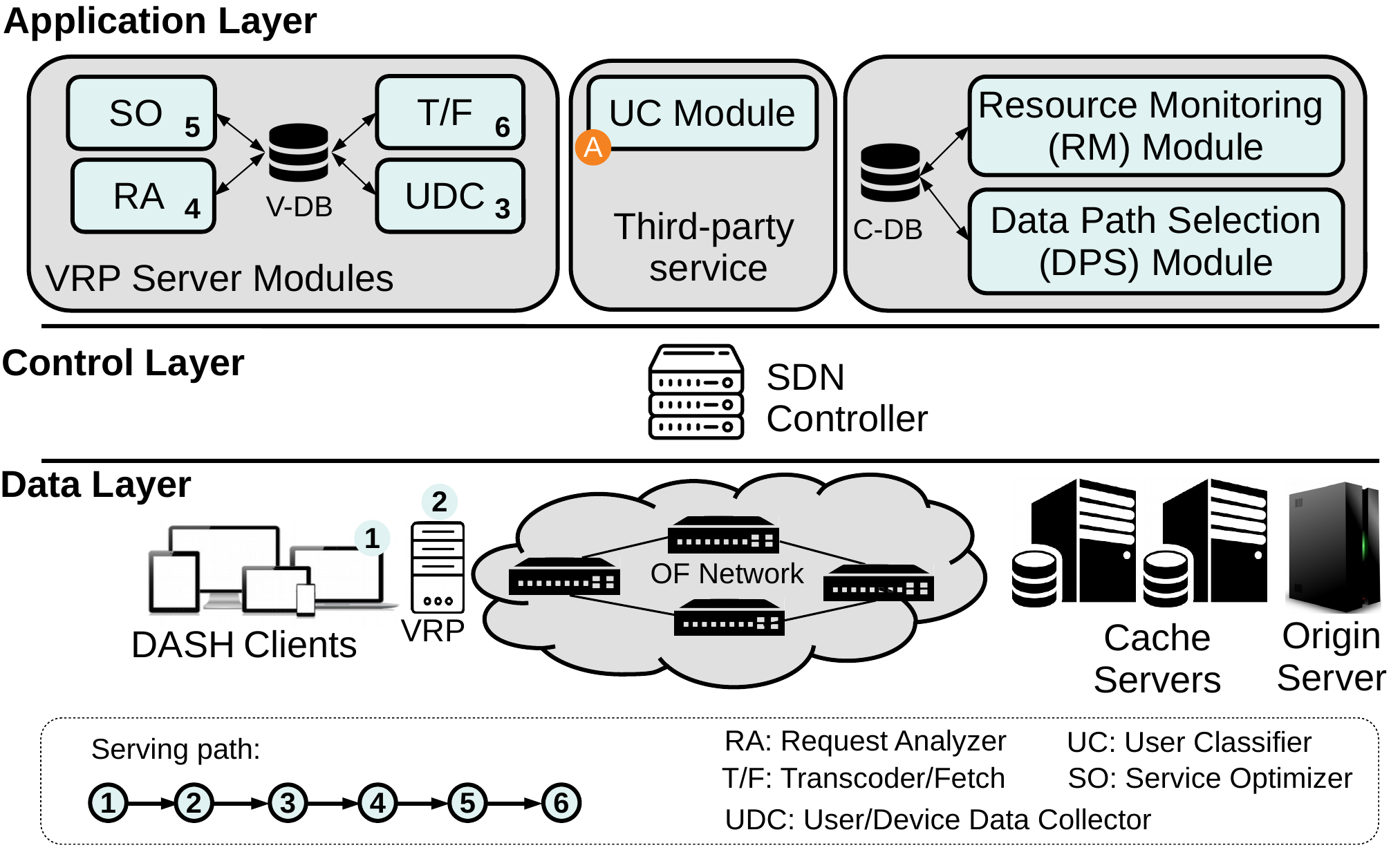}
	\caption{\small Proposed CSDN architecture.}
 \vspace{.5cm}
	\label{CSDN-Arch}
\end{figure}

Inspired by the overall architecture of an SDN-enabled network, we design the \texttt{CSDN} architecture in three main layers~(Figure~\ref{CSDN-Arch}), where each layer consists of a set of components that are described as follows:\\
\textbf{$(1)$~Data Layer:} This layer consists of four groups of entities, \ie \textit{(i)} DASH clients, \textit{(ii)} CDN components, including an origin server and multiple cache servers, \textit{(iii)} several VRP servers as gateways for DASH clients to the network and vice versa, and \textit{(iv)} OpenFlow (OF) switches that are connected to an SDN controller, CDN servers, and VRPs.\\
\textbf{$(2)$~Control Layer:} An SDN controller is placed in this layer.\\
\textbf{$(3)$~Application Layer:} The application layer comprises different components that are categorized into four groups regarding their functionalities: \textit{(i)} VRP server modules, \textit{(ii)} databases, \textit{(iii)} third-party service, and \textit{(iv)} controller modules.\\
These entities in the \texttt{CSDN} architecture collaborate as follows to serve DASH clients' requests efficiently. The SDN controller periodically monitors the cache servers' occupancy information (cache maps) and paths' available bandwidths and stores them in its database (C-DB). We define a \textit{Resource Monitoring Module} (RM) as one of the main controller modules to collect the aforementioned CDN- and network-level information from the OF switches. The C-DB data will be used by the controller's \textit{Data Path Selection Module} (DPS) in order to set up new routes in the network (\ie between VRP and CDN components), or by the VRP server modules in order to select the next segments/servers efficiently.
%%%%%%%%%%%%%
%%%%%%%%%%%%%
It is worth noting that, like our previous system \texttt{ES-HAS} does, \texttt{CSDN} operates in a time-slotted manner with an equal time slot duration of~$\theta$. 
Each time slot consists of two modified intervals: \textit{(i) Data Collecting} and \textit{(ii) Optimization Intervals}. 

In the first interval, user- and device-level information (\eg user identity, the associated devices, and users' video requests) are gathered and aggregated by the VRP's \textit{User/Device Data Collector Module} (UDC). This data is provided by the \textit{User Classifier Module} (UC) of a third-party service to which the user is subscribed (\eg Netflix or Amazon Prime). It is assumed that the aforementioned information will be available by the third-party service. Consequently, when clients request video segments during each time slot, the UDC fetches the associated profiles and collects the basic specifications (\eg display size, buffer size, type of device, and subscription plan), and assigns a user priority to each client. 

To prevent sending identical requests (issued by multiple clients in a given time slot), the VRP's \textit{Request Analyzer Module} (RA) identifies received user-level information from the UDC, considers only one request per segment, and stores the requests in the VRP's database (V-DB).  Moreover, using RESTful messages, the VRP periodically retrieves the required information (\ie cache maps plus available bandwidths between each cache server and the VRP that are available in the C-DB) from the SDN controller. In the second interval, the VRP's \textit{Service Optimizer Module} (SO) is executed to serve clients' requests optimally. 
% %%%%%%%
% \begin{figure}[!t]
% 	\centering
% 	\includegraphics[width=.8\linewidth]{Figures/Fig-Chap3-2--CSDN/CSDN-timeslot-structure.pdf}
% 	\caption{Proposed time slot structure}
% 	\label{CSDN-TS}
% \end{figure}
% %%%%%%%%

The data transmission will start in the first interval of the next time slot according to the SO's results. Actually, for each request the SO selects an appropriate cache server that hosts the requested quality by using the optimization model presented later in this section. However, in some situations, \eg when the requested quality is not available in a cache server, the SO chooses one of the following solutions based on users' preferences, then the VRP's \textit{Transcoder/Fetching Module} (T/F) prepares the optimal segment from an optimal server: \textit{(i)} determine the optimal replacement quality level on a cache server (with the minimum deviation from the original requested quality level) and forward it to the associated clients; \textit{(ii)} determine the optimal replacement quality level on a cache server and transcode it to the original requested quality level;\\ 
\textit{(iii)} fetch the original requested quality from the origin server. 

Like in \texttt{ES-HAS}, we utilize the notion that a requested quality level of a segment may be overwritten by a \textit{replacement quality}, based on Consumer Technology Association's \textit{Common Media Client Data} (CMCD) standard~\cite{cta-wave-cmcd}. To do that, we assume that a client can accept only better replacement quality levels instead of the actual requested quality of a segment. Furthermore, we employ the replacement quality levels determined by the SO as reference segments for the transcoding process in T/F. For the sake of simplification, we assume that each DASH client can request just one segment during each time slot. However, it is possible to take into account multiple requests from each client in each time slot without significant changes in the proposed model. We will demonstrate how to determine the time slot duration in Section~\ref{sec:EdgeSDN:CSDN:PerformanceEvaluation}.
%%%%%%%%%%%%%%%%%%%%%%%%%%%%%%%%%%%%%%%%%%%%%%%%%%%%%%%%%%%%%%%%%%%%TABLE%%%%%%%%%%%%%%%%%%%%%%%%%%%%%%%%%%%
\begin{table}[!b]
\centering
\caption {\small CSDN Notation.}
\label{tab:CSDN:notation}
\begin{tabular}{llllll}
\cline{1-2}
\multicolumn{2}{|c|}{\textbf{Input Parameters}}                                                                                  
&  &  &  &  \\ \cline{1-2}                                                                                         
\multicolumn{1}{|l|}{\begin{tabular}[c]{@{}l@{}}
$\mathcal{S}$\\ 
$\mathcal{C}$\\ 
$\mathcal{T}$\\ 
$\mathcal{A}$ \\
$a^{c,s}_q$\\ 
$\mu^c_{q,p}$\\ 
$\omega^{c}_{q,p}$\\ \\ 
$\mathcal{R}$ \\
$R_s$\\  
$i_c$\\ 
$m$\\ 
$\mathcal{K}_c$\\ \\ 
$\delta^c_q$\\ 
$\pi^c_q$\\ 
$\theta$\\ 
$\Omega$\\
\end{tabular}} 
& \multicolumn{1}{l|}{\begin{tabular}[c]{@{}l@{}}
Set of cache servers and origin server, $s\in \mathcal{S}$\\ 
Set of clients, $c\in \mathcal{C}$\\ 
Set of transcoding status, $t\in \mathcal{T}$\\
Set of available quality levels in $\mathcal{S}$\\
$a^{c,s}_q=1$ if quality $q$ requested by client $c$ is available in server $s$, $a^{c,s}_q=0$ otherwise\\ 
Required time to transcode quality $p$ requested by client $c$ from quality $q$ at VRP\\ 
Required computational resource (CPU usage in \%) to transcode quality $p$ \\requested by client $c$ from quality $q$ at VRP\\
Set of available bandwidth values \\
The available bandwidth between the VRP to server $s$\\
Quality level requested by client $c$\\ 
Integer number to limit the range of potential replacement quality levels for $i_c$\\ 
Set of eligible quality levels for a quality  requested by client $c$, where\\ $\mathcal{K}_c=\{i_c,{i_c}+1,...,min[{i_c}+m,q_{max}]\}$\\ 
Size of the segment of quality level $q$\\ 
Bitrate of the segment of quality level $q$ \\ 
Time slot duration \\ 
Compute resource threshold (max.~CPU usage in \%) for transcoding\end{tabular}} &  &  &  &  \\ \cline{1-2}
\multicolumn{2}{|c|}{\textbf{Variables}}                                                                                                                                 &  &  &  &  \\ \cline{1-2}                                                 
\multicolumn{1}{|l|}{\begin{tabular}[c]{@{}l@{}}$B^{c,s}_{q,t}$\\ \\ \\$D_c$\\ $T^{c,s}_q$\\ $\tau_c$\\ $\Gamma^{c,s}_q$ \end{tabular}}                                                                & \multicolumn{1}{l|}{\begin{tabular}[c]{@{}l@{}}$B^{c,s}_{q,0}$=1 if quality $q$ requested by $c$ is served from server $s$, $B^{c,s}_{q,1}$=1\\if quality $q$ requested by $c$ is served from server $s$ and transcoded \\at the VRP, $B^{c,s}_{q,1}$=0 otherwise\\
Deviation of quality level to serve client $c$ w.r.t. $i_c$\\ 
Required time to fetch quality $q$ for client $c$ from server $s$\\ 
Required transcoding time of quality level to serve client $c$ w.r.t. $i_c$\\
Serving time consisting of $T^{c,s}_q$ and $\tau_c$  \\ \end{tabular}}                                                    
&  &  &  &  \\ \cline{1-2}                                                                                                              
& &  &  &  &                                              
\end{tabular}
\end{table}
%%%%%%%%%%%%%%%%%%%%%%%%%%%%%%%%%%%%%%%%%%%%%%%%%%%%%%TABLE%%%%%%%%%%%%%%%%%%%%%%%%%%%%%%%%%%%
%%%%%%%%%%%%%%%%%%%%%%%%%%%%%%%%%%%%%%%%%%%%%%%%%%%%%%%%%%%%%%%%%%%%%%%%%%%%%%%%%%%%%%%%%%%%%%%%%%%%%%%%%%%%%%%%%%%%%%%%%%%%%%%%%%%%%%%%%%%%%%%%%%%%%%%%%%%%%%%%%%%%%%%%%%%%%%%%%%%%%%%%%%%%%%%%
\clearpage
\subsubsection{CSDN Optimization Problem Formulation}
\label{sec:EdgeSDN:CSDN:OM} 
As discussed earlier, the VRPs' SO module is responsible for executing a server/segment selection policy to serve clients' requests optimally, \ie \textit{(i)} determine an optimal server, and \textit{(ii)} select appropriate actions (\ie fetch-based or transcoding-based actions) based on users' preferences. For that, we introduce a mixed-integer linear programming (MILP) optimization model.
The proposed MILP model finds the optimal solution by minimizing the serving times, including server fetching time and transcoding time, and the segments' quality level deviations. 

Let set $\mathcal{A}$ denote the cache map (\ie availability of segments/bitrates) that a VRP receives from the SDN controller, where $a^{c,s}_q=1$ means that the quality level $q$ requested by client $c \in \mathcal{C}$ is available on the server $s\in \mathcal{S}$ (see Table~\ref{tab:CSDN:notation} for notations). Moreover, let a VRP's database (V-DB) host set $\mathcal{R}$ containing available bandwidth values between the cache servers and the VRP.  We derive the following constraints that must be satisfied to achieve an optimal solution. Let us define $i_c$ as the quality level requested by client $c\in\mathcal{C}$. We also define $\mathcal{K}_c=\{i_c,{i_c}+1,...,min[{i_c}+m,q^c_{max}]\}$ as the set of eligible (potentially, replacement) quality levels for the segment requested by client $c$, where $m$ and $q^c_{max}$ denote the maximum deviation from $i_c$ and the maximum quality level of the segment requested by $c$, respectively.

In each optimization interval, we select only one server to serve the client $c$. For this purpose, we introduce binary variable $B^{c,s}_{q,t}$, with the meaning defined in Table~\ref{tab:CSDN:notation}. As mentioned earlier, in the case of a cache miss, \ie when $i_c$ is not available in any cache server, the VRP can fetch it from the origin server, use other quality levels available in cache servers, or transcode it from the fetched replacement quality from a cache server. Thus, constraint ($1$) must be satisfied:
%const1
\begin{flalign}
&\sum_{t\in \mathcal{T}}\sum_{s\in \mathcal{S}}\sum_{q\in \mathcal{K}_c} B^{c,s}_{q,t}~.~{a}^{c,s}_q=1, &&\hspace{.4cm} \forall c \in \mathcal{C}\label{CSDN:eq:1}
\end{flalign}
Moreover, we should force the model not to select a replacement quality from the origin server by constraint ($2$): 
%const2
\begin{flalign} 
&\sum_{t\in \mathcal{T}}\sum_{q\in\mathcal{K}_c} B^{c,s}_{q,t}~.~q = i_c && \forall c\in \mathcal{C}, \text{where }s=Origin
\label{CSDN:eq:2}
\end{flalign}
The required time to fetch the quality level $q$ for client $c$ from server $s$, denoted by $T^{c,s}_q$, is determined by constraint ($3$): 
%cost3
\begin{flalign}
&\delta^c_q~.~\sum_{t \in \mathcal{T}}B^{c,s}_{q,t}\leq T^{c,s}_q~.~R_s,&&\forall c\in \mathcal{C},q\in \mathcal{K}_c,s\in \mathcal{S}, 
\label{CSDN:eq:3}
\end{flalign}
where $\delta^c_q$ is the size of the quality $q$ requested by client $c$ and $R_s$ is the available bandwidth between the VRP and cache server $s$. We also define $\tau_c$ as the transcoding time to serve client $c\in\mathcal{C}$ according to constraint ($4$):
%const4
\begin{flalign}
&\sum_{q\in \mathcal{K}_c} B^{c,s}_{q,t}~.~\mu^{c}_{q,p}\leq \tau_{c},&&\forall c\in\mathcal{C},s\in \mathcal{S}, p=i_c \text{ and } t=1
\label{CSDN:eq:4}
\end{flalign}
where $\mu^c_{q,p}$ is defined based on the meaning in Table~\ref{tab:CSDN:notation}.
Moreover, the model should consider the resource limitations, \ie available bandwidth on network paths and the available VRPs' computational resources. For this purpose, the selected quality should not violate $R_s$, the available bandwidth between the server $s$ to VRP, and $\Omega$, the VRP's computational resource threshold for transcoding (expressed as maximum CPU usage in \%). Thus the following constraints ($5-6$) must be satisfied:
%const5
\begin{flalign}
&\sum_{t\in \mathcal{T}}\sum_{c\in \mathcal{C}}\sum_{q\in \mathcal{K}_c} B^{c,s}_{q,t}~.~\pi^c_q \leq R_{s}&& \forall s\in \mathcal{S}\label{CSDN:eq:5}
\end{flalign}
%const6
\begin{flalign}
&\sum_{s\in \mathcal{S}}\sum_{c\in \mathcal{C}}\sum_{q\in \mathcal{K}_c}B^{c,s}_{q,t}~.~\omega^{c}_{q,p} \leq \Omega,&& \text{where }t=1 \text{ and } p=i_c 
\label{CSDN:eq:6}
\end{flalign}
where $\pi^c_q$ is the bitrate of the selected quality level $q$ for serving client $c$, and $\omega^{c}_{q,p}$ is the required computational resource (\ie CPU usage in \%) to transcode quality $p$ requested by client $c$ from quality $q$ at the VRP. To determine the quality deviation when the requested quality $i_c$ is not available in $\mathcal{K}_c$ or the requested quality has to be transcoded from the replacement quality, we introduce constraint ($7$):
%const7
\begin{flalign} 
&\sum_{s\in \mathcal{S}}\sum_{q\in \mathcal{K}_c} (B^{c,s}_{q,t}~.~q)-i_c \leq D_c&& \forall c\in \mathcal{C}, \text{where }t=0 \label{CSDN:eq:7}\end{flalign}

Finally, the optimization problem can be written as follows:
%objective
{\begin{flalign}
\textit{Minimize}&\hspace{0.5cm}\sum_{c\in \mathcal{C}}\sum_{s\in \mathcal{S}}\sum_{q\in \mathcal{K}_c}\alpha_c~.~\frac{\Gamma^{c,s}_q }{\mathbf{M}_\Gamma}+\sum_{c\in\mathcal{C}}\beta_c~.~\frac{D_c}{\mathbf{M}_D} \label{CSDN:eq:8}\\
  s.t.&\hspace{.5cm}\text{Constraints}\hspace{.5cm}\text{Eq.}(\ref{CSDN:eq:1})-\text{Eq.}(\ref{CSDN:eq:7})&&\nonumber\\
  &\hspace{.5cm}T^{c,s}_q+\tau_c = \Gamma^{c,s}_q, \hspace{1.5cm}\forall c\in \mathcal{C},q\in \mathcal{K}_c,s\in \mathcal{S}\nonumber\\
  vars.&\hspace{.5cm} T^{c,s}_q,\tau_c, \Gamma^{c,s}_q,D_c \geq 0, B^{c,s}_{q,t}\in\{0,1\}\nonumber
\end{flalign}
}
\normalsize where $\mathbf{M}_\Gamma$ and $\mathbf{M}_D$ are the maximum values for the serving time and the quality level deviation, respectively. The SO runs the above model for all clients' requested qualities in a time-slotted manner to minimize the serving times (\ie fetching time plus transcoding time) and the quality level deviations.  Moreover, we define two weight coefficients $\alpha_c$ and $\beta_c$  to set desirable priorities for $\Gamma^{c,s}_q $ and $D_c$. These weights can be adjusted based on each user's preference where $\alpha_c + \beta_c$ =1.
%%%%%%%%%%%%%%%%%%%%%%%%%%%%%%%%%%%%%%%%%%%%%%%%%%%%%%%%%%%%%%%%%%%%%%%%%%%%%%%%%%%%%%%%%%%%%%%%%%%%%%%%%%%%%%%%%%%%%%%%%%%%%%%%%%%%%%%%%%%%%%%%%%%%%%%%%%%%%%%%%%%%%%%%%%%%%%%%%%%%%%%%%%%%%%%
\subsection{CSDN Performance Evaluation}
\label{sec:EdgeSDN:CSDN:PerformanceEvaluation}
In this section, we first describe the evaluation setup and then discuss the performance evaluation of \texttt{CSDN} compared to \texttt{ES-HAS} (Section~\ref{chap:EdgeSDN:ES-HAS}), SABR~\cite{bhat2017network}, and a pure client-based approach.
%%%%%%%%%%%%%%%%%%%%%%%%%%%%%%%%%%%%%%%%%%%%%%%%%%%%%%%%%%%%%%%%%%%%%%%%%%%%%%%%%%%%%%%%%%%%%%%%%%%%%%%%%%%%%%%%%%%%%%%%%%%%%%%%%%%%%%%%%%%%%%%%%%%%%%%%%%%%%%%%%%%%%%%%%%%
\subsubsection{Evaluation Setup}
\label{sec:EdgeSDN:CSDN:EvaluationSetup}
Our testbed consists of 114 nodes running on Ubuntu 18.04 LTS inside Xen virtual machines. The VRPs use a 2.6 GHz, 6 Core CPU with 32 GB RAM. The proposed network topology is instantiated in the CloudLab~\cite{ricci2014introducing} environment. 

\begin{figure}[!t]
	\centering
	\includegraphics[width=.85\linewidth]{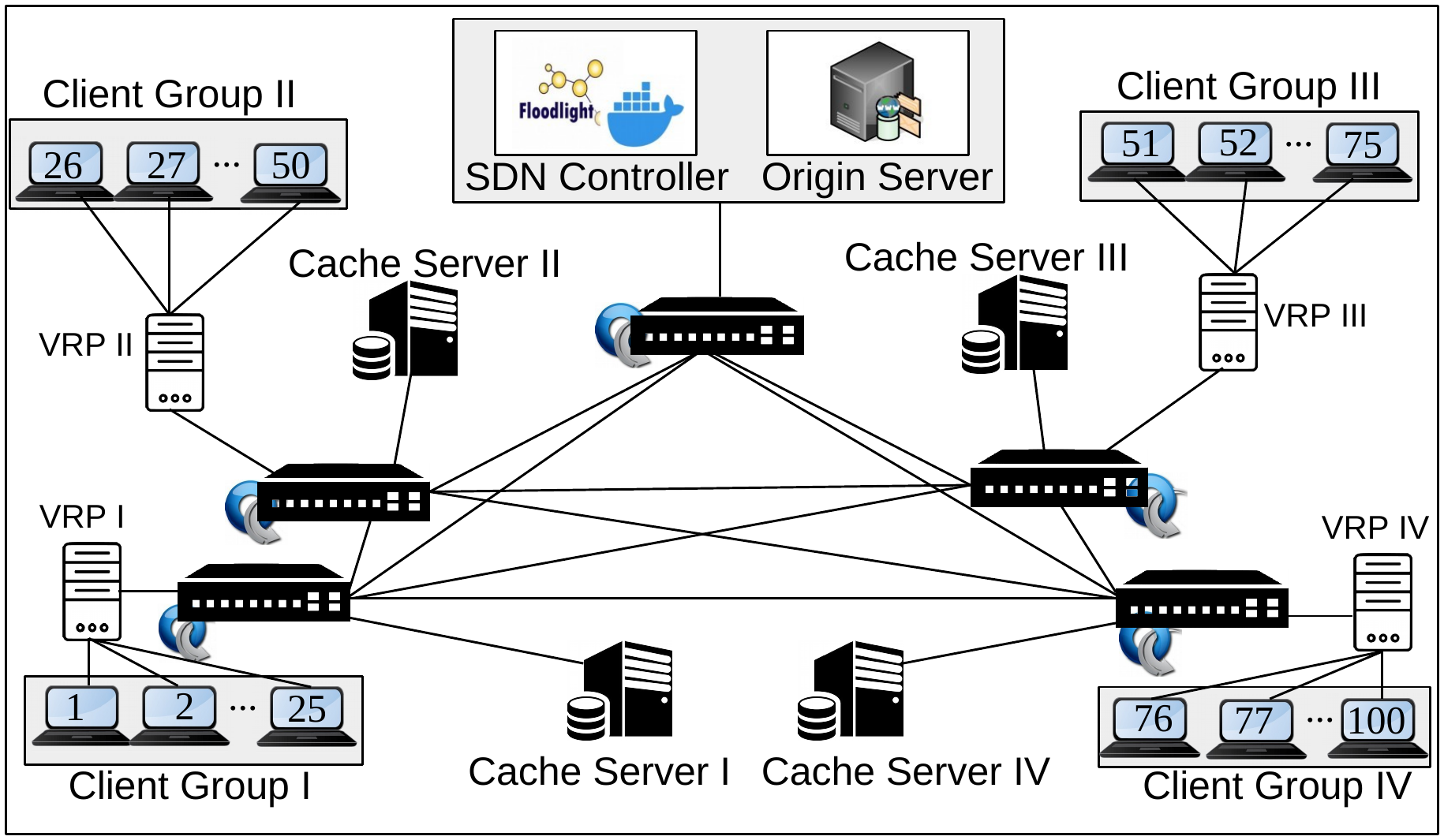}
	\caption{\small CSDN evaluation testbed.}
        \vspace{.5cm}
	\label{CSDN-Testbed}
\end{figure}

As illustrated in Figure~\ref{CSDN-Testbed}, it consists of five OpenFlow (OF) switches, 100 AStream DASH players~\cite{juluri2015sara,AStream} in headless mode, four VRP servers with the modules described in Section~\ref{sec:System Model}. Note that, in the current implementation, we assume homogeneous devices for all users, therefore, we use the client IP addresses as their identifiers. Moreover, we use four cache servers and one additional server that jointly hosts an origin server and a dockerized SDN controller. 
Cache servers I, II, III, IV are considered as local cache servers for client groups I, II, III, IV, respectively. The bandwidth values in different paths between cache servers I, II, III, IV, and the origin server to each VRP server are adjusted as follows:
VRP I:\{120, 100, 60, 80, 20\}, VRP II:\{100, 120, 80, 60, 20\}, VRP III:\{60, 100, 120, 80, 20\}, VRP IV:\{80, 60, 100, 120, 20\} (all in Mbps), which explicitly prioritize downloading segments from local cache servers. Apache~\cite{Apache} and MongoDB~\cite{Mongodb} with supporting RESTful APIs for cache map exchange are installed on all cache servers. Moreover, Least Recently Used (LRU) is considered in all cache servers as the cache replacement policy. 

The policy on a cache miss is that the requested quality will be fetched from the origin server to all cache servers. Floodlight~\cite{Floodlight} is utilized as an SDN controller. Among other tasks, it monitors the network to find paths’ available bandwidths (in one-second intervals) and then assigns the path with the highest available bandwidth between each cache server and a VRP server. For the sake of simplicity, we assume that all clients already joined the network and users' information is available in the third-party service (see Fig.~\ref{CSDN-Arch}).  
Ten test videos~\cite{lederer2012dynamic} with 300 seconds durations each are used in our experiments. These videos comprise two-second segments in five representations (0.089, 0.262, 0.791, 2.4, 4.2 Mbps). 

Although different strategies, \eg a popularity-based approach, could be used for the distribution of videos over CDN servers and for the clients' requests patterns, for the sake of simplicity we adopt the following scheme. 60\% of the videos’ segments are stored in each cache server randomly in most scenarios (only in one scenario, we store 30\% to 60\% videos segments on cache servers). Moreover, each client requests one video where $video_1$ is streamed to clients ({1,11,21,31,41,51, 61,71,81,91}), $video_2$ is streamed to clients ({2,12,22,32,42,52,62,72,82,92}), and so on.
The time slot duration is adjusted to 32 milliseconds in all experiments. (We will discuss how to obtain the time slot duration in the next subsection). Two different ABR algorithms, BOLA~\cite{spiteri2016bola} and SQUAD~\cite{wang2016squad}, representing buffer-based and hybrid approaches, are used in all experiments. Computational resource threshold (\ie $\Omega$) is set to 50\% of the overall VRPs' CPU usage. Python and the PuLP library~\cite{PuLP} are employed to implement and solve the \texttt{CSDN} MILP model. We use bitrate as the representative metric to evaluate the quality of segments, \ie the higher the bitrate, the higher the quality. The ``veryfast'' preset of FFmpeg~\cite{ffmpeg} is used for transcoding since we found it to induce less video distortion compared to ``ultrafast'' or ``superfast'' presets, at good transcoding speed.
%%%%%%%%%%%%%%%%%%%%%%%%%%%%%%%%%%%%%%%%%%%%%%%%%%%%%%%%%%%%%%%%%%%%%%%%%%%%%%%%%%%%%%%%%%%%%%%%%%%%%%%%%%%%%%%%%%%%%%%%%%%%%%%%%%%%%%%%%%%%%%%%%%%%%%%%%%%%%%%%%%%%%%%%%%%%%%%%%%%%%%%%%%%%%%%%
\subsubsection{Evaluation Results}
\label{sec:EdgeSDN:CSDN:EvaluationResult}
As mentioned in Section~\ref{sec:EdgeSDN:CSDN:System Model}, gathering different information items (in the data collecting interval) and running the MILP model (in the optimization interval) to determine the optimal servers/segments are done in a time-slotted manner. An inaccurate time slot value imposes an additional delay in the client's request-response interval (due to an overly long time slot duration in the VRP) and may convey an incorrect network situation to the client; consequently, the client may request a lower quality level for the next segment. Therefore, in the first experiment, we investigate various time slot durations for two-second segment durations in order to determine suitable values of $\theta$. 

\begin{figure}[!t]
	\centering
	\includegraphics[width=1\linewidth]{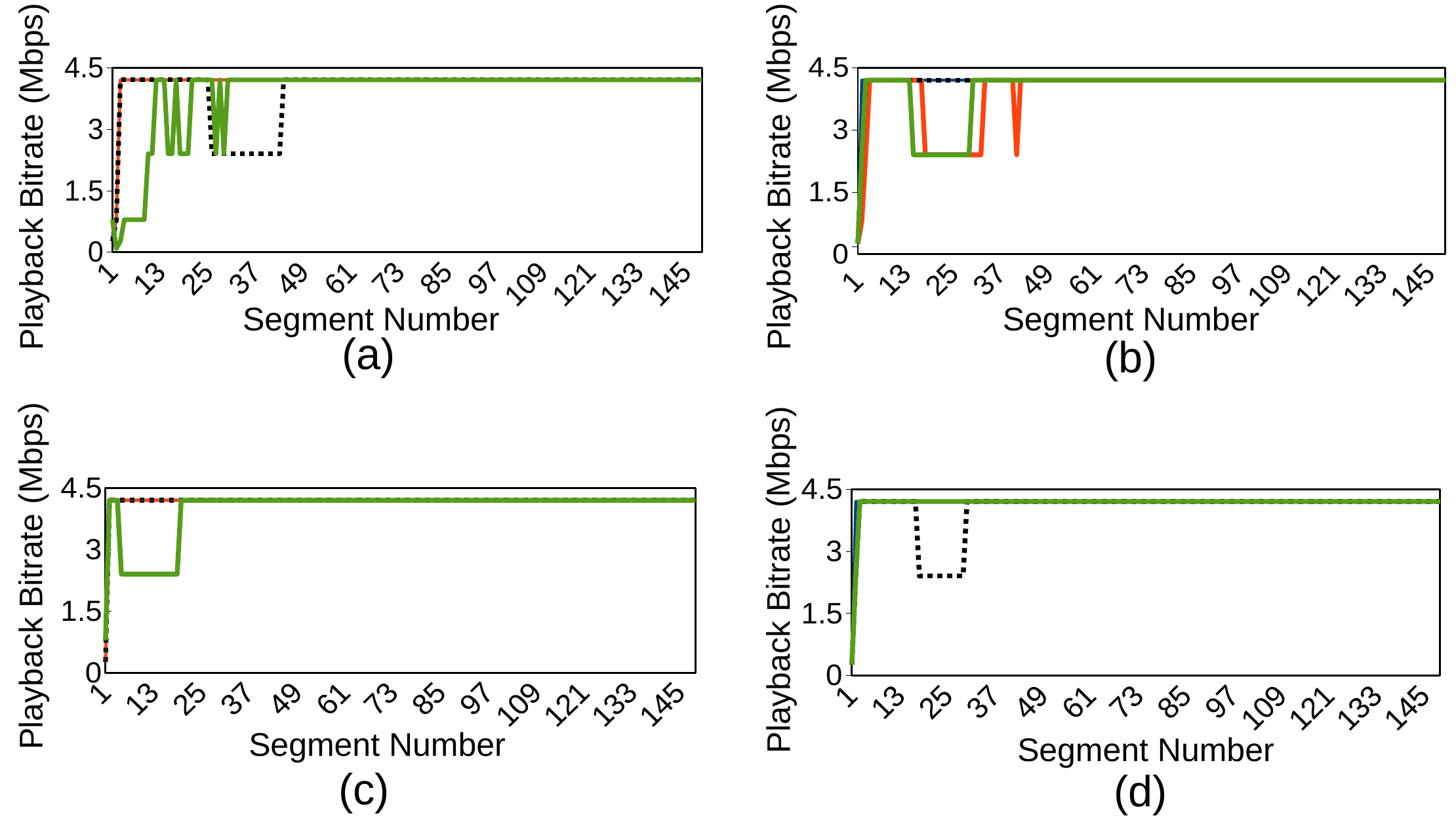}
	\caption{\small Impact of changing $\theta$ on playback bitrate: (a)~$\theta$ = 0.5s, (b)~$\theta$ = 50 milliseconds, (c)~$\theta$ = 32 milliseconds, and  (d)~$\theta$ = 0.}
        \vspace{.5cm}
	\label{timeslotDuration}
\end{figure}

We connect four clients to a server with limited bandwidth of 20 Mbps and start the experiment with the initial value of $\theta = 2$ sec and gradually decrease it in each run by 10 milliseconds. When we set $\theta = 0$, all four clients request the highest quality levels (\ie 4 x 4.2 Mbps) and receive the highest qualities (4 x 4.2 Mbps) from the server with 20 Mbps bandwidth.
The results show that clients start to request lower quality levels for time slot durations longer than 32 milliseconds (\eg the results for four $\theta$ values are shown in Fig. \ref{timeslotDuration}). Repeating the experiment shows that these values are 70 and 165 milliseconds for 4 and 6 sec.~segments, respectively. We also studied the time slot duration in more complex settings (like the one used for the experiments, Fig.~\ref{CSDN-Testbed}), with the same results. Let us consider the worst-case and imagine that clients' requests arrive at the beginning of the optimization interval; then, they will wait to be processed in the next optimization interval. In other words, a request could wait for two optimization intervals plus a data collecting interval. To avoid optimization interval overlapping, similar to \texttt{ES-HAS}, we assume the optimization interval should be less than or equal to $\frac{\theta}{2}$. Therefore, based on the first experiment, the optimization interval should be shorter than 16, 35, and 82 milliseconds for 2, 4, and 6 sec.~segments, respectively.  

In the second experiment, similar to \texttt{ES-HAS} evaluation strategy, we numerically study the scalability of the proposed MILP model. For that, we increase the number of clients' requests gradually (by four requests) and measure the proposed MILP execution times on the same server that runs the VRP's modules. 
The results show that \texttt{CSDN} MILP model can handle about 13, 35, and 72 different requests at the discussed optimization times (\ie $\frac{\theta}{2}$) for 2-, 4-, and 6-sec.~segments, respectively, in a topology including four cache servers. This is less than in \texttt{ES-HAS}, which could process about 60, 100, and 210 requests per optimization interval in the same condition. Now, let us calculate the maximum number of clients that can be handled by a VRP in the aforementioned topology. Each client sends 150 requests to download 150 segments with 2 seconds in our experiments. With a time slot duration of 32 milliseconds, we have 9375 ($\frac{300}{0.032}$) time slots in each experiment. Assuming a uniform distribution, each client sends a request in a given time slot with a probability of 0.016 ($\frac{150}{9375}$). Assuming that up to 13 distinct requests can be handled by the \texttt{CSDN} MILP model in each time slot, one VRP can serve up to a maximum of 812 clients ($812\leq\frac{13}{0.016}$); this number is 2300 in the \texttt{ES-HAS} approach. Utilizing transcoding-based actions leads to an increase of the MILP model execution time; in conjunction with different time slot durations, this decreases the number of managed requests/clients in \texttt{CSDN} as compared to \texttt{ES-HAS}. 
%%%%%%
\begin{figure}[!t]
	\centering
	\includegraphics[width=1\linewidth]{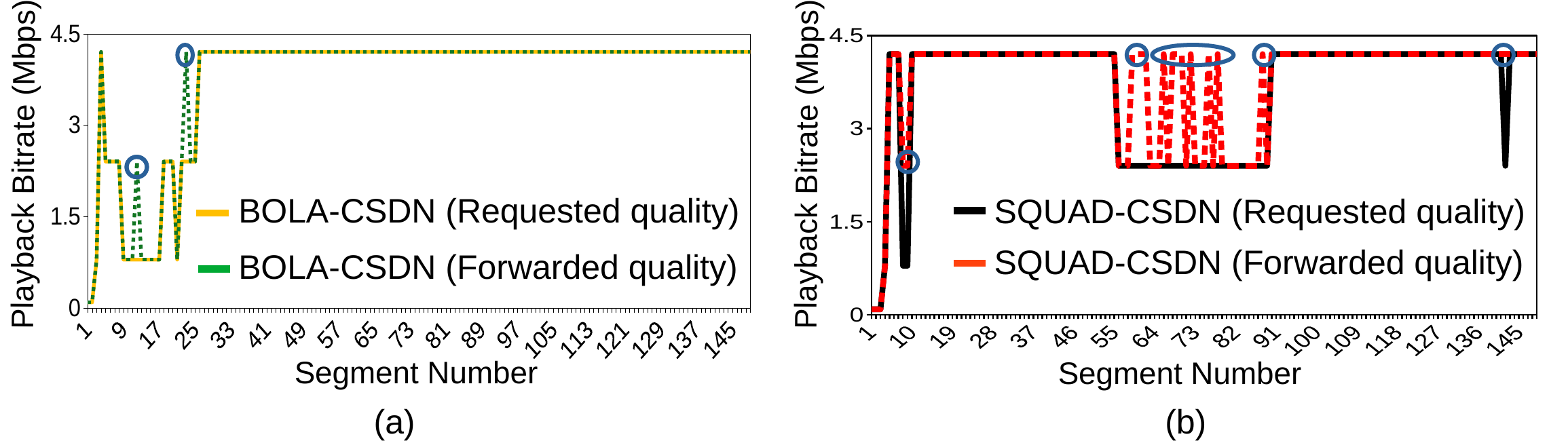}
	\caption{\small Requested quality levels by (a) CSDN-BOLA and (b) CSDN-SQUAD vs. forwarded quality levels for client 50.}
        \vspace{.5cm}
	\label{replacementQuality}
\end{figure}
%%%%%%

In the next experiment, we study the reactions of clients' ABR algorithms when replacement qualities overwrite their decisions. To do that, we set $\alpha$, $\beta$, and $m$ to 0.9, 0.1, and 4 for all clients to permit the MILP model transmitting replacement quality levels in case of cache misses. 
All events in the system, \eg requested qualities, forwarded qualities, selected servers are logged by the VRP. As shown in Figure~\ref{replacementQuality}, the requested quality levels of the BOLA and SQUAD algorithms are overwritten multiple times by replacement qualities (depicted by blue circles); however, the overall playback quality of subsequent segments is not significantly damaged. Other clients have a similar reaction to the replacement quality. 
%%%%%%
\begin{figure}[!t]
\centering
\includegraphics[width=1\linewidth]{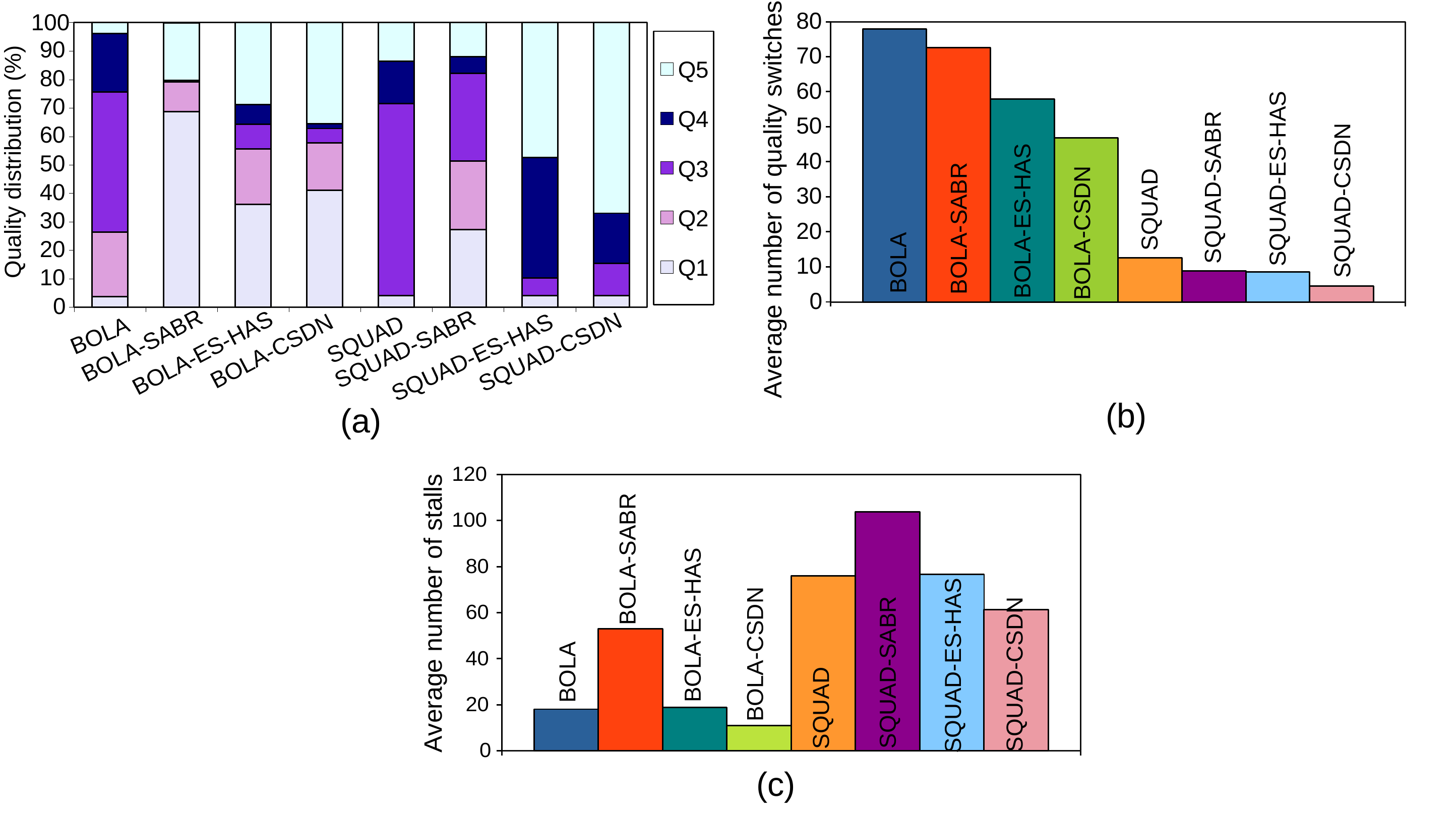}
\caption{\small (a) Average number of downloaded video quality levels, (b) average number of quality switches, and (c) average number of stalls for pure client-based ABR algorithms, SABR, ES-HAS, and CSDN for 100 clients.}
\vspace{.5cm}
\label{QS-Stall-Dist}	
\end{figure}
%%%%%%%

\begin{figure}[!t]
    \centering
    \includegraphics[width=1\linewidth]{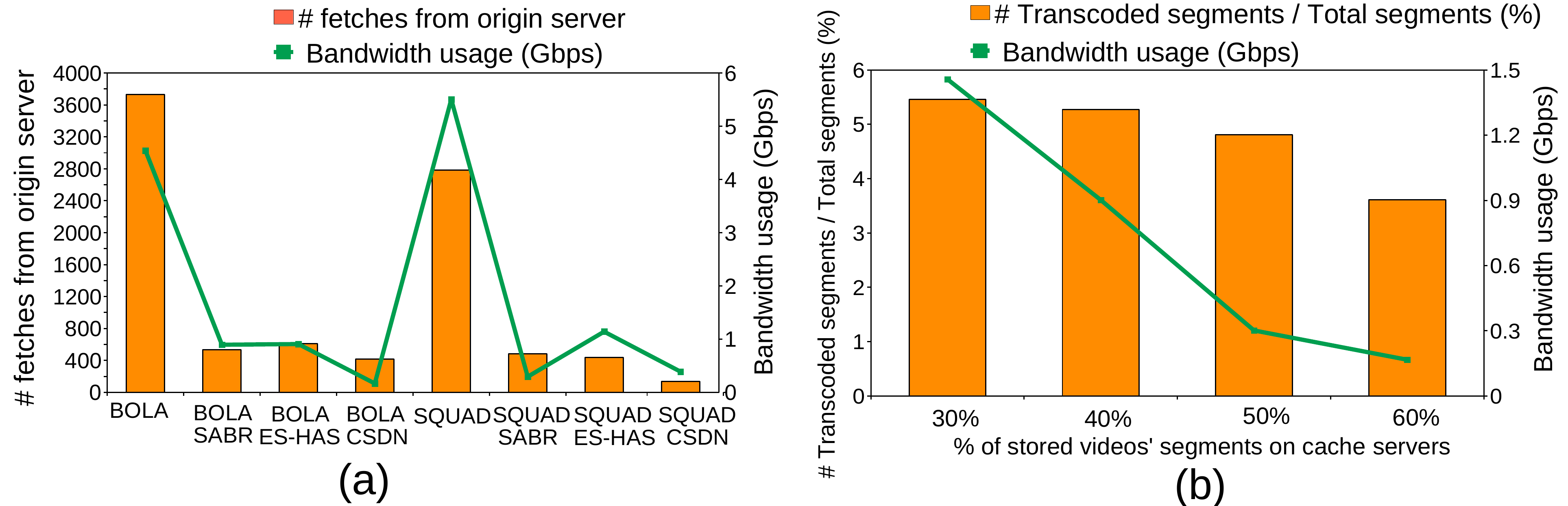}
    \caption{\small (a) Network backhaul bandwidth usage and number of fetches from the origin server for different approaches, and (b) network backhaul bandwidth usage and number of transcoded segments for different percentages of preloaded segments on cache servers for BOLA-CSDN for 100 clients.}
    \vspace{.5cm}
    \label{Bandwidth1}
\end{figure}

In another scenario, we compare the performance of \texttt{CSDN} with the pure client-based, SABR~\cite{bhat2017network}, and \texttt{ES-HAS} (Section~\ref{chap:EdgeSDN:ES-HAS}) approaches, when 60\% of the video dataset is uploaded on cache servers randomly. In the pure client-based approach, video players in the four client groups I to IV directly send their requests to their preset cache server (\ie cache server I for client group I, cache server II for client group II,  and so on). Because SABR employed different components than we have in our testbed (\eg the SDN controller, the monitoring module), we implemented the SABR system and reproduced its results without using its available source code. The \texttt{ES-HAS} model's parameters (\ie $\alpha_1$, $\alpha_2$, $\alpha_3$, and $m$) and the \texttt{CSDN} model's parameters ($\alpha$, $\beta$, and $m$) are set to (0.9, 0.1, 0, 4) and (0.9, 0.1, 4), respectively, to prioritize serving time and let both systems use the replacement quality levels on cache misses. 
Figures~\ref{QS-Stall-Dist} and \ref{Bandwidth1}(a) depict that the \texttt{CSDN} framework (for both ABR algorithms) outperforms the other methods in terms of the average bitrate distribution (Fig.~\ref{QS-Stall-Dist}(a)), the average number of quality switches (Fig.~\ref{QS-Stall-Dist}(b)), the average number of stalls (Fig.~\ref{QS-Stall-Dist}(c)), the number of fetches from the origin server, and backhaul bandwidth (Fig.~\ref{Bandwidth1}(a)).
In fact, \texttt{CSDN} outperforms the pure client-based method by fetching demanded segments from the cache server with the shortest serving time. Moreover, our proposed approach improves over SABR's results by \textit{(i)} identifying identical client requests (video/segment/quality),  \textit{(ii)} sending only one request instead of several requests to the selected cache server, and \textit{(iii)} applying an optimization-based approach for the server/segment selection policy. Employing an expanded action list (\ie transcoding-based actions besides fetch-based actions) and deciding accurately about server/segment selection helps \texttt{CSDN} fetch the requested segments from \textit{(i)} a cache server with higher available bandwidth (effective on reducing the number of fetches from the origin server, backhaul bandwidth usage, and the number of stalls), and \textit{(ii)} with minimum deviation (effective on reducing the number of quality switches), consequently improving over the \texttt{ES-HAS} results. 
%%%%%%%%%
%%%%%%%%

In the next scenario, we upload different amounts of the videos' segments (\ie from 30\% to 60\%) on cache servers randomly and measure the backhaul bandwidth usage and the number of transcoded segments in the \texttt{CSDN} system. The results for BOLA-\texttt{CSDN} (Fig.~\ref{Bandwidth1}(b)) show that, when cache servers contain a small number of video sequences/segments, the number of fetches from the origin server (to VRPs or cache servers) and the backhaul bandwidth consumption increase; moreover, \texttt{CSDN} needs to apply more transcoding actions to save backhaul traffic. However, less content distribution in the caches causes the number of transcoding not to increase significantly. An in-depth analysis shows that \texttt{CSDN} consumes less network bandwidth than other frameworks even with less content in the caches. For instance, \texttt{CSDN} consumes less than 900 Mbps of backhaul bandwidth by using only 40\% of cache storage~(Fig.~\ref{Bandwidth1}(b)), which is almost equal to the \texttt{ES-HAS} bandwidth consumption with 60\% of content in the caches (see BOLA-ES-HAS in Fig.~\ref{Bandwidth1}(a)). 
%%%%%%%%%
\begin{figure}[!t]
\centering
\includegraphics[width=1\textwidth]{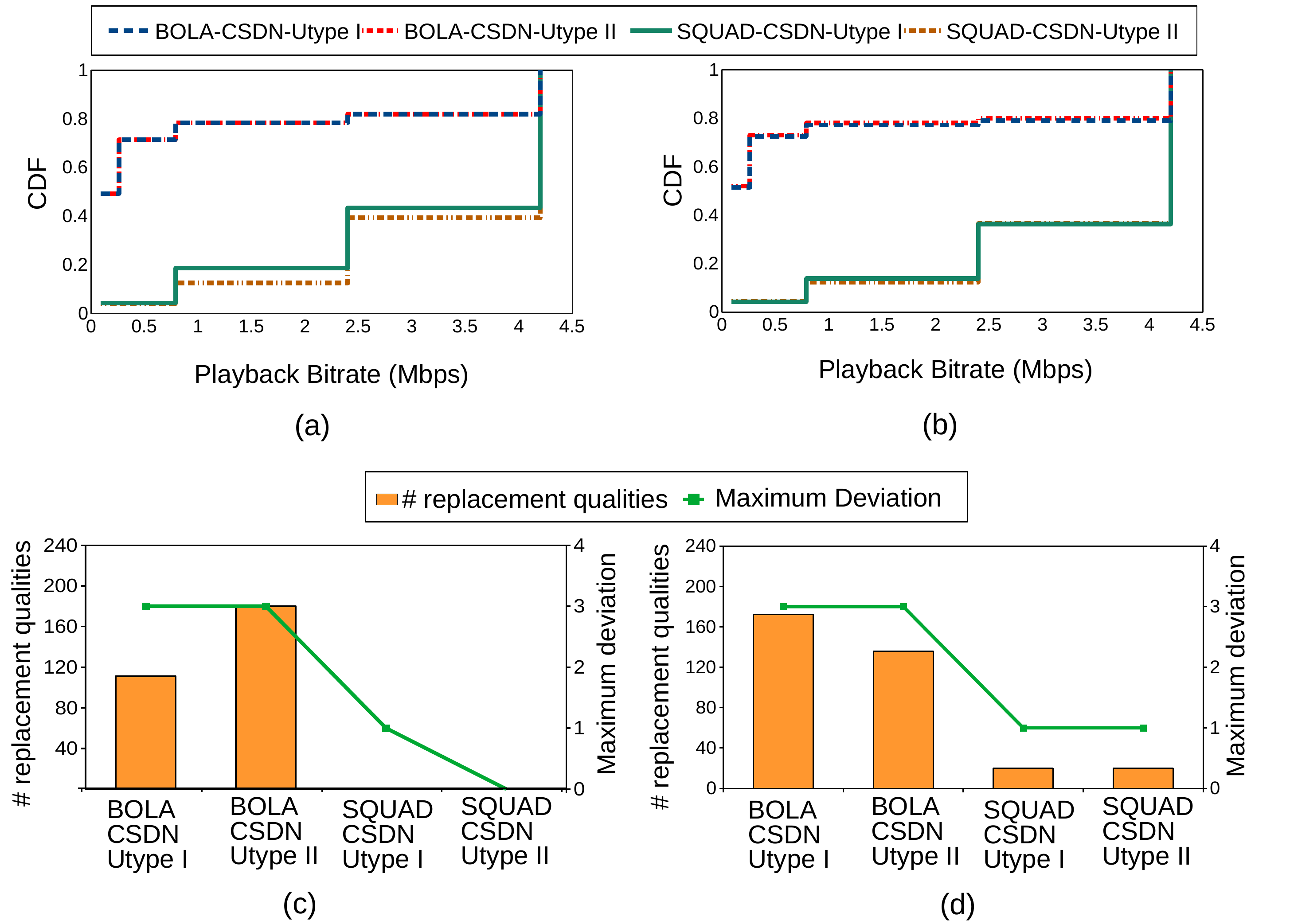}
\caption{\small CDF of average playback bitrate (a--b) and number of replacement quality levels and maximum quality deviation (c--d) for two types of users (I, II); parameters in (a) and (c): type I: $\alpha$=0.9, $\beta$=0.1; type II: $\alpha$=0.1, $\beta$=0.9; parameters in (b) and (d): type I: $\alpha$=0.1, $\beta$=0.9 and type II: $\alpha$=0.9, $\beta$=0.1.}
\vspace{.5cm}
\label{CDF-Diffusers}	
\end{figure} 
%%%%%%%%%

As a final scenario, we investigate the behavior of the \texttt{CSDN} system for two different types of users. For that, we set various values of $\alpha$ and $\beta$ as serving time and quality level deviation coefficients, respectively, to create two types of users. For instance, by setting ($\alpha_c=0.9$ and $\beta_c=0.1$) for users' type I, and ($\alpha_c=0.1$ and $\beta_c=0.9$) for users' type II, \texttt{CSDN}  serves their requests based on serving times, and quality level deviations, respectively. The results for user types I and II, including 40 and 60 clients, respectively, are illustrated in Figure~\ref{CDF-Diffusers}. As shown, SQUAD-\texttt{CSDN} has better performance for both user types compared to BOLA-\texttt{CSDN} in terms of playback bitrate, the number of replacement qualities, and maximum quality deviation. Moreover, the results show prioritizing $\beta$ does not necessarily force the model to fetch fewer replacement quality levels; in fact, the replacement quality can be used for transcoding actions too.
%%%%%%%%%
\section{Summary}
\label{chap:EdgeSDN:Conclusion}
This chapter proposed the \texttt{ES-HAS} framework as an edge- and SDN-assisted framework for HAS clients. In an \texttt{ES-HAS}-enabled system, cache server occupancy, available bandwidth values, and requested video quality levels are considered as feasible and useful information from the CDN/SDN and HAS clients. Virtualized Reverse Proxy (VRP) servers at the edge of the \texttt{ES-HAS} system are introduced to assist clients in receiving requested quality levels from the cache servers with the shortest fetching time. In case of a cache miss, a client's request is served by a replacement quality (only better quality levels with minimum deviation) from a cache server, or by the originally requested quality level from the origin server. This goal is achieved through a comprehensive network view provided by the SDN controller, collecting relevant information from the CDN (cache server occupancy) and from clients (requested video qualities) in a time-slotted manner. 
We formulated our problem as an MILP model considering resource constraints. We designed a midscale testbed, including 60 HAS clients and ran several experiments to evaluate the \texttt{ES-HAS} performance through QoE parameters and compared the obtained results with another state-of-the-art approach~\cite{bhat2017network} and two client-based schemes~\cite{spiteri2016bola,wang2016squad}. The experimental results showed that, although the SABR state-of-the-art system~\cite{bhat2017network} and \texttt{ES-HAS} performed (almost) identically in terms of the number of quality switches, \texttt{ES-HAS} outperformed SABR in terms of playback bitrate (by at least 70\%) and the number of stalls (by at least 40\%).

We equipped the \texttt{ES-HAS} VRPs with transcoding capability and introduced a CDN-Aware QoE Optimization in SDN-Assisted HTTP Adaptive Video Streaming framework called \texttt{CSDN}. We also introduced a new server/segment selection policy and formulated it as an MILP model to serve clients' requests from servers with an appropriate fetching time. This MILP model was enabled to determine a replacement quality (with the minimum quality deviation) instead of the originally requested quality or to decide on transcoding the requested quality from an available higher quality at the edge in case that quality is not available in the cache servers. We scaled up the \texttt{ES-HAS} architecture and deployed \texttt{CSDN} on a testbed, including 100 clients and four edge devices. The obtained results demonstrated that \texttt{CSDN} not only \textit{(i)} improved the users' QoE, \eg the overall video playback bitrate (7.5\%), the number of quality switches (19\%), or the number of stalls (19\%) at the client, but also \textit{(ii)} reduced backhaul bandwidth usage (63\%) compared to the SABR and \texttt{ES-HAS} systems.

%% file: Chapters/Chapter4/4-2-SARENA.tex
%************************************************
% \singlespacing
\chapter{SFC-Enabled Architecture for HAS}\label{chap:SFCEnabled}
%************************************************
\doublespacing
\vspace{-1cm}
\textit{5G} and \textit{6G} networks are expected to support various novel emerging adaptive video streaming services (\eg live, VoD, immersive media, and online gaming) with versatile QoE  requirements such as high bitrate, low latency, and sufficient reliability. It is widely agreed that these requirements can be satisfied by adopting emerging networking paradigms like SDN, NFV, and MEC. Previous studies have leveraged these paradigms to present NAVS frameworks, but mostly in isolation without devising chains of VNFs that consider the QoE requirements of \textit{various} types of multimedia services (MSs). To bridge the aforementioned gaps, in this chapter, we first introduce a set of multimedia VNFs, \ie \textit{Virtual Proxy Function} (VPF), \textit{Virtual Transcoding Function} (VTF), and \textit{Virtual Cache Function} (VCF) at the edge of an SDN-enabled network. We also form diverse \textit{Service Function Chains} (SFCs) based on the QoE requirements of different MSs. We then propose \texttt{SARENA} as an SFC-enabled architecture for adaptive video streaming applications. Next, we formulate the problem as a central scheduling optimization model executed at the SDN controller. We also present a lightweight heuristic solution consisting of \textit{two} phases that run on the SDN controller and edge servers to alleviate the time complexity of the optimization model in \rf{large-scale} scenarios. Finally, we design a large-scale cloud-based testbed including 250 HAS players requesting two popular MS applications (\ie live and VoD), conduct various experiments, and compare \texttt{SARENA}'s effectiveness with baseline systems. Experimental results illustrate that \texttt{SARENA} outperforms the baseline schemes in terms of users' QoE and network utilization in both MSs.\\
\noindent The contribution of this chapter, \ie ``\texttt{SARENA}: SFC-Enabled Architecture for Adaptive Video Streaming Applications'' has been accepted for publication at the IEEE International Conference on Communications (ICC) 2023~\cite{farahani2023sarena}.
%%%%%%%%%
\section{SARENA Framework}\label{chap:SFCEnabled:SARENA}
This section introduces \texttt{SARENA}~\cite{farahani2023sarena} as an SFC-enabled architecture for adaptive video streaming applications. We explain the details of the \texttt{SARENA} architecture and problem formulation in Section~\ref{sec:SARENADesign}. We then propose a lightweight heuristic method to solve the problem in Section~\ref{sec:SARENADesign:Heuristic}. The \texttt{SARENA} evaluation setup, methods, metrics, and obtained results are described in Section~\ref{sec:SFC:Performance Evaluation}.
%%%%%%%%%%%%%%%%%%%%%%%%%%%%%
\subsection{SARENA System Design}
\label{sec:SARENADesign}
In this section, we first delve into the \texttt{SARENA} architecture and then model the problem as an MILP optimization problem.
%%%%%%%%%%%%%%%%%%%%%%%%%%%%%
\subsubsection{SARENA Architecture}
\label{sec:SARENADesign:arch}
\texttt{SARENA} consists of three network layers as shown in Fig.~\ref{SARENA-arch}. These layers are:
\begin{enumerate}[noitemsep]
\item\textbf{Edge Layer (EL).} This layer includes edge servers close to base stations (\eg gNodeB in 5G) and clients. Inspired by the \textit{Consumer Technology Association} CTA-5004 standard~\cite{CTA}, each edge server periodically communicates with the SDN controller and shares its available computational resources' status, \eg CPU and RAM (in so-called \textit{comp stat} messages), the number of connected clients (\textit{client stat} messages), the number/type of multimedia service requests (\textit{MS stat} messages), and the cache occupancy (in \textit{edge stat} messages) with the SDN controller. All edge devices are equipped with the following VNFs: \\
\tiny{\circled[text=black,fill=pink,draw=black]{\scriptsize{\textbf{1}}}}\normalsize~\textit{Virtual Proxy Function} (VPF), which enables an edge server to play the role of a gateway between its associated clients and the network, \ie receiving clients' demands for different MS services, employing determined decisions by the SDN controller, and responding to them. \\
\tiny{\circled[text=black,fill=cyan!18!white,draw=black,draw=black]{\scriptsize{\textbf{2}}}}\normalsize~\textit{Virtual Cache Function} (VCF), utilized to store a limited number of popular segments for different MS services.\\
\tiny{\circled[text=black,fill=green!10!white,draw=black]{\scriptsize{\textbf{3}}}}\normalsize~\textit{Virtual Transcoding Function} (VTF), responsible for using computational resources and transcoding high quality segments to the required lower quality levels. These functions are \textit{placed} at the edge can collaborate and/or use other layer services to build SFCs and serve clients with desired service requirements.
%%%%%%%%%%%%%
\begin{figure}[!t]
	\centering
	\includegraphics[width=.7\textwidth]{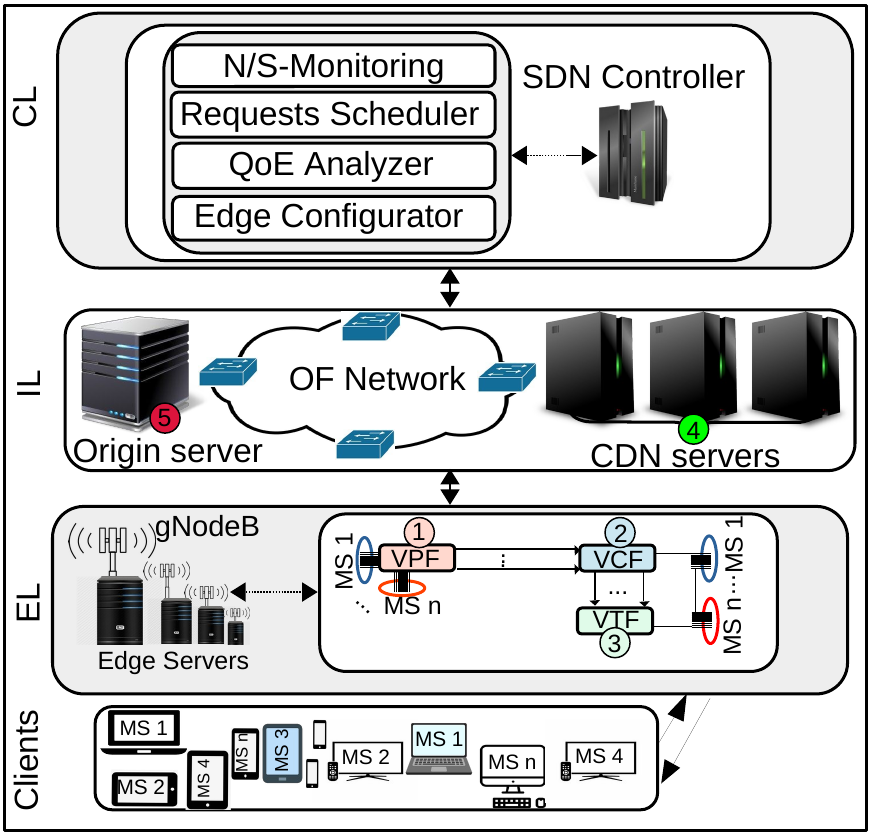}
	\caption{\small Proposed SARENA architecture.}
        \vspace{.5cm}
	\label{SARENA-arch}
\end{figure}
%%%%%%%%%%%%%
\item\textbf{Infrastructure Layer (IL).} This layer consists of a group of CDN servers (either OTT servers or purchased services from CDN providers) denoted by \tiny{\circled[text=black,fill=green,draw=black]{\scriptsize{\textbf{4}}}}\normalsize; each CDN contains various parts of video sequences. Like the EL layer, CDN servers periodically inform the SDN controller about their cache occupancy through \textit{CDN stat} messages. Moreover, an origin server (shown by \tiny{\circled[text=black,fill=red,draw=black]{\scriptsize{\textbf{5}}}}\normalsize) contains all video segments in multiple representations is placed in this layer. OpenFlow (OF) switches form an SDN-enabled backbone network.
\item\textbf{Core Layer (CL).} An SDN controller is placed in this layer, as shown in Fig.~\ref{SARENA-arch}, augmented with a \textit{Network/Service (N/S) Monitoring} module to monitor the EL and IL layers and collect the \textit{stat} information about the network and MS services. 
Operating in a \textit{time-slotted fashion}, the N/S monitoring module feeds a central decision-maker optimization module, called \textit{Requests Scheduler}. This module optimally serves different MS requests with diverse requirements invoked by all edge servers. For this purpose, the requests scheduler module must respond to the following crucial questions:
\begin{enumerate}[noitemsep]
    \item Where is the optimal place (\ie edge, CDN servers, or origin server) for fetching the content quality level requested by each client, while efficiently employing layers' available resources and satisfying service requirements (\eg service deadlines)? 
    \item How can we use the functions/services provided in the EL and IL layers to form MS function chains (SFCs)? 
    \item What is the optimal SFC for responding to the requested quality level with specific service requirements? 
\end{enumerate}    
\end{enumerate}

To respond to these questions, the requests scheduler considers all possible functions/services to make the following SFCs, then runs an \textit{optimization model} to determine the \textit{optimal node} and \textit{optimal SFC}, considering the service requirements. We will elaborate on the optimization model in more detail in Section~\ref{sec:SARENADesign:sysmodel}.
Possible SFC chains are:\\
\textbf{\textit{SFC\#1}}: Fetch the demanded quality level directly from the edge server (\tiny{\circled[text=black,fill=pink,draw=black]{\scriptsize{\textbf{1}}}}\normalsize$\rightarrow$\tiny{\circled[text=black,fill=cyan!18!white,draw=black]{\scriptsize{\textbf{2}}}}\normalsize).\\
\textbf{\textit{SFC\#2}}: Transcode the demanded quality level from a higher quality at the edge (\tiny{\circled[text=black,fill=pink,draw=black]{\scriptsize{\textbf{1}}}}\normalsize$\rightarrow$\tiny{\circled[text=black,fill=cyan!18!white,draw=black]{\scriptsize{\textbf{2}}}}\normalsize$\rightarrow$\tiny{\circled[text=black,fill=green!10!white,draw=black]{\scriptsize{\textbf{3}}}}\normalsize).\\ 
\textbf{\textit{SFC\#3}}: Fetch the requested quality level directly from the best CDN server (\ie in terms of the available bandwidth) (\tiny{\circled[text=black,fill=pink,draw=black]{\scriptsize{\textbf{1}}}}\normalsize$\rightarrow$\tiny{\circled[text=black,fill=green,draw=black]{\scriptsize{\textbf{4}}}}\normalsize).\\
\textbf{\textit{SFC\#4}}: Fetch the requested quality from the origin server (\tiny{\circled[text=black,fill=pink,draw=black]{\scriptsize{\textbf{1}}}}\normalsize$\rightarrow$\tiny{\circled[text=black,fill=red,draw=black]{\scriptsize{\textbf{5}}}}\normalsize).\\
\textbf{\textit{SFC\#5}}: Fetch a higher quality level from the best CDN server and transcode it into the demanded one at the edge (\tiny{\circled[text=black,fill=pink,draw=black]{\scriptsize{\textbf{1}}}}\normalsize$\rightarrow$\tiny{\circled[text=black,fill=green,draw=black]{\scriptsize{\textbf{4}}}}\normalsize$\rightarrow$\tiny{\circled[text=black,fill=green!10!white,draw=black]{\scriptsize{\textbf{3}}}}\normalsize). 

Notably, the SDN controller informs the VPF of edge servers about the optimal SFC and location; accordingly, edge servers serve their clients' requests from the determined server based on their service requirements. Furthermore, the SDN controller employs a \textit{QoE Analyzer} module to periodically analyze each edge server's served MS requests. Indeed, it uses information provided by the N/S monitoring and requests scheduler modules as inputs of a QoE model (see Section~\ref{sec:SFC:Performance Evaluation}), and if the calculated QoE value cannot meet the service threshold, it triggers the \textit{Edge Configurator} module to adjust the edge configurations automatically. The \textit{auto-scaling} feature utilized in the edge configurator, recently popular in both academia~\cite{aslanpour2020auto} and industry~\cite{aws-auto}, assists the edge servers in providing assured service for streaming applications requiring varying amounts of computing resources in response to dynamic client behavior over time.
%%%%table%%%%
\begin{table}[!t]
\centering
\caption {\small SARENA Notation.}
\label{tab:SARENA:notation}
\begin{tabular}{llllll}
\cline{1-2}
\multicolumn{2}{|c|}{\textbf{Input Parameters}}                                                                                  
&  &  &  &  \\ \cline{1-2}                                                                                         
\multicolumn{1}{|l|}{\begin{tabular}[c]{@{}l@{}}
$\mathcal{E}$\\ 
$\mathcal{S}$\\ 
$\mathcal{C}$\\ 
$\mathcal{R}$\\ 
$\mathcal{R}_{e}$\\  
$\mathcal{Q}^{r}$\\  \\ \\
$\mathcal{A}^{q,r}_{i}$\\ \\ \\
$\delta^{q,r}$\\ 
$\sigma^{q,r}$\\ \\
$\eta^{q,r}$\\ 
$\mu^{q,r}$\\ \\
$\pi_{e}$\\ 
$\omega_{i,e}$\\
$\theta^{r}$  \\ \end{tabular}} 
& \multicolumn{1}{l|}{\begin{tabular}[c]{@{}l@{}}
Set of $n$ edge servers \\ 
Set of $k$ cloud servers, including CDNs and an origin (i.e., $s=0$)\\
Set of SFC chains, where $c=\{1,2,3,4,5\}$\\
Set of $x$ various MS requests received by the SDN controller from $\mathcal{E}$\\
Set of various MS requests issued by edge server $e\in\mathcal{E}$\\
Set of possible quality levels for serving quality $q^*$ issued by $r\in\mathcal{R}$,\\ where $\mathcal{Q}^r=\{q^*,q^*+1,...,q^{*}_{max}\}$ and $q^{*}_{max}$ is the maximum quality\\ level for the requested segment\\
Available quality levels in edge VCF (i.e., $i\in\mathcal{E}$) or cloud servers \\(i.e., $i\in\mathcal{S}$) to serve $r\in\mathcal{R}$; $\alpha^{q,r}_{i}=1$ means $i$ hosts quality $q$ to serve \\$r\in\mathcal{R}$, otherwise  $\alpha^{q,r}_{i}=0$ \\
Segment size in quality $q$ requested by $r\in\mathcal{R}$\\
Required resources (i.e., CPU time in seconds) for transcoding quality\\ $q\in\mathcal{Q}^r$ requested by $r\in\mathcal{R}$\\
Bitrate associated to quality level $q\in\mathcal{Q}^{r}$ requested by $r\in\mathcal{R}$\\
Required time for transcoding quality $q\in\mathcal{Q}^{r}$ into the quality $q^*$\\ requested by $r\in\mathcal{R}$\\
Available computational resource (available CPU) of VTF in $e\in\mathcal{E}$\\
Available bandwidth on path between $i\in\mathcal{S}$ and $e\in\mathcal{E}$\\
Given deadline for delivering request $r\in\mathcal{R}$ based on its MS type\\
\end{tabular} }
&  &  &  &  \\ \cline{1-2}
\multicolumn{2}{|c|}{\textbf{Variables}} 
&  &  &  &  \\ \cline{1-2}                                                 
\multicolumn{1}{|l|}{\begin{tabular}[c]{@{}l@{}}
$D^{q,r}_{i,e,c}$\\ \\ \\ 
$\mathcal{P}^{q,r}_{e}$\\ \\
$\mathcal{T}^{q,r}_{i,e}$\\ \\
$\chi$
\end{tabular}}
&\multicolumn{1}{l|}{\begin{tabular}[c]{@{}l@{}}
Binary variable where $D^{q,r}_{i,e,c}=1$ indicates edge $i=e$  or $i\in\mathcal{S}$\\ serves request $r$ with quality level $q\in\mathcal{Q}^{r}$ via SFC chain $c\in\mathcal{C}$,\\ otherwise $D^{q,r}_{i,e,c}=0$\\
Required transcoding time at VTF in $e\in\mathcal{E}$ for serving $r\in\mathcal{R}$  with\\ quality level $q\in\mathcal{Q}^{r}$ \\
Required time for transmitting quality $q\in\mathcal{Q}^{r}$ in response to request\\ $r\in\mathcal{R}$ issued by $e\in\mathcal{E}$ from $i\in\mathcal{S}$ and $c\in\{3,4,5\}$\\
Total serving latency consisting of $\mathcal{P}^{q,r}_{e}$ and $\mathcal{T}^{q,r}_{i,e}$ for all MS services\\
\end{tabular}}                                                    
&  &  &  &  \\ \cline{1-2}                                                                                                              
& &  &  &  &                                              
\end{tabular}
\end{table}
%%%%table%%%%
%%%%%%%%%%%%%%%%%%%%%%%%%%%%%%%%%%%%%%%%%%%%%%%%%%%%%%%%%%%%%%%%%%%%%%%%%%%%%%%%%%%%%%%%%%%%%%%%%%%%%%%%%%%%%%%%%%%%%%%%%%%%%%%%%%%%%%%%%%%%%%%%%%%%%%%%%%%%%%%%%%%%%%%%%%%%%%%%%%%%%%%%%%%%%%%%
\clearpage
\subsubsection{SARENA Optimization Problem Formulation}
\label{sec:SARENADesign:sysmodel}
We introduce an MILP optimization model consisting of \textit{four} constraint groups: \textit{Chain Selection}, \textit{Latency Calculation}, \textit{Service/Policy}, and \textit{Resource Utilization} constraints. Table~\ref{tab:SARENA:notation} summarizes the notations used in this work.

\textbf{\textit{(i)} Chain Selection constraint.}  Let us define the binary decision-making variables $\mathcal{D}^{q,r}_{i,e,c}$ where $D^{q,r}_{i,e,c}=1$ shows edge $i=e$ or $i\in\mathcal{S}$ serves request $r$ with quality level $q\in\mathcal{Q}^{r}$ via SFC chain $c\in\mathcal{C}$, otherwise $D^{q,r}_{i,e,c}=0$. Therefore, Eq.~(\ref{SARENA:eq:1}) chooses the best SFC chain for each request $r$ issued by edge server $e\in\mathcal{E}$ by setting the $D^{q,r}_{i,e,c}$ to assure that each request is not parallelized, \ie split over multiple chains:
%%%%EQ1%%%%%%
\begin{flalign}
\label{SARENA:eq:1}
&\sum_{i\in\mathcal{S}\cup{e}}\sum_{c\in\mathcal{C}}\sum_{q\in\mathcal{Q}^{r}} \mathcal{D}^{q,r}_{i,e,c}~.~ \alpha^{q,r}_{i}=1,
&&\forall e\in\mathcal{E}, r\in \mathcal{R}_e
\end{flalign}
%%%%EQ1%%%%%%

\textbf{\textit{(ii)} Latency Calculation constraints.} Eq.~(\ref{SARENA:eq:2}) measures the transmission time $\mathcal{T}^{q,r}_{i,e}$, if cloud server $i$ in one of the SFC\#3, SFC\#4, or SFC\#5 is selected to transmit quality level $q\in\mathcal{Q}^{r}$ to edge server $e$:
%%%%EQ2%%%%%%
\begin{flalign}
\label{SARENA:eq:2}
&\sum_{c\in\mathcal{C}} \mathcal{D}^{q,r}_{i,e,c}~.~\delta^{q,r} \leq \mathcal{T}^{q,r}_{i,e}~.~\omega_{i,e},
&&\hspace{-1cm}\forall r\in\mathcal{R}_e, q\in\mathcal{Q}^{r},i\neq e
\end{flalign}
%%%%EQ2%%%%%%
Furthermore, Eq.~(\ref{SARENA:eq:3}) determines the required transcoding time $\mathcal{P}^{q,r}_{e}$ at edge $e$ in case of serving the quality demanded by $r$ from a higher quality $q$ in SFC\#2 or SFC\#5:
%%%%EQ3%%%%%%
\begin{flalign}
\label{SARENA:eq:3}
&\sum_{c\in\mathcal{C}\setminus{\{1,3,4}\}}\sum_{q\in \mathcal{Q}^{r}} \mathcal{D}^{q,r}_{i,e,c}~.~\mu^{q,r}\leq \mathcal{P}^{q,r}_{e},
&\forall e\in\mathcal{E}, i\in\mathcal{S}\cup e,r\in\mathcal{R}_e
\end{flalign}
%%%%EQ3%%%%%%

\textbf{\textit{(iii)} Service/Policy constraints.} The first constraint (Eq.~(\ref{SARENA:eq:4})) of this group guarantees that the total request's serving latency for preparing the requested quality $q$ of request $r$ must respect the request service deadline (denoted by $\theta^r$):
%%%%EQ4%%%%%%
\begin{flalign}
\label{SARENA:eq:4}
&{\sum_{i\in \mathcal{S}\cup e}\sum_{q\in \mathcal{Q}^{r}} {\mathcal{T}^{q,r}_{i,e}+\mathcal{P}^{q,r}_{e}}}\leq \theta^{r}, 
&&\forall e\in\mathcal{E}, r\in\mathcal{R}_e
\end{flalign}
%%%%EQ4%%%%%%
The total services' serving latency, namely $\chi$, \ie fetching time plus transcoding time, for all requests can be expressed as the following constraints (Eq.~(\ref{SARENA:eq:5})):
%%%%EQ5%%%%%%
\begin{flalign} 
\label{SARENA:eq:5}
&\sum_{r\in \mathcal{R}}\sum_{i\in \mathcal{S}\cup e}\sum_{e\in \mathcal{E}}\sum_{q\in \mathcal{Q}^{r}} \mathcal{T}^{q,r}_{i,e}+\mathcal{P}^{q,r}_{e}\leq \chi&&
\end{flalign}
%%%%EQ5%%%%%%
Moreover, Eq.~(\ref{SARENA:eq:6}) sets the IL policy by forcing the model to fetch the exact quality $q^*$ from the origin server when the origin server (\ie $s=0$) is selected to serve $r$.
%%%Const6%%%%
\begin{flalign}
&\sum_{q\in\mathcal{Q}^{r}} \mathcal{D}^{q,r}_{i=0,e,c=4}~.~q = q^*, && \forall e\in\mathcal{E},r\in\mathcal{R}_e
\label{SARENA:eq:6}
\end{flalign}
%%%Const6%%%%

\textbf{\textit{(iv)} Resource Utilization constraints.} Eq.~(\ref{SARENA:eq:7}) ensures that the required bandwidth for transmitting quality $q$ on the link between cloud server $i$ and edge $e$ must respect the available bandwidth (denoted by $\omega_{i,e}$): 
%%%%EQ7%%%%%%
\begin{flalign}
\label{SARENA:eq:7}
&\sum_{r\in\mathcal{R}_e}\sum_{c\in\mathcal{C}}\sum_{q\in \mathcal{Q}^{r}} \mathcal{D}^{q,r}_{i,e,c}~.~\eta^{q,r} \leq \omega_{i,e},
&&\hspace{-1cm}\forall e\in\mathcal{E},i\in\mathcal{S}, i\neq e
\end{flalign}
%%%%EQ7%%%%%%
Furthermore, Eq.~(\ref{SARENA:eq:8}) restricts the maximum required processing capacity for transcoding to the available computational resource on each edge server $e$ (denoted by $\pi_{e}$):
%%%%EQ8%%%%%%
\begin{flalign}
\label{SARENA:eq:8}
&\sum_{r\in\mathcal{R}_e}\sum_{i\in \mathcal{S}\cup e}\sum_{c\in\mathcal{C}\setminus{\{1,3,4}\}}\sum_{q\in \mathcal{Q}^{r}} D^{q,r}_{i,e,c}~.~ \sigma^{q,r} \leq \pi_{e}
&&\hspace{-1cm}\forall e\in\mathcal{E}
\end{flalign}
%%%%EQ8%%%%%%

\textbf{Central Scheduling Optimization Model.} An ABR algorithm embedded in a HAS player assesses the network's bandwidth by measuring the time between sending the request to download a segment and receiving the segment's last packet. Thus, minimizing the serving latency in the optimization model directly impacts the HAS clients' performance. To this end, the model for minimizing total requests' serving latency, denoted by $\chi$, can be expressed as follows: 
%%%%%%%%%%%objective%%%%%%%%
\begin{flalign}
\textit{Minimize}&\hspace{.3cm} \chi
\label{SARENA:eq:9}\\
  s.t.&\hspace{.5cm}\text{constraints}\hspace{.5cm}\text{Eq.}(\ref{SARENA:eq:1})-\text{Eq.}(\ref{SARENA:eq:8})&&\nonumber\\
  vars.&\hspace{.5cm} \mathcal{T}^{q,r}_{i,e},\mathcal{P}^{q,r}_{e}, \chi \geq 0, D^{q,r}_{i,e,c}\in\{0,1\}\nonumber 
\end{flalign}
%%%%%%%%%%%%

By running the MILP model (Eq.~(\ref{SARENA:eq:9})), an optimal chain will be selected for each request to minimize the total serving time. However, since the MILP model is NP-hard~\cite{lewis1983michael}, it suffers from high time complexity and is impractical for large-scale scenarios. Thus, we introduce a lightweight heuristic solution in the next section to cope with the aforementioned problems.
%%%%%%%%%%%%%%%%%%%%%%%%%%%%%%%%%%%%%%%%%%%%%%%%%%%%%%%%%%%%%%%%%%%%%%%%%%%%%%%%%%%%%%%%%%%%%%%%%%%%%%%%%%%%%%%%%%%%%%%%%%%%%%%%%%%%%%%%%%%%%%%%%%%%%%%%%%%%%%%%%%%%%%%%%%%%%%%%%%%%%%%%%%%%%%%%%%%%%%%%%%%%%%%%%%%
\subsection{SARENA Lightweight Heuristic Algorithms}
\label{sec:SARENADesign:Heuristic}
This section proposes two simple lightweight algorithms distributed on the SDN controller and edge servers. Our solution aims at satisfying the constraints (\ref{SARENA:eq:1})--(\ref{SARENA:eq:8}) by reducing the responsibilities of the SDN controller in terms of MS requests' scheduling and introducing a new VNF called \textit{Virtual Scheduler Function} (VSF) deployed at the edge. The VSF function utilizes constraints (\ref{SARENA:eq:1})--(\ref{SARENA:eq:8}) in the form of a lightweight heuristic algorithm to produce a nearly-optimal solution for its local edge server instead of running a central complex optimization model for all edge servers. Although we split the central MILP model (\ref{SARENA:eq:9}) into edge VSFs, the SDN controller, as a coordinator node, still collects \textit{stats} information from the layers (\ie by N/S monitoring), advertises this information to edge servers, and analyzes served requests' QoE (\ie by the QoE analyzer) and reconfigures edge servers if their associated MS services cannot meet the defined service thresholds (\ie by the edge configurator). %based on their requirements .

Like the centralized model, the VSF-based solution works in a time-slotted manner. A time slot consists of \textit{two} intervals: \textit{(i)}~\textit{Stats/Requests Collector} (SRC) interval and \textit{(ii)}~\textit{Requests Scheduler} (RES) interval. In the SRC interval, an edge server simultaneously receives \textit{stats} information and MS requests provided by the SDN controller and the VPF function, respectively. Considering the provided data, the VSF function in the RES interval runs Alg.~\ref{SARENA:Edge_Heu} to serve MS requests by choosing suitable SFCs, \ie 
\textit{SFC\#1}: \tiny{\circled[text=black,fill=cyan!18!white,draw=black]{\scriptsize{\textbf{{2}}}}\normalsize,
\textit{SFC\#2}: \tiny{\circled[text=black,fill=cyan!18!white,draw=black]{\scriptsize{\textbf{{2}}}}\normalsize$\rightarrow$\tiny{\circled[text=black,fill=green!10!white,draw=black]{\scriptsize{\textbf{{3}}}}\normalsize,
\textit{SFC\#3}: \tiny{\circled[text=black,fill=green,draw=black]{\scriptsize{\textbf{{4}}}}\normalsize,
\textit{SFC\#4}: \tiny{\circled[text=black,fill=red,draw=black]{\scriptsize{\textbf{{5}}}}\normalsize, or
\textit{SFC\#5}: \tiny{\circled[text=black,fill=green,draw=black]{\scriptsize{\textbf{{4}}}}\normalsize$\rightarrow$\tiny{\circled[text=black,fill=green!10!white,draw=black]{\scriptsize{\textbf{{3}}}}\normalsize, and minimizes the total serving latency w.r.t.~the service requirements and objective function of Eq.~(\ref{SARENA:eq:9}). 
We present the proposed algorithms separately in Alg.~\ref{SARENA:Edge_Heu} and Alg.~\ref{SARENA:SDN_Heu}, which are deployed on the edge servers and the SDN controller, respectively. 
%%%%%%%
\subsubsection{Edge-based Scheduling Heuristic Algorithm} 
As shown in Alg.~\ref{SARENA:Edge_Heu}, each edge server receives data provided by the SDN controller, \ie $stat\_info$ and information on bandwidth to CDN/origin servers (\ie $BW\_info$) in the SRC interval. The edge server calls the \textit{ExtractFeatures} function to extract some essential features of input requests, \eg MS types, requested video sequences/channels, requested qualities, request receiving time, and service deadlines (line 2). After that, based on the extracted features stored in the $features$ list, the \textit{MakeQueues} function is utilized to form different queues of requests (line 3). 
For instance, two popular MS applications, \ie VoD and live requests, are placed in separate queues based on MS type, requested live channel/VoD video IDs, and bitrate levels. 
Considering the system's current state, \ie available information on resources (\ie $stat\_info$) and queues of MS requests, the edge server in the RES interval must run multiple threads of a VSF function (one thread per MS type) to answer the questions mentioned in Section~\ref{sec:SARENADesign:arch}. 
%%%%%%%%%%%%%%%%%%%%%%%Edge_Heu%%%%%%%%%%%%%%%%%%%%%%%
	\begin{algorithm}[!t]
		\small
            \caption{\small SARENA edge server heuristic algorithm.}\label{SARENA:Edge_Heu}
		\begin{algorithmic}[1]
            \State \textbf{Input} $requests$, $stat\_info$, $BW\_info$, $on\_the\_fly$
            \State $features\leftarrow$ ExtractFeatures($requests$)
            \State $MS\_queues\leftarrow$ MakeQueues($requests,features$)
            \State *//Each $MS_m$ Thread: VSF()
            \State sort($MS\_queue_{m}, features_{m}$)
            \For{each $req$ in $MS\_queues_{m}$}
                \If{req $\in$ $on\_the\_fly$}
                    \State HoldReq($req$)
                \Else
                    \State $on\_the\_fly$.add($req$)
                    \State $SFC\_set\leftarrow$SFCDetector()
                    \State $SFC\_cost\leftarrow\text{CostFunction($SFC\_set$)}$
                    \State $opt\_SFC\leftarrow$OptimalSFC($SFC\_set$,$SFC\_cost$)
                    \State ServeRequest($req$,$opt\_SFC$)
                    \State UpdateVariables$()$
                \EndIf
            \EndFor
	   \end{algorithmic}
	\end{algorithm}
%%%%%%%%%%%%%%%%%%%%%%%Edge_Heu%%%%%%%%%%%%%%%%%%%%%%%

At the start of each VSF thread, associated MS queues are sorted based on the extracted information, like service deadline using the \textit{sort} function (line 5). 
Next, each MS request ($req$) is compared to $on\_the\_fly$ requests (\ie requests currently being served). If $req$ is in the $on\_the\_fly$ list, then it calls \textit{HoldReq} to hold the request and prevent network resource wastage and congestion (lines 7--8). Otherwise, the $on\_the\_fly$ list is updated by the $req$ to be processed (line 10). In the next step, the \textit{SFCDetector} function determines all feasible SFCs (\ie \textit{SFC\#1}--\textit{SFC\#5}) and stores them in the $SFC\_set$ list (line 11). 

Considering the objective function (\ref{SARENA:eq:9}) and defined constraints (\ref{SARENA:eq:1})--(\ref{SARENA:eq:8}), the serving latencies of the SFCs contained in $SFC\_set$ are calculated by calling the \textit{CostFunction} function and then the results are saved in $SFC\_cost$ (line 12). Next, the \textit{OptimalSFC} function calculates the minimum value (\ie serving latency) in the $SFC\_cost$ structure, retrieves its associated SFC from $SFC\_set$, and saves it as the optimal SFC (line 13). Finally, the \textit{ServeRequest} and \textit{UpdateVariables} functions are utilized to serve the clients' request with the optimal SFC and to upgrade \textit{stat} information and the $on\_the\_fly$ list, respectively (lines 14--15). Note that since more than one queue can proceed and might violate all/several resource constraints (\eg the bandwidth and/or computational limits), they are evaluated in a priority order where the queue with more requests and earlier service deadlines comes first. This process will be repeated in each RES interval until the MS session ends and all queues are served. Assume $\beta_1$, $\beta_2$, and $\beta_3$ indicate the number of MS services, number of channels/video sequences, and number of bitrates per channel/video. In the worst case, the time complexity of the multi-threaded Alg.~\ref{SARENA:Edge_Heu} employed by each edge server would be $O(\beta_1~.~\beta_2~.~\beta_3)$ in each time slot.
% %%%%%%%%%%%%%%%%%%%%%%%SDN_Heu%%%%%%%%%%%%%%%%%%%%%%%
\begin{center}
	\begin{algorithm}[!t]
		\small
            \caption{\small SARENA SDN controller heuristic algorithm.}\label{SARENA:SDN_Heu}
		\begin{algorithmic}[1]
            \While{$True$}
                \State $stat\_info, BW\_info\leftarrow$ N/S-Monitoring()
                \For{each $e$ in $\mathcal{E}$}
                    \State $MS\_QoE\_e\leftarrow QoEAnalyzer(e)$
                    \If{$MS\_QoE\_e  < MS\_QoE\_{th}$}
                        \State EdgeConfigurator($e$)
                    \EndIf
                \EndFor
                \State  Update($\mathcal{E}$)
                \State Wait ($\tau$)
            \EndWhile
	   \end{algorithmic}
	\end{algorithm}
\end{center}
% %%%%%%%%%%%%%%%%%%%%%%%SDN_Heu%%%%%%%%%%%%%%%%%%%%%%%
\subsubsection{SDN-based Management Heuristic Algorithm} 
In Alg.~\ref{SARENA:SDN_Heu}, the SDN controller uses the \textit{N/S-Monitoring} function to collect the \textit{stats} and bandwidth information from the network and servers, plus average QoE parameters for the served requests in the previous time slot (line 2). Based on the collected data, the \textit{QoE\_Analyzer} function is called to calculate the MS QoE scores for each edge server and store values in each edge $MS\_QoE\_e$ list (line 4) (see Section~\ref{sec:SFC:Performance Evaluation} for the used MS QoE model). Next, if the calculated values violate the service QoE thresholds (denoted by $MS\_QoE\_{th}$), which is adjusted by the network or video operator based on their business plans and services, the \textit{EdgeConfigurator} function is called to reconfigure the edge server (lines 5--6) and then updates all edge servers (line 9). We note that any policy on QoE analyzing and auto-scaling can be applied. This explained procedure will repeat periodically after $\tau$ seconds, where $\tau$ is the duration of the SRC interval within the while loop (line 10). The overall time complexity of Alg.~\ref{SARENA:SDN_Heu} can be given as $O(n)$, where $n$ is the number of edge servers.
%%%%%%%%%%%%%%%%%%%%%%%%%%%%%%%%%%%%%%%%%%%%%%%%%%%%%%%%%%%%%%%%%%%%%%%%%%%%%%%%%%%%%%%%%%%%%%%%%%%%%%%%%%%%%%%%%%%%%%%%%%%%%%%%%%%%%%%%%%%%%%%%%%%%%%%%%%%%%%%%%%%%%%%%%%%%%%%%%%%%%%%%%%%%%%%%
\subsection{SARENA Performance Evaluation}
\label{sec:SFC:Performance Evaluation}
In this section, we first describe \texttt{SARENA}'s evaluation setup, metrics, and methods and then compare \texttt{SARENA}'s performance against baseline solutions in various experiments.
\subsubsection{Evaluation Setup} 
To evaluate the performance of \texttt{SARENA} in a realistic large-scale environment, we instantiate our testbed on the CloudLab~\cite{ricci2014introducing} and use InternetMCI~\cite{zoo} as a real backbone network topology. Our testbed includes 280 components, \ie \textit{(i)} 250 \textit{AStream}~\cite{AStream,juluri2015sara} DASH players running the \textit{BOLA}~\cite{spiteri2016bola} ABR algorithm in headless modes (five groups of 50 peers); \textit{(ii)} five Apache HTTP servers (\ie four CDN servers with a total cache size of 40\% of the MS video datasets and an origin server, holding all MS video sequences); \textit{(iii)} 19 OpenFlow (OF) backbone switches and 45 backbone layer-2 links; \textit{(iv)} five edge servers, each of which is responsible for one group of clients and includes a partial cache size of only 5\% of the MS video sequences; and a FloodLight SDN controller. Note that each element is run on Ubuntu 18.04 LTS inside Xen virtual machines within separated Linux namespaces. 

To emulate the auto-scaling feature, a basic configuration (\ie 4 CPU cores and 6 GB RAM) of virtual machines supporting a maximum of 10 CPU cores and 16 GB RAM is assigned to all edge servers at the beginning of all experiments. Next, the edge servers' configurations scale up edge configurations by adding 2 CPU cores and 2 GB RAM whenever the SDN edge configurator module is triggered.  Although \texttt{SARENA} is independent of the caching policy, for the sake of simplicity, Least Recently Used (LRU) cache replacement policy is considered in all CDNs and VCFs. Note that the most popular MS VoD video sequences are pre-cached on the VCFs to avoid a slow startup of the system. Python 3.7 is used to implement all modules of the SDN controller and the VNFs. Moreover, the Python PuLP library with the CPLEX solver and Dockerimage~\textit{jrottenberg/ffmpeg}~\cite{ffmpeg} are employed to implement the MILP model and VTSs, respectively.

We evaluate the performance of \texttt{SARENA} through two prevalent MS services, \ie VoD and live streaming applications. ($MS\_{QoE}\_{th}$, $\theta_r$) values associated with live and VoD services are set to (4, 2\,s) and (3.5, 4\,s), respectively, to make different service requirements. We consider five live channels, where each of which plays a unique video~\cite{lederer2012dynamic} with 300\,s duration, comprising two-second segments of the following bitrate ladder \{(0.089,320p), (0.262,480p), (0.791,720p), (2.4,1080p), (4.2,1080p)\} [Mbps, resolution]. Moreover, 20 video sequences~\cite{lederer2012dynamic} with 300 seconds duration and four-second segments in seven representations \{(0.128,320p), (0.320,480p), (0.780,720p), (1.4,720p), (2.4,1080p), (3.3,1080), (3.9,1080)\} are employed for the VoD services. For simplicity, we assume each client requests only one MS type during its streaming session. 

The live channel and VoD video access probability are generated following a Zipf distribution with the skew parameter $\alpha=0.75$. The probability of an incoming request for the $i^{th}$ most popular live channel or VoD video are given as $prob(i)=\frac{1/i^{\alpha}}{\sum_{j=1}^{K}1/j^{\alpha}}$, where $K=5$ and $K=20$ for the live and VoD services, respectively. In the literature, the bandwidth value from the CDN servers to an edge server is assumed to be higher than from the origin server to an edge server.~\cite{al2019multi}. Therefore, the Linux \textit{Wondershaper} tool is employed to set 50 and 100 Mbps as a bottleneck bandwidth in different paths from the CDN and origin servers to edge servers, respectively. In addition, we use a real \textit{4G network trace}~\cite{raca2018beyond} (average bandwidth 3780 kbps and standard deviation 3190 kbps) collected on bus rides for links between clients to edge servers in all experiments. The SDN QoE analyzer module uses the \textit{ITU-T Rec. P.1203} model in mode 0~\cite{p1203}, as a standard comprehensive QoE model. The computational and bandwidth costs are set to $0.029\$$ per CPU per hour and $0.12\$$ per GB, respectively~\cite{aws-calc}.
%%%%%%%%%%%%%%%%%%%%%%%%%%%%%%%%%%%%%%%%%%%%%%%%%%%%%%%%%%%%%%%%%%%%%%%%%%%%%%%%%%%%%%%%%%%%%%%%%%%%%%%%%%%%%%%%%%%%%%%%%%%%%%%%%%%%%%%%%%%%%%%%%%%%%%%%%%%%%%%%%%%%%%%%%%%%%%%%%%%%%%%%%%%%%%%%%%%%%%%%%%%%%%%%%%%
\subsubsection{Evaluated Methods and Metrics} \label{sec:SFC:methodmetric}
Since there are no SFC-enabled frameworks supporting various HAS applications in the literature, we compare the results obtained by \texttt{SARENA} with the following baseline methods: 
\begin{enumerate}[noitemsep]
\item \textbf{CDN-assisted} (CDA): this is a regular CDN-based streaming method. 
\item \textbf{Non VNF-assisted} (NVA): edge devices without VCF and VTF functions are used to find the best CDN server with maximum available bandwidth.
\item \textbf{Non VTF-enabled} (NTE): edge servers are equipped with the VCFs, but do not support VTFs. 
\item \textbf{Non Reconfiguration-enabled} (NRE): this approach employs a simple version of \texttt{SARENA} without the SDN edge configurator module.  
\end{enumerate}
For fair comparisons, the results of all systems are derived with respect to the same settings and the same topology in our testbed. Moreover, each experiment is run 20 times, and the average and standard deviation values are reported in the experimental results. We use the following evaluation metrics to compare \texttt{SARENA} results with the above-mentioned methods: 
\begin{enumerate}[noitemsep]
\item\textbf{Average Segment Bitrate} (ASB) of all the downloaded segments.
\item\textbf{Average Number of Quality Switches} (AQS), the average number of segments whose bitrate level changed compared to the previous one.
\item\textbf{Average Stall Duration} (ASD), the average of total video freeze time of all clients.
\item\textbf{Average Number of Stalls} (ANS), the average number of rebuffering events.
\item\textbf{Average Perceived Overall QoE} (APQ) calculated by the ITU-T Rec. P.1203 model in mode 0~\cite{p1203}.
\item\textbf{Average Serving Latency} (ASL), defined as the overall time for serving all clients, including fetching latency plus transcoding latency. 
\item\textbf{Backhaul Traffic Load} (BTL), the  volume of segments downloaded from the origin server.
\item\textbf{Edge Transcoding Ratio} (ETR), the fraction of segments reconstructed by the VTFs. 
\item\textbf{Network Cost Value} (NCV), consisting of computational and bandwidth costs. 
\end{enumerate}

Note that the arrows in each metric plot~(Figs.~\ref{SARENA-results1} (a)--(f) and \ref{SARENA-results2} (a)--(c)) indicate which direction (up or down) represents a better result. \rf{We use the following method to calculate the results reported in the next section to show whether \texttt{SARENA} outperforms a specific baseline method in terms of the aforementioned metrics:}\\
\rf{$\%~increase~or~\%~decrease=(\frac{SARENA\_obtained\_metric~-~Baseline\_obtained\_metric}{Baseline\_obtained\_metric}) \times 100$}
%%%%%%%%%%%%%%%%%%%%%%%%%%%%%%%%%%%%%%%%%%%%%%%%%%%%%%%%%%%%%%%%%%%%%%%%%%%%%%%%%%%%%%%%%%%%%%%%%%%%%%%%%%%%%%%%%%%%%%%%%%%%%%%%%%%%%%%%%%%%%%%%%%%%%%%%%%%%%%%%%%%%%%%%%%%%%%%%%%%%%%%%%%%%%%%%%%%%%%%%%%%%%%%%%%%
\subsubsection{Evaluation Results} 
In the first scenario, we conduct experiments to assess \texttt{SARENA}'s performance in terms of the QoE metrics and compare the results with described baseline systems. As plotted in Fig.~\ref{SARENA-results1}(a--d), \texttt{SARENA} downloads VoD and live streaming segments with higher ASBs, improves AQSs and ANSs, and shortens ASDs in both VoD and live streaming applications. Although enhancing the aforementioned common QoE parameters can generally improve the users' satisfaction, the standard comprehensive QoE model is utlized by the APQ metric to evaluate overall users' QoE. Note that stalling events (measured by ASD and ANS) seriously impact APQ compared to other parameters. Utilizing the VCFs and VTFs in SFCs enables the \texttt{SARENA} system to download MS segments with improved QoE metrics, especially ASD, and leads to achieving better APQ by at least 53\% and 39.6\% for VoD and live streaming, respectively, compared to the baseline approaches (Fig.~\ref{SARENA-results1}(e)). Moreover, as illustrated in Fig.~\ref{SARENA-results1}(f), the average serving latency values (ASL) obtained by \texttt{SARENA} in VoD and live applications are reduced by at least 32.5\% and 29.3\% compared to other methods. This is because \texttt{SARENA} \textit{(i)} uses latency- and QoE-sensitive policies to meet $\theta_{r}$ and $MS\_{QoE}\_{th}$ values, \textit{(ii)} utilizes all layers of possible resources employed in SFCs, and \textit{(iii)} operates an auto-scaling policy provided by the SDN controller, for serving MS requests.  
%%%%%%%%%%%%%%
\begin{figure}[!t]
	\centering
	\includegraphics[width=.85\textwidth]{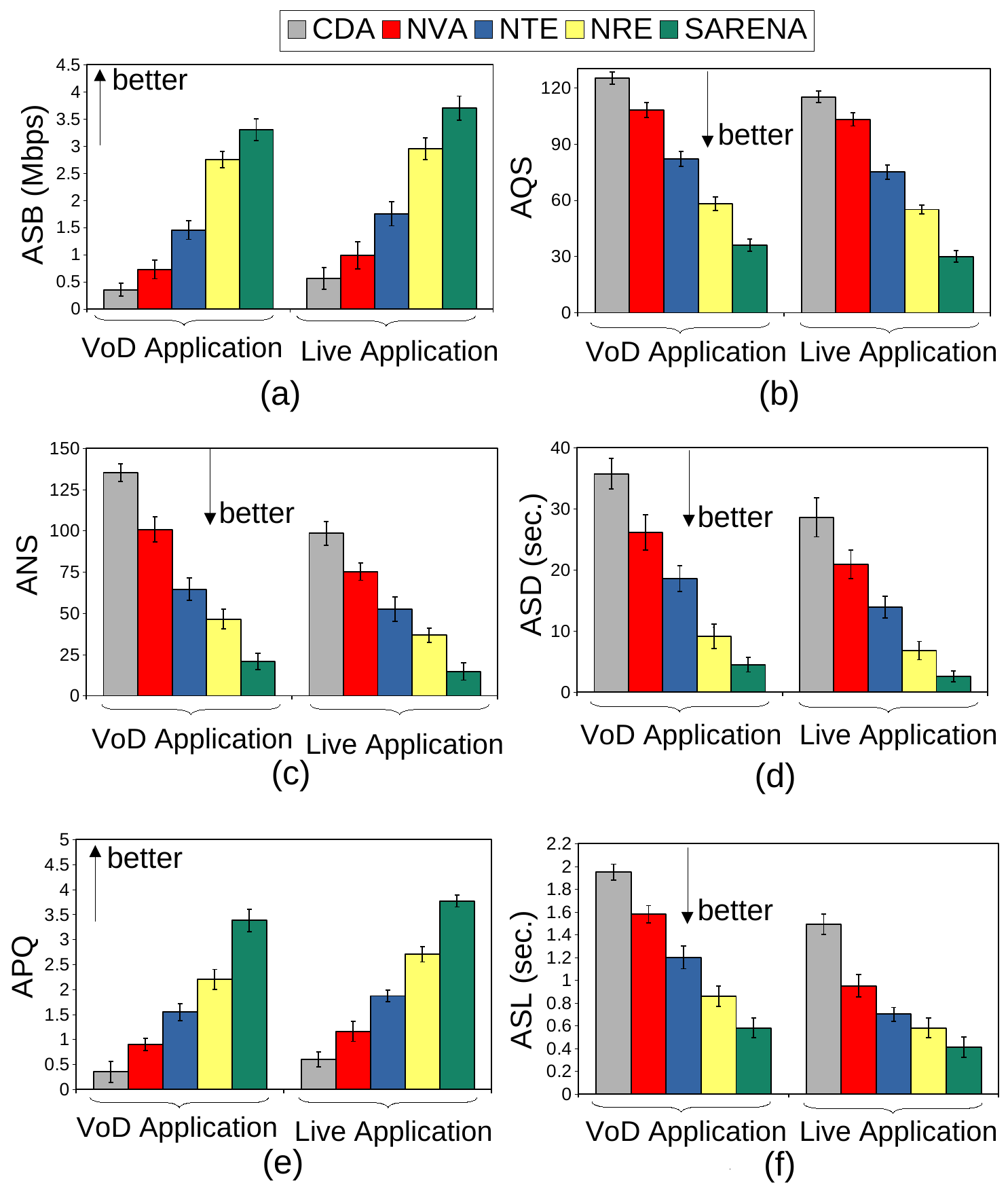}
	\caption{\small QoE results for the CDA, NVA, NTE, NRE, and SARENA systems for 250 clients.}
        \vspace{.5cm}
	\label{SARENA-results1}
\end{figure}
%%%%%%%%%%%%%%
\begin{figure}[!t]
	\centering
	\includegraphics[width=.85\textwidth]{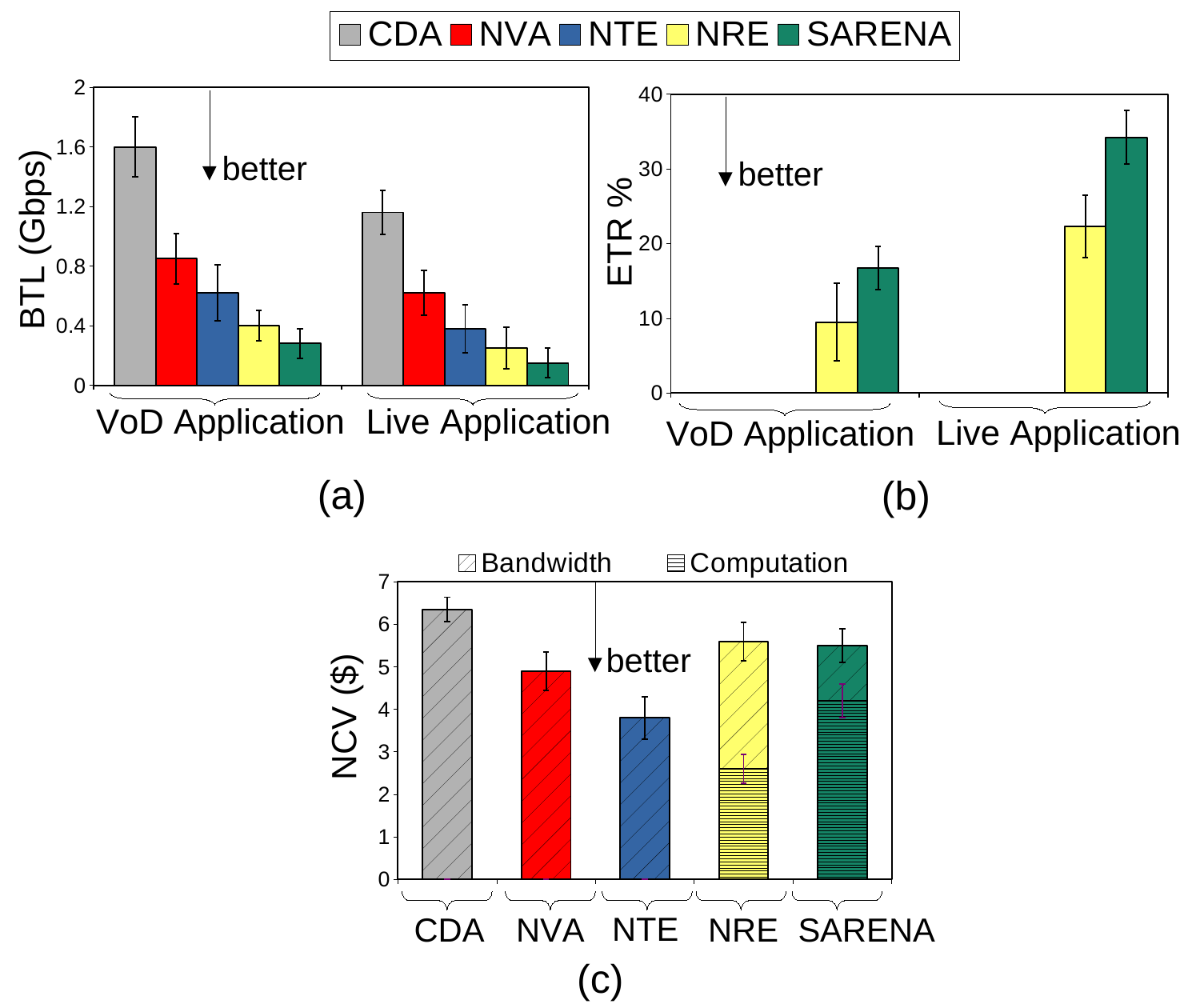}
	\caption{\small Network utilization results for the CDA, NVA, NTE, NRE, and \texttt{SARENA} systems for 250 clients.}
        \vspace{.5cm}
	\label{SARENA-results2}
\end{figure}
%%%%%%%%%%%%%%

The second scenario studies the effectiveness of \texttt{SARENA} in terms of network utilization (\ie BTL and ETR) and network cost metrics (\ie NCV). Since \texttt{SARENA} \textit{(i)} fetches requested or higher MS quality levels from the VCF or CDN servers and \textit{(ii)} transcodes requested MS quality levels by VTF, it downloads fewer segments from the origin server, consequently outperforms the CDA, NVA, and NTE systems in terms of BTL (Fig.~\ref{SARENA-results2}(a)). Moreover, the auto-scaling capability provided by the SDN controller assists \texttt{SARENA} in yielding a better BTL (40\% and 30\% for VoD and live applications, respectively) compared to the NRE system. Moreover, the ETR metric indicates that the auto-scaling feature allows \texttt{SARENA} to operate more computational resources of the edge servers to transcode MS applications, especially for serving live streaming requests, as latency-sensitive demands (Fig.~\ref{SARENA-results2}(b)). Note that the ETR metrics for not transcoding-enabled systems, \ie CDA, NVA, and NTE, are zero. Although the \texttt{SARENA} uses more edge computational resources, our final experiment (Fig~\ref{SARENA-results2}(c)) demonstrates that its NCV metric is still lower than for CDA and (almost) identical to the NRE system. This is because \texttt{SARENA} improves the costly BTL metric.
%%%%%%%%%%%
\clearpage
\section{Summary}
\label{chap:SFCEnabled:Conclusion}
In this chapter, we introduced an SFC-enabled architecture for various adaptive video streaming applications, named \texttt{SARENA}. We virtualized several multimedia functions as VNFs, designed chains of VNFs, and utilized all possible CDN and edge servers' resources for serving different multimedia service requests, \eg VoD and live video streaming, in an SDN-enabled network. We formulated the problem as a central MILP optimization model to provide a scheduling policy for various multimedia services. \texttt{SARENA}'s edge servers were augmented with lightweight scheduling functions to improve the scalability and alleviate time complexity problems of the MILP model and make it deployable in practical environments. We also utilized the SDN controller's capability to scale edge server resources based on service requirements dynamically. The performance evaluation of \texttt{SARENA} over a large-scale cloud-based testbed, including 250 HAS clients, confirmed that \texttt{SARENA} outperformed baseline systems in terms of users' QoE (by at least 39.6\%), latency (by at least 29.3\%), and network utilization (by at least 30\%) for both live and VoD services.

%% file: Chapters/Chapter5/5-1-Intro.tex
%************************************************
% \singlespacing
\chapter{Collaborative Edge-Assisted Frameworks for HAS}\label{chap:CollaborativeEdge}
%************************************************
\doublespacing
\vspace{-1cm}
Multi-Access Edge Computing (MEC) is considered one of the leading networking paradigms for designing NAVS systems by providing video processing and caching close to the end users. Despite the wide usage of this technology, designing NAVS architectures for HAS-based clients that support low-latency and high-quality video streaming, including \textit{edge collaboration}, is still a challenge. To address these issues, this chapter leverages the SDN, NFV, and MEC paradigms to propose two collaborative edge-assisted systems, named \texttt{LEADER} and \texttt{ARARAT} \rf{(\texttt{LEADER}'s extension)}, \rf{that work for live and VoD applications}. Aiming at minimizing HAS clients' serving time and delivery cost \rf{(in the \texttt{ARARAT} framework)}, besides considering available resources and all possible serving actions, we design multi-layer architectures and formulate the problems as a centralized optimization model executed by the SDN controller. However, to cope with the high time complexity of the centralized models, we introduce heuristic approaches that produce lightweight solutions through efficient collaboration between the SDN controller and edge servers. 
\rf{The proposed heuristics are equipped with fair bandwidth allocation strategies to provide sufficient bandwidth for each edge server based on its requirements.}
Finally, we implement the \texttt{LEADER} and \texttt{ARARAT} frameworks, conduct our experiments on a large-scale cloud-based testbed including 250 HAS players, and compare their effectiveness with state-of-the-art systems within comprehensive scenarios. The experimental results illustrate that the proposed solutions improve users' QoE, decrease the streaming cost, and enhance network utilization compared to their competitors. \rf{Moreover, the \texttt{ARARAT} results indicate that network fairness provided by the bandwidth allocation strategy indirectly improves users' QoE fairness.}\\
\noindent The first contribution of this chapter, \ie ``\texttt{LEADER}: A Collaborative Edge-and SDN-Assisted Framework for HTTP Adaptive Video Streaming'' was presented at the IEEE International Conference on Communications (ICC) 2022~\cite{farahani2022leader} and the second one, \ie ``\texttt{ARARAT}: A Collaborative Edge-Assisted Framework for HTTP Adaptive Video Streaming'' was published in the IEEE Transactions on Network and Service Management (TNSM) journal~\cite{farahani2022ararat}.

%% file: Chapters/Chapter5/5-2-LEADER.tex
\doublespacing
\section{LEADER Framework}\label{chap:CollaborativeEdge:LEADER}
This section presents \texttt{LEADER}~\cite{farahani2022leader}, a collaborative edge- and SDN-assisted framework for HTTP adaptive video streaming. The details of the \texttt{LEADER} architecture, action tree, and optimization MILP model are explained in Section~\ref{sec:LEADERDesign}. A distributed heuristic approach is introduced in Section~\ref{sec:LEADER:heuristic} to solve the proposed optimization model. The \texttt{LEADER} evaluation setup, methods, metrics, and results are presented in Section~\ref{sec:LEADER:PerformanceEvaluation}.
\subsection{LEADER System Design}
\label{sec:LEADERDesign}
This section explains the \texttt{LEADER} architecture and then formulates the problem as an MILP optimization model.
%%%%%%%%%%%%%%%%%%%%%%%
\subsubsection{LEADER Architecture}
\label{sec:LEADER:sysmodel}
The proposed hierarchical architecture of \texttt{LEADER}, which includes three layers, is illustrated in Fig.~\ref{LEADER-Arch}. These layers are:
\begin{enumerate}[noitemsep]
\item \textbf{Edge Layer.} Edge servers close to base stations (gNodeB) are located in this layer. On behalf of clients, each edge server periodically communicates with the SDN controller, sends the status of its available resources, \eg CPU and RAM (in so-called \textit{comp map} messages), and its cache occupancy (in \textit{edge map} messages) to the SDN controller, and responds to the clients. In our design, edge servers are categorized into \textit{Local Edge Servers} (LES) and \textit{Neighbor Edge Servers} (NES). LESs serve their clients' requests directly, while NESs can collaborate with LESs for transcoding and caching during cache miss situations or lack of processing resources. 
%%%%%%%%%%%%%%%%%%%%%%%
\begin{figure}[!t]
	\centering
	\includegraphics[width=.8\columnwidth]{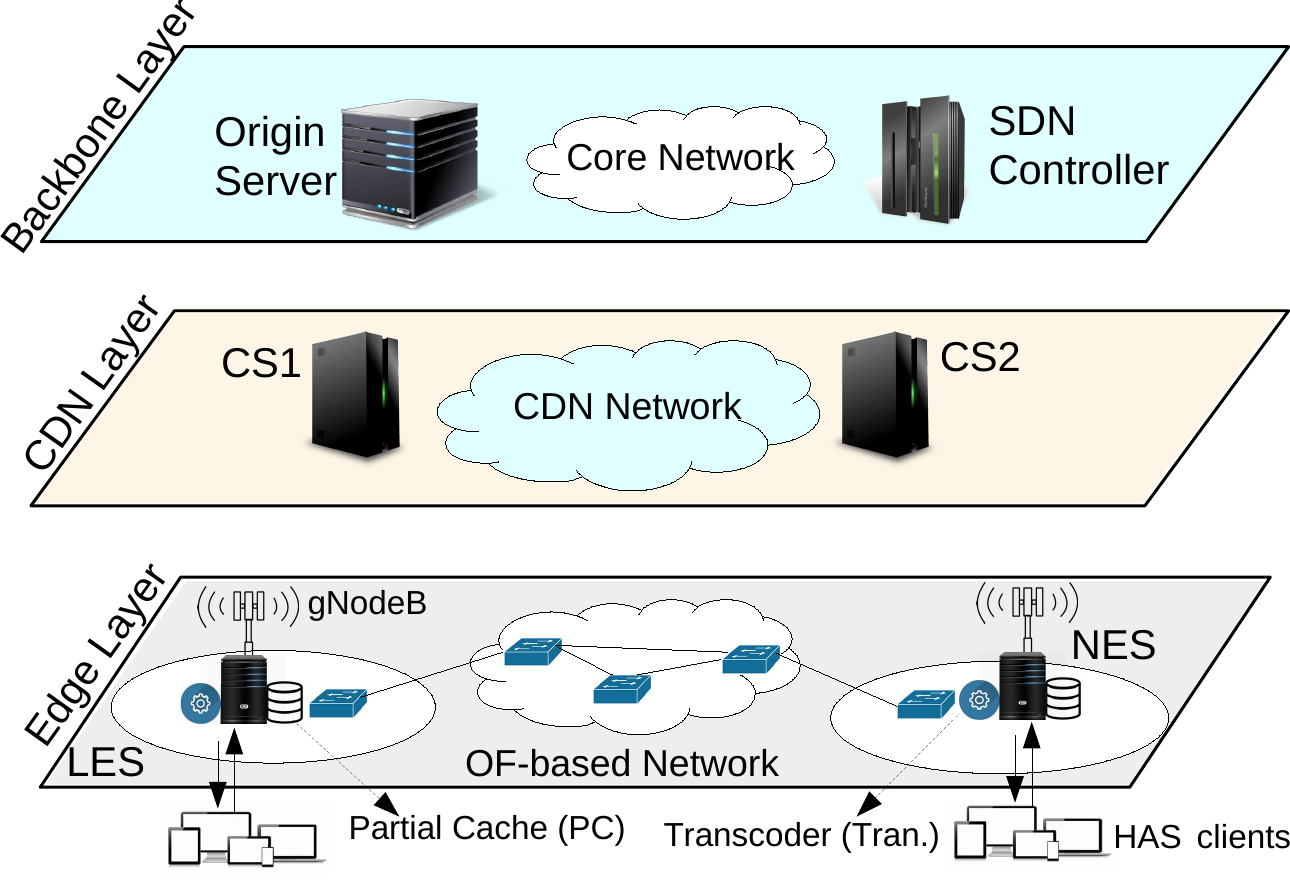}
	\caption{\small LEADER architecture.}
        \vspace{.5cm}
	\label{LEADER-Arch}
\end{figure}
%%%%%%%%%%%%%%%%%%%%%%
Each edge server includes the following components: \textit{(i)} \textit{Transcoding Module} (Tran.), which is responsible for transcoding segments' higher quality levels to the demanded quality levels; \textit{(ii)} \textit{Partial Cache} (PC), which is utilized to store a limited number of segments.
\item\textbf{CDN Layer.} The CDN Layer is constructed by multiple CDN servers (CS) where each of them contains various parts of video sequences and periodically 
informs the SDN controller about its cache occupancy through \textit{cache map} messages.
\item\textbf{Backbone Layer.} An origin server that contains all video segments in multiple representations and an SDN controller are placed in this 
layer. The SDN controller consists of the following modules:
\begin{enumerate}[noitemsep]
\item\textit{Bandwidth Monitoring Module} (BMM): It monitors the available bandwidth of different paths between each LES and other servers, \ie NESs, CSs, or the origin server. The output of the BMM will be stored in a repository called \textit{bandwidth map}. 
\item\textit{Path Selection Module} (PSM): It selects the best path with maximum available bandwidth  between each LES and NESs, CSs, or the origin server. 
\item\textit{Central Optimization Module} (COM): COM runs an MILP optimization model to respond to the following key questions:
\begin{enumerate}[noitemsep]
    \item Where is the optimal place (\ie LES, NESs, CSs, or the origin server) in terms of the minimum serving time for fetching each client's requested content quality level from? 
    \item What is the optimal approach for answering to the requested quality level, \ie fetch or transcode? 
    \item What is the optimal action to reach the requested quality? 
\end{enumerate}
\end{enumerate}
\end{enumerate}
%%%%%
\begin{figure}[!t]
	\centering
	\includegraphics[width=.6\columnwidth]{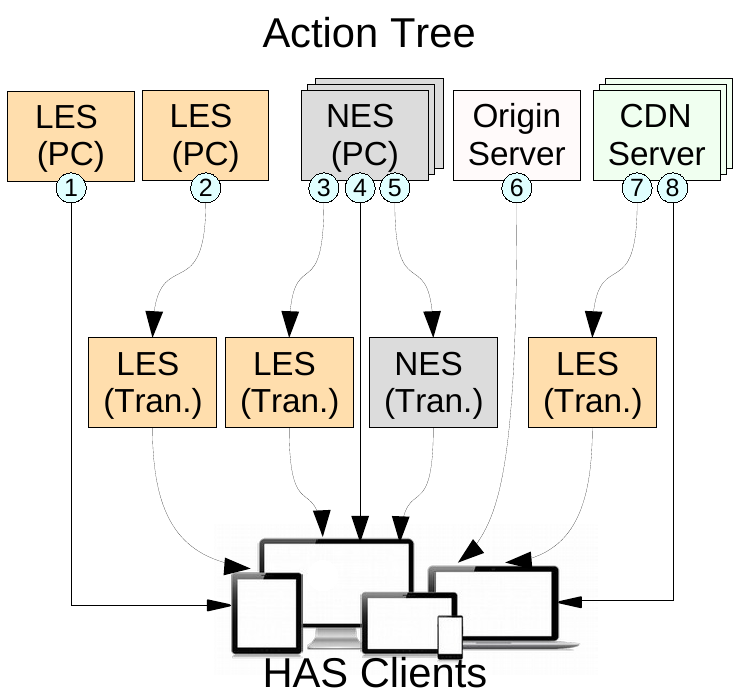}
	\caption{\small LEADER action tree.}
        \vspace{.5cm}
	\label{LEADER-tree}
\end{figure}
%%%%%%

To do that, COM utilizes the received items, \ie cache map, edge map, comp map, and bandwidth map and finds an optimal solution (\ie in terms of lowest serving time) from the \textit{Action Tree} (Fig.~\ref{LEADER-tree}): 
\begin{enumerate}[leftmargin=*,label=\protect\circledd{\arabic*},noitemsep]
\item Fetch the requested quality level directly from the LES (action $1$ in the action tree). 
\item Transcode the requested quality from a higher quality at the LES. 
\item Fetch a higher quality from the best adjacent NES and transcode it at the LES. 
\item Fetch the requested quality directly from the best NES server. 
\item Transcode the requested quality from a higher quality at the best NES and then transmit it from the NES to the LES. 
\item Fetch the requested quality from the origin server. 
\item Fetch a higher quality from the best CDN server and transcode it at the LES. 
\item Fetch the requested quality from the best CDN server. 
\end{enumerate}
Finally, the SDN controller informs edge servers about optimal actions; consequently, they can serve their clients' requests from the determined server in minimum serving time.
%%%%%%%%%%%%%%%%%%%%%%%%%%%%%%%%%%%%%%%%%%%%%%%%%%%%%%%%%%%%%%%%%%%%%%%%%%%%%%%%%%%%%%%%%%%%%%%%%%%%%%%%%%%%%%%%%%%%%%%%%%%%%%%%%%%%%%%%%%%%%%%%%%%%%%%%%%%%%%%%%%%%%%%%%%%%%%%%%%%%%%%%%%%%%%%%%%%%%%%%%%%%%%%%%%%
%%%%table%%%%%%%%%%%%%%%%%%%%%%%%%
\begin{table}[!b]
\centering
\caption {LEADER Notation.}
\label{tab:LEADER:notation}
\begin{tabular}{llllll}
\cline{1-2}
\multicolumn{2}{|c|}{\textbf{Input Parameters}}                                                                                  
&  &  &  &  \\ \cline{1-2}                                                                                         
\multicolumn{1}{|l|}{\begin{tabular}[c]{@{}l@{}}
$\mathcal{C}$\\ 
$\mathcal{V}$\\ 
$\mathcal{R}$\\ 
$R_i$\\
$\mathcal{Q}^r$\\ \\ \\
$\mathcal{T}$\\  \\ \\
$\alpha^{p,r}_{i}$\\ \\ 
$\omega_{i,j}$\\
$\delta^{r}_{p}$\\ 
$\pi^{r}_{p}$\\ \\
$\Omega_i$\\
$\mu^{r}_{p}$\\ \\
$\eta_p$\\
\end{tabular}} 
& \multicolumn{1}{l|}{\begin{tabular}[c]{@{}l@{}}
Set of $k$ CDN servers and an origin server (i.e., $c=0$)\\
Set of $n$ edge servers \\
Set of $m$ requests received by the SDN controller from $\mathcal{V}$\\
Set of requests issued by edge server $i\in\mathcal{V}$\\
Set of possible quality levels for serving quality $p^*$ requested by\\ $r\in\mathcal{R}$, where $\mathcal{Q}^r=\{p^*,p^*+1,...,p^{*}_{max}\}$ and $p^{*}_{max}$ is the maximum\\ quality level for the requested segment\\
Set of possible transcoding statuses: $\mathcal{T}=\{0,1,2\}$; $t$ is 1 or 2 if\\ the requested quality is transcoded from a higher quality $p\in\mathcal{Q}^r$\\ at the NES or at the LES, respectively; otherwise t=0\\
Available quality levels in server $i\in\mathcal{V}\cup\mathcal{C}$; $\alpha^{p,r}_{i}=1$ means server\\ $i$ hosts quality $p$ to serve $r\in\mathcal{R}$, otherwise  $\alpha^{p,r}_{i}=0$ \\
Available bandwidth on path between $i\in\mathcal{V}$ and $j\in\{\mathcal{V}\cup \mathcal{C}\} \setminus i$\\
Size of segment in quality $p$ requested by $r\in\mathcal{R}$ [bytes]\\
Computational cost (CPU usage in \%) for transcoding quality $p\in\mathcal{Q}^r$\\ into the quality requested by $r\in\mathcal{R}$\\
Available processing capacity (available CPU in \%) of $i\in \mathcal{V}$\\
Required time for transcoding quality $p\in\mathcal{Q}^r$ into the quality \\requested by $r\in\mathcal{R}$\\
Bitrate associated to quality level $p \in \mathcal{Q}^r$\\
\end{tabular} }
&  &  &  &  \\ \cline{1-2}
\multicolumn{2}{|c|}{\textbf{Variables}} 
&  &  &  &  \\ \cline{1-2}                                                 
\multicolumn{1}{|l|}{\begin{tabular}[c]{@{}l@{}}
$B^{p,r,t}_{i,j}$\\ \\ \\ 
$\tau^{r}_{i}$\\ 
$T^{p,r}_{i,j}$\\ \\
$\Gamma^{p,r}_{i,j}$\\
\end{tabular}}
&\multicolumn{1}{l|}{\begin{tabular}[c]{@{}l@{}}
Binary variable where $B^{p,r,t}_{i,j}=1$ shows server $i\in\mathcal{V}\cup\mathcal{C}$ transmits\\  quality $p\in\mathcal{Q}^r$ to server $j\in\mathcal{V}$ for request $r$ with transcoding status $t$,\\ otherwise $B^{p,r,t}_{i,j}=0$\\
Required transcoding time at server $i\in\mathcal{V}$ for serving $r\in\mathcal{R}$  \\
Required time of transmitting quality level $p\in\mathcal{Q}^r$ in response to\\ request $r\in\mathcal{R}$ from server $i\in\mathcal{V}\cup\mathcal{C}$ to $j\in\mathcal{V}$\\
Serving time consisting of $\tau^{r}_i$ and $T^{p,r}_{i,j}$\\
\end{tabular}}                                                    
&  &  &  &  \\ \cline{1-2}                                                                                                              
& &  &  &  &                                              
\end{tabular}
\end{table}
%%%%table%%%%%%%%%%%%%%%%%%%%%%%%%
\clearpage
\subsubsection{LEADER Optimization Problem Formulation}
Let set $\mathcal{C}$ consist of $k$ CDN servers and an origin server, where $c=0$ indicates the origin server. (See Table~\ref{tab:LEADER:notation} for a summary of the notation.) Moreover, let us define $\mathcal{R}$ as the set of $m$ requests received by the SDN controller from the set $\mathcal{V}$ of $n$ edge servers. The subset $R_{i}\subseteq\mathcal{R}$ shows all requests issued by edge server $i\in\mathcal{V}$. Furthermore, we assume that the SDN controller periodically receives the cache map and the edge map from cache and edge servers, respectively; thus, let $\alpha^{p,r}_{i}$ denote the availability of quality level $p$ of a segment demanded by request $r$ at cache/edge server $i$. In \texttt{LEADER}, it is assumed that the SDN controller defines a data path between two servers $i\in\mathcal{V}$ and $j\in\{\mathcal{C}\cup\mathcal{V}\} \setminus i$ where $\omega_{i,j}$ shows its available bandwidth accordingly. As mentioned earlier, when a LES requests quality $p^*$, it is possible to receive the requested quality or a higher one since all edge servers can run the transcoding function; therefore, for each request $r$, we define $\mathcal{Q}^r=\{p^*,p^*+1,...,p^{*}_{max}\}$ as the set of possible quality levels for serving requested quality $p^*$, where $p^{*}_{max}$ is the maximum quality level for the requested segment. In the following, we introduce an MILP model with four groups of constraints: \textit{Action Selection} (AS), \textit{Serving Time} (ST), \textit{CDN/Origin} (CO), and \textit{Resource Consumption} (RC) constraints.

\textbf{\textit{(i)} AS Constraint.} The first constraint determines a suitable action from the introduced action tree (Fig.~\ref{LEADER-tree}) for each request $r$ issued by LES server $j$. It selects appropriate values of the binary variable $B^{p,r,t}_{i,j}$ (see Table~\ref{tab:LEADER:notation} for the definition):
\begin{flalign}
&\sum_{i\in\mathcal{C}\cup\mathcal{V}}\sum_{t\in\mathcal{T}}\sum_{p\in\mathcal{Q}^r} B^{p,r,t}_{i,j}~.~\alpha^{p,r}_{i}
=1,&&\forall j\in\mathcal{V}, r\in R_{j}
\label{LEADER:eq:1}
\end{flalign}

\textbf{\textit{(ii)} ST Constraints.} The first ST constraint determines the required time (denoted by $T^{p,r}_{i,j}$) for transmitting  quality level $p\in\mathcal{Q}^r$ from each server $i\in\mathcal{V}\cup\mathcal{C}$ to each LES $j\in\mathcal{V}$:
%%%Const2%%%%%
\begin{flalign}
&\frac{\sum_{t\in\mathcal{T}} B^{p,r,t}_{i,j}~.~\delta^{r}_p}{\omega_{i,j}}\leq T^{p,r}_{i,j},&&\forall r\in R_j, p\in\mathcal{Q}^r,i\neq j
\label{LEADER:eq:2}
\end{flalign}
The next constraint measures the required transcoding time $\tau_{i}^{r}$ at server $i\in\mathcal{V}$ in case of serving the quality requested by $r\in R$ from a higher quality $p$ by means of transcoding:
%%%Const3%%%%%
\begin{flalign}
&\sum_{j\in \mathcal{V}}\sum_{p\in \mathcal{Q}^r} B^{p,r,t=1}_{i,j}~.~\mu^{r}_{p}\leq \tau_{i}^{r}&&\forall i\in\mathcal{V},r\in\mathcal{R}
\label{LEADER:eq:3}
\end{flalign}

\textbf{\textit{(iii)} CO Constraints.} This group of constraints forces the model to fetch the exact quality $p^*$ from the origin server when the origin server is selected to serve the LES:
%%%Const4%%%%%
\begin{flalign} 
&\sum_{t\in \mathcal{T}}\sum_{p\in\mathcal{Q}^r} B^{p,r,t}_{i=0,j}~.~p = p^*, && \forall j\in\mathcal{V},r\in R_j
\label{LEADER:eq:4}
\end{flalign}
Note that $i=0$ in Eq.~(\ref{LEADER:eq:4}) means that the origin server is serving requests. We also should prevent CSs from performing transcoding functions, expressed by the following equation:
\begin{flalign} 
&\sum_{i\in\mathcal{C}}\sum_{r\in R_j}\sum_{p\in\mathcal{Q}^r} B^{p,r,t=1}_{i,j}~.~p = 0 && \forall j\in\mathcal{V}
\label{LEADER:eq:5}
\end{flalign}

\textbf{\textit{(iv) }RC Constraints.} The first constraint in this group guarantees that the required bandwidth for transmitting segments on the link between two servers $j\in\mathcal{V}$ and $i\in\mathcal{C}\cup\mathcal{V}$ should not exceed the available bandwidth:
\begin{flalign}
&\hspace{-.1cm}\sum_{t\in \mathcal{T}}\sum_{r\in R_j}\sum_{p\in \mathcal{Q}^r} \hspace{-.1cm}B^{p,r,t}_{i,j}~.~\eta_p \leq \omega_{i,j}&& \forall j\in\mathcal{V},i\in\mathcal{C}\cup\mathcal{V}
\label{LEADER:eq:6}
\end{flalign}
Similarly, the next constraint limits the maximum required processing capacity for the transcoding operation to the available computational resource.
\begin{flalign}
&\sum_{r\in \mathcal{R}}\sum_{p\in \mathcal{Q}^r}(\sum_{i\in \mathcal{C}\cup\mathcal{V}}B^{p,r,t=2}_{i,j}+\sum_{i\in\mathcal{V} \setminus j} B^{p,r,t=1}_{j,i})~.~\pi^{r}_{p} \leq \Omega_j
&&\forall j\in\mathcal{V}&&
\label{LEADER:eq:7}
\end{flalign}

\textbf{Central MILP Optimization Model.} The model minimizes the serving time (\ie fetching time plus transcoding time), denoted by $\Gamma^{p,r}_{i,j}$, and can be formulated as follows:
\begin{flalign}
\textit{Minimize}&\hspace{0cm}\sum_{r\in \mathcal{R}}\sum_{i\in \mathcal{V}\cup\mathcal{C}}\sum_{j\in \mathcal{V}}\sum_{p\in \mathcal{Q}^r}{\Gamma^{p,r}_{i,j}}
\label{LEADER:eq:8}\\
  s.t.&\hspace{0cm}\text{constraints}\hspace{.5cm}\text{Eq.}(\ref{LEADER:eq:1})-\text{Eq.}(\ref{LEADER:eq:7})&&\nonumber\\
  &\hspace{-.5cm}T^{p,r}_{i,j}+\tau^{r}_i = \Gamma^{p,r}_{i,j}, \forall r\in\mathcal{R},i\in\mathcal{V}\cup\mathcal{C},j\in\mathcal{V},p\in\mathcal{Q}^r\nonumber\\
  vars.~&\hspace{0cm} T^{p,r}_{i,j},\tau^{r}_i, \Gamma^{p,r}_{i,j} \geq 0, B^{p,r,t}_{i,j}\in\{0,1\}\nonumber 
\end{flalign}
By running the MILP model, the SDN controller should select an optimal action for each request $r \in \mathcal{R}$ such that the total serving time is minimized.

\textbf{Preposition.} Considering shared links in Eq.~(\ref{LEADER:eq:8}) changes the model to a mixed integer non-linear programming (MINLP) model.
The proposed model~(\ref{LEADER:eq:8}) minimizes the total serving time assuming disjoint paths from edge servers in $\mathcal{V}$ to other servers in $\mathcal{V}\cup\mathcal{C}$. Let $\omega^{a,b}_{i,j}$ be an optimally allocated bandwidth between server $i$ and $j$ over the link $(a,b)$. Thus, the required time $T^{p,r}_{i,j}$ for transmitting  quality level $p\in\mathcal{Q}^r$ from server $i\in\mathcal{V}\cup\mathcal{C}$ to each LES $j\in\mathcal{V}$ for each $r\in R_j, p\in\mathcal{Q}^r$ is obtained as follows:
\begin{flalign}
\label{LEADER:eq:9}
&\frac{\sum_{t\in\mathcal{T}} B^{p,r,t}_{i,j}~.~\delta^{r}_p}{\bar{\omega}_{i,j}}\leq T^{p,r}_{i,j},&&\\
&\text{where }\bar{\omega}_{i,j}\leq\omega^{a,b}_{i,j}\hspace{1.5cm}\forall\text{ links }(a,b),i\in\mathcal{V}\cup\mathcal{C},j\in\mathcal{V}\nonumber
\end{flalign}
Although we should also apply the change to constraint (\ref{LEADER:eq:6}), replacing (\ref{LEADER:eq:2}) by (\ref{LEADER:eq:9}) turns the proposed MILP model (\ref{LEADER:eq:8}) into an MINLP model that is NP-complete~\cite{lee2011mixed} and suffers from high time complexity. In the following, we will propose a heuristic approach to remedy this time complexity problem.
%%%%%%%%%%%%%%%%%%%%%%%%%%%%%%%%%%%%%%%%%%%%%%%%%%%%%%%%%%%%%%%%%%%%%%%%%%%%%%%%%%%%%%%%%%%%%%%%%%%%%%%%%%%%%%%%%%%%%%%%%%%%%%%%%%%%%%%%%%%%%%%%%%%%%%%%%%%%%%%%%%%%%%%%%%%
\subsection{LEADER Distributed Heuristic Approach}
\label{sec:LEADER:heuristic}
Motivated by the characteristics of the objective function (\ref{LEADER:eq:8}), we propose a lightweight heuristic algorithm.
Our method aims at satisfying the proposed constraints~(\ref{LEADER:eq:1})--(\ref{LEADER:eq:7}) by considering shared links among servers in $\mathcal{C}$ and $\mathcal{V}$. First, let us define the time slot structure shown in Fig.~\ref{LEADER-TS}(a) that consists of two intervals: \textit{(i)} \textit{Edge Update} (EU) interval and \textit{(ii)} \textit{Serving Requests} (SR) interval. In the EU interval, the SDN controller sends some important resource data from the network and servers (\ie bandwidth, cache, edge, and comp maps) to the edge servers in $\mathcal{V}$; moreover, the SDN controller efficiently assigns appropriate bandwidth to each edge server on shared links. Having the provided information, edge servers in the SR interval serve their requests by selecting appropriate actions that minimize serving time according to the objective function in (\ref{LEADER:eq:8}). 
Let us consider a simple topology shown in Fig.~\ref{LEADER-TS}(b) where five data paths from each edge server (LES and NES) to the neighboring edge server, three cache servers (CS1, CS2, and CS3), and an origin server are illustrated. The paths from LES and NES to other servers are depicted by the blue and dashed red lines, respectively.

\subsubsection{SDN Controller Heuristic Algorithm}
\begin{figure}[!t]
	\centering
	\includegraphics[width=1\columnwidth]{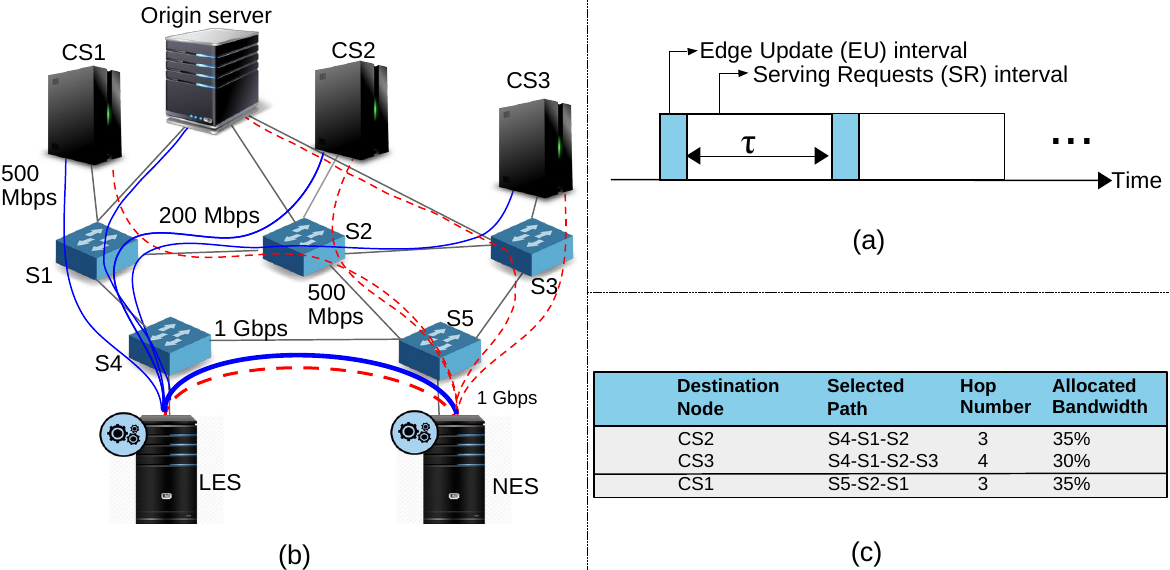}  
	\caption{\small Proposed time slot structure (a),
	and example of bandwidth allocation strategy (b-c).}
        \vspace{.5cm}
	\label{LEADER-TS}
\end{figure}

As shown in Fig.~\ref{LEADER-TS}(b), LES uses the shared link connecting OpenFlow (OF) switches S1 and S2 to reach CS2 and CS3, while NES utilizes it to fetch data from CS1. Let us follow the proposed algorithm in Alg.~\ref{LEADER:heu-SDN}. In an infinite loop, the SDN controller first collects resource data from the network and servers (line 2), it then determines one path from each edge server in $\mathcal{V}$ to each server in $\mathcal{C}$ (line 3). We note that any policy on selecting the paths can be applied. We here used the shortest path policy. 
Let $\mathcal{L}$ be the set of all links in the considered topology. To assign an appropriate bandwidth to each edge server on a shared link (lines 4--10), the SDN controller considers two main criteria: \textit{($i$)} the number of reachable cache servers and \textit{($ii$)} their hop counts.
%%%%%%%%%
%%%%%%%%%%

Suppose edge server $i\in\mathcal{V}$ is reaching server $j\in\mathcal{C}$ through link $l\in\mathcal{L}$ that is shared with a subset of edge servers denoted by $V^i_l$. Let $b^{i,j}_{l}$ be the amount of allocated bandwidth to server $i\in\mathcal{V}$ for reaching server $j\in\mathcal{C}$ through shared link $l$ among edge servers in $V^i_l$; thus, regarding the considered criteria, $b^{i,j}_{l}$ can be computed as follows:
\begin{flalign} 
&b_{l}^{i,j}= \frac{H_l - h^{i,j}_l}{\sum_{k\in V^i_l} (H_l - h^{k,j}_l)}~.~\omega_{l},&& \forall i\in \mathcal{V},j\in\mathcal{C}, l\in\mathcal{L}
\label{heu-eq}
\end{flalign}
where $H_l=\sum_{k\in {V^i_l}}\sum_{j\in \mathcal{C}} h^{k,j}_l$ and $h^{i,j}_l$ indicates the hop counts of the path containing link $l$ from $i\in\mathcal{V}$ to server $j\in\mathcal{C}$. 
Using this strategy, as shown in Fig.~\ref{LEADER-TS}(c), the bandwidth values allocated to shared link S1--S2 for LES and NES to reach CS2 and CS1, respectively, are more than the amount allocated to LES for accessing CS3.
Next, the SDN controller finds the available bandwidth in each path by determining the minimum bandwidth of used links as a bottleneck (lines 11--13), and then updates all edge servers (line 14). This described procedure will repeat after each $\tau+\xi$ seconds, where $\xi$ is the duration of the EU interval, within the while loop (line 15). Assume $n$, $s$, and $h$ indicate the number of edge servers, number of all shared links, and maximum hop counts of selected paths. The overall time complexity of Alg.~\ref{LEADER:heu-SDN} can be given as $O(s^{2}~.~n)+O(n~.~h)$.
% %%%%%%%%%%%%%%%%%%%%%%%SDN_Heu%%%%%%%%%%%%%%%%%%%%%%%
\begin{center}
	\begin{algorithm}[!t]
		\small
            \caption{\small LEADER SDN controller heuristic algorithm.}\label{LEADER:heu-SDN}
		\begin{algorithmic}[1]
            \While{$True$}
                \State UpdateSet($cache\_map$, $edge\_map$, $bw\_map$, $comp\_map$)
                \State SelectPath()
                \For{each $l$ in $\mathcal{L}$}
                    \For{each $i\in\mathcal{V}$ use $l$}
                        \For{each $j\in V^i_l$}
                             \State AllocateBW()
                        \EndFor
                    \EndFor
                \EndFor
                \For{each (LES, $\mathcal{V}\cup\mathcal{C}$, path)}
                    \State AvailableBw()
                \EndFor
                \State Update $\mathcal{V}$ in EU interval
                \State Wait ($\tau$)
            \EndWhile
        \end{algorithmic}
      \end{algorithm}          
\end{center}                    
%%%%%%%%%%%%%%%%%%%%%%%%SDN_Heu%%%%%%%%%%%%%%%%%%%%%%%
\subsubsection{Edge Server Heuristic Algorithm}
Each edge server runs Alg.~\ref{LEADER:heu-Edge} in the \textit{SR} interval to serve its own clients' requests with minimum serving times. For this purpose, each LES server receives cache and edge maps ($cache\_map$, $edge\_map$), bandwidth maps to other servers $\in\mathcal{V}\cup\mathcal{C}$ ($bw\_map$), and the NESs' available computational information ($comp\_map$) from the SDN controller. Next, it compares the received request ($req$) to $on\_the\_fly$ requests (\ie  requests currently being served). If $req$ is in the $on\_the\_fly$ set, then it calls \textit{HoldReq} to hold the request and prevent network resource wastage and congestion (lines 2--3). Otherwise, this set is updated by the new request to be processed (line 5). All feasible actions (Fig.~\ref{LEADER-tree}) are calculated by the \textit{ValidAction} function and stored in the $action\_set$ repository (line $6$). After that, based on the proposed objective function in (\ref{LEADER:eq:8}), the serving times of actions in $action\_set$ are calculated by utilizing the \textit{CostFunction} function and the results are saved in $action\_cost$ (line 7). Next, the minimum value (\ie serving time) in the $action\_cost$ is determined by the \textit{OptimalAction} function, and its associated action is retrieved from $action\_set$ and saved as the optimal action, $opt\_action$ (line 8). Finally, the edge server utilizes the \textit{ServeRequest} function to serve the client's request (line 9), and upgrades cache, edge, bandwidth, and comp maps, plus the $on\_the\_fly$ repository by the \textit{UpdateVariables} function (line 10). 
Assume $a$ denotes the number of discussed actions. Employing the Quicksort approach, the time complexity of Alg.~\ref{LEADER:heu-Edge} in the worst case would be $O(a*log(a))$.
% %%%%%%%%%%%%%%%%%%%%%%%LEADER:heu-Edge%%%%%%%%%%%%%%%%%%%%%%%
\begin{center}
	\begin{algorithm}[!t]
		\small
            \caption{LEADER edge server heuristic algorithm.}\label{LEADER:heu-Edge}
		\begin{algorithmic}[1]
            \State \textbf{Input} $req$, $cache\_map$, $edge\_map$, $bw\_map$, $comp\_map$, $on\_the\_fly$
            \If{req $\in$ $on\_the\_fly$}
                \State HoldReq($req$)
            \Else
            \State $on\_the\_fly$.add($req$)
            \State $action\_set\leftarrow$ValidAction()
            \State $action\_cost\leftarrow\text{CostFunction($action\_set$)}$
            \State $opt\_action\leftarrow$OptimalAction($action\_set$,$action\_cost$)
            \State ServeRequest($req$,$opt\_action$)
            \State UpdateVariables$()$
	\EndIf   
        \end{algorithmic}
	\end{algorithm}
\end{center}
% %%%%%%%%%%%%%%%%%%%%%%%LEADER:heu-Edge%%%%%%%%%%%%%%%%%%%%%%%
%%%%%%%%%%%%%%%%%%%%%%%%%%%%%%%%%%%%%%%%%%%%%%%%%%%%%%%%%%%%%%%%%%%%%%%%%%%%%%%%%%%%%%%%%%%%%%%%%%%%%%%%%%%%%%%%%%%%%%%%%%%%%%%%%%%%%%%%%%%%%%%%%%%%%%%%%%%%%%%%%%%%%%%%%%%%%%%%%%%%%%%%%%%%%%%%%%%%%%%%%%%%%%%%%%%
\subsection{LEADER Performance Evaluation}
\label{sec:LEADER:PerformanceEvaluation}
In this section, we describe \texttt{LEADER}'s evaluation setup, metrics, and methods and evaluate \texttt{LEADER}'s performance in two scenarios.
\subsubsection{Evaluation Setup}
\label{sec:LEADER:eval-setting}
To evaluate the \texttt{LEADER} framework in a realistic large-scale environment, we consider a real network topology called Geant~\cite{zoo} (40 OF switches, 61 layer-2 links) and instantiate our testbed including 301 components (each of them runs on Ubuntu 18.04 LTS inside Xen virtual machines) on the CloudLab~\cite{ricci2014introducing} environment.
We implement all modules of five considered edge servers and an SDN controller (\ie FloodLight) in Python to serve requests received from 250 AStream~\cite{juluri2015sara} DASH players. For simplicity, we assume that all clients already joined the network. 

The \textit{BOLA}~\cite{spiteri2016bola} quality adaptation algorithm is used in our testbed. Each edge server decides for its own clients' group (50 clients) and collaborates with other edge servers. Moreover, we employ four CDN servers (with a total cache size of 40\% of the video dataset) and an origin server (containing the entire video sequences) that run an Apache HTTP server and MongoDB. Each edge server contains a partial cache with only 5\% of the video dataset. The most popular video sequences are pre-cached on the edge
servers to avoid a slow startup of the system. 
Note that \texttt{LEADER} is independent of the caching policy and could be compatible with any type of caching strategies. However, for simplicity, Least Recently Used (LRU) is considered in all CDN and partial caches as the cache  replacement policy. Fifty video sequences~\cite{lederer2012dynamic} with 300 seconds duration, comprising two-second segments in five representations (0.089, 0.262, 0.791, 2.4, 4.2 Mbps) are used in our experiments. 

For the sake of simplicity, we assume that the popularity of each video is known in advance and sorted in descending order. The video access probability is generated following a Zipf distribution~\cite{cherkasova2004analysis} with the skew parameter $\alpha=0.75$, \ie the probability of an incoming request for the $i^{th}$ most popular video is given as $prob(i)=\frac{1/i^{\alpha}}{\sum_{j=1}^{K}1/j^{\alpha}}$, where $K=50$. The Docker image \textit{jrottenberg/ffmpeg}~\cite{ffmpeg} is utilized to measure the segment transcoding time on edge servers. 
We set the bandwidth values of all links in different paths from each LES to the origin server, cache servers, and other NES servers to 50, 100, and 200 Mbps, respectively.
%%%%%%%%%%%%%%%%%%%%%%%%%%%%%
\subsubsection{Evaluation Methods and Metrics}
\label{LEADER:Comparison Methods}
In our performance evaluation, we compare the acquired results by \texttt{LEADER} with the following methods:
\begin{enumerate}[noitemsep]
\item\textbf{Non Edge Collaborative} (NECOL): Like in most of the existing works, there is not any edge collaboration possibility in this approach. In a NECOL-based system, each LES serves its associated clients' requests separately via one of the actions $1,2,6,7$, or $8$ (Fig.~\ref{LEADER-tree}).
\item\textbf{Default Edge Collaborative} (DECOL): In this method, all edge servers collaborate in both caching and transcoding. However, the SDN controller does not take into account any bandwidth allocation policy for the shared links.
\end{enumerate}
Note that our testbed with a similar setup is used for all schemes to produce fair and robust comparisons. The performance of the aforementioned approaches is evaluated with common QoE-related parameters and network utilization metrics as follows:
\begin{enumerate}[noitemsep]
\item\textbf{Average Segment Bitrate} (ASB) of all the downloaded segments, 
\item\textbf{Average Number of Quality Switches} (AQS) as the average number of segments whose bitrate levels are changed compared to their previous one, 
\item\textbf{Average Number of Stalls} (ANS) that is equal to the average number of rebuffering events, 
\item\textbf{Average Stall Duration} (ASD) that is the average of total video freeze time in all clients, 
\item\textbf{Average Perceived QoE} (APQ) calculated by ITU-T Rec.P.1203 mode 0~\cite{p1203},
\item\textbf{Cache Hit Ratio} (CHR) defined as the fraction of segments fetched from the CDNs or edge servers, 
\item\textbf{Edge Transcoding Ratio} (ETR) that is the fraction of segments transcoded at the edge servers,
\item\textbf{Backhaul Traffic Load} (BTL) as the  volume of segments downloaded from the origin server.
\end{enumerate}
\rf{Note that we use the method discussed in Section~\ref{sec:SFC:Performance Evaluation} to calculate the results reported in the next section.}
%%%%%%%%%%%%%%%%%%%%%%%%%%%%%%%%%%%
\subsubsection{Evaluation Results}
\label{LEADER:results}
%%%%%%%
\begin{figure}[t]
\centering
\includegraphics[width=.9\textwidth]{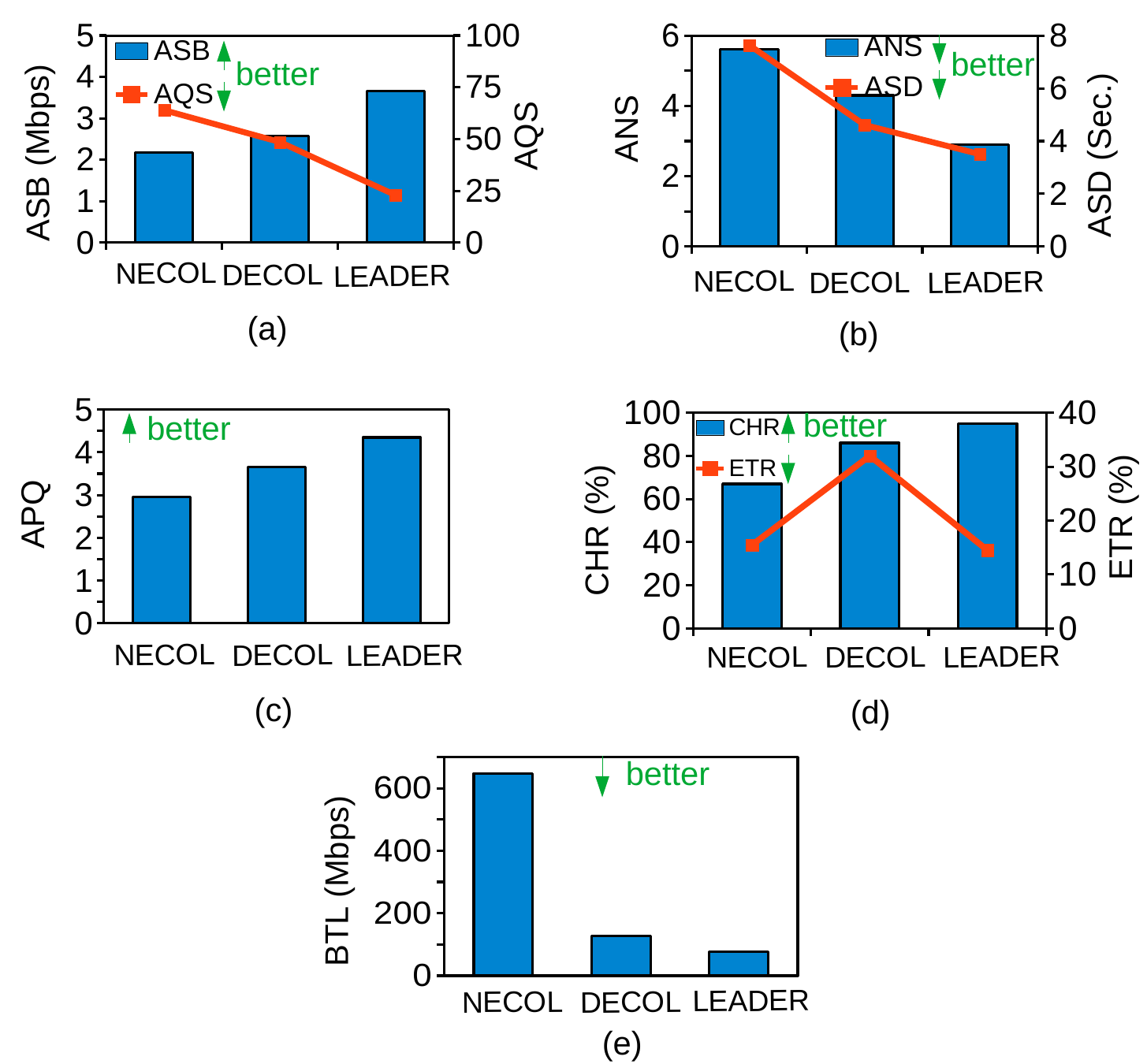}
\caption{\small QoE (a--c), and network utilization (d--e) results for the NECOL, DECOL, and \texttt{LEADER} systems for 250 clients.}
\vspace{.5cm}
\label{res-testbed}	
\end{figure}
%%%%%%%
In the first scenario, we evaluate \texttt{LEADER}'s performance in terms of the QoE parameters and compare the results with NECOL and DECOL. As shown in Fig.~\ref{res-testbed}(a--c), DECOL serves clients with higher ASB, decreases AQS, improves ANS, and shortens ASD, consequently increases APQ by up to 20\% compared to NECOL (Fig.~\ref{res-testbed}(c)), since it uses NESs' cooperation and can shorten clients' serving time through fetching/transcoding actions by NESs. Because \texttt{LEADER} employs a bandwidth allocation strategy, each LES has sufficient bandwidth for accessing other servers, specifically when competing with other NESs in shared links. Consequently, \texttt{LEADER}'s LESs download segments with higher ASB, decrease AQS, ANS, ASD, and thus improve APQ by about 47\% and 22\% compared to NECOL and DECOL, respectively. 

The effectiveness of \texttt{LEADER} regarding the network utilization aspect is evaluated in the second scenario.
Note that the NECOL-enabled system encounters a cache miss event when \textit{(i)} the requested or higher quality levels do not exist in the LES cache or CDNs, \textit{(ii)} available bandwidth values are insufficient to fetch the requested or higher quality levels from CDNs, or \textit{(iii)} the available LES's processing capability is not enough to transcode the requested quality from a higher quality. In addition, a cache miss occurs in DECOL and \texttt{LEADER} if the LES cannot fetch or transcode the requested quality by NESs' assistance. 

The CHR metric shows that DECOL outperforms the NECOL system due to its capability for fetching requested or higher quality levels from NESs or using NESs' idle resources for performing transcoding (see Fig.~\ref{res-testbed}(d)). Therefore, it downloads fewer segments from the origin server and improves backhaul bandwidth usage (Fig.~\ref{res-testbed}(e)), but uses more computational resources at the edge servers due to performing transcoding tasks, as compared to NECOL (Fig.~\ref{res-testbed}(d)).  
Although \texttt{LEADER} and DECOL have a similar behavior of using NESs' resources for fetching/transcoding, the bandwidth strategy helps \texttt{LEADER} to provide sufficient bandwidth between LESs and cache servers and fetch more segments directly from CDNs with less serving time (\ie without transcoding); \texttt{LEADER} thus outperforms the DECOL system in terms of CHR, BTL, and ETR. However, \texttt{LEADER} and DECOL as collaborative systems increase the number of communicated messages (cache, edge, bandwidth, comp maps) to/from the SDN controller compared to the NECOL system.
\clearpage

%% file: Chapters/Chapter5/5-3-ARARAT.tex
\doublespacing
\section{ARARAT Framework}
\label{chap:CollaborativeEdge:ARARAT}
This section introduces the \texttt{ARARAT} system~\cite{farahani2022ararat} as another collaborative edge-assisted framework for HTTP adaptive video streaming. We first discuss the motivation and problem definition for designing the \texttt{ARARAT} framework in Section~\ref{sec:ARARAT:design}. We also explain the proposed hierarchical architecture, formulate the problem as a central optimization model, and analyze its time complexity. % in Section~\ref{sec:ARARAT:design}. 
Section~\ref{sec:ARARAT:Heuristic} describes our proposed heuristic approaches. The evaluation setup and results are presented in Section~\ref{sec:ARARAT:Performance Evaluation}.

%%%%%%%%%%%%%%%%%%%%%%%%%%%%%%%%%%%%%%%%%%%%%%%%%%%%%%%%%%%%%%%%%%%%%%%%%%%%%%%%%%%%%%%%%%%%%%%%%%%%%%%%%%%%%%%%%%%%%%%%%%%%%%%%%%%%%%%%%%%%%%%%%%%%%%%%%%%%%%%%%%%%%%%%%%%%%%%%%%%%%%%%%%%%%%%%%%%%%%%%%%%%%%%%%%%
\subsection{ARARAT System Design}
\label{sec:ARARAT:design}
%This section first discusses the problem statement and introduces the \texttt{ARARAT} architecture. The problem formulation and its time complexity analysis are then presented.
%%%%%%%%%%%%%%%%%%%%%%%%%%%%%%%%%%%%%%%%%%%%%%%%%%%%%%%%%%%%%%%%%%%%%%%%%%%%%%%%%%%%%%%%%%%%%%%%%%%%%%%%%%%%%%%%%%%%%%%%%%%%%%%%%%%%%%%%%%%%%%%%%%%%%%%%%%%%%%%%%%%%%%%%%%%%%%%%%%%%%%%%%%%%%%%%%%%%%
\subsubsection{ARARAT Problem Statement}
\label{sec:PROBLEM STATEMENT}
As discussed in Section~\ref{chap:SOTA}, there are many NAVS systems essentially studying a single-edge server in an SDN-enabled environment. However, a single-edge server does not include enough cache and computing resources to satisfy different users' demands. 
Edge servers collaborating on caching and transcoding can potentially improve edge servers' resource utilization, decrease backhaul bandwidth usage, and improve users' QoE. Let us consider a simple example including \textit{two} types of edge servers, \ie \textit{Local Edge Server} (LES) and \textit{Neighboring Edge Servers} (NESs), which are adjacent to the LES. Both the LES and the NESs are equipped with transcoding (Tran) and partial caching (PC) functions. (We will elaborate on this in more detail in Section~\ref{sec:System Model:arch}).

A LES serves its clients' requests directly, while NESs collaborate with the LES on transcoding and caching in case of lack of processing resources or on cache misses. Note that a LES concurrently can play the NES role for other NESs. Moreover, we consider multiple CDN servers containing various parts of video sequences and an origin server including all video segments in various representations. 
In such systems, like illustrated in Fig.~\ref{ARARAT:actionTree}, a bunch of requests is collected by each LES in specified time slots (intervals). Then, the SDN controller runs a centralized approach (\eg an optimization model) to come up with appropriate decisions for serving all clients' requests. The \texttt{ARARAT} system aims at achieving the best solutions to respond to the following key questions:
%%%%%
\begin{figure}[!t]
	\centering
	\includegraphics[width=.7\columnwidth]{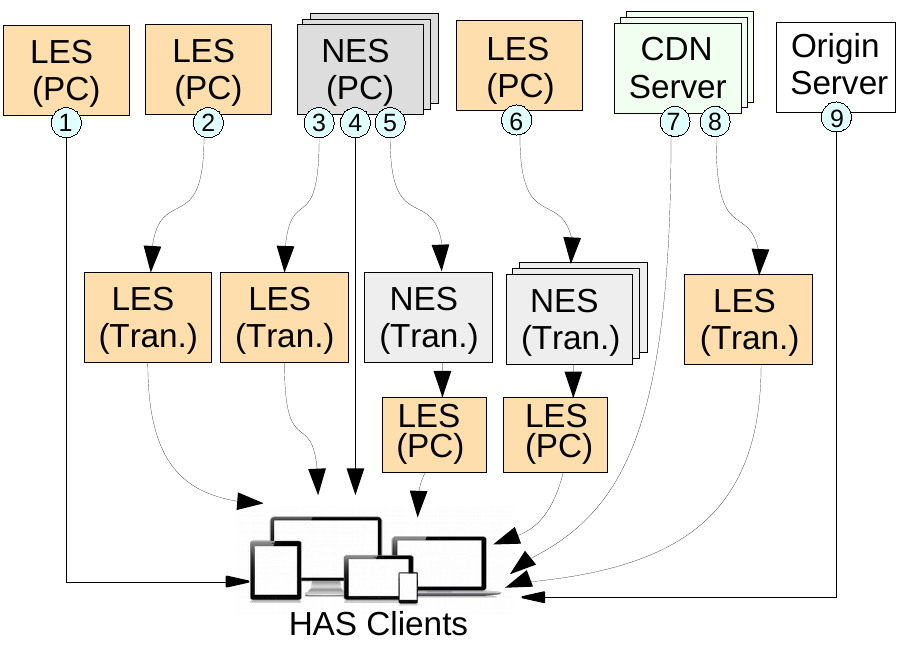}
	\caption{\small ARARAT action tree.}
         \vspace{.5cm}
	\label{ARARAT:actionTree}
\end{figure}
%%%%%
\begin{enumerate}[noitemsep]
\item Where is the optimal server (\ie LES, NES, CDN node, or the origin) in terms of the minimum serving time (\ie fetching, potentially plus transcoding) for acquiring each client's requested content quality level from? 
\item What is the optimal approach for delivering each client's requested quality level, \ie fetch or transcode it from a higher quality, considering edge servers' bandwidth and computational limitations? 
\item What is the optimal action to reach each client's requested quality?
\end{enumerate}

To this end, \texttt{ARARAT} considers clients' serving times, edge resource limitations, and network costs to find the solution by utilizing actions from the \textit{Action Tree} depicted in Fig.~\ref{ARARAT:actionTree} (action numbering as in the figure):
\begin{enumerate}[leftmargin=*,label=\protect\circledd{\arabic*},noitemsep]
 \item Fetch the requested quality level directly from the LES. 
 \item Transcode the requested quality from a higher quality at the LES.
 \item Fetch a higher quality from the best adjacent NES and transcode it at the LES.  
 \item Fetch the requested quality directly from the best NES.
 \item Transcode the requested quality from a higher quality at the best NES and then transmit it from the NES to the LES.
 \item Send a higher quality from LES to the best NES (``best'' in terms of available processing capability), transcode it at the NES, and then forward it back from the NES to the LES.
 \item Fetch the requested quality from the best CDN server. 
 \item Fetch a higher quality from the best CDN server and transcode it at the LES. 
 \item Fetch the requested quality from the origin server. 
\end{enumerate}
Sections~\ref{sec:System Model:arch} and \ref{sec:System Model:MILP} will describe how the SDN controller communicates with edge and CDN servers to determine the optimal actions in terms of serving time and network cost.
%%%%%%%%%%%%%%%%%%%%%%%%%%%%%%%%%%%%%%%%%%%%%%%%%%%%%%%%%%%%%%%%%%%%%%%%%%%%%%%%%%%%%%%%%%%%%%%%%%%%%%%%%%%%%%%%%%%%%%%%%%%%%%%%%%%%%%%%%%%%%%%%%%%%%%%%%%%%%%%%%%%%%%%%%%%%%%%%%%%%%%%%%%%%%%%%%%%%%%%%%%%%%%%%%%%
\label{sec:System Model} 
\subsubsection{ARARAT Architecture}
\label{sec:System Model:arch}
Regarding our discussion above and the problem statement, we here propose a multi-layer architecture for \texttt{ARARAT} consisting of \textit{three} core layers: \textit{edge layer}, \textit{CDN/origin layer}, and \textit{control layer} (see Fig.~\ref{ARARAT-arch}).
\begin{enumerate}[noitemsep]
\item \textbf{Edge Layer.} As introduced in Section~\ref{sec:PROBLEM STATEMENT}, edge servers are categorized into LES and NESs and located close to base stations (\eg gNodeB in 5G). Each of them periodically communicates with the SDN controller, sends the status of its available resources, \eg CPU and RAM (in so-called \textit{comp\_map} messages), and its cache occupancy (in \textit{edge\_map} messages) to the SDN controller, and responds to the clients. Each edge server consists of the following components:
\textit{(i)} \textit{Transcoding Module} (Tran) to transcode segments from a higher quality level to the demanded quality level; \textit{(ii)} \textit{Partial Cache} (PC) that stores a limited number of segments; and \textit{(iii) }\textit{Request Analyzer Module} (RAM), which is designed to aggregate identical requests (issued by several clients) and consider only one request for each segment. Note that the network slicing method is employed to design an independent logical network-assisted video streaming instance over the common edge layer; therefore, the edge layer can be utilized to serve various network services, \eg video streaming services with different requirements.
%%%%%%%%%%%%%%%%%%%%%%%%%%%%%%%%%%%%%%%%%%%
\begin{figure}[!t]
	\centering
	\includegraphics[width=1\columnwidth]{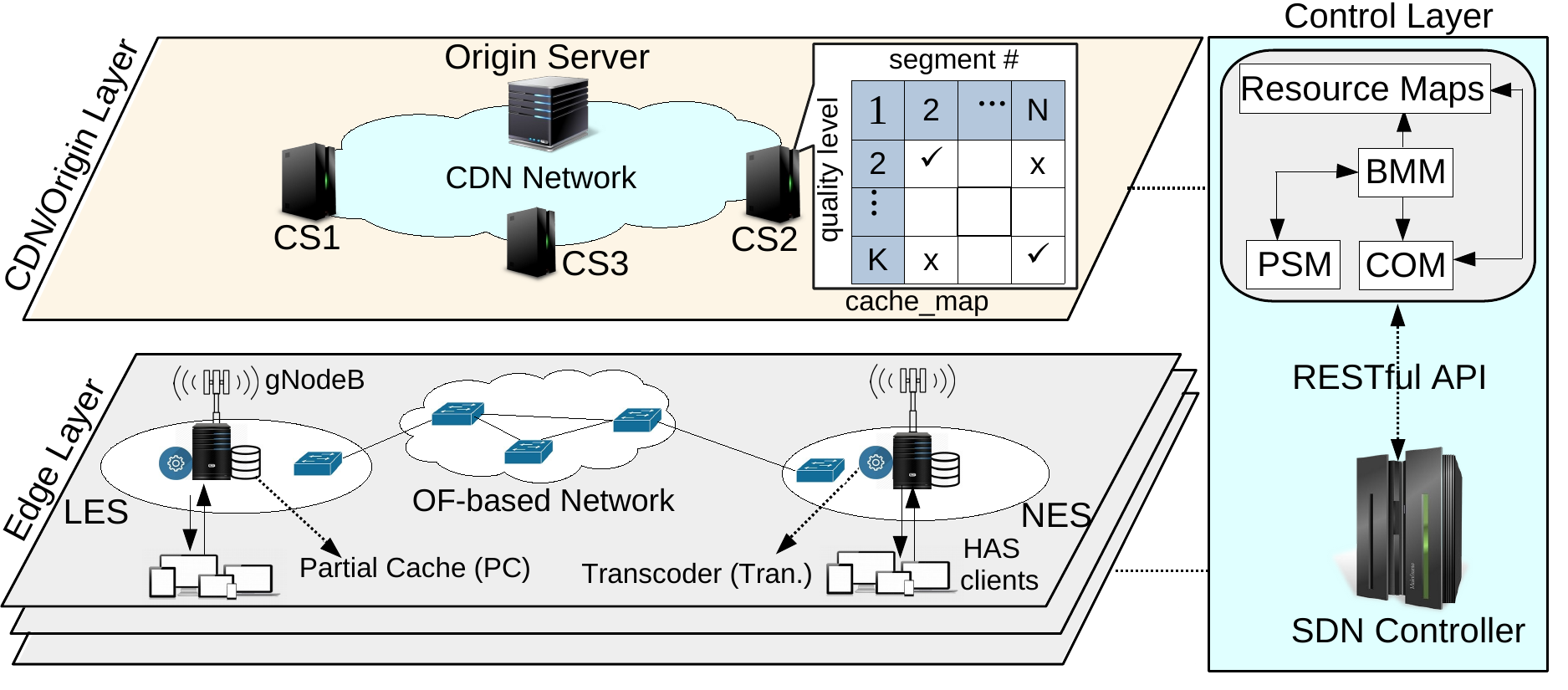}
	\caption{\small ARARAT architecture.}
        \vspace{.5cm}
	\label{ARARAT-arch}
\end{figure}
%%%%%%%%%%%%%%%%%%%%%%%%%%%%%%%%%%%%%%%%%%%
\item \textbf{CDN/Origin Layer.} 
% The CDN/origin layer is constructed by multiple CDN servers (CSs) where each of them contains various parts of video sequences and periodically informs the SDN controller about its cache occupancy through \textit{cache\_map} messages.
The CDN/origin layer is constructed by multiple CDN servers (CSs) where each of them contains various parts of video sequences and periodically informs the SDN controller about the availability of the stored segments and their quality levels through \textit{cache\_map} messages. The \textit{cache\_map} structure is depicted in Fig.~\ref{ARARAT-arch}. Moreover, an origin server that contains all video segments in multiple representations is placed in this layer.

\item \textbf{Control Layer.} An SDN controller is placed in this layer. The SDN controller consists of the following modules: \textit{(i)} \textit{Bandwidth Monitoring Module} (BMM): It monitors the available bandwidth of different paths between each LES and other servers, \ie NESs, CSs, and the origin server. The output of the BMM will be stored in a sub-repository called \textit{bandwidth\_map} which is part of a large repository named \textit{resource maps}. \textit{(ii)} \textit{Path Selection Module} (PSM): It uses BMM results and selects the shortest paths with maximum available bandwidth between each LES and NESs, CSs, and the origin server. \textit{(iii)} \textit{Central Optimization Module} (COM): COM runs an optimization model to determine an appropriate solution for the questions listed in Section~\ref{sec:PROBLEM STATEMENT}. For this purpose, COM utilizes the received items that are stored in \textit{resource maps}, \ie \textit{cache\_map, edge\_map, comp\_map}, and \textit{bandwidth\_map}, and finds an optimal solution (\ie in terms of lowest serving time, and network cost) from the \textit{Action Tree} (Fig.~\ref{ARARAT:actionTree}).
\end{enumerate}
%%%%%%%%%%%%%%%%%%%%%%%%%%%%%%%%%%%%%%%%%%%%%%%%%%%%%%%%%%%%%%%%%%%%%%%%%%%%%%%%%%%%%%%%%%%%%%%%%%%%%%%%%%%%%%%%%%%%%%%%%%%%%%%%%%%%%%%%%%%%%%%%%%%%%%%%%%%%%%%%%%%%%%%%%%%%%%%%%%%%%%%%%%%%%%%%%%%%%%%%%%%%%%%%%%%
%%%%table%%%%
\begin{table}[!t]
\centering
\caption {\small ARARAT Notation.}
\label{tab:ARARAT:notation}
\begin{tabular}{llllll}
\cline{1-2}
\multicolumn{2}{|c|}{\textbf{Input Parameters}}                                                                                  
&  &  &  &  \\ \cline{1-2}                                                                                         
\multicolumn{1}{|l|}{\begin{tabular}[c]{@{}l@{}}
$\mathcal{C}$\\ 
$\mathcal{V}$\\ 
$\mathcal{R}$\\ 
$R_i$\\
$\mathcal{Q}^r$\\ \\ \\
$\mathcal{T}$\\  \\ \\
$\alpha^{p,r}_{i}$\\ \\ 
$\omega_{i,j}$\\
$l$\\ \\
$\delta^{r}_{p}$\\  
$\pi^{r}_{p}$\\ \\
$\Omega_i$\\ 
$\mu^{r}_{p}$\\ \\
$\eta_p$\\
$\Delta_{tr}$\\
$\Delta_{bw}$\\ 
$\beta_1, \beta_2$\\ \\
\end{tabular}} 
& \multicolumn{1}{l|}{\begin{tabular}[c]{@{}l@{}}
Set of $k$ CDN servers and an origin server (i.e., $c=0$)\\
Set of $n$ edge servers \\
Set of $m$ requests received by the SDN controller from $\mathcal{V}$\\
Set of requests issued by edge server $i\in\mathcal{V}$\\
Set of possible quality levels for serving quality $p^*$ requested\\by $r\in\mathcal{R}$, where $\mathcal{Q}^r=\{p^*,p^*+1,...,p^{*}_{max}\}$ and $p^{*}_{max}$\\ is the maximum quality level for the demanded segment\\
Set of possible transcoding statuses: $\mathcal{T}=\{0,1,2\}$; $t$ is 1 or 2 \\if the requested quality is transcoded from a higher quality\\ $p\in\mathcal{Q}^r$ at the NES or at the LES, respectively; otherwise t=0\\
Availability of quality level $p$ in server $i\in\mathcal{V}\cup\mathcal{C}$; $\alpha^{p,r}_{i}=1$\\means server $i$ hosts quality $p$ to serve $r\in\mathcal{R}$, otherwise  $\alpha^{p,r}_{i}=0$ \\
Available bandwidth on path between $i\in\mathcal{V}$ and $j\in\{\mathcal{V}\cup \mathcal{C}\} \setminus i$\\
Action index, where $l=1$, if the selected action from the \\action tree is action $6$, otherwise $l=0$\\
Size of segment in quality $p$ requested by $r\in\mathcal{R}$ (in bytes)\\
Required resources (i.e., CPU time in seconds) for transcoding \\quality $p\in\mathcal{Q}^r$ into the requested quality by $r\in\mathcal{R}$\\
Available computational resources (available CPU) of $i\in \mathcal{V}$\\
Required time for transcoding quality $p\in\mathcal{Q}^r$ into the quality\\ requested by $r\in\mathcal{R}$\\
Bitrate associated to quality level $p \in \mathcal{Q}^r$\\
Computational cost per CPU core per second\\
Bandwidth cost per bit per second\\
Adjustable weighting coefficients for the serving time  and the\\ network cost, respectively \\
\end{tabular} }
&  &  &  &  \\ \cline{1-2}
\multicolumn{2}{|c|}{\textbf{Variables}} 
&  &  &  &  \\ \cline{1-2}                                                 
\multicolumn{1}{|l|}{\begin{tabular}[c]{@{}l@{}}
$B^{p,r,t}_{i,j,l}$\\ \\ \\ 
$\tau^{r}_{i}$\\ 
$T^{p,r}_{i,j}$\\ \\
$\Gamma$\\
 $\Pi, \Lambda$\\ \\
 $\Psi$\\
\end{tabular}}
&\multicolumn{1}{l|}{\begin{tabular}[c]{@{}l@{}}
Binary variable where $B^{p,r,t}_{i,j,l}=1$ shows server $i\in\mathcal{V}\cup\mathcal{C}$\\ transmits quality $p\in\mathcal{Q}^r$ to server $j\in\mathcal{V}$ for request $r$ with\\ transcoding status $t$ and action index $l$, otherwise $B^{p,r,t}_{i,j,l}=0$\\
Required transcoding time at server $i\in\mathcal{V}$ for serving $r\in\mathcal{R}$ \\
Required time of transmitting quality level $p\in\mathcal{Q}^r$ in response \\ to request $r\in\mathcal{R}$ from server $i\in\mathcal{V}\cup\mathcal{C}$ to $j\in\mathcal{V}$\\
Serving time consisting of $\tau^{r}_i$ and $T^{p,r}_{i,j}$\\
Costs of computation and bandwidth for serving clients' requests,\\ respectively\\
Total network cost consisting of $\Pi$ and $\Lambda$\\
\end{tabular}}                                                    
&  &  &  &  \\ \cline{1-2}                                                                                                              
& &  &  &  &                                              
\end{tabular}
\end{table}
%%%%table%%%%
\clearpage
\subsubsection{ARARAT Centralized Optimization Model}
\label{sec:System Model:MILP}
Let set $\mathcal{C}$ consist of $k$ CDN servers and an origin server, where $c=0$ indicates the origin server (see Table~\ref{tab:ARARAT:notation} for a summary of the notation). Moreover, let us define $\mathcal{R}$ as the set of $m$ requests received by the SDN controller from the set $\mathcal{V}$ of $n$ edge servers. The subset $R_{i}\subseteq\mathcal{R}$ shows all requests issued by edge server $i\in\mathcal{V}$. Furthermore, we assume that the SDN controller periodically receives the \textit{cache\_map} and the \textit{edge\_map} from cache and edge servers, respectively. $\alpha^{p,r}_{i}$ denotes the availability of quality level $p$ of a segment demanded by request $r$ at cache/edge server $i$. 

We assume that the SDN controller defines a data path between two servers $i\in\mathcal{V}$ and $j\in\{\mathcal{C}\cup\mathcal{V}\} \setminus i$ where $\omega_{i,j}$ shows its available bandwidth between $i$ and $j$ accordingly. As mentioned earlier, when a LES $i$ requests quality $p$ from server $j\in\{\mathcal{C}\cup\mathcal{V}\} \setminus i$, it is possible to receive the requested quality or a higher one from $j$.
In fact, if $p$ is available in the cache of $j$, $p$ can be directly forwarded to $i$; otherwise, since all edge servers can run the transcoding function, $j$ can send a higher quality to $i$;
therefore, for each request $r$, we define $\mathcal{Q}^r=\{p^*,p^*+1,...,p^{*}_{max}\}$ as the set of possible quality levels for serving requested quality $p^*$, where $p^{*}_{max}$ is the maximum quality level for the requested segment. In the following, we introduce an MILP optimization model that decides an optimal action for each request $r\in\mathcal{R}$ based on the given \textit{resource maps} information in such a way that the total serving time and network cost are minimized. For this purpose, the following aspects are required to be taken into account to force the model to:
\begin{enumerate}[noitemsep]
\item select only one action for each request $r\in\mathcal{R}$.
\item determine the best action in terms of serving time (\ie fetching and transcoding) among all feasible actions for each request $r\in\mathcal{R}$.
\item not violate the given available network resources by the selected action.
\item not violate essential regulations defined by stakeholders.
\item determine the best action in terms of network cost (\ie bandwidth and computation) among all feasible actions for each request.
\end{enumerate}
Therefore, the model should satisfy five groups of constraints: \textit{Action Selection} (AS), \textit{Serving Time} (ST), \textit{CDN/Origin} (CO), \textit{Resource Consumption} (RC), and \textit{Network Cost} (NC) constraints, which are described as follows.

\textbf{\textit{(i)} AS Constraint.} 
The first constraint determines a suitable action from the introduced action tree (Fig.~\ref{ARARAT:actionTree}) for each request $r$ issued by LES $j$. As explained in Section~\ref{sec:PROBLEM STATEMENT}, action 6 of the action tree forces \textit{(i)} the LES to send a higher quality to the NES with highest available computational resources, and \textit{(ii)} the NES to send back the requested quality to the LES after completing the transcoding process. However, the LES does not need to transmit a higher quality to other servers in other actions. 
To distinguish action 6, we define an action index $l$, where $l=1$ if the selected action is action 6, otherwise $l=0$. Moreover, $t\in\{0,1,2\}$ defines the following transcoding statuses: $t=0$ indicates that the requested quality is transferred without transcoding, and $t=1$ or $t=2$  means the requested quality is transcoded from a higher quality at the NES or the requesting LES, respectively. Therefore, the first constraint selects appropriate values of the binary variable $B^{p,r,t}_{i,j,l}$, where $B^{p,r,t}_{i,j,l}=1$ indicates server $i$ (\ie CDN, origin, or NES) replies to request $r$ issued by LES $j$ through sending quality $p$ with transcoding status $t$, and action index $l$ (refer to Table~\ref{tab:ARARAT:notation} for notations):
%%%%%Const1%%%%%%%
\begin{flalign}
1\leq &\sum_{i\in\mathcal{C}\cup\mathcal{V}}\sum_{t\in\mathcal{T}}\sum_{p\in\mathcal{Q}^r} (B^{p,r,t}_{i,j,l=0}~.~\alpha^{p,r}_{i})
+\hspace{-.3cm}\sum_{i\in\mathcal{V}\setminus j}\sum_{p\in\mathcal{Q}^r} (B^{p,r,t=0}_{j,i,l=1}~.~\alpha^{p,r}_{j}+ B^{p,r,t=1}_{i,j,l=1}~.~(1-\alpha^{p,r}_{i}))
\leq 2,
\nonumber&&\\
&\forall j\in\mathcal{V}, r\in\mathcal{R}_j
\label{ARARAT:eq:1}
\end{flalign}
%%%%%Const1%%%%%%%
Note that since the 6th action of the action tree forces the LES to send a higher quality to the NES with the highest available computational resources, and, it receives the requested quality from that NES, the MILP model needs to set both binary variables $B^{p,r,t=0}_{i,j,l=1}$ and $B^{p,r,t=1}_{i,j,l=1}$. Therefore, the upper bound will be equal to two when MILP selects action 6. Moreover, $\alpha^{p,r}_i$ is used to assure the quality $p$ requested by request $r$ is available in server $i$.

\textbf{\textit{(ii)} ST constraints.} The first ST constraint determines the required time, denoted by $T^{p,r}_{i,j}$, for transmitting quality level $p\in\mathcal{Q}^r$ in response to request $r$ from each server $i$ to each LES $j\in\mathcal{V}$:
%%%%%Const2%%%%%%%
\begin{flalign}
&T^{p,r}_{i,j} =
\begin{cases}
\frac{B^{p,r,t=0}_{j,i,l=1}~.~\delta^r_{p}+B^{p,r,t=1}_{i,j,l=1}~.~\delta^r_{p^{*}}}{\omega_{i,j}}
\hspace{.4cm}\forall r\in\mathcal{R}_j, i\in\mathcal{V}\setminus j
\\
\\
\frac{\sum_{t\in\mathcal{T}} B^{p,r,t}_{i,j,l=0}~.~\delta^{r}_p}{\omega_{i,j}} 
\hspace{.4cm}\forall r\in\mathcal{R}_j,  i\in\{\mathcal{C}\cup\mathcal{V}\}\setminus j
\end{cases} 
\label{ARARAT:eq:2}
\end{flalign}
%%%%%Const2%%%%%%%
where $\delta^{r}_p$ and $\delta^{r}_{p^*}$ show the size (in bytes) of quality $p$ and $p^*$ of the segment requested by $r\in\mathcal{R}$, respectively, where $p^*\leq p$. The next constraint measures the required transcoding time $\tau_{i}^{r}$ at server $i\in\mathcal{V}$ in case of serving the quality requested by $r\in R$ from a higher quality $p$ by transcoding:
%%%%%Const3%%%%%%%
\begin{flalign}
&\sum_{j\in\mathcal{V}}\sum_{l\in \{0,1\}}\sum_{p\in \mathcal{Q}^r} B^{p,r,t=1}_{i,j,l}~.~\mu^{r}_{p}\leq \tau_{i}^{r},&\hspace{.7cm}\forall  i\in\mathcal{V},r\in\mathcal{R}
\label{ARARAT:eq:3}
\end{flalign}
%%%%%Const3%%%%%%%
Therefore, the serving time, namely $\Gamma$, \ie fetching time plus transcoding  time can be expressed as follows:
%%%%%Const4%%%%%%%
\begin{flalign}
&\sum_{r\in \mathcal{R}}\sum_{i\in\mathcal{V}\cup\mathcal{C}}\sum_{j\in \mathcal{V}}\sum_{p\in \mathcal{Q}^r}T^{p,r}_{i,j}+\tau^{r}_i\leq \Gamma&&
\label{ARARAT:eq:4}
\end{flalign}
%%%%%Const4%%%%%%%

\textbf{\textit{(iii)} CO constraints.} This group of constraints forces the model to fetch the exact quality $p^*$ from the origin server when the origin server is selected to serve the LES; so, we have:
%%%%%Const5%%%%%%%
\begin{flalign} 
&\sum_{t\in \mathcal{T}}\sum_{p\in\mathcal{Q}^r} B^{p,r,t}_{i=0,j,l=0}~.~p = p^*, && \forall j\in\mathcal{V},r\in R_j
\label{ARARAT:eq:5}
\end{flalign}
%%%%%Const5%%%%%%%
Note that $i=0$ in Eq.~(\ref{ARARAT:eq:5}) means that the origin server is serving requests. We also should prevent CSs from performing transcoding functions by the following equation:
%%%%%Const6%%%%%%%
\begin{flalign} 
&\sum_{i\in\mathcal{C}}\sum_{r\in R_j}\sum_{p\in\mathcal{Q}^r} B^{p,r,t=1}_{i,j,l=0}~.~p = 0, && \forall j\in\mathcal{V}
\label{ARARAT:eq:6}
\end{flalign}
%%%%%Const6%%%%%%%

\textbf{\textit{(iv) }RC constraints.} This group of constraints guarantees that the required bandwidth for transmitting segments on the link between two servers $j\in\mathcal{V}$ and $i\in\mathcal{C}\cup\mathcal{V}$ should not exceed the available bandwidth:
%%%%%Const7%%%%%%%
\begin{flalign}
&\sum_{t\in \mathcal{T}}\sum_{l\in \{0,1\}}\sum_{r\in R_j}\sum_{p\in \mathcal{Q}^r}B^{p,r,t}_{i,j,l}~.~\eta_p \leq \omega_{i,j},&\forall j\in\mathcal{V},i\in\{\mathcal{C}\cup\mathcal{V}\}\setminus{j}
\label{ARARAT:eq:7}
\end{flalign}
%%%%%Const7%%%%%%%
where $\eta_p$ is the bitrate associated to quality level $p$. Similarly, the next constraint limits the maximum required processing capacity for the transcoding operation to the available computational resource on each server $j\in\mathcal{V}$ (denoted by $\Omega_j$):
%%%%%Const8%%%%%%%
\begin{flalign}
&\sum_{r\in \mathcal{R}}\sum_{p\in \mathcal{Q}^r}\sum_{l\in\{0,1\}}(\sum_{i\in \mathcal{C}\cup\mathcal{V}}B^{p,r,t=2}_{i,j,l} + \sum_{i\in\mathcal{V} \setminus j} \hspace{-.3cm}B^{p,r,t=1}_{j,i,l})~.~\pi^{r}_{p} \leq \Omega_j
,&\forall j\in\mathcal{V}
\label{ARARAT:eq:8}
\end{flalign}
where $\pi_p^r$ is the required computational resources (\ie CPU time in seconds) for transcoding a higher bitrate $p$ into bitrate $p^*$ requested by $r$.
%%%%%Const8%%%%%%%

\textbf{\textit{(v) }NC constraints.} The last group of constraints formulates the total bandwidth cost $\Lambda$ for transmitting segments on the link between two servers $j\in\mathcal{V}$ and $i\in\{\mathcal{C}\cup\mathcal{V}\}\setminus{j}$  as follows:
%%%%%Const9%%%%%%%
\begin{flalign} 
&\Delta_{bw}~.~\sum_{t\in \mathcal{T}}\sum_{l\in \{0,1\}}\sum_{r\in R_j}\sum_{p\in \mathcal{Q}^r} \hspace{-.1cm}B^{p,r,t}_{i,j,l}~.~ \eta_p \leq \Lambda,&\forall j\in\mathcal{V},i\in\{\mathcal{C}\cup\mathcal{V}\}\setminus{j}
\label{ARARAT:eq:9}
\end{flalign}
%%%%%Const9%%%%%%%
where $\Delta_{bw}$ refers to the bandwidth cost per bit per second. Similarly, the next constraint specifies 
the total required computational cost $\Pi$ for transcoding processes:
%%%%%Const10%%%%%%%
\begin{flalign}
&\Delta_{tr}~.~\sum_{r\in \mathcal{R}}\sum_{p\in \mathcal{Q}^r}\sum_{l\in\{0,1\}}(\sum_{i\in \mathcal{C}\cup\mathcal{V}}B^{p,r,t=2}_{i,j,l}+ \sum_{i\in\mathcal{V} \setminus j} B^{p,r,t=1}_{j,i,l})~.~\pi^{r}_{p} \leq \Pi
,&\forall j\in\mathcal{V}
\label{ARARAT:eq:10}
\end{flalign}
%%%%%Const10%%%%%%%
where $\Delta_{tr}$ indicates resource computational cost per CPU core per second. Therefore, the total network cost, \ie $\Psi$ can be obtained as follows:
%%%%%Const11%%%%%%%
\begin{flalign}
&\Pi+\Lambda \leq \Psi&&
\label{ARARAT:eq:11}
\end{flalign}
%%%%%Const11%%%%%%%

\textbf{Central optimization model.} As mentioned earlier, a HAS client runs an ABR algorithm to estimate the network's bandwidth by measuring the time between sending the request to download a segment and receiving the segment's last packet. Thus, minimizing the serving time in the optimization model directly affects the HAS clients' performance. On the other hand, the optimization model must have the ability to be adjusted by the network operator based on their business plans and desired network costs.
Therefore, making a trade-off between serving time and network cost in the objective enables the model to select actions based on the operators' desired policies. To this end, the following model, in a multi-objective fashion, minimizes the clients' total serving time as well as the total network cost: 
%%%%%Const12%%%%%%%
\begin{flalign}
\textit{Minimize}&\hspace{.3cm}\beta_1~.~ \frac{\Gamma}{\Gamma^*} + \beta_2~.~\frac{\Psi}{\Psi^*}
\label{ARARAT:eq:12}\\
  s.t.&\hspace{.5cm}\text{constraints}\hspace{.5cm}\text{Eq. }(\ref{ARARAT:eq:1})-\text{Eq. }(\ref{ARARAT:eq:11})&&\nonumber\\
  vars.&\hspace{.5cm} T^{p,r}_{i,j},\tau^{r}_i, \Gamma, \Psi, \Pi,\Lambda\geq 0, B^{p,r,t}_{i,j,l}\in\{0,1\}\nonumber 
\end{flalign}
%%%%%Const12%%%%%%%

The first term in the objective function is the total serving time for all clients' requests in $\mathcal{R}$ while the second term determines the total network cost. By minimizing the objective function~(\ref{ARARAT:eq:12}), the SDN controller can determine an optimal action for each request $r\in\mathcal{R}$. Two weighting coefficients ($\beta_1$ and $\beta_2$) are defined to set desirable priorities for serving time and network cost, respectively. These weights can be adjusted by operators based on network policies where $\beta_1 + \beta_2 =1$. Since $\Gamma$ and $\Psi$ have different units (\ie seconds and dollars, respectively), a normalization strategy is required to form the objective function. Therefore, we normalize each term by dividing it by its associated maximum calculated value (\ie $\Gamma^*$ and $\Psi^*$, respectively). %determined over all feasible actions, multiplied by the number of requests in $R$. 
To determine $\Gamma^*$, we run the MILP model with $\beta_{1}=0$ to achieve the minimum network cost, in other words, the maximum delay. Similarly, we set $\beta_{2}=0$ to get $\Psi^*$. The obtained $\Gamma^*$ and $\Psi^*$ are then used to run the model for different sets of $(\beta_1, \beta_2)$. 
%%%%%%%%%%%%%%%%%%%%%%%%%%%%%%%%%%%%%%%%%%%%%%%%%%%%%%%%%%%%%%%%%%%%%%%%%%%%%%%%%%%%%%%%%%%%%%%%%%%%%%%%%%%%%%%%%%%%%%%%%%%%%%%%%%%%%%%%%%%%%%%%%%%%%%%%%%%%%%%%%%%%%%%%%%%%%%%%%%%%%%%%%%%%%%%%%%%%%%%%%%%%%%%
\subsubsection{Complexity Analysis of the Optimization Model}
\label{sec:System Model:analysis}
\textbf{{\textit{Theorem I}}}: The proposed MILP model (\ref{ARARAT:eq:12}) is an NP-hard problem.\\
\textbf{Proof.} The combination of introduced binary and non-binary variables makes the optimization model (12) a MILP model. Using methods introduced by \cite{conforti2014integer}, the proposed MILP model can be reduced to the Multiple Knapsack Problem (MKP), which is a well-known NP-hard problem~\cite{lewis1983michael}, in polynomial time.

\textbf{{\textit{Theorem II}}}: Considering shared links between edge servers to reach other servers (in $\mathcal{V}\cup\mathcal{C}$) in Eq.~(\ref{ARARAT:eq:12}) changes the model to a mixed integer non-linear programming (MINLP) model.\\
\textbf{Proof:} The proposed model~(\ref{ARARAT:eq:12}) minimizes the clients' serving time plus network cost, assuming disjoint paths from edge servers in $\mathcal{V}$ to other servers in $\mathcal{V}\cup\mathcal{C}$. Let us define two variables, $\bar{\omega}_{i,j}$ and $\omega^{a,b}_{i,j}$ as optimal allocated bandwidth between server $i$ and $j$ and optimal allocated bandwidth between server $i$ and $j$ over the link $(a,b)$, where $\bar{\omega}_{i,j}\leq\omega^{a,b}_{i,j}, \forall$ links $(a,b),i\in\{\mathcal{V}\cup\mathcal{C}\}\setminus j,j\in\mathcal{V}$ for considering bottleneck links. 
Thus, the required time $T^{p,r}_{i,j}$ for transmitting  quality level $p\in\mathcal{Q}^r$ from server $i$ to each LES $j\in\mathcal{V}$ for each request $r\in R_j$ and quality $ p\in\mathcal{Q}^r$ is obtained as follows:
%%%%%Const13%%%%%%%
\begin{flalign}
&T^{p,r}_{i,j} =
\begin{cases}
\frac{B^{p,r,t=0}_{j,i,l=1}~.~\delta^r_{p}+B^{p,r,t=1}_{i,j,l=1}~.~\delta^r_{p^{*}}}{\bar{\omega}_{i,j}} 
\hspace{1.4cm}\forall  i\in\mathcal{V}\setminus j
\\
\\
\frac{\sum_{t\in\{0,1,2\}} B^{p,r,t}_{i,j,l=0}~.~\delta^{r}_p}{\bar{\omega}_{i,j}} 
\hspace{1.4cm}\forall  i\in\{\mathcal{C}\cup\mathcal{V}\}\setminus j
\end{cases} 
\label{ARARAT:eq:13}
\end{flalign}
%%%%%Const13%%%%%%%
Moreover, the formulation of required bandwidth for transmitting segments on the link between two servers $j\in\mathcal{V}$ and $i\in\{\mathcal{C}\cup\mathcal{V}\}\setminus j$ should be updated as follows:
%%%%%Const14%%%%%%%
\begin{flalign}
&\sum_{l\in \{0,1\}}\sum_{t\in \{0,1,2\}}\sum_{r\in R_j}\sum_{p\in \mathcal{Q}^r} \hspace{-.1cm}B^{p,r,t}_{i,j,l}~.~\eta_p \leq \bar{\omega}_{i,j},&\forall j\in\mathcal{V},i\in\{\mathcal{C}\cup\mathcal{V}\}\setminus j
\label{ARARAT:eq:14}
\end{flalign}
%%%%%Const14%%%%%%%

Therefore, replacing Eqs.~(\ref{ARARAT:eq:2}) and (\ref{ARARAT:eq:7}) by Eqs.~(\ref{ARARAT:eq:13}) and (\ref{ARARAT:eq:14}), respectively, where the multiplication of two variables (\ie ${\bar{\omega}_{i,j}}~.~T^{p,r}_{i,j}$) occurs, turns the proposed MILP model (\ref{ARARAT:eq:12}) into an MINLP model that is NP-complete~\cite{lee2011mixed}. This model suffers from high time complexity. To mitigate high time complexity issues, we will propose two heuristic approaches in the next section.
%%%%%%%%%%%%%%%%%%%%%%%%%%%%%%%%%%%%%%%%%%%%%%%%%%%%%%%%%%%%%%%%%%%%%%%%%%%%%%%%%%%%%%%%%%%%%%%%%%%%%%%%%%%%%%%%%%%%%%%%%%%%%%%%%%%%%%%%%%%%%%%%%%%%%%%%%%%%%%%%%%%%%%%%%%%%%%%%%%%%%%%%%%%%%%%%%%%%%%%%%%%%%%%%%%%

\subsection{ARARAT Heuristic Approaches}
\label{sec:ARARAT:Heuristic} 
%%%%%%%%%%%%%%
\begin{figure}[t]
\centering
\includegraphics[width=1\textwidth]{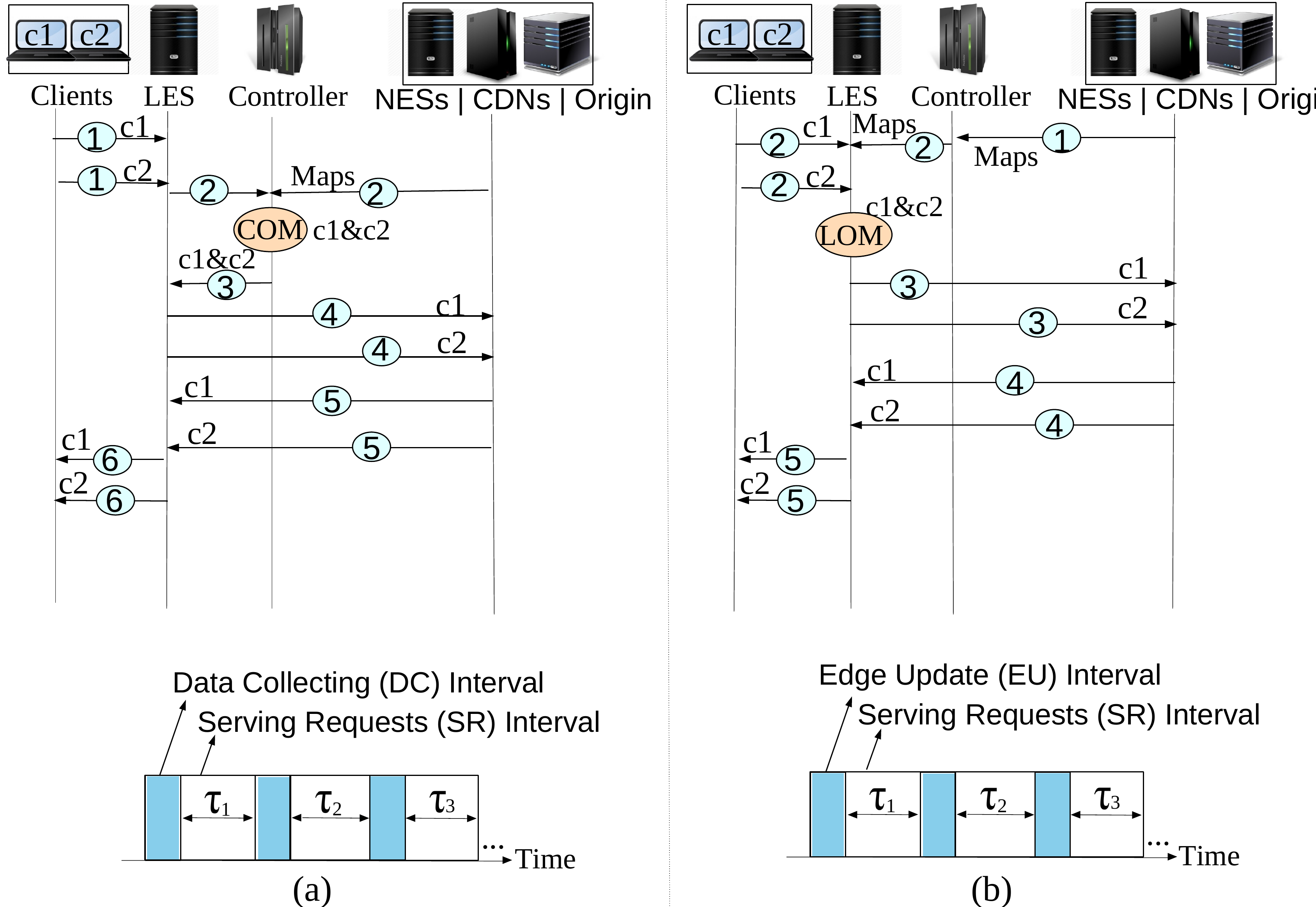}
\caption{\small Workflows and proposed time slot structures: (a) Centralized, (b) Coarse-Grained approaches.}
\vspace{.5cm}
\label{ARARAT-workflows1}	
\end{figure}
%%%%%%%%%%%%%%
Motivated by the characteristics of the objective function (\ref{ARARAT:eq:12}), this section proposes two heuristic approaches called \textit{Coarse-Grained} (CG) and \textit{Fine-Grained} (FG) methods that aim at satisfying the proposed constraints Eqs.~(\ref{ARARAT:eq:1})--(\ref{ARARAT:eq:11}). In the CG approach, each edge server runs a reduced version of the MILP model (\ref{ARARAT:eq:12}) in terms of the number of variables. Indeed, the reduced MILP model inherits the defined constraints Eqs.~(\ref{ARARAT:eq:1})-(\ref{ARARAT:eq:11}), but it produces only an optimal solution for its host edge server. Although we split the MILP model (\ref{ARARAT:eq:12}) in the CG model, the SDN controller is still required to work as a coordinator node for collecting/advertising parameters from/to edge servers. Subsequently, we will show that the CG approach suffers from high time complexity when the number of requests goes up. To cope with this challenge and also take shared links between edge servers and CDN/origin servers (Theorem II) into account, we then propose two Fine-Grained (FG) heuristic approaches, namely FG I and FG II. In the FG approaches, each edge server runs a lightweight heuristic algorithm upon receiving a request.

Before describing the details of each approach, let us consider the simple example introduced in Figs.~\ref{ARARAT-workflows1} and \ref{ARARAT-workflows2} and then define the workflow and time slot structure of the discussed centralized approach (Section~\ref{sec:System Model:MILP}) as a baseline method (see Fig.~\ref{ARARAT-workflows1}(a)). The time slot structure in the centralized model (Fig.~\ref{ARARAT-workflows1}(a)) consists of two intervals: \textit{(i)} \textit{Data Collecting} (DC) interval and \textit{(ii)} \textit{Serving Requests} (SR) interval. 
In the DC interval, the SDN controller collects clients' requests (\eg $c1$ and $c2$ in Fig.~\ref{ARARAT-workflows1}(a)) from the edge servers and some important resource maps from the network and servers, \ie \textit{bandwidth\_map, cache\_map, edge\_map, comp\_map} (step 2). For simplicity in displaying the workflows, zero propagation delay is assumed. Having the provided information, the SDN controller in the SR interval runs the Centralized MILP Optimization Model (COM) (\ref{ARARAT:eq:12}), which considers serving delay and network cost as its objective and informs LES and NESs about the optimal actions (3). In the next step, each edge server employs the determined action (4), and after fetching/preparing the requested qualities (5), it serves clients (6). This time slot structure continues as long as the controller receives the requests from the edge servers. 
%%%%%%%%%%%%%%%%%%%%%%%%%%%%%%%%%%%%%%%%%%%%%%%%%%%%%%%%%%%%%%%%%%%%%%%%%%%%%%%%%%%%%%%%%%%%%%%%%%%%%%%%%%%%%%%%%%%%%%%%%%%%%%%%%%%%%%%%%%%%%%%%%%%%%%%%%%%%%%%%%%%%%%%%%%%%%%%%%%%%%%%%%%%%%%%%%%%%%%%%%%%%%%%
\subsubsection{ARARAT Coarse-Grained Heuristic Approach}
\label{sec:CGA}
Inspired by the frameworks introduced in chapter \ref{chap:EdgeSDN}, we propose a coarse-grained approach; its workflow is shown in Fig.~\ref{ARARAT-workflows1}(b). This approach again uses a time-slot structure which includes two intervals: \textit{(i)} \textit{Edge Update} (EU) interval and \textit{(ii)} \textit{Serving Requests} (SR) interval. In the first interval, the SDN controller updates each edge server (\eg LES in Fig.~\ref{ARARAT-workflows1}(b)) via the \textit{resource maps} (step 2) that are received periodically from the network and servers (1). Moreover, a number of clients' requests are collected by the associated edge servers in the EU interval as well. Let us define binary variables $X^{p,r}_{i,t}$, and $Y^{p,r}_{i,t}$ for each LES $j$, where $X^{p,r}_{i,t}=1$ shows LES $j$ receives quality $p$ in response to request $r$ through server $i$ (\ie CDN, origin, or NES)
with transcoding status $t$, while $Y^{p,r}_{i,t}=1$ indicates LES $j$ sends quality $p$ in response to request $r$ to server $i$ (\ie NES) with transcoding status $t$. 

Furthermore, to simplify the notation, let us change the introduced variable $T^{p,r}_{i,j}$ into $T^{p,r}_{i}$ and define $\Lambda_x$ and $\Lambda_y$ as bandwidth costs when receiving $p$ from or sending $p$ to other servers, respectively (refer to the notation in Table~\ref{tab:ARARAT:notation}). 
Therefore, each edge server $j$ in the SR interval runs a \textit{Local MILP Optimization Model} (LOM) (Eq.~(29)), which considers serving delay and network cost as its objective for the collected requests to answer the key questions mentioned in Section~\ref{sec:PROBLEM STATEMENT} (steps 3 and 4 in Fig.~\ref{ARARAT-workflows1}(b)) and finally serves clients' requests (5). In other words, in the CG approach, the centralized MILP model is split into multiple LOM models, each of which is executed on an edge server individually by coordination of the SDN controller.
To do that, by having minor modifications in the objective function (\ie Eq. (\ref{ARARAT:eq:12})) and constraints (\ie Eqs.~(\ref{ARARAT:eq:1})-(\ref{ARARAT:eq:11})), we introduce the coarse-grained optimization model as follows:
%%%%CONS15

\textbf{\textit{(i) }AS Constraint}
\begin{flalign}
&1\leq\hspace{-.3cm}\sum_{i\in\mathcal{C}\cup\mathcal{V}}\sum_{t\in\mathcal{T}}\sum_{p\in\mathcal{Q}^r} (X^{p,r}_{i,t}~.~\alpha^{p,r}_{i})
+\hspace{-.2cm}\sum_{i\in\mathcal{V}\setminus j}\sum_{p\in\mathcal{Q}^r} (Y^{p,r}_{i,t=0}~.~ \alpha^{p,r}_{j}+ X^{p,r}_{i,t=1}~.~ (1-\alpha^{p,r}_{i}))
\leq 2,&\hspace{-.3cm}\forall r\in\mathcal{R}_j
\label{ARARAT:eq:15}
\end{flalign}
%%%%CONS16

\textbf{\textit{(ii) }ST Constraints}
\begin{flalign}
&T^{p,r}_{i} =
\begin{cases}
\frac{Y^{p,r}_{i,t=0}~.~\delta^r_{p}+X^{p,r}_{i,,t=1}~.~\delta^r_{p^{*}}}{\omega_{i,j}}
\hspace{.4cm}\forall r\in\mathcal{R}_j, i\in\mathcal{V} \setminus j
\\
\\
\frac{\sum_{t\in\mathcal{T}} X^{p,r}_{i,t}~.~\delta^{r}_p}{\omega_{i,j}} 
\hspace{.4cm}\forall r\in\mathcal{R}_j,  i\in\{\mathcal{C}\cup\mathcal{V}\}\setminus j
\end{cases} 
\label{ARARAT:eq:16}
\end{flalign}
%%%%CONS17
\begin{flalign}
&\sum_{p\in \mathcal{Q}^r} X^{p,r}_{i,t=2}~.~\mu^{r}_{p}\leq \tau_{j}^{r},&\forall  i\in\mathcal{V}\cup\mathcal{C},r\in\mathcal{R}
\label{ARARAT:eq:17}
\end{flalign}
%%%%CONS18
\begin{flalign}
&\sum_{r\in \mathcal{R}}\sum_{i\in\mathcal{V}\cup\mathcal{C}}\sum_{p\in \mathcal{Q}^r}T^{p,r}_{i}+\tau^{r}_j\leq \Gamma&&
\label{ARARAT:eq:18}
\end{flalign}
%%%%CONS19

\textbf{\textit{(iii) }CO Constraints}
\begin{flalign} 
&\sum_{t\in \mathcal{T}}\sum_{p\in\mathcal{Q}^r} X^{p,r}_{i=0,t}~.~p = p^*, && \forall r\in R_j
\label{ARARAT:eq:19}
\end{flalign}
%%%%CONS20
\begin{flalign} 
&\sum_{i\in\mathcal{C}}\sum_{r\in R_j}\sum_{p\in\mathcal{Q}^r} X^{p,r}_{i,t=1}~.~p = 0 && &&
\label{ARARAT:eq:20}
\end{flalign}
%%%%CON21

\textbf{\textit{(iv) }RC Constraints}
\begin{flalign}
&\sum_{t\in \mathcal{T}}\sum_{r\in R_j}\sum_{p\in \mathcal{Q}^r} \hspace{-.1cm}X^{p,r}_{i,t}~.~\eta_p \leq \omega_{i,j},&&\forall i\in\mathcal{C}\cup\mathcal{V}
\label{ARARAT:eq:21}
\end{flalign}
%%%%CONS22
\begin{flalign}
&\sum_{t\in \mathcal{T}}\sum_{r\in R_j}\sum_{p\in \mathcal{Q}^r} \hspace{-.1cm}Y^{p,r}_{i,t}~.~\eta_p \leq \omega_{i,j},&&\forall i\in\mathcal {V} \setminus j
\label{ARARAT:eq:22}
\end{flalign}
%%%%CONS23
\begin{flalign}
&\sum_{r\in \mathcal{R}}\sum_{p\in \mathcal{Q}^r}(\sum_{i\in\mathcal{C}\cup\mathcal{V}}X^{p,r}_{i,t=2} + \sum_{i\in\mathcal{V} \setminus j} Y^{p,r}_{j,t=1})~.~\pi^{r}_{p} \leq \Omega_j&&
\label{ARARAT:eq:23}
\end{flalign}
%%%%CONS24

\textbf{\textit{(v) }NC Constraints}
\begin{flalign} 
&\Delta_{bw}~.~\sum_{t\in \mathcal{T}}\sum_{r\in R_j}\sum_{p\in \mathcal{Q}^r} \hspace{-.1cm}X^{p,r}_{i,t}~.~\eta_p \leq \Lambda_{x} ,&&\hspace{-.2cm}\forall i\in\{\mathcal{C}\cup\mathcal{V}\}\setminus j
\label{ARARAT:eq:24}
\end{flalign}
%%%%%CONS25
\begin{flalign} 
&\Delta_{bw}~.~\sum_{t\in \mathcal{T}}\sum_{r\in R_j}\sum_{p\in \mathcal{Q}^r} \hspace{-.1cm}Y^{p,r}_{i,t}~.~ \eta_p \leq \Lambda_{y} ,&\hspace{-.4cm}\forall i\in\mathcal {V}\setminus j
\label{ARARAT:eq:25}
\end{flalign}
%%%%%CONS26
\begin{flalign}
&\Lambda_x+\Lambda_y \leq \Lambda&&
\label{ARARAT:eq:26}
\end{flalign}
%%%%%CONS27
\begin{flalign}
&\pi^{r}_{p}~.~\sum_{r\in \mathcal{R}}\sum_{p\in \mathcal{Q}^r}(\sum_{i\in\mathcal{C}\cup\mathcal{V}}X^{p,r}_{i,t=2}+ \sum_{i\in\mathcal{V} \setminus j} Y^{p,r}_{i,t=1})~.~\Delta_{tr} \leq \Pi&&
\label{ARARAT:eq:27}
\end{flalign}
%%%%%CONS28
\begin{flalign}
&\Pi+\Lambda \leq \Psi&&
\label{ARARAT:eq:28}
\end{flalign}
%%%%%OBJ

\textbf{Local MILP Optimization Model}
\begin{flalign}
\textit{Minimize}&\hspace{.3cm}\beta_1~.~ \frac{\Gamma}{\Gamma^*} + \beta_2~.~\frac{\Psi}{\Psi^*}
\label{ARARAT:eq:29}\\
  s.t.&\hspace{.2cm}\text{constraints}\hspace{.5cm}\text{Eq.}(\ref{ARARAT:eq:15})-\text{Eq.}(\ref{ARARAT:eq:28})&&\nonumber\\
  vars.&\hspace{.2cm} T^{p,r}_{i},\tau^{r}_j, \Gamma, \Psi, \Pi,\Lambda_x, \Lambda_y, \Lambda \geq 0, X^{p,r}_{i,t}, Y^{p,r}_{i,t}\in\{0,1\}\nonumber 
\end{flalign}

It is worth noting that each edge server can individually set coefficients $\beta_1$ and $\beta_2$ based on its policy. Although using the localization technique in LOM can mitigate the time complexity problem of the centralized optimization approach (\ie COM), applying the analysis used in Section~\ref{sec:System Model:analysis} demonstrates that the LOM is still suffering from high time complexity and is not practical at scale. Consequently, we propose two fine-grained approaches that can be employed in practical large-scale scenarios.
%%%%%%%%%%%%%%%%%%%%%%%%%%%%%%%%%%%%%%%%%%%%%%%%%%%%%%%%%%%%%%%%%%%%%%%%%%%%%%%%%%%%%%%%%%%%%%%%%%%%%%%%%%%%%%%%%%%%%%%%%%%%%%%%%%%%%%%%%%%%%%%%%%%%%%%%%%%%%%%%%%%%%%%%%%%%%%%%%%%%%%%%%%%%%%%%%%%%%%%%%%%%%%%
\subsubsection{ARARAT Fine-Grained Heuristic Approach I}
\label{sec:FGAI}
%%%%%%%%%%%%%%
\begin{figure}[t]
\centering
\includegraphics[width=1\textwidth]{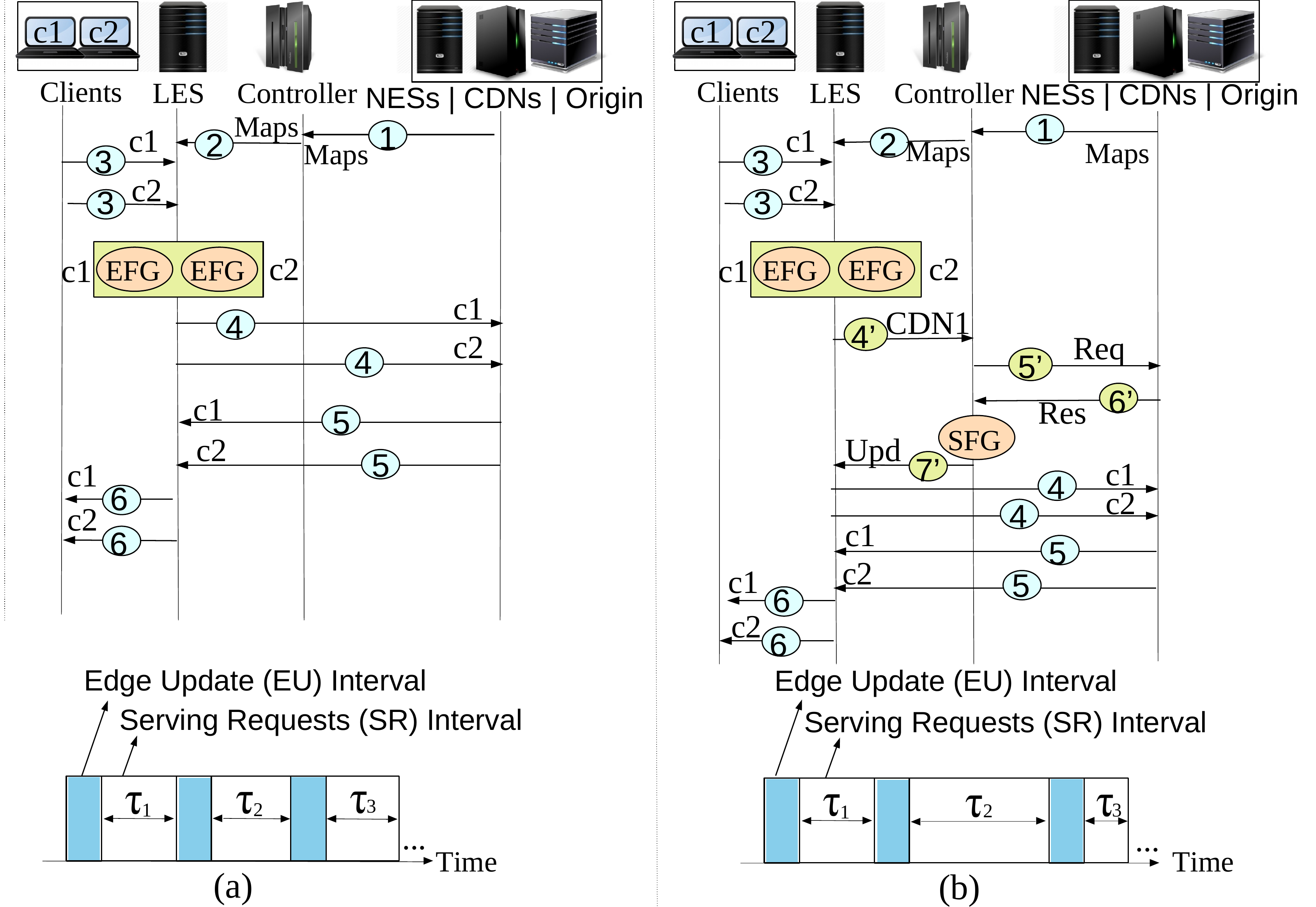}
\caption{\small Workflows and proposed time slot structures: (a) Fine-Grained I, (b) Fine-Grained II approaches.}
\vspace{.5cm}
\label{ARARAT-workflows2}	
\end{figure}
%%%%%%%%%%%%%%
Inspired by the CG method, the proposed Fine-Grained approach I is designed in two main steps: edge servers \textit{(i)} receive essential resource maps from the SDN controller in the EU interval (step 2 in Fig.~\ref{ARARAT-workflows2}(a)) and \textit{(ii)} execute an \textit{Edge Fine-Grained} (EFG) algorithm in the SR interval. 
Due to the lightweight property of the proposed approach, we employ the EFG algorithm \textit{per request} in lieu of executing it for a number of collected requests. The EFG algorithm, shown in Alg.~\ref{EFGH1}, begins with the $resource\_maps$ (\ie $cache\_map$, $edge\_map$, map of bandwidths to other servers in $\mathcal{V}\cup\mathcal{C}$ ($bw\_map$), the NESs' available compute capacity ($comp\_map$)), the client's request $r$ and an $on\_the\_fly$ list (holding requests currently being served) as input parameters. As a result, it serves $r$ with an optimal action $opt\_action$ from the action tree (Fig.~\ref{ARARAT:actionTree}). 
Whenever the new request $r$ is received, $r$ will be compared to the requests in the $on\_the\_fly$ list. The \textit{HoldReq} function is called when $r$ is in the $on\_the\_fly$ list (i.e., $r$ is requested by another client and still is in progress) to hold the request in order to prevent sending identical requests (lines 2--3). 

When $r$ is not found in the $on\_the\_fly$ list, that means $r$ is a completely new request and the $on\_the\_fly$ list must be updated by $r$ to be processed (line 5). Therefore, the \textit{HoldReq} function with the $on\_the\_fly$ list enables edge servers to send exactly one request instead of multiple identical requests and thus saves network resources, \eg bandwidth and computation, and prevents network congestion. By extracting the requested quality $p^*$ from request $r$, considering the received \textit{resource maps}, the set of CDN servers ($\mathcal{C}$) and NESs ($\mathcal{V}$), and coefficient values $\beta_1$, $\beta_2$, the proposed algorithm (Alg.~\ref{EFGH1}) calls the \textit{CostFunction} function, described as Alg.~\ref{EFGH1-COST}, to determine the optimal action based on the objective function in Eq.~(\ref{ARARAT:eq:29}) and adjusted weighting coefficients (\ie $\beta_1$ and $\beta_2$) for the request $r$. The result is stored in $opt\_ action$ (line 6). Finally, the edge server calls the \textit{ServeRequest} function to serve the client's request $r$ (line 7) and updates the \textit{resource maps} as well as the $on\_the\_fly$ list via the \textit{UpdateVariables} function (line 8). 
\raggedbottom
% %%%%%%%%%%%%%%%%%%%%%%%ARARAT:EFGH1%%%%%%%%%%%%%%%%%%%%%%%
\begin{center}
	\begin{algorithm}[!t]
		\small
            \caption{\small ARARAT Edge Fine-Grained (EFG) heuristic algorithm I.}\label{EFGH1}
		\begin{algorithmic}[1]
            \State \textbf{Input} $r$, $resource\_maps$, $on\_the\_fly$
            \If{$r\in on\_the\_fly$}
                \State HoldReq($r$)
            \Else
            \State $on\_the\_fly$.add($r$)
            \State $opt\_action\leftarrow$CostFunction($r$, $p^*$, $\mathcal{C}$, $\mathcal{V}$, $\beta_1$, $\beta_2$, $resource\_maps$)
            \State ServeRequest($r$,$opt\_action$)
            \State UpdateVariables$()$
            \EndIf
	   \end{algorithmic}
	\end{algorithm}
\end{center}
% %%%%%%%%%%%%%%%%%%%%%%%ARARAT:EFGH1%%%%%%%%%%%%%%%%%%%%%%%
\raggedbottom

The \textit{CostFunction} function (Alg.~\ref{EFGH1-COST}) consists of three main parts: \textit{(i)} calculating the $serving\_time$ and $network\_cost$ of all feasible actions (lines 3--77), \textit{(ii)} calculating the objective values (Eq.~\ref{ARARAT:eq:29}) for the determined actions (lines 78--80), and  \textit{(iii)} finding the minimum objective value and returning its associated action as the optimal action $opt\_action$ (lines 81--83). More precisely, the possible quality levels for serving $p^*$ requested by $r$ (\ie $\mathcal{Q}^{r}$) are extracted from $edge\_map[0]$ (\ie the LES $edge\_map$) (line 3). If $\mathcal{Q}^{r}$ includes at least one $p \geq p^*$, the minimum quality will be selected as $p$ (line 4). If the selected quality $p$ is identical to the desired quality $p^*$, the  
$network\_cost[1]$ and $serving\_time[1]$ associated with action 1 of the action tree (Fig.~\ref{ARARAT:actionTree}) will be set to zero (lines 5--7). Otherwise, if $p>p^*$, and the available resources in the LES (\ie $comp\_map[0]$) are enough for running the transcoding task, action 2 will be selected to transcode $p^*$ from $p$ at the LES with $\mu^{r}_{p}$ and $\pi^{r}_{p}~.~\Delta_{tr}$ as $serving\_time$ and $network\_cost$, respectively (lines 9--11). If the transcodable version $p$ exists in $\mathcal{Q}^{r}$, but $comp\_map[0]$ is not available at the LES, the algorithm can select the best NES (denoted by $SS$) in terms of the available bandwidth and the available computational resources for running action 6 (lines 13--20). 

Note that the threshold value $thr\_comp$ is defined as forcing NESs not to use more than $thr\_comp$ percent of their computational resources to run transcoding processes requested by their neighboring edge servers. In fact, each edge server allocates the desired amount of its computational resources to serve its own requests by setting the $thr\_comp$ value. If $SS$ can be selected by the described policy, $serving\_time$ and $network\_cost$ associated with action 6 will be set (lines 24--26). If $\mathcal{Q}^{r}$ is empty or the computational resources are insufficient at the LES, other servers (\ie $\mathcal{C}\cup\mathcal{V}\setminus{V[0]}$) can assist the LES to serve $p$ (line 31). To accomplish this, all servers that include $p\geq p^*$ are determined, and their available bandwidths to the LES are stored in the auxiliary data structure $dict1$. Moreover, the available computational resources of all servers containing at least one $p>p^*$ are stored in the $dict2$ data structure (lines 32--38). If $dict1$ is not empty and the selected server $SS$ with maximum available bandwidth includes the  quality $p=p^*$, the $serving\_time$ and $network\_cost$ entries for action 4, action 7 or action 9, respectively, are set (lines 39--55). If there is not a selected quality $p$ equal to $p^*$ and the selected server $SS$ is an edge server, the location of transcoding will be specified by the maximum available computing resources, and then the $SS$ value will be updated (lines 57--60). Therefore, based on the $SS$ value and  $comp\_map$, the transcoding function will be run at the best NES or the LES server through action 5 or action 3 (lines 56--75); otherwise, $p$ is retrieved from the CDN server with the highest available bandwidth and transcoded at the LES (\ie action 8) (lines 71--75).
Then $max\_cost$ and $max\_time$ are determined from $network\_cost$ and $serving\_time$, respectively (lines 76--77). In the subsequent loop, the objective values for all determined feasible actions are calculated and saved to the $obj$ data structure (lines 78--80). Finally, the minimum objective value is determined and its action is saved and returned as $opt\_action$ (line 81--83).

Assume $k$ denotes the number of discussed feasible actions. The time complexity of this approach is $O(k)$. Note that bandwidth allocation for the shared links between edge servers is still not taken into account in this approach. The following approach completes the introduced method by a bandwidth allocation strategy.
\clearpage
%%%%%%%%%%%%%%%%%%%%%%%Alg2-CostFunction
\begin{center}
 \begin{algorithm}[!t]
            \small
            \caption{\small ARARAT CostFunction}\label{EFGH1-COST}
		\begin{algorithmic}[1]
            \State \textbf{Input} $r$, $p^*$, $\mathcal{C}$, $\mathcal{V}$ $\beta_1$, $\beta_2$, $resource\_maps$
            \State \textbf{Output} $opt\_action$
            \If{($\mathcal{Q}^{r}\subset edge\_map[0]$)}
                \State $p\leftarrow$ min($\mathcal{Q}^{r}$)
                \If{$p==p^*$}
                     \State $network\_cost[1]\leftarrow$ 0
                     \State $serving\_time[1]\leftarrow$ 0
                \Else
                    \If{$comp\_map[0]$}
                        \State $network\_cost[2]\leftarrow \pi^{r}_{p}~.~\Delta_{tr}$
                        \State $serving\_time[2]\leftarrow \mu^{r}_{p}$ 
                    \Else
                        \For{$i\in\mathcal{V}\setminus{\mathcal{V}[0]}$}
                            \State $min\_bw\leftarrow \eta_p$
                            \State $min\_comp\leftarrow0$
                            \If{$comp\_map[i]\geq thr_{comp}$}
                                \If{$bw\_map[i]>min\_bw$ or $comp\_map[i]>min\_comp$}
                                    \State $min\_bw\leftarrow bw\_map[i]$
                                    \State $min\_comp\leftarrow comp\_map[i]$
                                    \State $SS\leftarrow i$
                                \EndIf
                            \EndIf
                        \EndFor
                        \If{$SS$}
                            \State $network\_cost[6]\leftarrow \pi^{r}_{p}~.~\Delta_{tr} + (\eta_p + \eta_{p^*})~.~\Delta_{bw}$
                            \State $serving\_time[6]\leftarrow \mu^{r}_{p} + \frac{\eta_p + \eta_{p^*}}{bw\_map[SS]}$
                        \EndIf
                    \EndIf
                \EndIf
            \EndIf
            \If{not ($\mathcal{Q}^{r}\subset edge\_map[0]$ or $comp\_map[0]$)}
                \For{$i\in\{\mathcal{V}\cup\mathcal{C}\}\setminus{\mathcal{V}[0]}$}
                    \If{($\mathcal{Q}^{r}\subset edge\_map[i] or cache\_map[i]$)}
                        \State $dict1[i]\leftarrow bw\_map[i]$
                        \State $dict2[i]\leftarrow comp\_map[i]$
                    \EndIf
                \EndFor
            \EndIf\label{alg:last-step}
            \end{algorithmic}
    \end{algorithm}
\end{center}
\clearpage
%%%%%%%%%%%%%%%%%%%
\begin{center}
 \begin{algorithm}[!t]
    % \caption{CostFunction()}\label{EFGH1-COST2}
    \small
    \addtocounter{ALG@line}{-1}
    \begin{algorithmic}[1]
    \setcounterref{ALG@line}{alg:last-step}
        \If{dict1}
            \State $max\_bw\leftarrow$  max($dict1$.values())
             \State $SS\leftarrow$ $dict1$.index($max\_bw$)
             \State $p\leftarrow$ min($\mathcal{Q}^r \subset edge\_map[SS]$)
             \If{$p==p^*$}
                 \If{$SS\in \mathcal{V}\setminus{\mathcal{V}[0]} and $max\_bw$>\eta_p$}
                    \State $network\_cost[4]\leftarrow$ $\eta_{p^*}~.~ \Delta_{bw}$
                    \State $serving\_time[4]\leftarrow \frac{\eta_{p^*}}{bw\_map[SS]}$
                \EndIf
                \If{$SS != \mathcal{C}[0] and $max\_bw$>\eta_p$}
                    \State $network\_cost[7]\leftarrow$ $\eta_{p^*}~.~\Delta_{bw}$
                    \State $serving\_time[7]\leftarrow \frac{\eta_{p^*}}{bw\_map[SS]}$
                \EndIf
                \If{$max\_bw >\eta_p$}
                    \State $network\_cost[9]\leftarrow$ $\eta_{p^*}~.~ \Delta_{bw}$
                    \State $serving\_time[9]\leftarrow \frac{\eta_{p^*}}{bw\_map[SS]}$
                \EndIf
            \Else
                \If{$SS \in \mathcal{V}\setminus{\mathcal{V}[0]}$}
                    \State $max\_comp\leftarrow$  max($dict2$.values())
                   \State $SS\leftarrow$$ dict2$.index($max\_comp$)
                   \State $p\leftarrow$ min($\mathcal{Q}^r \subset edge\_map[SS]$)
                   \If{$SS \in \mathcal{V}\setminus{\mathcal{V}[0]} and max\_comp>thr_{comp}$}
                        \State $network\_cost[5]\leftarrow \pi^{r}_{p}~.~\Delta_{tr} + (\eta_{p^*})~.~\Delta_{bw}$
                        \State $serving\_time[5]\leftarrow \mu^{r}_{p} + \frac{\eta_{p^*}}{bw\_map[SS]}$
                    \EndIf
                    \If{$comp\_map[0]$}
                        \State $network\_cost[3]\leftarrow \pi^{r}_{p}~.~\Delta_{tr} + (\eta_p) ~.~ \Delta_{bw}$
                        \State $serving\_time[3]\leftarrow \mu^{r}_{p} + \frac{\eta_p}{bw\_map[SS]}$
                    \EndIf
                \EndIf
             \EndIf
             \If{$comp\_map[0]$}
                \State $network\_cost[8]\leftarrow \pi^{r}_{p}~.~\Delta_{tr} + (\eta_p)~.~\Delta_{bw}$
                \State$serving\_time[8]\leftarrow \mu^{r}_{p} + \frac{\eta_p}{bw\_map[SS]}$
            \EndIf   
        \EndIf
    \State $max\_cost\leftarrow$max($network\_cost$.values())
    \State $max\_time\leftarrow$max($serving\_time$.values())
    \For{$a \in network\_cost$}
        \State $obj[a]\leftarrow$ $\beta_1 ~.~ serving\_time[a]$/ max($max\_time$) + $\beta_2~.~Network\_cost[a] $ / max($max\_time$)
    \EndFor
    \State $min\_obj\leftarrow$ min($obj$.values())
    \State $opt\_action\leftarrow  obj$.index($min\_obj$)
    \State \textbf{Return} $opt\_action$
     \end{algorithmic}
     \end{algorithm}
\end{center}
%%%%%%%%%%%%%%%%%%%%%%%Alg2-CostFunction
%%%%%%%%%%%%%%%%%%%%%%%%%%%%%%%%%%%%%%%%%%%%%%%%%%%%%%%%%%%%%%%%%%%%%%%%%%%%%%%%%%%%%%%%%%%%%%%%%%%%%%%%%%%%%%%%%%%%%%%%%%%%%%%%%%%%%%%%%%%%%%%%%%%%%%%%%%%%%%%%%%%%%%%%%%%%%%%%%%%%%%%%%%%%%%%%%%%%%%%%%%%%%%%
\subsubsection{ARARAT Fine-Grained Heuristic Approach II}
\label{sec:FGA2}
In this method, clients' requests are served by edge servers again in fine-granular form (\ie per request) in a time-slotted fashion (Fig.~\ref{ARARAT-workflows2}(b), steps 1--6). As shown in Fig.~\ref{ARARAT-workflows2}(b), each edge server, in addition to running the \textit{Edge Fine-Grained} (EFG) algorithm II, can inform the SDN controller to run the \textit{SDN Fine-Grained} (SFG) algorithm (step 4') to allocate a new bandwidth value to the other servers (\eg CDN1). Note that informing the SDN controller and running EFG for serving clients' requests is executed separately in a distinct thread not to harm the performance. When the SDN controller receives a bandwidth allocation request from one or more edge servers, it collects bandwidth information from all edge servers (steps 5' and 6'), runs the SFG algorithm, and updates all edges with new bandwidth values. Therefore, the SR intervals (\ie $\tau_1, \tau_2$, and $\tau_3$ in Fig.~\ref{ARARAT-workflows2}(b)) could have different lengths based on asynchronous updating of the bandwidth values by the SDN controller. We describe the details of this approach by means of the EFG and SFG algorithms indicated in Alg.~\ref{EFGH2} and Alg.~\ref{SFGH2}, respectively.

\textbf{Edge Fine-Grained (EFG) Algorithm II.}
As mentioned earlier, this algorithm executes two functions in separate threads, named \textit{TriggerController} and \textit{ServingRequests}. 
Since each edge device receives resource maps inputs (\eg $cache\_map, edge\_map$ or $bw\_map$) from the SDN controller in the EU interval, it realizes the reason for cache miss events for the requested quality $p^*$. Therefore, if $p^*$ exists in the server $i\in\mathcal{C}\cup\mathcal{V}$, but 
the bandwidth between $i$ and $V[0]$ (\ie the LES) is not sufficient for the requested quality transmission, $i$ can not serve $p^*$ and then the counter $miss\_counter[i]$ will be updated at the end of the \textit{ServingRequests} thread by calling the \textit{UpdateVariables} function (line 17).
The value of the $miss\_counter[i]$ is compared with a predefined $thr\_miss$ threshold in a loop frequently by the \textit{TriggerController} thread. If the $miss\_counter[i]$ exceeds $thr\_miss$, then \textit{SendSDNMSG} function is called to inform the SDN controller about updating bandwidth values between $i$ and $V[0]$ by sending \textit{miss}\_\textit{bitrates}$[i]$ and \textit{hit}\_\textit{bitrates}$[i]$ information (lines 2--9).   \textit{miss}\_\textit{bitrates}$[i]$ and \textit{hit}\_\textit{bitrates}$[i]$ denote the amounts of total missed bitrates (\ie due to the bandwidth insufficiency) and served bitrates (in Mbps) from server $i$ to the LES, respectively. Moreover, it should be noted that $thr\_miss$ is an input parameter that can be set based on the edge layer policies. In the \textit{ServingRequests} thread, the requested quality $p^*$ is served similar to Alg.~\ref{EFGH1}. 
%%%%%%%%%%%%%%%%%%%%%%%Alg3%%%%%%%%%%%%%%%%%%%%%%%
\begin{center}
	\begin{algorithm}[!t]
		\small
            \caption{\small ARARAT Edge Fine-Grained (EFG) heuristic algorithm II.}\label{EFGH2}
		\begin{algorithmic}[1]
            \State \textbf{Input} $r$, $resource\_maps$, $on\_the\_fly$
            \State *//Thread (I): TriggerController()            
            \While{$True$}
                \For{$i\in \mathcal{\{C}\cup\mathcal{V\}}\setminus{\mathcal{V}[0]}$}
                    \If{$miss\_counter[i]> thr\_miss$}
                        \State SendSDNMSG(\textit{miss}\_\textit{bitrates}$[i]$, \textit{hit}\_\textit{bitrates}$[i]$
                    \EndIf               
                \EndFor
            \EndWhile
            \State *//Thread (II): ServingRequests()
            \If{$r\in on\_the\_fly$}
                \State HoldReq($r$)
            \EndIf
            \State $on\_the\_fly$.add($r$)
            \State $opt\_action\leftarrow$CostFunction($r$,$p^*$, $\mathcal{C}$, $\mathcal{V}$, $\beta_1$, $\beta_2$, $resource\_maps$)
            \State ServeRequest($r$,$opt\_action$)
            \State UpdateVariables$()$            
        \end{algorithmic}
      \end{algorithm}          
\end{center}                    
%%%%%%%%%%%%%%%%%%%%%%%Alg3%%%%%%%%%%%%%%%%%%%%%%%

\textbf{SDN Fine-Grained (SFG) Algorithm.}
In the first step, the SDN controller uses $bw\_map$ information, extracted from $resource\_maps$ and $net\_map$, the global information of the entire network topology, to determine one path from each edge server $j\in\mathcal{V}$ to other servers in $\{\mathcal{C}\cup\mathcal{V}\}\setminus j$ (line 2). We note that any policy for selecting the paths can be applied. We here use the maximum value of $\frac{available\_bandwidth}{hop\_count}$ as a path selection metric. It is clear that, for paths with the same hop count, the maximum available bandwidth is considered as a path selection metric. The bandwidth on a link shared by multiple paths between edge and/or other servers is allocated equally at the beginning of the system. In other words, a uniform bandwidth allocation strategy is utilized at system initiation time. However, a key question is: \textit{``What will happen if edge servers are unequally loaded and always receive unequal traffic?"} 
Therefore, the SFG algorithm with the \textit{SynUpdate} function updates all edge servers with all collected resource maps from the network and servers (\ie $cache\_map, edge\_map, bw\_map, comp\_map$) in the first EU interval (line 3). It is worth mentioning that the \textit{SynUpdate} function updates edge servers in a predefined time. 

In an infinite loop, the SDN controller employs \textit{receiveEdgeMSG} to check if any edge server requests a bandwidth update or not. If at least one bandwidth update request is detected, the \textit{EdgeServerStates} function is called to collect \textit{miss}\_\textit{bitrates}$[i]$ and \textit{hit}\_\textit{bitrates}$[i]$ from all servers $i\in\{\mathcal{C}\cup\mathcal{V}\}\setminus j$ to all $j\in\mathcal{V}$ and store results in the $all\_edge\_states$ data structure (lines 4-6). 
Next, based on the $all\_edge\_states$ information, the \textit{BWAllocation} function is called to determine a new amount of bandwidth ($new\_rate$) to each link in each route detected by the SDN controller (line 7). Subsequently, the new rates are configured by the \textit{ConfigRate} function on routes (line 8) and then all edge servers are updated by the \textit{AsynUpdate} function (line 9). Note that the described procedure will be repeated after detecting each bandwidth update request; therefore, the new bandwidth values are updated by \textit{AsynUpdate}. If no edge server demands bandwidth updates, \textit{SynUpdate} updates edge servers on a regular basis (line 11).
\raggedbottom
%%%%%%%%%%%%%%%%%%%%%%%Alg4%%%%%%%%%%%%%%%%%%%%%%%
\begin{center}
	\begin{algorithm}[!t]
		\small
            \caption{\small ARARAT SDN Fine-Grained (SFG) heuristic algorithm.}\label{SFGH2}
		\begin{algorithmic}[1]
            \State \textbf{Input} $net\_map, resource\_maps$
            \State $routes\leftarrow$SelectPath()           
            \State SynUpdate($\mathcal{V}$)            
            \While{$True$}
                \If{receiveEdgeMSG()}
                    \State $all\_edge\_states\leftarrow$EdgeServerStates()
                    \State $new\_rate\leftarrow$BWAllocation($routes, all\_edge\_states$)
                    \State ConfigRate($new\_rate$)
                    \State AsynUpdate($\mathcal{V}$)  % with miss\_counter,miss\_traffic,hit\_traffic update
                \Else 
                    \State SynUpdate($\mathcal{V}$)  % without miss\_counter,miss\_traffic,hit\_traffic update
                \EndIf
              
            \EndWhile         
        \end{algorithmic}
      \end{algorithm}          
\end{center}                    
%%%%%%%%%%%%%%%%%%%%%%%Alg4%%%%%%%%%%%%%%%%%%%%%%%

\raggedbottom
As mentioned earlier, to answer the aforementioned question, the \textit{BWAllocation} function receives the determined routes from each edge server to other servers (in $\mathcal{V}\cup\mathcal{C}$), plus $all\_edge\_states$, and then allocates new bandwidth values to the links of each route based on edge servers' traffic load in a fair way.
Let us define $d_{i,j}$ as the total amount of bitrates including \textit{hit}\_\textit{bitrates}$[i]$ and \textit{miss}\_\textit{bitrates}$[i]$ 
from server $i\in\{\mathcal{C}\cup\mathcal{V}\}\setminus j$ to server $j\in\mathcal{V}$.
Moreover, let us consider $x^{a,b}_{i,j}$ and $f^{a,b}_{i,j}$ as the allocated bandwidth and the fairness coefficient to the shared link $(a,b)$ in the route between $i$ and $j$.
In the following, we introduce a \textit{FairnessModel} as a linear programming (LP) optimization model that is employed by the \textit{BWAllocation} function to calculate the fairness coefficients and new allocated bandwidth values and store them in the $frn\_coef$ and $new\_rate$ lists, respectively.
This model should satisfy two groups of constraints:

\textbf{\textit{(i) }Fairness Constraints.} Based on the $all\_edge\_states$ information, the model should calculate fairness for each link in the path between $i$ to $j$. Therefore, the fairness can be expressed as follows:
\begin{flalign}
&d_{i,j}-x^{a,b}_{i,j}\leq (1-f^{a,b}_{i,j})~.~d_{i,j},&
\forall\text{ links }(a,b),i\in\mathcal{V}\cup\mathcal{C},j\in\mathcal{V}&\label{ARARAT:eq:30}
\end{flalign}
%%%%%%%%%%%%%%%%%%
\begin{flalign}
&F\leq f^{a,b}_{i,j}&&
&\forall\text{ links }(a,b),i\in\mathcal{V}\cup\mathcal{C},j\in\mathcal{V}
&\label{ARARAT:eq:31}
\end{flalign}
where $F$ is the minimum value of fairness among all fairness coefficients. 
%%%%%%%%%%%%%%%%%%

\textbf{\textit{(ii) }Bandwidth Constraint.} The bandwidth allocated to link $(a,b)$ in the path between $i$ and $j$, \ie $x^{a,b}_{i,j}$, should not violate the whole capacity $\omega^{a,b}_{i,j}$. Therefore the following constraint should be satisfied: 
\begin{flalign}
&\sum_{i\in\mathcal{V}\cup\mathcal{C}}\sum_{j\in\mathcal{V}}x^{a,b}_{i,j}\leq \omega^{a,b}_{i,j}&&
&\forall \text{ links }(a,b) \label{ARARAT:eq:32}
\end{flalign}
%%%%%%

\textbf{Fairness LP Optimization Model.} The following model maximizes the  links' fairness:  
\begin{flalign}
\textit{Maximize}&\hspace{.3cm}F
\label{ARARAT:eq:33}\\
&s.t.&\hspace{-.2cm}\text{constraints}\hspace{.2cm}
\text{Eq. }(\ref{ARARAT:eq:30})-\text{Eq. }(\ref{ARARAT:eq:32})&&\nonumber&&\\
&var.&\hspace{.2cm}F,f^{a,b}_{i,j},x^{a,b}_{i,j}\geq 0\nonumber&&
\end{flalign}
\begin{figure}[!t]
	\centering
	\includegraphics[width=.7\columnwidth]{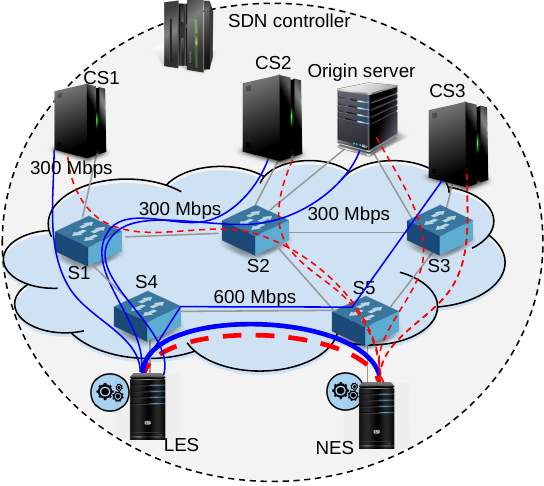}
	\caption{\small An example topology for bandwidth allocation strategy.}
        \vspace{.5cm}
	\label{bw-allocation-fig}
\end{figure}

To better illustrate the proposed bandwidth allocation strategy, a simple network example is depicted in Fig.~\ref{bw-allocation-fig} where five data paths from each edge server (LES and NES) to the neighboring edge server, cache servers (CS1, CS2, and CS3), and an origin server are depicted. The paths from LES and NES to other servers are shown by the blue and dashed red lines, respectively. Moreover, imagine all links' capacities between edge servers to CDN/origin servers and NES are 300 Mbps and 600 Mbps, respectively. As mentioned, The SDN controller allocates the bandwidth on a link shared by multiple paths (\eg S1--S2) between edge and/or other servers equally at the beginning of the system (\ie 100 Mbps for three depicted paths). For instance,  Table~\ref{bw-allocation-tbl} shows that the bandwidth values assigned to a shared link S1--S2 for LES to reach CS2 and origin, or NES to CS1, are calculated based on their requests, in a fair way.
%%%%%%%%%%%%
\begin{table}[!t]
\centering
\caption{Calculated fairness coefficients and allocated bandwidth for the example topology of Fig.~\ref{bw-allocation-fig}.}
\label{bw-allocation-tbl}
\begin{tabular}{|l|l|c|c|c|c|}
\hline
\textbf{Edge}                 & \textbf{DST} & \multicolumn{1}{l|}{\textbf{Selected Path}} & $d_{i,j}$ & $x^{1,2}_{i,j}$ & \multicolumn{1}{l|}{$\textbf{F}$} \\ \hline
\multirow{2}{*}{LES} & CS2 & S4-\color{red}{S1-S2}                           & 600       & 174.75          & \multirow{3}{*}{.291}   \\ \cline{2-5}
                     & Origin & S4-\color{red}{S1-S2}                        & 180       & 52.42           &                          \\ \cline{1-5}
NES                 & CS1 & S5-\color{red}{S2-S1}                           & 250       & 72.81           &                          \\ \hline
\end{tabular}
\vspace{.5cm}
\end{table}
%%%%%%%%%%%%

Assuming $k$ as the total number of actions, the time complexity of the FG algorithm II is similar to FG I (\ie $O(k)$) since, for each request, the edge server should select an optimal action regarding the value of the cost function. 
Moreover, assume $c$, $n$, and $l$ indicate the number of CDN/origin servers, edge servers, the maximum number of shared links in each path. The overall time complexity of the LP optimization model~(Eq.~\ref{ARARAT:eq:33}) in the worst case would be $O(c~.~ n ~.~l)$.
%%%%%%%%%%%%%%%%%%%%%%%%%%%%%%%%%%%%%%%%%%%%%%%%%%%%%%%%%%%%%%%%%%%%%%%%%%%%%%%%%%%%%%%%%%%%%%%%%%%%%%%%%%%%%%%%%%%%%%%%%%%%%%%%%%%%%%%%%%%%%%%%%%%%%%%%%%%%%%%%%%%%%%%%%%%%%%%%%%%%%%%%%%%%%%%%%%%%%%%%%%%%%%%%%%%
\subsection{ARARAT Performance Evaluation}
\label{sec:ARARAT:Performance Evaluation}
This section explains the evaluation setup, metrics, and methods and evaluates the performance of \texttt{ARARAT} compared to baseline and \sota methods.
%%%%%%%%%%%%%%%%%%%%%%%%%%%%%%%%%%%%%%%%%%%%%%%%%%%%%%%%%%%%%%%%%%%%%%%%%%%%%%%%%%%%%%%%%%%%%%%%%%%%%%%%%%%%%%%%%%%%%%%%%%%%%%%%%%%%%%%%%%%%%%%%%%%%%%%%%%%%%%%%%%%%%%%%%%%%%%%%%%%%%%%%%%%%%%%%%%%%%%%%%%%%%%%%%%%
\subsubsection{Evaluation Setup}
\label{sec:setup}
\begin{figure}[!t]
	\centering
	\includegraphics[width=.9\columnwidth]{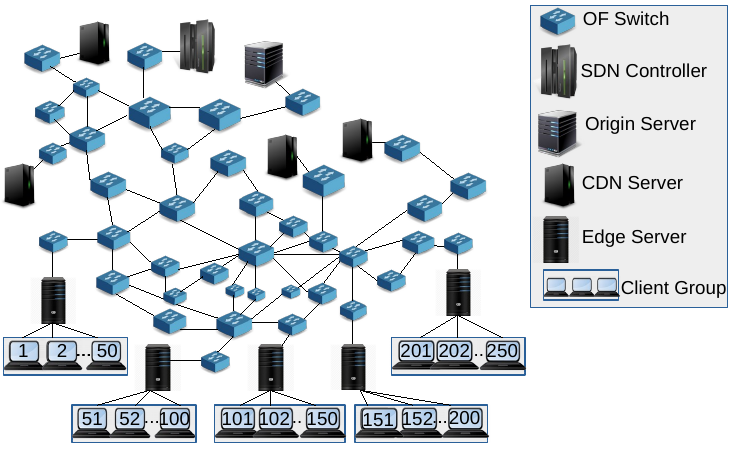}
	\caption{ARARAT evaluation testbed.}
        \vspace{.5cm}
	\label{ARARAT:testbed-topo}
\end{figure}

We employ a real network topology named Geant~\cite{zoo} to evaluate the \texttt{ARARAT} framework in a realistic large-scale setting. Our testbed is instantiated on the CloudLab~\cite{ricci2014introducing} environment. As depicted in Fig~\ref{ARARAT:testbed-topo}, we utilize 301 components (each of them runs on Ubuntu 18.04 LTS inside Xen virtual machines), \ie 40 OpenFlow (OF) switches, five edge servers, a FloodLight~\cite{Floodlight} SDN controller, four CDN servers, an origin server (running an Apache HTTP server~\cite{Apache} and MongoDB~\cite{Mongodb}), and 250 open-source AStream~\cite{juluri2015sara, AStream} DASH players working in
a headless mode. Python-based HTTP servers and the Python PuLP library~\cite{PuLP} with the CPLEX solver library~\cite{PuLP} are employed to implement edge servers and the proposed optimization models (\ie MILPs and LP), respectively. Furthermore, all modules of edge servers, as well as the SDN controller, are implemented in Python to serve requests received from DASH players. For the sake of simplicity, we assume that all clients have joined the network.
%%%%%%%%%%%%
%%%%%%%%%%%%

\textit{BOLA}~\cite{spiteri2016bola} and \textit{SQUAD}~\cite{wang2016squad} are used in DASH clients as buffer-based and hybrid quality adaptation algorithms, respectively. Fifty video sequences~\cite{lederer2012dynamic} with 300 seconds duration, comprising two-second segments in bitrate ladder (representation set) \{(0.089,240), (0.262,360), (0.791,720), (2.4,1080), (4.2,1080)\} (Mbps, vertical resolution) are used in all experiments except in the third experiment of scenario I (see Section~\ref{sec:Methods}). For that experiment, we use the same video sequences/segments in the following bitrate ladders: \{(0.089,240), (0.131,240), (0.262,360), (0.595,480), (0.791,720), (1.5,720), (2.4,1080), (4.2,1080)\} (8 representations); and
\{(0.089,240), (0.131,240), (0.262,360), (0.595,480), (0.791,720), (1,720), (1.5,720), (2.4,1080), (3,1080), (3.5,1080), (3.8,1080) , (4.2,1080)\} (12 representations).
Our work~on \texttt{CSDN}~\ref{chap:EdgeSDN:CSDN} demonstrates that when cache servers contain a small number of video segments, the number of fetches from the origin server and the backhaul bandwidth consumption increase and the system tends to apply more transcoding-based actions to save backhaul traffic. However, for the sake of simplicity, the total cache size of CDN servers is set to 40\% of the video dataset size, while the origin server holds all video sequences. Each edge server contains a partial cache with only 5\% capacity of the video dataset size. To avoid a slow startup of the system, the most popular video sequences are pre-cached on the edge servers. 

It is worth mentioning that \texttt{ARARAT} is entirely independent of the caching policies and compatible with any type of caching strategies. However, for simplicity, Least Recently Used (LRU) is used in all CDN and edge servers' partial caches as the cache replacement policy. 
Our experiments assume that the number of source video sequences and the video sequences' popularities will remain unchanged, and clients can request any bitrate of a segment from the mentioned bitrate ladder. For simplicity, we also assume that the access popularity of each video is known in advance and sorted in descending order. The video access probability is generated following a Zipf distribution~\cite{cherkasova2004analysis} with the skew parameter $\alpha=0.75$, \ie the probability of an incoming request for the $i^{th}$ most popular video is given as $prob(i)=\frac{1/i^{\alpha}}{\sum_{j=1}^{K}1/j^{\alpha}}$, where $K=50$. The Docker image \textit{jrottenberg/ffmpeg}~\cite{ffmpeg} is utilized to measure the segment transcoding time over edge servers. Moreover, the SDN controller employs the Linux \textit{Wondershaper} tool~\cite{wondershaper} to set $new\_rate$ values for the shared links in each selected path.

In \texttt{ARARAT} and described \sota systems, most costs consist of bandwidth and computational (\ie for running transcoding tasks) costs. Therefore, the fraction of other service costs are negligible compared to the above two kinds of costs. The computational and bandwidth costs (\ie $\Delta_{tr}$ and $\Delta_{bw}$) are set to $0.029\$$ per CPU per hour and $0.12\$$ per GB, respectively~\cite{aws-calc}. In the literature, the bandwidth values from an edge server to another edge server are assumed greater than to CDNs as well as to the origin server~\cite{yi2017lavea, al2019multi}. Therefore, we set the bandwidth values of all links in different paths from each LES to the origin server, cache servers, and other NESs to 50, 100, and 200 Mbps, respectively. To this aim, we limit a link of each path from each LES to the NES, CDN, and origin servers with the aforementioned bandwidth values as a bottleneck bandwidth. A 4G network trace~\cite{raca2018beyond} collected on bus rides is used for links between clients to edge servers to emulate the mobile network conditions in all experiments. The thresholds $thr\_comp$ and $thr\_miss$ are set to 50\% of the available computational resources (\ie $\Omega_i$) and 100, respectively. Furthermore, we set the weighting parameters $\beta_1=0.5, \beta_2=0.5$ and the SDN monitoring interval to one second in all experiments.
%%%%%%%%%%%%%%%%%%%%%%%%%%%%%%%%%%%%%%%%%%%%%%%%%%%%%%%%%%%%%%%%%%%%%%%%%%%%%%%%%%%%%%%%%%%%%%%%%%%%%%%%%%%%%%%%%%%%%%%%%%%%%%%%%%%%%%%%%%%%%%%%%%%%%%%%%%%%%%%%%%%%%%%%%%%
\subsubsection{Evaluation Methods, Scenarios, and Metrics}
\label{sec:Methods}
In our performance evaluation, we first investigate the performance of the proposed optimization models and heuristic approaches. Next, we refer to our proposed heuristic approaches and provide practical results to evaluate \texttt{ARARAT} performance compared with the following \sota and baseline schemes: 
\begin{enumerate}[noitemsep]
\item \textbf{SABR}~\cite{bhat2018sabr}: The edge computing paradigm is not used in this method. Customized DASH players utilize some important resource data (\ie \textit{cache\_map and bandwidth\_map}) provided by the SDN controller to make decisions about the next segment requests. We use modified AStream DASH players to receive the aforementioned resource data from the SDN controller. 
\item \textbf{ES-HAS}~(Section~\ref{chap:EdgeSDN:ES-HAS}): Like in most existing works, there is no edge collaboration possibility in this approach. ES-HAS utilizes edge servers without video transcoding capability where each of them runs an MILP model on the collected client requests to serve them via one of the actions $1, 7$ or $9$ (Fig.~\ref{ARARAT:actionTree}). Note that the time slot structure introduced in Fig.~\ref{ARARAT-workflows1}(b) is employed to reproduce ES-HAS results.
\item \textbf{CSDN}~(Section~\ref{chap:EdgeSDN:CSDN}): In this non-collaborative approach, each edge server runs a simplified version of the proposed Coarse-Grained approach (\ie an MILP model) for the collected client requests and serves them separately via one of the actions $1,2,7,8$ or $9$ (Fig.~\ref{ARARAT:actionTree}). Note that we use the time slot structure proposed in Fig.~\ref{ARARAT-workflows1}(b) for this method too. 
\item \textbf{NECOL}: The Non Edge Collaborative (NECOL)-based baseline system does not support an edge collaboration possibility. Each NECOL edge server executes a simplified version of the proposed Fined-Grained approach I for each client request to serve it through one of the actions $1,2,7,8$ or $9$ (Fig.~\ref{ARARAT:actionTree}). We use the time slot structure presented in Fig.~\ref{ARARAT-workflows2}(a) for this approach.
\end{enumerate}

For fair and robust comparisons, we implement all aforementioned \sota and baseline schemes without using available source codes, and the results are derived with respect to the same settings and the same topology in our testbed.
The performance of the aforementioned approaches is evaluated in three scenarios. In \textit{scenario I}, we study the impact of changing the bitrate ladder, scaling the network topology, and increasing the number of arriving requests on the \textit{algorithm performance} metrics. \textit{Scenario II} conducts experiments to evaluate the performance of the aforementioned systems in terms of common \textit{QoE-related} parameters. Finally, \textit{scenario III} investigates the behavior of different frameworks in terms of \textit{network utilization and cost} metrics.
Algorithm performance, QoE-related, network utilization and cost metrics are defined as follows:
\begin{enumerate}[noitemsep]
\item \textbf{Algorithm performance metrics}:
\begin{itemize}[noitemsep]
\item \textbf{ETV}: Execution Time Values for the different \texttt{ARARAT} schemes.
\item \textbf{NOV}: Normalized Objective Value for the \texttt{ARARAT} schemes.
\end{itemize}
\item \textbf{QoE-related metrics}:
\begin{itemize}[noitemsep]
\item \textbf{ASB}: Average Segment Bitrate of all downloaded segments.
\item \textbf{AQS}: Average Quality Switches, \ie the number of segments whose bitrate levels change compared to the previous ones.
\item \textbf{ASD}: Average Stall Duration, \ie the average of total video freeze times in all clients.
\item \textbf{ANS}: Average Number of Stalls, \ie the average number of rebuffering events.
\item \textbf{APQ}: Average Perceived QoE, calculated by ITU-T Rec. P.1203 mode 0~\cite{p1203}.
\end{itemize}
\item \textbf{Network utilization and cost metrics}:
\begin{itemize}[noitemsep]
\item \textbf{CHR}: Cache Hit Ratio, defined as the fraction of segments fetched from the CDN or edge servers.
\item \textbf{ETR}: Edge Transcoding Ratio, \ie the fraction of segments transcoded at the edge servers.
\item \textbf{BTL}: Backhaul Traffic Load, the volume of segments downloaded from the origin server.
\item \textbf{ANU}: Average Network Utilization per link, \ie $\frac{\kappa_l}{\mathcal{K}_l}$ where  $\kappa_l$ and ${\mathcal{K}_l}$ represent the measured traffic (in bit/s) on link $l$ and the total allocated bandwidth to link $l$, respectively.
\item \textbf{AST}: Average Serving Time for all clients, including fetching time plus transcoding time. 
\item \textbf{NCV}: Network Cost Values, including computational and bandwidth costs.
\item \textbf{ANC}: Average Number of Communicated messages from/to the SDN controller to/from all clients (in the SABR method) or all edge servers (in other frameworks), including OF and HTTP messages.
\end{itemize}
\end{enumerate}
The performance of \texttt{ARARAT} is presented in the next section, where each experiment is executed 25 times to ensure accuracy, and the average values and standard deviations are reported in the experimental results. 
\rf{Note that we use the method discussed in Section~\ref{sec:SFC:Performance Evaluation} to calculate the results reported in the next section.}
%%%%%%%%%%%%%%%%%%%%%%%%%%%%%%%%%%%%%%%%%%%%%%%%%%%%%%%%%%%%%%%%%%%%%%%%%%%%%%%%%%%%%%%%%%%%%%%%%%%%%%%%%%%%%%%%%%%%%%%%%%%%%%%%%%%%%%%%%%%%%%%%%%%%%%%%%%%%%%%%%%%%%%%%%%%%%%%%%%%%%%%%%%%%%%%%%%%%%%%%%%%%%%%%%%%
\subsubsection{Evaluation Results}
\label{sec:Evaluation Results}
\begin{figure}[!t]
\centering
\includegraphics[width=1\textwidth]{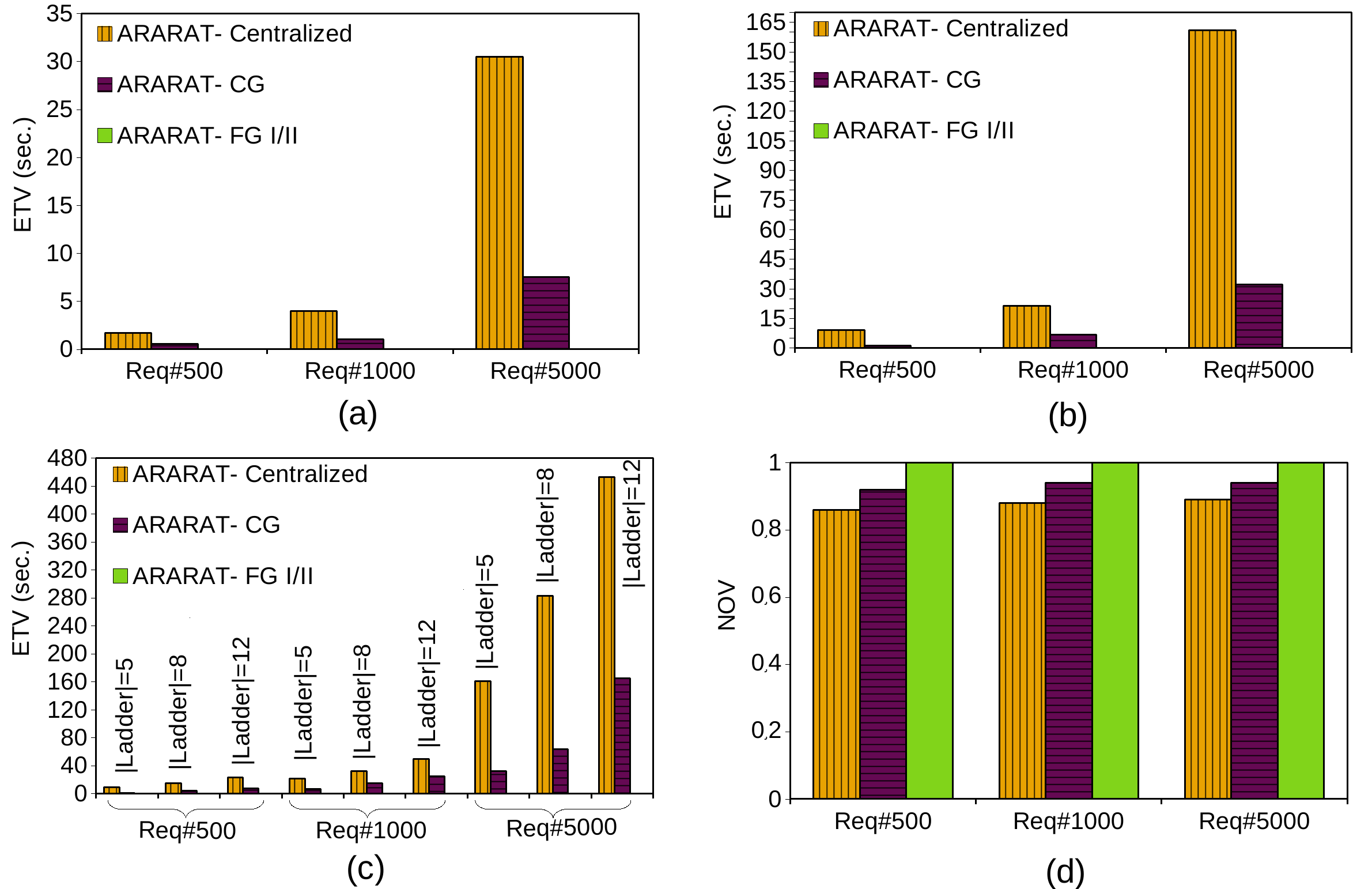}
\caption{Impact of increasing the number of arriving requests (req\#) in (a) a small topology, (b) in the Geant toplogy, and (c) changing the bitrate ladder on the performance of the centralized MILP model, CG, and FG  approaches in terms of ETV and (d) NOV.}
\vspace{.5cm}
\label{res-ETV}	
\end{figure}

\textbf{Scenario I: }
As mentioned earlier, in this scenario, we compare the performance of the proposed centralized optimization MILP model and the CG and FG approaches in terms of the EVT and NOV metrics (Fig.~\ref{res-ETV}).
%%%%%%%%%%%%%%%
\begin{figure}[!t]
\centering
\includegraphics[width=.8\columnwidth]{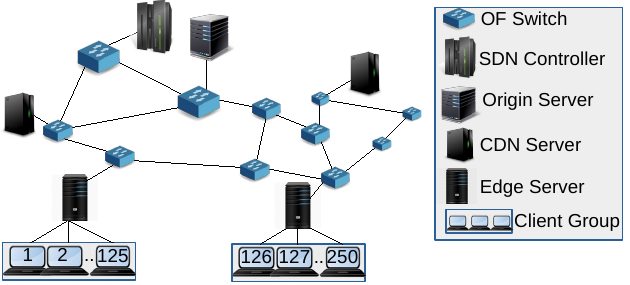}
\caption{\small Small-scale topology used in scenario I.}
\vspace{.5cm}
\label{small-topo}	
\end{figure}
%%%%%%%%%%%%%%

To this aim, we first calculate the ETV metric for each approach by increasing some vital criteria, \eg numbers of simultaneous requests (\ie Req\#), of the CDN/edge servers, and of representations in the bitrate ladder to evaluate the \texttt{ARARAT} schemes' scalability and practicality. Fig.~\ref{res-ETV}(a) highlights the ETV comparison of different \texttt{ARARAT} schemes for different numbers of simultaneous clients' requests (\ie Req\# = 500, 1000, and 5000) in a real small-scale network topology named Abilene~\cite{zoo}, including two CDN servers and two edge servers, depicted in Fig.~\ref{small-topo}. It is worth mentioning that all experimental settings are the same as discussed in Section~\ref{sec:setup} for this topology. The results show that although the CG approach outperforms the centralized scheme in terms of ETV, it still suffers from high execution times due to running the LOM model, and its run times are increased significantly by increasing the Req\#. 

In the second experiment, we replace the first experiment's topology with the described Geant topology (Fig.~\ref{ARARAT:testbed-topo}) and investigate the behavior of the schemes when the number of the CDN and edge servers increases. As depicted in Fig.~\ref{res-ETV}(b), the ETV values for the centralized and CG approaches for serving the same number of Req\# are increased compared to the small-scale scenario. In fact, by expanding the topology, COM and 
LOM need more time to select best servers, consequently best actions.
%%%%%%
%%%%%%

In the third experiment, the impact of changing the bitrate ladder on the values of ETV is studied. As shown in Fig.~\ref{res-ETV}(c), increasing the number of representations in the bitrate ladders (\ie $\mid$Ladder$\mid$=5, 8, 12), increases ETV values noticeably.
In fact, having a ladder with more representations causes COM and LOM to spend much more time for deciding on the best action with the best approach (transcoding, fetching). In all experiments, the FG approaches outperform the other schemes significantly in terms of the ETV, showing very small growth with increasing numbers of requests, servers, and representations (see Fig.~\ref{res-ETV}(a--c)).

In the final experiment of this scenario, we measure the normalized objective value (NOV) metric by changing the number of clients' requests. As illustrated in Fig.~\ref{res-ETV}(d), the results of this experiment for the topology of Fig.~\ref{ARARAT:testbed-topo} indicate that the centralized optimization MILP model surpasses the CG and FG approaches in terms of normalized objective value for all Req\#. Note that changing the topology and replacing the bitrate ladder show similar trends for the NOV. As mentioned earlier, in the centralized and the CG approaches, COM and LOM are run for all clients' requests received from all LESs and clients' requests obtained by each LES, respectively, while the FG approaches execute the EFG algorithms on each LES to specify an optimal solution for each request. Running LOM or EFG on each LES server in the CG and FG approaches, respectively, leads to a sub-optimal solution, but as shown in Fig.~\ref{res-ETV}(d), they are quite close to the centralized MILP model. 
%%%%%%%%%%%%%%%%%%%%%%%%%%%%%%%%%%%%%%%%%%%%%%%%%%%%%%%%%%%%%%%%%%%%%%%%%%%%%%%%%%%%%%%%%%%%%%%%%%%%%%%%%%%%%%%%%%%%%%%%%%%%%%%%%%%%%%%%%%%%%%%%%%%%%%%%%%%%%%%%%%%%%%%%%%%%%%%%%%%%%%%%%%%%%%%%%%%%%%%%%%%%%%%%%%%

\begin{figure}[!t]
\centering
\includegraphics[width=1\textwidth]{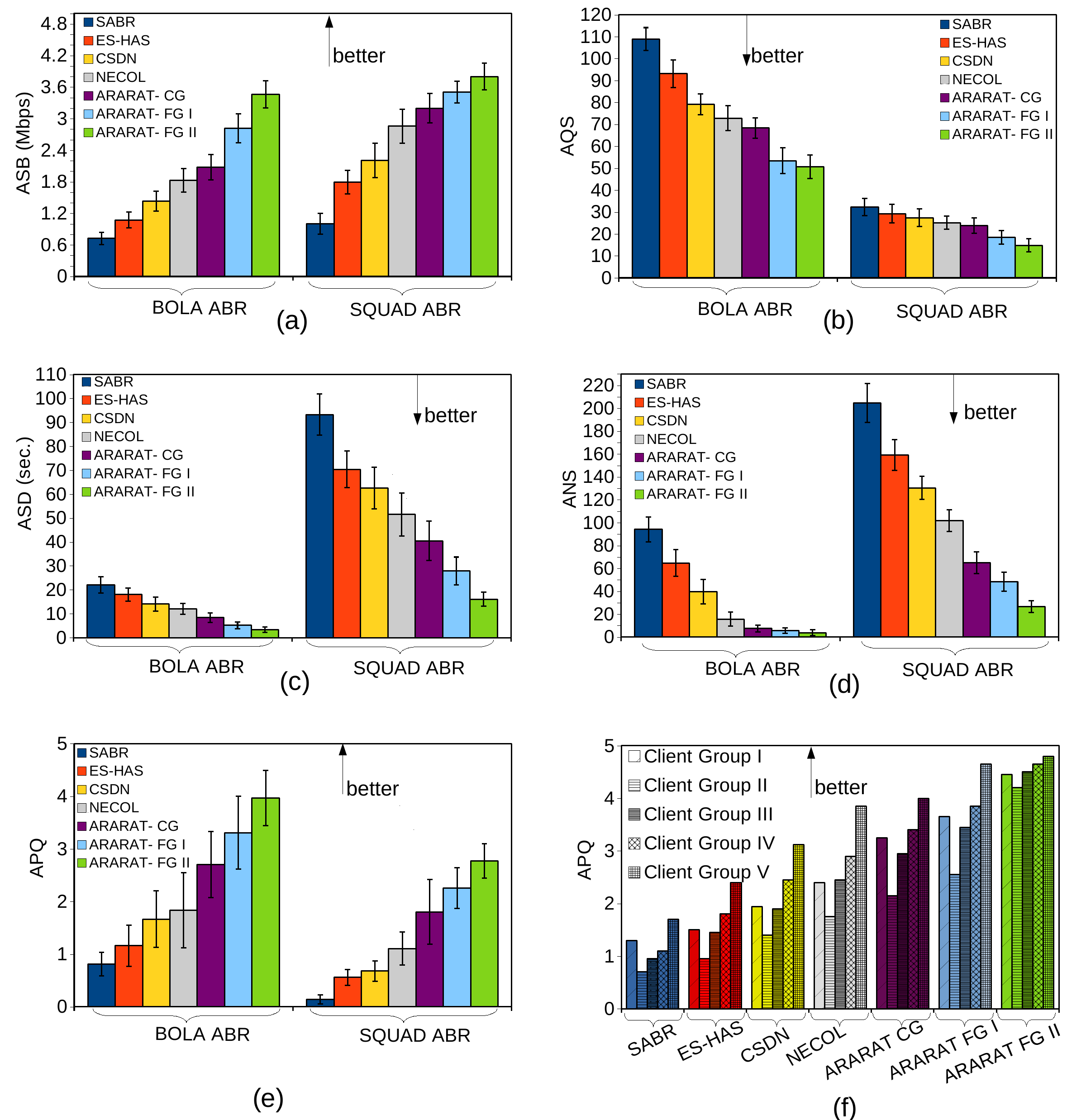}
\caption{Performance of the \texttt{ARARAT} approaches compared with \sota methods in terms of (a) ASB, (b) AQS, (c) ASD, (d) ANS, (e) APQ for 250 clients running BOLA and SQUAD ABR algorithms, and (f) APQ for each client group for the BOLA ABR algorithm.}
\vspace{.5cm}
\label{res-QoE}	
\end{figure}

\begin{figure}[!t]
\centering
\includegraphics[width=.85\textwidth]{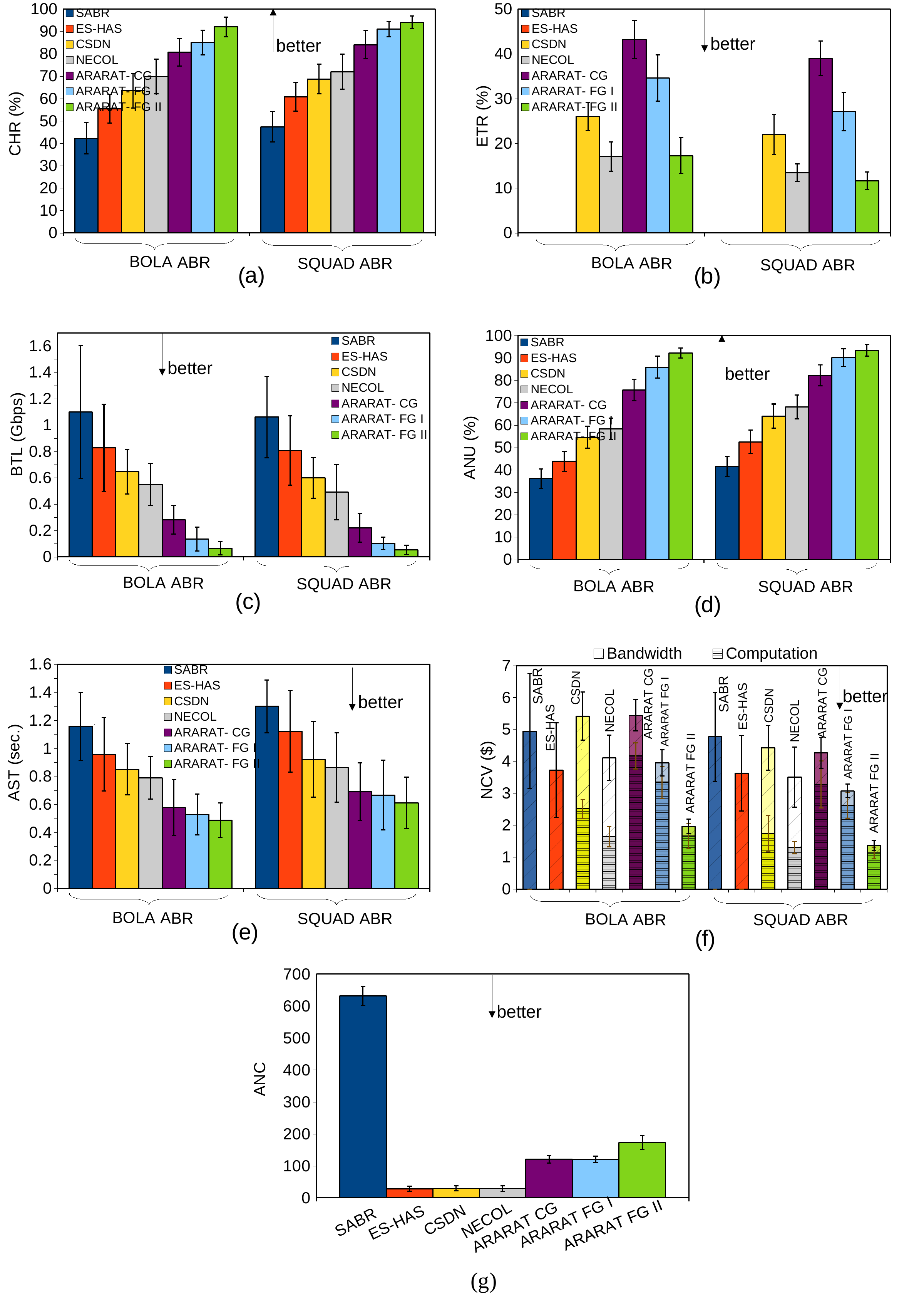}
\caption{\small Performance of the \texttt{ARARAT} approaches compared with \sota methods in terms of (a) CHR, (b) ETR, (c) BTL, (d) ANU, (e) AST, (f) NCV, and (g) ANC for 250 clients running BOLA and SQUAD ABR algorithms.}
\label{res-Netutil}	
\end{figure}

\textbf{Scenario II:}
In the \textit{second scenario}, we study the effectiveness of the proposed \texttt{ARARAT} CG and FG schemes on the testbed  and compare the QoE results with \sota methods. The \texttt{ARARAT} centralized method is not used in this scenario due to its high time complexity and impracticality in large-scale environments. The results are shown in two groups, \ie QoE and network utilization, in Fig.~\ref{res-QoE} and Fig.~\ref{res-Netutil}, respectively. 
%%%%%%%%%%%%%%%%%%%%%%%%%%%%%%%%%%%%%%%%%%%%
%%%%%%%%%%%%%%%%%%%%%%%%%%%%%%%%%%%%%%%%%%%

As depicted in Fig.~\ref{res-QoE}(a), the \texttt{ARARAT} approaches (\ie CG, FG I, and FG II) serve clients with higher ASB for both BOLA- and SQUAD-based adaptation due to using NESs' cooperation. More precisely, the \texttt{ARARAT} FG II scheme outperforms the other approaches in terms of ASB since it utilizes the bandwidth allocation strategy.
The \texttt{ARARAT} FG I algorithm has the second-best ASB results and surpasses the CG scheme despite its higher time complexity. It is worth noting that, although the \texttt{ARARAT} CG scheme employs NESs' collaboration, it only improves over the NECOL system, a non-collaborative system, by about 13.5\% (for both ABR algorithms) in terms of ASB, while these improvements are at least 25\% and 35\% for the \texttt{ARARAT} FG I and FG II approaches, respectively. Moreover, the \texttt{ARARAT} approaches demonstrate the best performance in terms of AQS for both ABR algorithms (Fig.~\ref{res-QoE}(b)). Similar to ASB, the \texttt{ARARAT} FG II method results in the best AQS results and outperforms the \sota methods by at least 31\%. This is again because \texttt{ARARAT} FG II employs the bandwidth allocation strategy that allows each LES to access other servers with sufficient bandwidth, specifically when competing with other NESs on shared links. Also, \texttt{ARARAT} FG II outperforms the \texttt{ARARAT} CG system in terms of AQS by at least 26\%, while it slightly improves over \texttt{ARARAT} FG I by at least 5\%. 

The performance of different frameworks in terms of ASD and ANS are plotted in Fig.~\ref{res-QoE}(c) and (d), respectively. In contrast to ASB and AQS, BOLA-based players, by focusing more on the buffer status, work better than SQUAD-based players in terms of ASD and ANS. Edge collaboration, selecting the best server with minimum delay and employing the best approach (\ie fetching or transcoding), allows the \texttt{ARARAT} CG system to serve BOLA-based players' requests with about 30\% and 52\% fewer ASD and ANS values, respectively, compared with \sota frameworks. However, the \texttt{ARARAT} FG approaches utilizing fast EFG algorithms instead of an optimization model reduce ASD and ANS obtained by the \texttt{ARARAT} CG system by at least 38\% and 26\%, respectively. Again, \texttt{ARARAT} FG II, including a bandwidth allocation strategy, achieves the best stalling results among all methods, improving over ASD and ANS results of \texttt{ARARAT} FG I by at least 35\% and 31\%, respectively.

Although improving the aforementioned common QoE parameters (\ie ASB, AQS, ASD, and ANS) can generally improve the users' satisfaction, a standard comprehensive QoE model is required to assess the users' QoE. Therefore, we use a standard QoE model~\cite{p1203} to evaluate the performance of the proposed system in terms of the APQ metric. As shown in Fig.~\ref{res-QoE}(e), the value of the APQ metric is between zero and five, which means the worst- and best-perceived quality. It should be noted that stalling events (measured by ASD and ANS) significantly impact APQ compared to other parameters. Therefore, BOLA-enabled players with better stalling behavior acquire better APQ compared to SQUAD-enabled players. 

It is obvious from the results that the \texttt{ARARAT} approaches demonstrate superior performance for both ABR algorithms. More precisely, \texttt{ARARAT} FG II improves  APQ by at least 104\% compared to \sota techniques. Also, by improving the ASB, AQS, ASD, and ANS results, \texttt{ARARAT} FG II increases APQ by about 46\% and 20\% over \texttt{ARARAT} CG and FG I. \rf{In the next experiment, we investigate the indirect impact of network bandwidth fairness on users' QoE fairness. Therefore, to} show the SFG algorithm's impact on QoE in more detail, we measured the APQ metric for each group of clients (\ie client groups I--V). As illustrated in Fig.~\ref{res-QoE}(f), APQ variations between different client groups are decreased between 40\% and 70\% by the \texttt{ARARAT} FG II method compared to other approaches. \rf{It can be deduced that using the SFG algorithm for allocating bandwidth to each LES based on their requests results in better average QoE fairness among different client groups.}

\textbf{Scenario III: }
In the \textit{third scenario}, we investigate the performance of the proposed \texttt{ARARAT} CG and FG methods on the testbed and compare the network utilization and cost results with \sota methods. Similar to Scenario II, we do not use the \texttt{ARARAT} centralized method in this scenario.

The effectiveness of \texttt{ARARAT} regarding the CHR and ETR metrics are plotted in Fig's.~\ref{res-Netutil} (a) and (b), respectively. Note that the SABR- and ES-HAS-enabled systems encounter a cache miss event when \textit{(i)} the requested or higher quality levels do not exist in CDN servers or \textit{(ii)} available bandwidth is insufficient to fetch the requested quality from the selected CDN server. In addition, a cache miss occurs in CSDN and NECOL if the available edge's processing capability is not sufficient to transcode the requested quality from a higher quality. Furthermore, a cache miss is experienced by \texttt{ARARAT} approaches when the LES cannot fetch or transcode the demanded quality by NESs' assistance. The CHR metric shows that \texttt{ARARAT} CG outperforms \sota systems by at least 15\% due to its capability for fetching requested or higher quality levels from NESs or using NESs' resources for transcoding tasks (Fig.~\ref{res-Netutil}(a)). However, \texttt{ARARAT} CG increases ETR by at least 114\% due to using a collaborative system (Fig.~\ref{res-Netutil}(b)). Since SABR and ES-HAS are not transcoding-enabled systems, their ETR metrics are zero. 
%%%%%%%%%%%%%%%%%%%%%%%%%%%%%%%%%%%%%%%%%%%%
%%%%%%%%%%%%%%%%%%%%%%%%%%%%%%%%%%%%%%%%%%%
\clearpage
As depicted in Fig's.~\ref{res-Netutil}(a) and (b), the \texttt{ARARAT} FG I approach improves over \texttt{ARARAT} CG slightly, by 5\% and 19\% in terms of CHR and ETR, respectively. Although \texttt{ARARAT} FG I and FG II  have a similar behavior of using NESs' resources for fetching/transcoding, the bandwidth allocation strategy helps \texttt{ARARAT} FG II to provide sufficient bandwidth between LESs and cache servers and fetch more segments directly from CDN servers with less transcoding compared to \texttt{ARARAT} FG I. Improving the cache hit ratio (\ie CHR) directly affects the BTL metric since the system downloads fewer segments from the origin server; \texttt{ARARAT} FG II thus outperforms other systems and significantly reduces fetching bitrates from the origin server (Fig.~\ref{res-Netutil}(c)).

In the next experiment, the ANU is studied to assess the network's performance for the different methods. Collaboration between SDN and LESs for providing adequate bandwidth to edge servers allows for a better ANU for the \texttt{ARARAT} FG II method, as is evident from Fig.~\ref{res-Netutil}(d). \texttt{ARARAT} FG II has better ANU results than the \sota methods and the two other \texttt{ARARAT} approaches by at least 37\% and 4\%, respectively. Moreover, we measure the AST, which is considered an essential part of the end-to-end delay. As shown in Fig.~\ref{res-Netutil}(e), using \texttt{ARARAT} FG II reduces the AST by at least 30\% and 7\% compared to NECOL and \texttt{ARARAT} FG I, respectively, since it fetches more segments directly from best CDN servers with sufficient bandwidth between LESs and cache servers (\ie without transcoding). Note that due to more high-quality segments downloaded by SQUAD-based players (Fig.\ref{res-QoE}(a)), their AST values are higher than those of BOLA-based players. Our final experiment measures the NCV based on bandwidth and computational costs for the different approaches (Fig.~\ref{res-Netutil}(f)). Although the proposed \texttt{ARARAT} FG approaches run transcoding functions on all LESs, their NCVs are lower than those of other methods. This is because they improve the CHR and ETR metrics. \texttt{ARARAT} FG II achieves the best performance among all the methods and reduces network costs.

In the final experiment of this scenario, the ANC is investigated to evaluate the SDN controller's overhead for the different methods. As shown in Fig.~\ref{res-Netutil}(g), on the downside, the \texttt{ARARAT} approaches, especially \texttt{ARARAT} FG II, increase the ANC, including OF and HTTP messages for receiving resource maps (\ie cache, edge, bandwidth, comp.~maps) and for allocating sufficient bandwidth to edge servers compared to the \sota approaches. However, since the \texttt{ARARAT} approaches employ the aggregation technique at edge servers and do not require establishing direct communication between the SDN controller and clients, they still perform much better than the SABR system in terms of the ANC.
%%%%%%%%%
% \clearpage
\section{Summary}
\label{chap:CollaborativeEdge:Conclusion}
This chapter proposed a collaborative edge- and SDN-assisted framework for HAS clients called \texttt{LEADER}. Similar to \texttt{CSDN} (Section~\ref{chap:EdgeSDN:CSDN}) and \texttt{SARENA} (Section~\ref{chap:SFCEnabled}), \texttt{LEADER} employed VNFs with transcoding and partial caching capabilities at the edge of an SDN-enabled network. In contrast to others, \texttt{LEADER} leveraged the SDN controller's abilities to establish a collaboration between a Local Edge Server (LES) and other Neighboring Edge Servers (NESs) by using idle resources (computation, bandwidth, storage) for running video caching/transcoding functions. In addition, \texttt{LEADER}'s SDN controller employed a set of actions (of a so-called \textit{Action Tree}), plus a bandwidth allocation strategy to assist edge servers in achieving appropriate bandwidth resources to serve their requests from the most suitable servers in terms of time. We formulated \texttt{LEADER}'s problem as a central MILP optimization model and then proposed a distributed lightweight heuristic algorithm to alleviate the overheads of the proposed model. Deploying \texttt{LEADER} over a large-scale testbed, consisting of 250 DASH clients revealed that \texttt{LEADER} outperformed baseline schemes in terms of users' QoE by at least 22\% and network utilization by at least 13\%.

Inspired by the core idea of the \texttt{LEADER} framework, we also introduced an extended version of \texttt{LEADER}, called \texttt{ARARAT}. Similar to \texttt{LEADER}, \texttt{ARARAT} proposed a hierarchical architecture for supporting edge collaboration for serving clients' requests from the servers with the shortest delay, utilizing the best policies (\ie fetching or transcoding) by considering all possible resource constraints. However, \texttt{ARARAT} considered delivery costs as important criteria and took into account all feasible resources provided by different resource providers for serving client requests, and then extended \texttt{LEADER}'s \textit{Action Tree}. The problem was formulated as a multi-objective MILP model to jointly optimize serving time (\ie fetching and transcoding) and overall network cost (\ie bandwidth and computation). \texttt{ARARAT} also proposed coarse-grained and fine-grained heuristic approaches to solve the optimization problem. Additionally, the bandwidth allocation approach utilized by the SDN controller in \texttt{ARARAT} was distinct from the bandwidth allocation strategy employed by \texttt{LEADER}. The experimental results of comprehensive scenarios executed on a large-scale cloud-based testbed and comparison with those of our previous works and other state-of-the-art schemes showed that the \texttt{ARARAT} approaches provided significant benefits to both OTT service providers and network operators in terms of users' QoE, network cost, and network utilization by at least 47\%, 47\%, and 48\%, respectively.

%% file: Chapters/Chapter6/6-1-Intro.tex
%************************************************
% \singlespacing
\chapter{Hybrid P2P-CDN Architectures for HAS}\label{chap:Hybrid-P2PCDN}
%************************************************
\doublespacing
\vspace{-1cm}
Designing a cost-effective, scalable, and flexible NAVS architecture that supports the high-definition and low-latency video streaming requirements for live streaming services is still an open challenge. \rf{P2P content delivery providers employ several incentives to motivate peers to share their resources (\eg upload bandwidth) to benefit the entire network. These incentives can vary depending on the policies of each P2P provider. For instance, the ``tit-for-tat'' principle can be offered, where active peers receive faster download speeds when they contribute their upload bandwidth to other peers. Another example is the ``reward-based'' policy, where active peers receive compensation in terms of monetary rewards or free Internet services for their resource contributions. However, addressing or even designing such incentive methods is out of the scope of the thesis.} Considering all advantages provided by CDN and P2P networks, this chapter leverages novel networking paradigms, \ie MEC and NFV, and promising video processing solutions, \ie \textit{Video Super-Resolution} (SR), and \textit{Distributed Video Transcoding} (TR), to introduce hybrid P2P-CDN systems for HAS, namely, \texttt{RICHTER} and \texttt{ALIVE}.
We introduce two \textit{multi-layer} architectures and design \textit{action trees} that consider all feasible resources (\ie storage, computation, and bandwidth) provided by peers, edge, and CDN servers for serving peer requests with acceptable latency and quality. We formulate the problems as optimization models executed at the edge of the network. To alleviate the high time complexity of the optimization models, we propose the \texttt{RICHTER} online learning model and \texttt{ALIVE} lightweight heuristic scheme. Finally, we design and instantiate large-scale cloud-based testbeds, including 350 HAS players, deploy the proposed solutions on them, and conduct a series of experiments to evaluate the performance of the proposed solutions in comprehensive scenarios. Experimental results express that the proposed solutions improve users' QoE, decrease the incurred cost of the streaming service provider, shorten clients' serving latency, enhance edge server energy consumption, and reduce backhaul bandwidth usage compared to baseline approaches.\\
\noindent The first contribution of this chapter, \ie ``Hybrid P2P-CDN Architecture for Live Video Streaming: An Online Learning Approach'' was presented at the IEEE Global Communications Conference (GLOBECOM) 2022~\cite{farahani2022hybrid} and the second one, \ie ``\texttt{ALIVE}: A Latency- and Cost-Aware Hybrid P2P-CDN Framework for Live Video Streaming'' has been submitted to the IEEE Transactions on Network and Service Management (TNSM) journal.

%% file: Chapters/Chapter6/6-2-RICHTER.tex
\doublespacing
\section{RICHTER Framework}\label{chap:Hybrid-P2PCDN:RICHTER}
This section presents \texttt{RICHTER}~\cite{farahani2022hybrid}, an online-learning (OL)-based hybrid P2P-CDN architecture for live video streaming. Section~\ref{sec:RICHTERDesign} describes the details of the \texttt{RICHTER} architecture and formulation. We explain our proposed OL-enabled method in Section~\ref{sec:RICHTER:OL}. \texttt{RICHTER}'s evaluation setup, methods, metrics, and results are explained in Section~\ref{sec:RICHTER:Performance Evaluation}. 
%%%%%%%%%%%%%%%%%%%%%%%%%%%%%%%%%%%%%%%%%%%%%%%%%%%%%%%%%%%%%%%%%%%%%%%%%%%%%%%%%%%%%%%%%%%%%%%%%%%%%%%%%%%%%%%%%%%%%%%%%%%%%%%%%%%%%%%%%%%%%%%%%%%%%%%%%%%%%%%%%%%%%%%%%%%
\subsection{RICHTER System Design}
\label{sec:RICHTERDesign}
 This section discusses the \texttt{RICHTER} multi-layer architecture and shows how the problem is formulated as an optimization model.
%%%%%%%%%%%%%%%%%%%%%%%%%%%%%%%%%%%%%%%%%%%%%%%%%%%%%%%%%%%%%%%%%%%%%%%%%%%%%%%%%%%%%%%%%%%%%%%%%%%%%%%%%%%%%%%%%%%%%%%%%%%%%%%%%%%%%%%%%%%%%%%%%%%%%%%%%%%%%%%%%%%%%%%%%%%
\subsubsection{RICHTER Architecture}
\label{sec:RICHTER:sysmodel}
The proposed architecture of \texttt{RICHTER} includes \textit{four} core layers and is shown in Fig.~\ref{RICHTER-arch}. These layers are:
\begin{enumerate}[noitemsep]
\item\textbf{Media Organization Layer.} In this layer, raw live video sequences are encoded and packaged into DASH format, then stored on the origin server. Note that this layer is able to package the encoded video seuences to other formats like HLS~\cite{HLS} or \textit{Common Media Application Format} (CMAF)~\cite{CMAF}.
\item\textbf{CDN Layer.} This layer is constructed by a group of CDN servers (either OTT servers or a purchased service from CDN providers), each of which contains various parts of video sequences. Inspired by the \textit{Consumer Technology Association} CTA-5004 standard~\cite{CTA,CTA-wave}, CDN servers periodically inform the edge layer about their cache occupancy via \textit{Common Media Server Data} (CMSD) messages.  
\item\textbf{P2P Layer.} Given the continuous increases in smartphone capabilities, \eg high-bandwidth access to the Internet, energy resources, and hardware-accelerated video transcoding, \texttt{RICHTER} utilizes the peers' idle resources to provide a distributed video transcoding approach besides video transmission. Like most hybrid P2P-CDN schemes, we construct the P2P layer based on the tree-mesh structure, including two types of peers: \textit{Seeders} and \textit{Leechers}. In this scheme, seeders' requests can be served by all nodes (\ie CDNs, origin, edge, or other seeders) except leechers, while leechers' requests can be served by all nodes. Inspired by the CTA-5004 standard~\cite{CTA}, peers periodically inform the edge layer about their cache occupancies through \textit{Common Media Client Data} (CMCD) messages and receive updates from the edge layer via CMSD messages. 
%%%%
\begin{figure}[!t]
	\centering
	\includegraphics[width=.9\columnwidth]{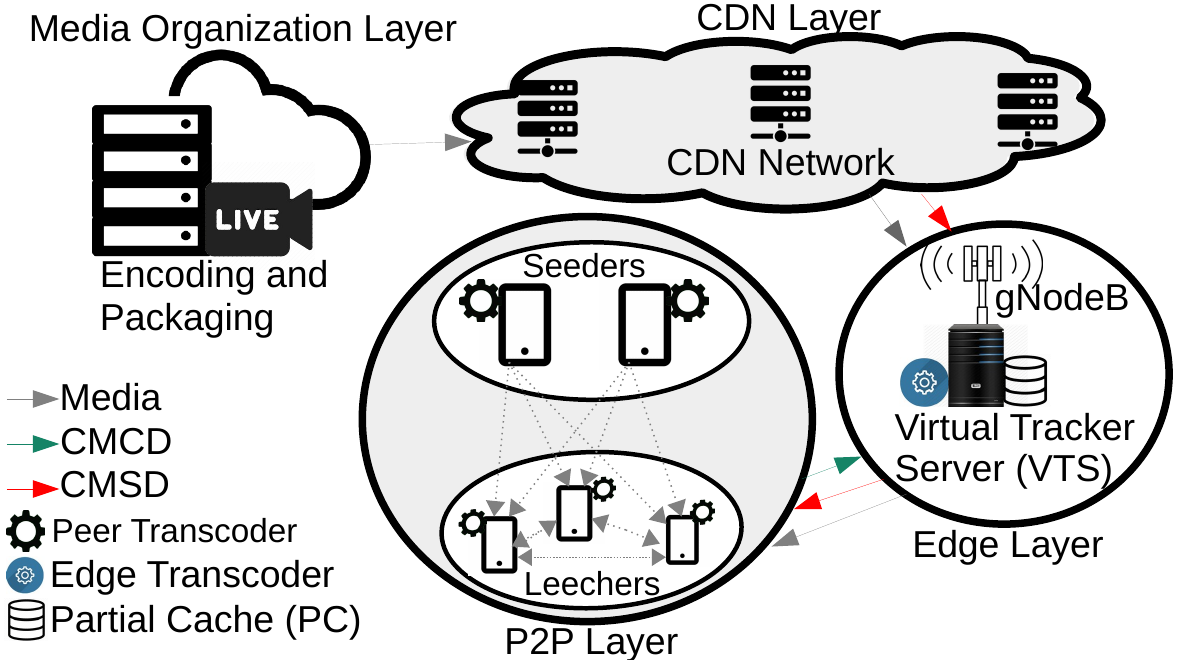}
	\caption{\small RICHTER system architecture.}
        \vspace{.5cm}
	\label{RICHTER-arch}
\end{figure}
%%%%
\item\textbf{Edge Layer.} This layer leverages the capabilities of NFV and edge computing and presents virtualized edge components called \textit{Virtual Tracker Servers} (VTSs) close to base stations (\eg gNodeB in 5G). Note that, in the proposed system, during a live session, clients' requests are directed to a VTS, and then they get responses based on the VTS's decisions.
As shown in Fig.~\ref{RICHTER-arch}, a VTS is equipped with \textit{transcoding} and \textit{partial cache} functions to serve clients' requests from existing higher content qualities (by transcoding) or directly from cached qualities, respectively. Note that because the VTS has a broader view of both P2P and CDN layers (based on the received CMCD/CMSD messages and monitored information), it can track clients' requests and store a mapping between all transmitted content and all served clients in its peer-map lists.
\end{enumerate}
Thus, it must respond to the following vital questions whenever it needs to decide to serve received requests:
\begin{enumerate}[noitemsep]
\item Where is the optimal place (\ie adjacent peers, VTS, CDN servers, or origin server) in terms of lowest latency for fetching each client's requested content quality level from, while efficiently utilizing the available resources?
\item What is the optimal approach for responding to the requested quality level (\ie fetch or transcode)?
\end{enumerate}
%%%%%
\begin{figure}[t]
	\centering
	\includegraphics[width=.55\columnwidth]{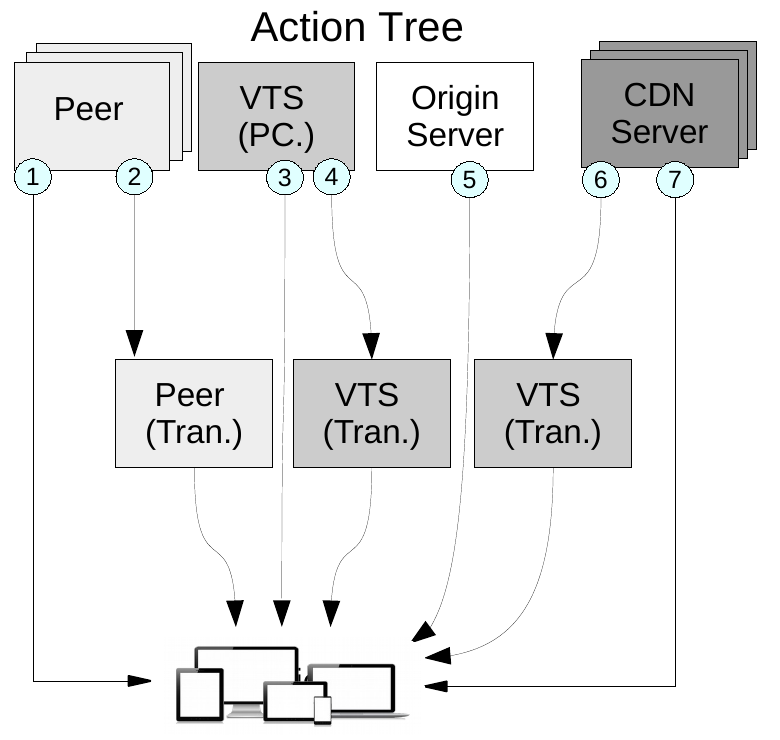}
	\caption{\small RICHTER action tree.}
        \vspace{.5cm}
	\label{AT}
\end{figure}
%%%%%%

Among other tasks, a VTS monitors the system frequently to obtain precise information about the available resources (\eg bandwidth, peers' computational and power resources), and peers' joining/leaving times. Therefore, when a VTS receives a new request, it can find the optimal solution (\ie in terms of minimum latency) from the \textit{action tree} (Fig.~\ref{AT}) (action numbering as in the figure):
\begin{enumerate}[leftmargin=*,label=\protect\circledd{\arabic*},noitemsep]
\item Use the P2P network and transmit the requested quality directly from the best adjacent peer with maximum stability (\ie the least recent joining time). 
\item Transcode the requested quality from a higher quality at the most stable adjacent peer and transmit it through the P2P network. 
\item Fetch the requested quality directly from the edge, \ie the VTS. 
\item Transcode the requested quality from a higher quality at the VTS. 
\item Fetch the requested quality from the origin server. 
\item Fetch a higher quality from the best CDN server and transcode it at the VTS. 
\item Fetch the requested quality from the best CDN server. 
\end{enumerate}
%%%%%%%%%%%%%%%%%%%%%%%%%%%%%%%%%%%%%%%%%%%%%%%%%%%%%%%%%%%%%%%%%%%%%%%%%%%%%%%%%%%%%%%%%%%%%%%%%%%%%%%%%%%%%%%%%%%%%%%%%%%%%%%%%%%%%%%%%%%%%%%%%%%%%%%%%%%%%%%%%%%%%%%%%%%
%%%%table%%%%
\begin{table}[!t]
\centering
\caption {\small RICHTER Notation.}
\label{tab:RICHTER:notation}
\begin{tabular}{llllll}
\cline{1-2}
\multicolumn{2}{|c|}{\textbf{Input Parameters}}                                                                                  
&  &  &  &  \\ \cline{1-2}                                                                                         
\multicolumn{1}{|l|}{\begin{tabular}[c]{@{}l@{}}
$\mathcal{C}$\\ 
$\mathcal{P}$ \\ \\
$\mathcal{V}$\\ 
$\mathcal{Q}_j$\\ \\ \\
$\mathcal{T}$\\  \\ \\
$\mathcal{R}$\\
$\overline{\mathcal{Q}}$\\
$\alpha^{q}_{i,j}$\\ \\
$\omega_{i,j}$\\
$\theta^{q}_{i,j}$\\ \\
$\eta^{q}_{i,j}$\\ \\
$\mu^{q}_{i,j}$\\ \\
$\delta^{q}_j$\\
$\lambda^q_j$\\
$\Omega_i$\\
$\phi_i$\\
\end{tabular}} 
& \multicolumn{1}{l|}{\begin{tabular}[c]{@{}l@{}}
Set of $k$ CDN servers and an origin server (i.e., $c=0$)\\
Set of $n$ peers including $s$ seeders and $l$ leechers in subsets $\mathcal{P}_1$ and $\mathcal{P}_2$,\\ respectively\\
Virtual Tracer Servers (VTSs) \\
Set of possible quality levels for serving quality $q^*$ requested by $j\in\mathcal{P}$,\\ where $\mathcal{Q}_j=\{q^*,q^*+1,...,q^{*}_{max}\}$ and $q^{*}_{max}$ is the maximum quality\\ level for the demanded segment\\
Set of possible transcoding statuses, where $\mathcal{T}=\{0,1,2\}$ and $t=1$ or\\ $t=2$ if the requested quality is transcoded from a higher quality $q\in\mathcal{Q}_j$\\ at a VTS $i\in\mathcal{V}$ or a peer $i\in\mathcal{P}\setminus{j}$, respectively; otherwise t=0\\
Set of $\rho$ peer regions\\
Set of quality level queues in the VTS\\
Available quality levels in  $i\in\mathcal{P}\cup\mathcal{V}\cup\mathcal{C}$; $\alpha^{q}_{i,j}=1$ means node $i$ hosts\\ quality $q$ requested by peer $j\in{P}$; otherwise $\alpha^{q}_{i,j}=0$\\
Available bandwidth on path between $i\in\{\mathcal{P}\cup\mathcal{V}\cup\mathcal{C}\}\setminus{j}$, $j\in\mathcal{P}$\\
Required resources (i.e., CPU usage in \%) for transcoding quality $q\in\mathcal{Q}_j$ \\into the quality requested by $j\in\mathcal{P}$ in $i\in\{\mathcal{P}\cup\mathcal{V}\}\setminus{j}$\\
Required power (in milliampere-hour) for transcoding quality $q\in\mathcal{Q}_j$ \\into the quality requested by $j\in\mathcal{P}$ in $i\in\mathcal{P}\setminus{j}$\\
Required time (in seconds) for transcoding quality $q\in\mathcal{Q}_j$ into the\\ quality requested by $j\in\mathcal{P}$ in $i\in\{\mathcal{P}\cup\mathcal{V}\}\setminus{j}$\\
Size of segment in quality $q$ (in bytes) requested by $j\in\mathcal{P}$ \\
Bitrate for quality level $q\in \mathcal{Q}_j$ requested by $j\in\mathcal{P}$\\
Available computation resources (available CPU) of $i\in\mathcal{P}\cup\mathcal{V}$\\
Available power resources of $i\in\mathcal{P}$\\
\end{tabular} }
&  &  &  &  \\ \cline{1-2}
\multicolumn{2}{|c|}{\textbf{Variables}} 
&  &  &  &  \\ \cline{1-2}                                                 
\multicolumn{1}{|l|}{\begin{tabular}[c]{@{}l@{}}
 $X^{q,t}_{i,j}$\\ \\ \\
$\tau^{q}_{i,j}$\\ \\
$T^{q}_{i,j}$\\ \\
$\Psi$\\
\end{tabular}}
&\multicolumn{1}{l|}{\begin{tabular}[c]{@{}l@{}}
Binary variable where $X^{q,t}_{i,j}=1$ indicates source $i\in\{\mathcal{P}\cup\mathcal{V}\cup\mathcal{C}\}\setminus{j}$ \\transmits quality $q\in\mathcal{Q}_j$ requested by peer $j\in\mathcal{P}$ with transcoding status \\$t$, otherwise $X^{q,r,t}_{i,j}=0$\\ 
Required transcoding time at source $i\in\{\mathcal{P}\cup\mathcal{V}\}\setminus{j}$ to serve quality\\ $q\in\mathcal{Q}_j$ requested by $j\in\mathcal{P}$\\
Required time of transmitting quality level $q\in\mathcal{Q}_j$ in response to peer \\ $j\in\mathcal{P}$ from server $i\in\{\mathcal{P}\cup\mathcal{V}\cup\mathcal{C}\}\setminus{j}$\\
Serving time consisting of $\tau^{q}_{i,j}$ and $T^{q}_{i,j}$\\
\end{tabular}}                                                    
&  &  &  &  \\ \cline{1-2}                                                                                                              
& &  &  &  &                                              
\end{tabular}
\end{table}

%%%%%%
\clearpage
\subsubsection{RICHTER Optimization Problem Formulation}
\label{sec:RICHTER:Formulation}
We introduce an MILP optimization model that includes four groups of constraints: \textit{Action Selection} (AS), \textit{Serving Time} (ST), \textit{Origin/Peer} (CP), and \textit{Resource Usage} (RU).

\textbf{\textit{(i)} AS constraint.} Constraint~(\ref{RICHTER:eq:1}) selects an appropriate action from the proposed action tree (Fig.~\ref{AT}) for the request issued by peer $j$. It chooses a suitable value of the binary variable $X^{q,t}_{i,j}$ (refer to Table~\ref{tab:RICHTER:notation} for the definition):
%%%%%%%%%%%Const1%%%%%%%%
\begin{flalign}
\label{RICHTER:eq:1}
&\sum_{i\in\{\mathcal{P}\cup\mathcal{V}\cup\mathcal{C}\}\setminus{j}}\sum_{t\in\mathcal{T}}\sum_{q\in\mathcal{Q}_j} X^{q,t}_{i,j}~.~\alpha^{q}_{i,j}
=1,&&\forall j\in\mathcal{P}
\end{flalign}
%%%%%%%%%%%%%%%%%%%%%%%%%%

\textbf{\textit{(ii)} ST constraint.} 
Constraint~(\ref{RICHTER:eq:2}) determines transmitting time $T^{q}_{i,j}$ to transmit quality level $q\in\mathcal{Q}_j$ from source node $i$ to peer $j$:  
%%%%%%%%%%%Const2%%%%%%%%
\begin{flalign}
\label{RICHTER:eq:2}
&\frac{\sum_{t\in\mathcal{T}} X^{q,t}_{i,j}~.~\delta^{q}_j}{\omega_{i,j}}\leq T^{q}_{i,j},&\forall j\in\mathcal{P}, i\in\{\mathcal{P}\cup\mathcal{V}\cup\mathcal{C}\}\setminus{j}, q\in\mathcal{Q}_j
\end{flalign}
%%%%%%%%%%%%%%%%%%%%%%%%%
Constraint~(\ref{RICHTER:eq:3}) determines the required transcoding time $\tau^{q}_{i,j}$ at node $i$ in case of serving the quality requested by peer $j$ from a higher quality $q$ by transcoding:
%%%%%%%%%%%Const3%%%%%%%%
\begin{flalign}
\label{RICHTER:eq:3}
&\sum_{t\in\mathcal{T}\setminus{0}}\sum_{q\in \mathcal{Q}_j} X^{q,t}_{i,j}~.~\mu^{q}_{i,j}\leq \tau^{q}_{i,j},&&\forall j\in\mathcal{P}, i\in\{\mathcal{P}\cup\mathcal{V}\}\setminus{j}
\end{flalign}
%%%%%%%%%%%%%%%%%%%%%%%%%
Therefore, the serving time, namely $\Psi$, \ie fetching time plus transcoding time, can be expressed as follows:
%%%%%%%%%%%Const4%%%%%%%%
\begin{flalign}
\label{RICHTER:eq:4}
&\sum_{i\in \{\mathcal{P}\cup\mathcal{V}\cup\mathcal{C}\}\setminus{j}}\sum_{j\in \mathcal{P}}\sum_{q\in \mathcal{Q}_j}T^{q}_{i,j}+\tau^{q}_{i,j}\leq \Psi&&
\end{flalign}
%%%%%%%%%%%%%%%%%%%%%%%%%

\textbf{\textit{(iii)} CP constraint.} Constraint~(\ref{RICHTER:eq:5}) forces the model to fetch the exact quality $q^*$ from the origin or CDNs when one of them is chosen to serve peer $j\in\mathcal{P}$. 
%%%%%%%%%%%Const5%%%%%%%%
\begin{flalign} 
\label{RICHTER:eq:5}
&\sum_{i\in\mathcal{C}}\sum_{q\in\mathcal{Q}_j} X^{q,t=0}_{i,j}~.~q = q^*, && \forall j\in\mathcal{P}
\end{flalign}
%%%%%%%%%%%%%%%%%%%%%%%%%
Note that $i=0$ in Eq.~(\ref{RICHTER:eq:5}) denotes the origin server. Moreover, we should prevent seeders from fetching requested qualities from leechers, expressed in Eq. (\ref{RICHTER:eq:6}):
%%%%%%%%%%%Const6%%%%%%%%
\begin{flalign} 
\label{RICHTER:eq:6}
&\sum_{i\in\mathcal{P}_2}\sum_{t\in\mathcal{T}\setminus{1}}\sum_{q\in\mathcal{Q}_j} X^{q,t}_{i,j}~.~q = 0, && \forall j\in\mathcal{P}_1
\end{flalign}

\textbf{\textit{(iv)} RU constraint.} Constraint~(\ref{RICHTER:eq:7}) guarantees that the required bandwidth for transmitting segments on the link between nodes $i$ and $j$ must respect the available bandwidth: 
%%%%%%%%%%%Const7%%%%%%%%
\begin{flalign}
\label{RICHTER:eq:7}
&\sum_{t\in\mathcal{T}}\sum_{q\in \mathcal{Q}_j} \hspace{-.1cm}X^{q,t}_{i,j}~.~\lambda^{q}_j\leq \omega_{i,j},&& \hspace{-.5cm}\forall j\in\mathcal{P},i\in\{\mathcal{P}\cup\mathcal{V}\cup\mathcal{C}\}\setminus{j}
\end{flalign}
%%%%%%%%%%%%%%%%%%%%%%%%%
Constraint~(\ref{RICHTER:eq:8}) limits the maximum required processing capacity for the transcoding operation to the available computational resource.
%%%%%%%%%%%Const8%%%%%%%%
\begin{flalign}
\label{RICHTER:eq:8}
&\sum_{j\in\mathcal{P}}\sum_{q\in \mathcal{Q}_j}(X^{q,t=1}_{i\in\mathcal{V},j}+X^{q,t=2}_{i\in\mathcal{P}\setminus{j},j})~.~\theta^{q}_{i,j}\leq\Omega_i&&\forall i\in\mathcal{P}\cup\mathcal{V}
\end{flalign}
%%%%%%%%%%%%%%%%%%%%%%%%%
Similarly, constraint~(\ref{RICHTER:eq:9}) limits the maximum required peers' power resources for running the transcoding function to the available power resource.
%%%%%%%%%%%Const9%%%%%%%%
\begin{flalign}
\label{RICHTER:eq:9}
&\sum_{j\in\mathcal{P}}\sum_{q\in \mathcal{Q}_j}X^{q,t=2}_{i,j}~.~\eta^{q}_{i,j}\leq\phi_{i}&&\forall i\in\mathcal{P}\setminus{j}
\end{flalign}
%%%%%%%%%%%%%%%%%%%%%%%%%

\textbf{MILP Optimization Model.} The following model minimizes the requests' serving times (\ie fetching  time  plus  transcoding  time), denoted by $\Psi$: 
%%%%%%%%%%%objective%%%%%%%%
\begin{flalign}
\textit{Minimize}&\hspace{.3cm} \Psi
\label{RICHTER:eq:10}\\
  s.t.&\hspace{.5cm}\text{constraints}\hspace{.5cm}\text{Eq.}(\ref{RICHTER:eq:1})-\text{Eq.}(\ref{RICHTER:eq:9})&&\nonumber\\
  vars.&\hspace{.5cm} T^{q}_{i,j},\tau^{q}_{i,j},\Psi \geq 0, X^{q,t}_{i,j}\in\{0,1\}\nonumber 
\end{flalign}
%%%%%%%%%%%%
By running the MILP model~(\ref{RICHTER:eq:10}), an optimal action will be selected for each request issued by peer $j \in \mathcal{P}$ such that the total serving time is minimized. However, the MILP model (\ref{RICHTER:eq:10}) is NP-hard~\cite{lewis1983michael}, and suffers from high time complexity. The next section introduces an OL-based approach based on SOM~\cite{kohonen-som} to remedy this issue.
%%%%%%%%%%%%%%%%%%%%%%%%%%%%%%%%%%%%%%%%%%%%%%%%%%%%%%%%%%%%%%%%%%%%%%%%%%%%%%%%%%%%%%%%%%%%%%%%%%%%%%%%%%%%%%%%%%%%%%%%%%%%%%%%%%%%%%%%%%%%%%%%%%%%%%%%%%%%%%%%%%%%%%%%%%%%%%%%%%%%%%%%%%%%%%%%%%%%%%%%%%%%%%%%%%%
\subsection{RICHTER Online Learning Approach}
\label{sec:RICHTER:OL}
We design an OL-based solution depicted in Fig.~\ref{OL-fig1}, which works in a time-slotted fashion. As shown in Fig.~\ref{OL-fig2}(a), the proposed time slot structure consists of two intervals: \textit{(i)} \textit{Collecting Data} (CD) and
\textit{(ii)} \textit{Serving Requests} (SR). In addition to the transcoding and partial cache functions, a VTS is equipped with the four following modules: \textit{Resource monitoring Module} (RM), \textit{Manager Module} (MM), \textit{Queuing Module} (QM), and \textit{OL Agent}. Moreover, the VTS hosts multiple queues, one each per peer region, live video channel, and bitrate in each channel. In the CD interval, the following modules are called to prepare inputs for the OL agent:
\begin{enumerate}[noitemsep]
\item \textbf{RM.} This module is responsible for collecting received CMCD and CMSD messages, monitoring available resources (\ie bandwidth, power, computation, joining/leaving times), queues, and notifying the MM module.  
%%%%
\begin{figure}[!t]
\centering
\includegraphics[width=1\textwidth]{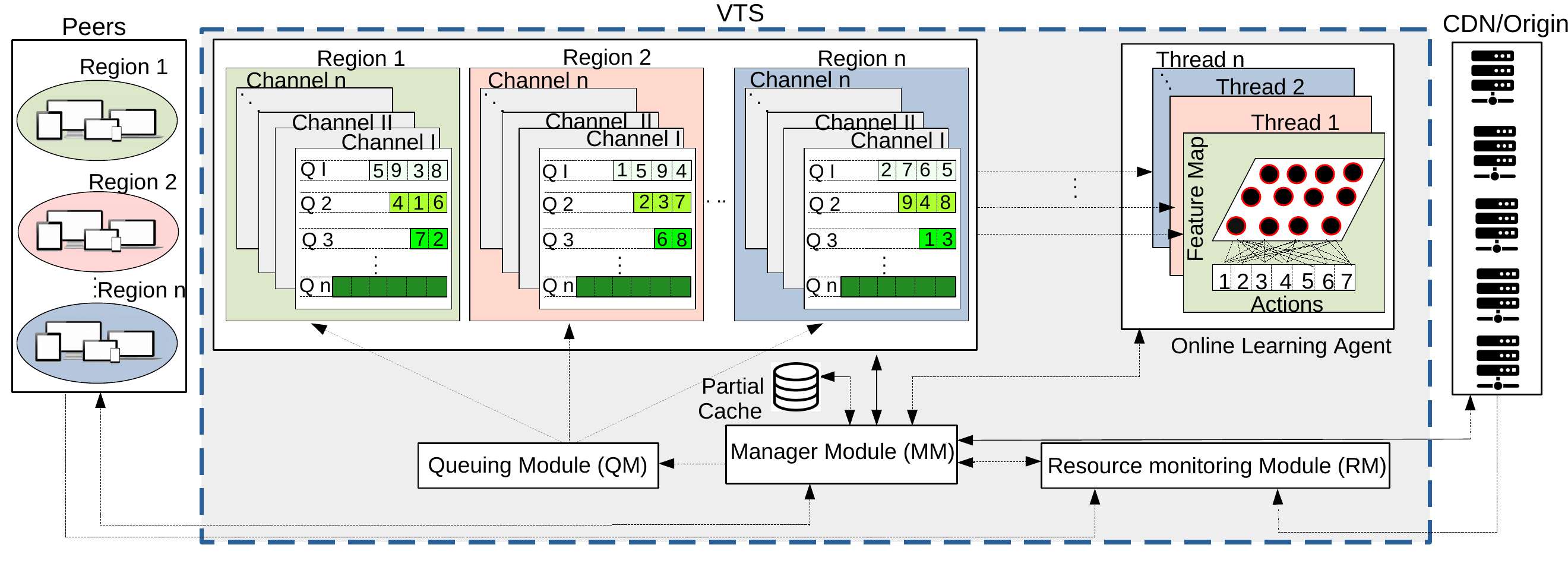}
\caption{\small Proposed online learning  structure.}
\vspace{.5cm}
\label{OL-fig1}	
\end{figure}
%%%%
\item \textbf{MM.} This module is used to \textit{(i)} receive HTTP requests from players, \textit{(ii)} extract regions based on IP addresses, requested channels, and bitrates from the incoming HTTP requests, \textit{(iii)} aggregate and forward the incoming HTTP requests and the extracted information (\ie region/channel/bitrate) to the QM module, \textit{(iv)} update the OL agent based on the items received by the RM, \textit{(v)} control the correctness of decisions made by the OL agent before fetching or transcoding qualities from nodes, \textit{(vi)} communicate with the peers, CDN and/or origin server regarding the decisions made by the OL agent, and \textit{(vii)} store popular segments fetched from CDN/origin server into the partial cache. Note that the MM module immediately responds to a requested segment that exists in the partial cache. Furthermore, it includes an \textit{on the fly} list that is used to prevent delivering a request to the QM module if a response to the request is in flight from the CDN/origin server.
\item \textbf{QM.} This module receives extracted features of requests from the MM module and places requests in separate queues based on peer regions, requested channel IDs, and bitrates. 
\end{enumerate}

Considering the system's current state, \ie available information on resources and queues of requests provided by the MM module, the \textbf{OL agent} in the SR interval must run multiple threads of an OL algorithm (one thread per peer region) to answer the questions mentioned in Section~\ref{sec:RICHTER:sysmodel}. Since SOM~\cite{kohonen-som} \textit{(i)} is one of the widely used techniques for unsupervised classification problems, \textit{(ii)} can be applied to solve NP-hard problems~\cite{bentaleb2021catching}, \textit{(iii)} does not require a prepared dataset for supervised model training, \textit{(iv)} allows online real-time decision making, and \textit{(v)} evolves its model quickly over time, it is adopted as the request management solution in the OL agent. 
For each queue $\overline{\mathcal{Q}}_b\in\overline{\mathcal{Q}}$ with requested bitrate level $b$, a set of SOM neurons (black circles in Fig.~\ref{OL-fig1}) is created, each of which is a feasible node holding the requested quality (\ie same-region peer, VTS, CDN/origin server) or a higher quality (\ie same-region peer, VTS) for serving $b$ through fetching or running transcoding, respectively. Note that since more than one queue can proceed and might violate all/several resource constraints (\eg the bandwidth, computation, or power limitations), they are evaluated in a priority order where the queue with a higher number of requests comes first. 
%%%%
\begin{figure}[!t]
\centering
\includegraphics[width=1\textwidth]{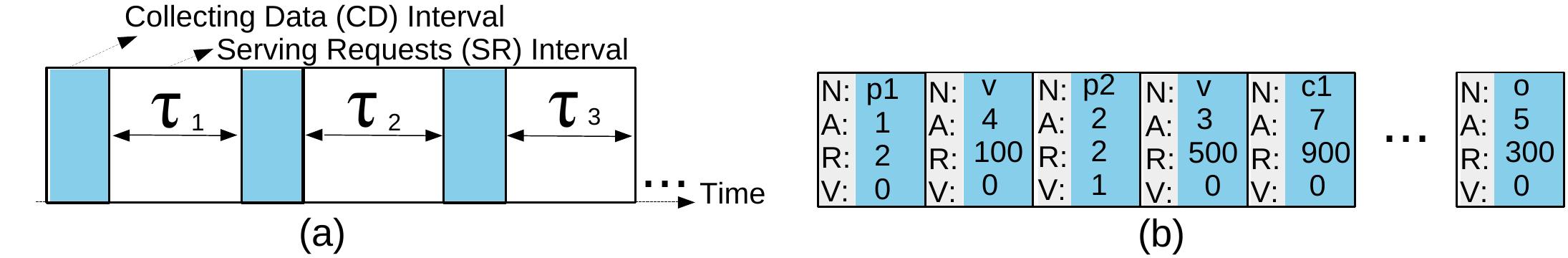}
\caption{\small (a) time slot structure, and (b) a sample of the OL agent output.}
\vspace{.5cm}
\label{OL-fig2}	
\end{figure}
%%%%

Each SOM's neuron has two features (\ie feature map) that are defined as a $<$latency, penalty$>$ tuple. The latency feature indicates fetching plus transcoding times, while the penalty feature is used to penalize the neuron whenever the agent makes an incorrect decision (due to violating one/several constraints (\ref{RICHTER:eq:1})-(\ref{RICHTER:eq:9})). For the sake of simplicity, we assume that each violating action increases the sum of penalties by one. Moreover, in order to represent the SOM features in the same space, we use normalized features in the range between 0 and 1. 
When the SOM thread is executed, it will consider the neurons' feature map and classify neurons to find the \textit{best matching unit} (BMU) with the maximum reward, \ie minimum $<$latency, penalty$>$ values. 

The Euclidean distance function $D_{\overline{\mathcal{Q}}_b}(i,j)=\sqrt{\sum_{n=1}^{2}w_{\overline{\mathcal{Q}}_b}^n(i[n]-j[n])^2}$ as a simple discriminant function is used to calculate the best matching of the features used in each neuron $j$ compared to BMU $i$, where $w_{\overline{\mathcal{Q}}_b}^n$ in weighting matrix $w_{\overline{\mathcal{Q}}_b}$ is used for the $n^{th}$ feature of each feature list. Usually, after selecting the BMU, the corresponding neuron and its neighbors must be updated. Note that the neighborhood function employed in the SOM is the Gaussian distribution function $H_{\overline{\mathcal{Q}}_b}(i,j)=e^{\frac{-D_{\overline{\mathcal{Q}}_b}(i,j)^2}{2\sigma^2}}$, where $\sigma$ is the learning rate. Finally, an output list of tuples ($N, A, R, V$) sorted in ascending order (in terms of latency) is sent to the MM module, where each tuple indicates the determined node $N$, action $A$, the maximum number of requests $R$ that can be served via that node/action, and a violation signal $V$, respectively (Fig.~\ref{OL-fig2}(b)). 

For instance, tuple (p1,1,2,0) of the output list shows that \textit{$peer_1$} using \textit{$action_1$} can serve \textit{two} requests without violating the defined constraints. The MM module follows the SOM decisions for serving requests with tuples with $V=0$, while it ignores tuples with $V\geq1$. Note that the MM module updates the inputs of the OL agent (\ie available resources) regarding the accepted outputs of the OL agent since the SOM threads might execute several times during two consecutive CD intervals.
This process will be repeated in each SR interval until the live streaming session ends and all queues are served. Assume $\rho$, $\beta$, and $\gamma$ indicate the number of peer regions, number of live channels, and number of bitrates per channel, respectively. In the worst case, the time complexity of the multi-thread SOM method employed by the OL agent would be $O(\rho~.~\beta~.~\gamma)$ in each time slot.
%%%%%%%%%%%%%%%%%%%%%%%%%%%%%%%%%%%%%%%%%%%%%%%%%%%%%%%%%%%%%%%%%%%%%%%%%%%%%%%%%%%%%%%%%%%%%%%%%%%%%%%%%%%%%%%%%%%%%%%%%%%%%%%%%%%%%%%%%%%%%%%%%%%%%%%%%%%%%%%%%%%%%%%%%%%%%%%%%%%%%%%%%%%%%%%%%%%%%%%%%%%%%%%%%%%
\subsection{RICHTER Performance Evaluation}
\label{sec:RICHTER:Performance Evaluation}
This section explains the evaluation setup, metrics, and methods and evaluates the performance of \texttt{RICHTER} in two scenarios.
%%%%%%%%%%%%%%%%%%%%%%%%%%%%%%%%%%%%%%%%%%%%%%%%%%%%%%%%%%%%%%%%%%%%%%%%%%%%%%%%%%%%%%%%%%%%%%%%%%%%%%%%%%%%%%%%%%%%%%%%%%%%%%%%%%%%%%%%%%%%%%%%%%%%%%%%%%%%%%%%%%%%%%%%%%%%%%%%%%%%%%%%%%%%%%%%%%%%%%%%%%%%%%%%%%%
\subsubsection{Evaluation Setup}
To assess the effectiveness of \texttt{RICHTER} in a realistic large-scale environment, InternetMCI \cite{zoo} is considered as a real backbone network topology. We instantiate our testbed including 375 elements, \ie 350 \textit{AStream}~\cite{AStream,juluri2015sara} DASH players running the \textit{BOLA}~\cite{spiteri2016bola} adaptive bitrate (ABR) algorithm (seven groups of 50 peers), five Apache HTTP servers (\ie four CDN servers with a total cache size of 40\% of the video dataset and an origin server, containing all video sequences), 19 OpenFlow (OF) backbone switches, 45 backbone layer-2 links, and a VTS server (with a partial cache size of only 5\% of the video sequences) on the CloudLab~\cite{ricci2014introducing} environment. Each element is run on Ubuntu 18.04 LTS inside Xen virtual machines. \texttt{RICHTER} is independent of the caching policy and is compatible with various caching strategies. For simplicity, Least Recently Used (LRU) is considered in all CDN and partial caches as the cache replacement policy. Note that we assume each peer can cache five segments of the video sequences at most. We implement all modules of VTS in Python to serve clients' requests for five live channels (\ie CH I--CH V). Each live channel plays a unique video~\cite{lederer2012dynamic} with 300 seconds duration, comprising two-second segments in bitrate ladder \{(0.089,320p), (0.262,480p), (0.791,720p), (2.4,1080p), (4.2,1080p)\} [Mbps, content resolution]. 

The Docker image \textit{jrottenberg/ffmpeg}~\cite{ffmpeg} is utilized to measure the segment transcoding time on the VTS. To measure the transcoding time on the heterogeneous P2P network, we run the transcoding function via \textit{FFmpegKit}~\cite{ffmpeg-mobile} on an iPhone 11 (Apple A13 Bionic, iOS 15.3), a Xiaomi Mi11 (Snapdragon 888, Android 11), and a PC (Apple M1, MacOS 12.0.1). Moreover, power consumption is measured via device tools, such as \textit{Android Energy Profiler} and \textit{Android Battery Manager}. 
The bandwidth of all links in different paths from the CDN and origin servers to the VTS are set to 50 and 100 Mbps, respectively. To emulate the mobile network conditions, we assume 250 peers initiate the experiments, and then, every three seconds, a new peer joins the sessions. The VTS directs the first peer to the best CDN server (in terms of lowest latency), while other participating peers can be connected on both CDN and P2P links. A real 4G network trace~\cite{raca2018beyond} collected on bus rides is employed for links between peers to edge servers in all experiments. The average bandwidth of this trace is approximately \SI{3780}{kbps} with a standard deviation of {3190}{kbps}. The channel access probability is generated following a Zipf distribution with the skew parameter $\alpha=0.7$, \ie the probability of an incoming request for the $i^{th}$ channel in each peer group is given as $prob(i)=\frac{1/i^{\alpha}}{\sum_{j=1}^{K}1/j^{\alpha}}$, where $K=5$. The learning rate and weighting parameters associated with latency and penalty are set to 0.01, 0.5, and 0.5, respectively. 
%%%%%%%%%%%%%%%%%%%%%%%%%%%%%%%%%%%%%%%%%%%%%%%%%%%%%%%%%%%%%%%%%%%%%%%%%%%%%%%%%%%%%%%%%%%%%%%%%%%%%%%%%%%%%%%%%%%%%%%%%%%%%%%%%%%%%%%%%%%%%%%%%%%%%%%%%%%%%%%%%%%%%%%%%%%%%%%%%%%%%%%%%%%%%%%%%%%%%%%%%%%%%%%%%%%
\subsubsection{Evaluation Methods and Metrics}
The results achieved by the \texttt{RICHTER} will be compared with the following baseline methods:
\begin{enumerate}[noitemsep]
\item\textbf{Non Hybrid (NOH)}: regular CDN-based streaming with no P2P support. 
\item\textbf{Non Transcoding-enabled Hybrid (NTH)}: Like in most works, there is no transcoding capability in this approach. In an NTH-based system, peers can only be served via one of actions 1, 5 or 7 (Fig.~\ref{AT}). 
\item\textbf{Edge Caching/Transcoding Hybrid (ECT)}: In this approach, transcoding at the peer side is not considered, and requests can be served via all actions except action 2. 
\end{enumerate}

For fair comparisons, our testbed with a similar setup is used in all systems. Moreover, the NOH, NTH, and ECT systems employ lightweight heuristic approaches to answer the questions mentioned in Section~\ref{sec:RICHTER:sysmodel} by considering Eqs.~(\ref{RICHTER:eq:1})--(\ref{RICHTER:eq:10}).
The performance of these systems is evaluated through the following metrics: 
\begin{enumerate}[noitemsep]
\item\textbf{Average Segment Bitrate (ASB)} of all the downloaded segments. 
\item\textbf{Average Number of Quality Switches (AQS)}, the average number of segments whose bitrate level changed compared to the previous one. 
\item\textbf{Average Stall Duration (ASD)}, the average of total video freeze time of all clients.
\item\textbf{Average Number of Stalls (ANS)}, the average number of rebuffering events. 
\item\textbf{Average Perceived Overall QoE (APQ)} calculated by the ITU-T Rec. P.1203 model in mode 0~\cite{p1203}. 
\item\textbf{Average Serving Time (AST)}, defined as the overall time for serving all clients, including fetching time plus transcoding time. 
\item\textbf{Backhaul Traffic Load (BTL)}, the  volume of segments downloaded from the origin server. 
\item\textbf{Edge Transcoding Ratio (ETR)}, the fraction of segments transcoded at the VTS or peers. 
\item\textbf{Cache Hit Ratio (CHR)}, defined as the fraction of segments fetched from the CDN or edge servers or peers. 
\end{enumerate}
Each experiment is executed 20 times, and the average and standard deviation values are reported in the experimental results. 
\rf{Note that we use the method discussed in Section~\ref{sec:SFC:Performance Evaluation} to calculate the results reported in the next section.}
%%%%%%%%%%%%%%%%%%%%%%%%%%%%%%%%%%%%%%%%%%%%%%%%%%%%%%%%%%%%%%%%%%%%%%%%%%%%%%%%%%%%%%%%%%%%%%%%%%%%%%%%%%%%%%%%%%%%%%%%%%%%%%%%%%%%%%%%%%%%%%%%%%%%%%%%%%%%%%%%%%%%%%%%%%%%%%%%%%%%%%%%%%%%%%%%%%%%%
\subsubsection{Evaluation Results}
Running transcoding on peers must be fast enough, not significantly impose a delay to the live system, and not consume much battery; otherwise, the clients' requests may use other actions that congest the network and edge server. In the first scenario, we run experiments to investigate the latency and energy overheads of running transcoding tasks on peers. To evaluate the latency overheads, we measure transcoding times for a five-minute video in different resolutions/bitrates on the mobile devices. In fact, transcoding demands decoding video into raw frames and then re-encoding those frames into new frames. Thus, transcoding time at the peer-side is equal to the encoding time due to leveraging the video processing that is already being done to capture or view video.

As shown in Table~\ref{tab:RICHTER:P2P-Trans}, running transcoding for the whole video takes 8.5--254.2 seconds on these devices (0.056--1.69 seconds per segment) and is fast enough in action 2.
%%%%%%%%%%%%%%%%%%%table%%%%%%%%%%%%%%%
\begin{table}[!t]
\caption{\small Average transcoding times for a 5-min. video on peers.}
\label{tab:RICHTER:P2P-Trans}
\centering
\begin{tabular}{|l|l|c|c|c|}
\hline
\multicolumn{1}{|c|}{\textbf{Resolution}} & \multicolumn{1}{c|}{\textbf{Bitrate}} & \multicolumn{1}{c|}{\textbf{iOS}} & \multicolumn{1}{c|}{\textbf{Android}} & \multicolumn{1}{c|}{\textbf{PC}} \\ \hline
1080p $\rightarrow$ 240p        & 4219k$\rightarrow$ 89k      & 34.2                     & 53.25                        & 18.85                   \\ \hline
1080p $\rightarrow$ 360p        & 4219k $\rightarrow$ 262k    & 42.7                     & 61                           & 24.9                    \\ \hline
1080p $\rightarrow$ 720p        & 4219k $\rightarrow$ 791k    & 166.5                    & 130.9                        & 53.1                    \\ \hline
1080p $\rightarrow$ 1080p       & 4219k $\rightarrow$ 2484k   & 254.2                    & 249.6                        & 87.2                    \\ \hline
1080p $\rightarrow$ 240p        & 2484k $\rightarrow$ 89k     & 35                       & 55.1                         & 18.9                    \\ \hline
1080p $\rightarrow$ 360p        & 2484k $\rightarrow$ 262k    & 45                       & 62.7                         & 21.8                    \\ \hline
1080p $\rightarrow$ 720p        & 2484k $\rightarrow$ 791k    & 172.2                    & 132.5                        & 52                      \\ \hline
720p $\rightarrow$ 240p         & 791k $\rightarrow$ 89k      & 16.25                    & 34.75                        & 10.8                    \\ \hline
720p $\rightarrow$ 360p         & 791k $\rightarrow$ 262k     & 25.1                     & 49.3                         & 15.4                    \\ \hline
360p $\rightarrow$ 240p         & 262k $\rightarrow$ 89k      & 8.5                      & 19.5                         & 8.8                     \\ \hline
\end{tabular}
\vspace{.5cm}
\end{table}
%%%%%%%
\begin{figure}[!t]
\centering
\includegraphics[width=1\textwidth]{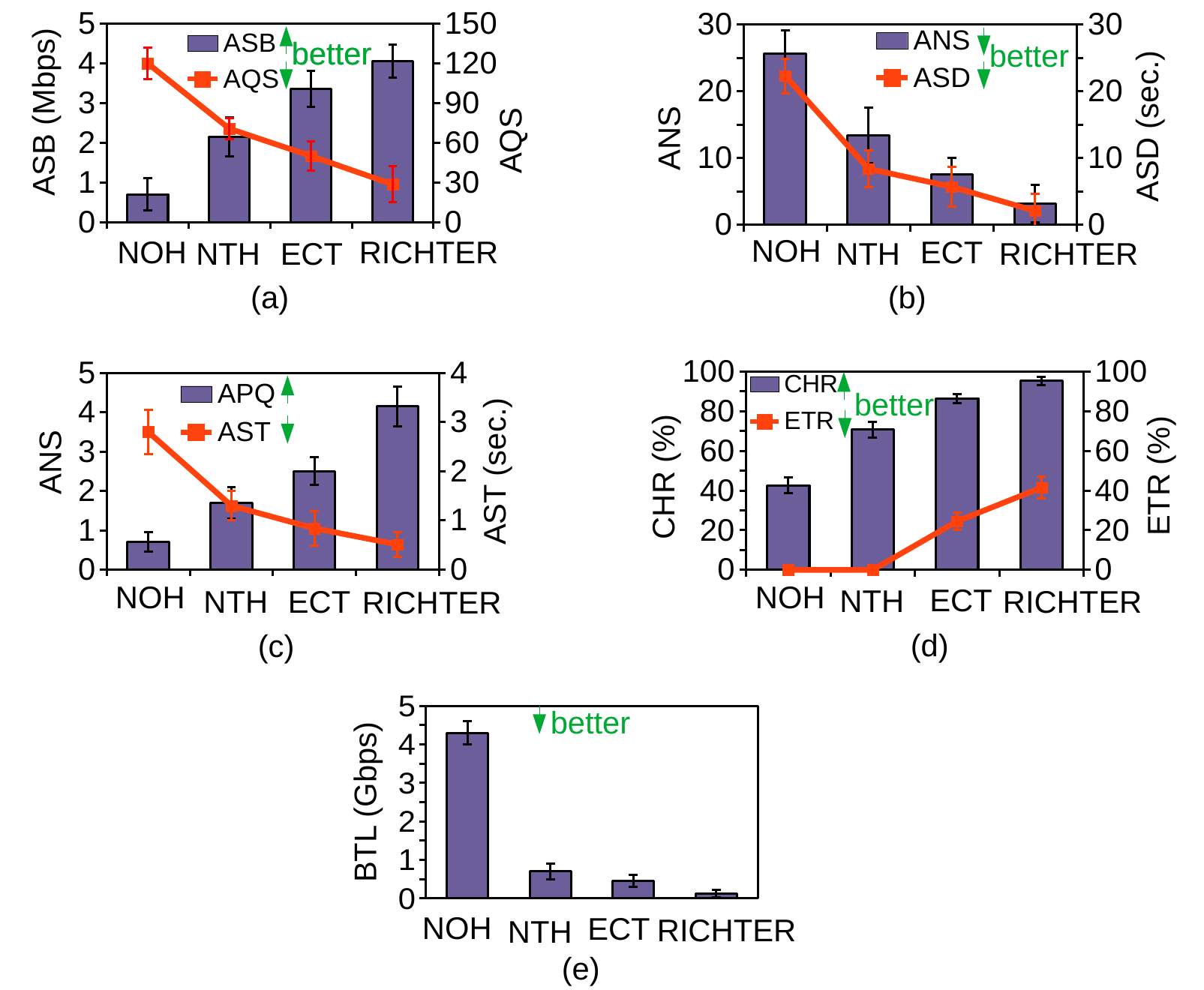}
\caption{\small Evaluation results for the NOH, NTH, ECT, and \texttt{RICHTER} systems for 350 clients.}
\vspace{.5cm}
\label{RICHTER-Result}	
\end{figure}
%%%%%%%%%%%%%%%%%%%table%%%%%%%%%%%%%%%
In another experiment, we measure the battery consumption of peers when they \textit{(i)} play a video, \textit{(ii)} transcode a video from a higher quality, or \textit{(iii)} play video I, transcode video II, and transmit video III, simultaneously. The average values for five-minute video sequences are approximately 0.8\%, 0.4\%, and 1.3\% of peers' battery usage, respectively. Thus, a combination of playing, transcoding, and transmitting tasks does not put a significant burden on the peers' batteries compared to the energy used to play or transcode video.
%%%%%%%%%%%%%%%%%%%%%%%%%%%%%%%%%

In the second scenario, we evaluate \texttt{RICHTER}'s effectiveness in terms of the aforementioned metrics and compare the results with the baseline systems. As illustrated in Fig.~\ref{RICHTER-Result}(a--c), \texttt{RICHTER} downloads segments with higher ASB, decreases AQS and ANS, shortens ASD, and thus improves APQ and AST by at least 59\% and 39\% compared to the baseline approaches (Fig.~\ref{RICHTER-Result}(c)), respectively. Thus, the average latency can be significantly reduced due to shortening the ASD and AST values. This is because \texttt{RICHTER} utilizes all peers' possible resources for serving clients.  

The performance of \texttt{RICHTER} regarding the CHR, BTL, and ETR metrics is shown in Fig.~\ref{RICHTER-Result}(d--e). Note that a cache miss event occurs when \textit{(i)} the requested or higher quality levels are not available in the partial caches or on CDN servers, \textit{(ii)} available bandwidth values are insufficient to fetch the requested or higher quality levels from CDN servers or peers, \textit{(iii)} the available edge or peers' processing capabilities are not sufficient to transcode the requested quality from a higher quality. 
The CHR and BTL metrics indicate that \texttt{RICHTER} outperforms other systems due to its ability to fetch requested or higher quality levels in a hybrid system or using distributed transcoded. Although \texttt{RICHTER} downloads fewer segments from the origin server and improves backhaul bandwidth usage (by about 70\%) compared to ECT, it uses more computation resources of the edge and P2P layer due to employing a distributed transcoding approach.  
\clearpage

%% file: Chapters/Chapter6/6-3-ALIVE.tex
\doublespacing
\section{ALIVE Framework}\label{chap:Hybrid-P2PCDN:ALIVE}
This section presents the \texttt{ALIVE} system~\cite{farahani2023alive}, a latency- and cost-aware hybrid P2P-CDN framework for live video streaming. We first discuss the problem statement, introduce the \texttt{ALIVE} multi-layer architecture, and then formulate the problem as an MILP optimization model in Section~\ref{sec:ALIVE:Design}. Section~\ref{subsec:ALIVE:Heuristic} explains the proposed greedy-based lightweight heuristic algorithm. The evaluation setup, the details of the used super-resolution (SR) models, and obtained results are presented in Section~\ref{sec:ALIVE:Performance Evaluation}.
%%%%%%%%%%%%%%%%%%%%%%%%%%%%%%%%%%%%%%%%%%%%%%%%%%%%%%%%%%%%%%%%%%%%%%%%%%%%%%%%%%%%%%%%%%%%%%%%%%%%%%%%%%%%%%%%%%%%%%%%%%%%%%%%%%%%%%%%%%%%%%%%%%%%%%%%%%%%%%%%%%%%%%%%%%%%%%%%%%%%%%%%%%%
\subsection{ALIVE System Design}
\label{sec:ALIVE:Design}
%This section first presents the problem statement and then introduces the \texttt{ALIVE} architecture. Next, we formulate the problem as an optimization model and analyze its time complexity.
%%%%%%%%%%%%%%%%%%%%%%%%%%%%%%%%%%%%%%%%%%%%%%%%%%%%%%%%%%%%%%%%%%%%%%%%%%%%%%%%%%%%%%%%%%%%%%%%%%%%%%%%%%%%%%%%%%%%%%%%%%%%%%%%%%%%%%%%%%%%%%%%%%%%%%%%%%%%%%%%%%%%%%%%%%%%%%%%%%%%%%%%%%%
\subsubsection{ALIVE Problem Statement}
\label{sec:ALIVE:Design:PROBLEMSTATEMENT}
As discussed in Section~\ref{chap:SOTA}, most existing hybrid-based NAVS systems use peers' bandwidth and caching resources without employing edge server assistance. However, utilizing all possible peers' resources (\ie computational, caching, and bandwidth) and controlling such hybrid systems through edge servers augmented with caching and transcoding can satisfy different users' demands and requirements, improve network utilization, and decrease network costs.
Imagine a hybrid P2P-CDN system, including multiple peers, some of which are enabled to run peer transcoding (PTR) and/or peer super-resolution (PSR) functions. Moreover, a \textit{Virtual Tracker Server} (VTS) is placed at the edge of the network and augmented with edge transcoding (ETR) and partial caching (EPC) functions. (We will elaborate on this in more detail in Section~\ref{sec:Design:Architecture}). Moreover, consider multiple CDN servers containing various parts of video sequences and an origin server including all video segments in various representations. 

In such a system, as depicted in Fig.~\ref{ALIVE:actionTree}, a bunch of live requests is controlled by the VTS (by running a decision-making strategy, \ie an optimization model or a heuristic algorithm), and appropriate decisions are taken to serve all clients' requests with suitable policies. The \texttt{ALIVE} system aims at achieving the most suitable solutions to respond to the following vital questions:
\begin{enumerate}[noitemsep]
\item Where is the best place (\ie adjacent peers, VTS, CDN servers, or origin server) in terms of the lowest latency for fetching each client's requested content quality level from, while efficiently utilizing the available resources?
\item What is the best approach for responding to the requested quality level, \ie fetch, transcode it from a higher quality (using PTR or ETR), or upscale it from a lower resolution (running PSR) considering peers, edge, and CDN servers' resource limitations?
\end{enumerate}

To this end, \texttt{ALIVE} considers various network nodes' (\ie peer, edge, CDN) resource limitations and peers' stabilities (in the sense that the least recent joining time is interpreted as the highest stability), and answers the above questions by utilizing actions from the \textit{action tree} shown in Fig.~\ref{ALIVE:actionTree} (action numbering as in the figure):
\begin{enumerate}[leftmargin=*,label=\protect\circledd{\arabic*},noitemsep]
\item Use the P2P network and transmit the requested quality representation directly from the best adjacent peer with maximum stability (\ie the least recent joining time).
\item Use the PTR function to transcode the requested quality representation from a higher quality at the peer or the best adjacent peer (``best'' in terms of stability, available processing, and available bandwidth) and then forward it to the peer.
 \item Use the PSR function to upscale the requested resolution representation from a lower resolution received from the most stable adjacent peer in the P2P network. 
\item Use the EPC function to fetch the requested representation directly from the VTS.
\item Use the ETR function to transcode the requested quality representation from a higher quality at the VTS and then transmit it from the VTS to the peer.
\item Fetch a higher quality representation from the best CDN server (\ie in terms of bandwidth), run the ETR function to transcode it at the VTS, and then transmit it from the VTS to the peer.
 \item Fetch the requested representation directly from the best CDN server (\ie highest available bandwidth). 
 \item Fetch the requested representation from the origin server. 
 \end{enumerate}

As discussed in Section~\ref{chap:SOTA}, the SR output is an uncompressed video representation; thus, this uncompressed representation produced by SR must be recompressed, and then an extra encoding time must be imposed to make a transferable representation. To prevent additional compression and transmission latencies for  reconstructed representations, the proposed action tree always uses the PSR function at the last mile, where the peer plays out the uncompressed representation immediately after reconstruction (see Section~\ref{sec:SRTR} for further explanation).
%%%%%%%%%%%%%%%%%%%%%%%%%%%%%%%%%%%%%%%%%%%%%%%%%%%%%%%%%%%%%%%%%%%%%%%%%%%%%%%%%%%%%%%%%%%%%%%%%%%%%%%%%%%%%%%%%%%%%%%%%%%%%%%%%%%%%%%%%%%%%%%%%%%%%%%%%%%%%%%%%%%%%%%%%%%
\subsubsection{ALIVE Architecture}
\label{sec:Design:Architecture} 
%%%%%%%%%%%%%Fig1%%%%%%%%%%%%%%%%%
\begin{figure}[!t]
	\centering
	\includegraphics[width=.7\textwidth]{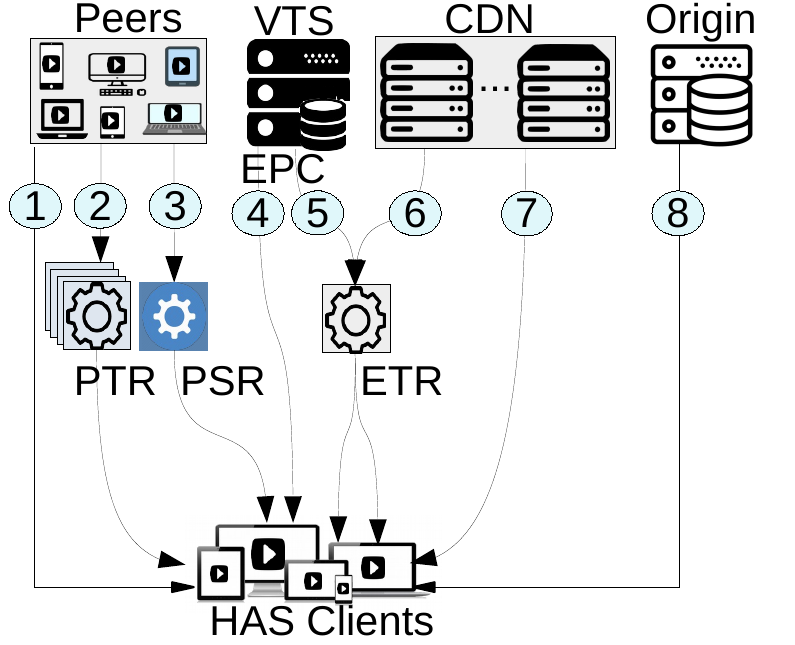}
	\caption{ ALIVE action tree; ETR:= edge Transcoding; PTR := Peer Transcoding; PSR := Peer Super-resolution.}
 \vspace{.5cm}
	\label{ALIVE:actionTree}
\end{figure}
%%%%%%%%%%%%%Fig1%%%%%%%%%%%%%%%%%
%%%%%%%%%%%%%Fig2%%%%%%%%%%%%%%%%%
\begin{figure}[t]
	\centering
	\includegraphics[width=1\textwidth]{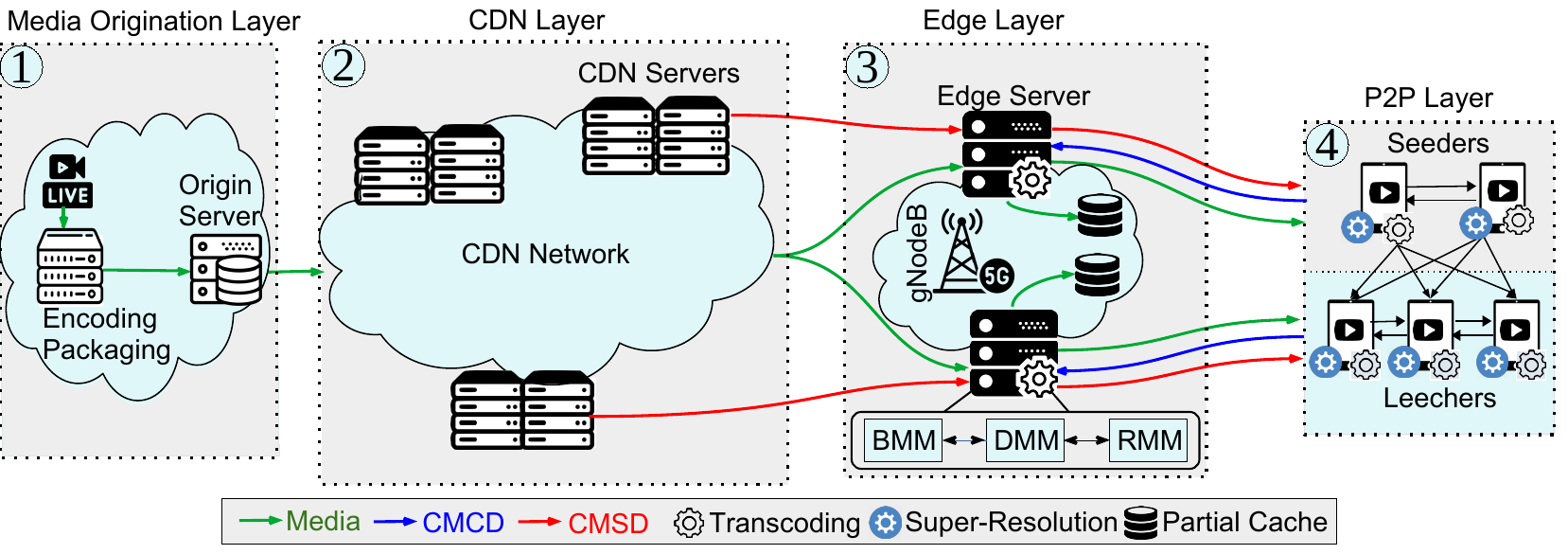}
	\caption{ALIVE multi-layers architecture.}
 \vspace{.5cm}
	\label{ALIVE-arch}
\end{figure}
%%%%%%%%%%%%%Fig2%%%%%%%%%%%%%%%%%
Regarding our discussion above and the problem statement, we propose the \texttt{ALIVE} multi-layer architecture, depicted in Fig.~\ref{ALIVE-arch}, including \textit{four} core layers: \textit{Media Organization}, \textit{CDN}, \textit{Edge}, and \textit{P2P}.
\begin{enumerate}[noitemsep]
\item \textbf{Media Organization Layer. }In this layer, the raw live video sequences are encoded and packaged into DASH format, then are stored on the origin server. Note that this layer is able to package the encoded video sequences to other formats like HLS or \textit{Common Media Application Format} (CMAF).
\item \textbf{CDN Layer. }This layer includes a group of CDN servers (either OTT servers or a purchased service from CDN providers), each of which holds various parts of video sequences. Inspired by the \textit{Consumer Technology Association} CTA-5004 standard~\cite{CTA,CTA-wave}, CDN servers periodically inform the edge layer about their cache occupancy (\ie availability of the stored segments and their available representations) via \textit{Common Media Server Data} (CMSD) messages.
\item \textbf{Edge Layer. }This layer leverages NFV and edge computing to present \textit{Virtual Tracker Servers} (VTSs) positioned close to base stations (\eg gNodeB in 5G). Note that, in the proposed system, during a live session, clients' requests are directed to a VTS, and then they get responses based on the VTS's decisions. As shown in Fig.~\ref{ALIVE-arch}, a VTS is equipped with the following modules:\\
\textit{(i)} \textit{Edge Transcoding Module} (ETR) to transcode segments from an existing higher quality representation to the demanded quality representation; \\
\textit{(ii)} \textit{Edge Partial Cache Module} (EPC) that stores a limited number of segments; \\
\text{(iii)} \textit{Bandwidth Monitoring Module} (BMM) that monitors the available bandwidth between the VTS and other servers (\ie CDN and origin server) and then feeds the DMM module;\\ \text{(iv)} \textit{Resources Monitoring Module} (RMM) which is designed to monitor the system frequently, collect CMCD/CMSD messages, obtain precise information about the peers' available resources (\eg available bandwidth, computational, and power resources), find peers' joining/leaving times, and notify the DMM module; \\
\textit{(v)} \textit{Decision-Making Module} (DMM) that operates an optimization model or a heuristic algorithm to determine appropriate solutions for the three questions listed in 
Section~\ref{sec:ALIVE:Design:PROBLEMSTATEMENT}. 

For this purpose, DMM employs the information provided by BMM and RMM and finds the best action from the \textit{action tree} (Fig.~\ref{ALIVE:actionTree}). 
It is worth noting that the network slicing method is utilized to design an independent logical network-assisted video streaming instance over the common edge layer; therefore, the edge layer can be used to serve various network services, \eg video streaming services with various requirements.
\item \textbf{P2P Layer. }Given the continuous growth in smart devices' capabilities, \eg high-bandwidth access to the Internet, energy resources, and hardware-accelerated methods for running some computation-intensive tasks, \texttt{ALIVE} employs the peers' idle resources to provide \textit{(i)} video \textit{Peer Transcoding} (PTR) and \textit{(ii)} \textit{Peer Super-Resolution} (PSR) approaches, besides their tiny cache and idle upload bandwidth for video transmission. Like most hybrid P2P-CDN schemes, we construct the P2P layer based on the tree-mesh structure, including two types of peers: \textit{Seeders} and \textit{Leechers}. In this scheme, seeders' requests can be served by all nodes (\ie CDNs, origin, edge, or other seeders) except leechers, while all nodes can serve leechers' requests. Inspired by the CTA-5004 standard~\cite{CTA}, peers periodically inform the edge layer about their cache occupancy and available resources (\ie battery, computational, and bandwidth) through \textit{Common Media Client Data} (CMCD) messages and receive updates from the edge layer via CMSD messages. 
\end{enumerate}
%%%%%%%%%%%%%%%%%%%%%%%%%%%NotationTable%%%%%%%%%%%%%%%%%%%%%%%%%%%%%%
\begin{table}[!t]
\centering
\caption {\small ALIVE Notation.}
\label{tab:ALIVE:notation}
\begin{tabular}{llllll}
\cline{1-2}
\multicolumn{2}{|c|}{\textbf{Input Parameters}}                                                                                  
&  &  &  &  \\ \cline{1-2}                                                                                         
\multicolumn{1}{|l|}{\begin{tabular}[c]{@{}l@{}}
$\mathcal{C}$\\ 
$\mathcal{P}$ \\ 
$\mathcal{V}$\\ 
$\mathcal{R}_j$\\ \\ \\ \\ \\
$\mathcal{T}$\\  \\ \\ \\
$\mathcal{S}$\\ \\
$\chi$\\
$\overline{\mathcal{Q}}$\\ \\
$\alpha^{r}_{i,j}$\\ \\
$\omega_{i,j}$\\
$\theta^{r}_{i,j}$\\ \\ \\
$\eta^{r}_{i,j}$\\ \\
$\mu^{r}_{i,j}$\\ \\
$\delta^{r}_j$\\
$\lambda^r_j$\\
$\Omega_i,\phi_i$\\ \\
$\Delta_{co},\Delta_{bw}$\\ \\
 $\beta$\\ \\
\end{tabular}} 
& \multicolumn{1}{l|}{\begin{tabular}[c]{@{}l@{}}
Set of $n$ CDN servers and an origin server (i.e., $c=0$)\\                                     
Set of $m$ peers including $s$ seeders and $l$ leechers in subsets $\mathcal{P}_1$ and $\mathcal{P}_2$\\           
Set of Virtual Tracer Servers (VTSs)\\             
Set of possible representations (i.e., (quality, resolution)) for serving\\  representation $r^{*}=(x^*,y^*)$ requested by $j\in\mathcal{P}$, where \\$\mathcal{R}_j=\{(x^{*},y^{*}_{min}),...,(x^*,y^*),...,(x^{*}_{max},y^{*})\}$ and $y^{*}_{min}$, $x^{*}_{max}$ are the\\ available minimum resolution and maximum quality for the demanded\\  segment, respectively\\
Set of possible transcoding statuses, where $\mathcal{T}=\{0,1,2,3\}$; $t=1$,\\  $t=2$, or $t=3$ if the requested representation is transcoded from\\ a higher quality representation $r\in\mathcal{R}_j$ by the ETR, by the peer's\\ PTR, or by an adjacent peer's PTR, respectively; otherwise $t=0$\\ 
Set of possible SR statuses; $\mathcal{S}=\{0,1\}$; $s=1$ if a peer runs the\\ PSR function to build the requested representation; otherwise $s=0$\\
Set of $\rho$ peer regions\\ 
Set of queues of clients' requests (i.e., representations) in the VTS;\\ $\overline{\mathcal{Q}}$ = $\{\overline{\mathcal{Q}}_{r_{min}}$,...,$\overline{\mathcal{Q}}_{r_{max}}\}$\\
Available representations in $i\in\mathcal{C}\cup\mathcal{V}\cup\mathcal{P}$; $\alpha^{r}_{i,j}=1$ means node\\ $i$ hosts representation $r$ requested by peer $j\in{P}$; otherwise $\alpha^{r}_{i,j}=0$\\
Available bandwidth on path between $i\in\{\mathcal{P}\cup\mathcal{V}\cup\mathcal{C}\}\setminus{j}$, and $j\in\mathcal{P}$\\
Required computational resources (i.e., CPU time in seconds) for \\building $r\in\mathcal{R}_j$ requested by $j\in\mathcal{P}$ through TR in $i\in\mathcal{V}\cup\mathcal{P}$ or via\\ PSR in $i=j$, respectively \\
Required power (in milliampere-hour) for constructing $r\in\mathcal{R}_j$ \\requested by $j\in\mathcal{P}$ through TR in $i\in\mathcal{P}$ or SR in $i=j$\\
Required time (in seconds) for building $r\in\mathcal{R}_j$ requested by\\ $j\in\mathcal{P}$ through TR in $i\in\mathcal{V}\cup\mathcal{P}$ or via SR in $i=j$ \\
Size of segment in representation $r$ (in bytes) requested by $j\in\mathcal{P}$ \\ 
Bitrate for representation $r\in \mathcal{R}_j$ requested by $j\in\mathcal{P}$ \\     
Available computation resources (available CPU) of $i\in\mathcal{V}\cup\mathcal{P}$ and\\ available power resources of $i\in\mathcal{P}$, respectively \\     
Computational cost per CPU core per second and bandwidth cost\\ per bit per second, respectively\\  
Adjustable weighting coefficient for the latency, where $1-\beta$ is the \\ network costs coefficient\\
\end{tabular} }
&  &  &  &  \\ \cline{1-2}
\multicolumn{2}{|c|}{\textbf{Variables}} 
&  &  &  &  \\ \cline{1-2}                                                 
\multicolumn{1}{|l|}{\begin{tabular}[c]{@{}l@{}}
$D^{r,t,s}_{i,j}$\\ \\ \\
$\tau_{i,j}^{r}$\\ \\
$\mathcal{L}^{r}_{i,j}$\\ \\
$\Psi$\\
$\Pi,\Lambda$\\
$\xi$\\
\end{tabular}}
&\multicolumn{1}{l|}{\begin{tabular}[c]{@{}l@{}}
Binary variable where $D^{r,t,s}_{i,j}=1$ shows node $i\in\mathcal{C}\cup\mathcal{V}\cup\mathcal{P}$ transmits\\ representation $r\in\mathcal{R}_j$ requested by peer $j\in\mathcal{P}$
with TR status $t$ and\\ SR status $s$, otherwise $D^{r,t,s}_{i,j}=0$\\
Required computation time to build the representation $r\in\mathcal{R}_j$ requested \\ by $j\in\mathcal{P}$ via TR at node $i\in\mathcal{V}\cup\mathcal{P}$ or via PSR at $i=j$ \\
Required transmitting time of representation $r\in\mathcal{R}_j$ in response to peer\\ $j\in\mathcal{P}$ from node $i\in\{\mathcal{C}\cup\mathcal{V}\cup\mathcal{P}\}\setminus{j}$\\
Total serving latency, including $\tau^{r}_{i,j}$ and $\mathcal{T}^{r}_{i,j}$\\
Network computational and bandwidth costs, respectively\\
Total network costs consisting of $\Pi$ and $\Lambda$\\
\end{tabular}}                                                    
&  &  &  &  \\ \cline{1-2}                                                                                                              
& &  &  &  &                                              
\end{tabular}
\end{table}
\clearpage
%%%%%%%%%%%%%%%%%%%%%%%%%%%%%%%%%%%%%%%%%%%%%%%%%%%%%%
%%%%%%%%%%%%%%%%%%%%%%%%%%%%%%%%%%%%%%%%%%%%%%%%%%%%%%%%%%%%%%%%%%%%%%%%%%%%%%%%%%%%%%%%%%%%%%%%%%%%%%%%%%%%%%%%%%%%%%%%%%%%%%%%%%%%%%%%%%%%%%%%%%%%%%%%%%%%%%%%%%%%%%%%%%%
\subsubsection{ALIVE Optimization Problem Formulation}
\label{sec:ALIVE:Design:MILP}
Let set $\mathcal{C}$ denote $n$ CDN servers and an origin server, where $c=0$ means the origin server (see Table~\ref{tab:ALIVE:notation} for a summary of the notation). Moreover, let us consider $\mathcal{P}$ as the set of $m$ peers, consisting of $s$ seeders and $l$ leechers in two subsets $\mathcal{P}_1$ and $\mathcal{P}_2$, respectively. As discussed earlier, the DMM frequently receives CMCD/CMSD and bandwidth updates from the RMM and BMM modules. Considering this information, let $\alpha^{r}_{i,j}$ denote the availability of representation $r$ of a segment demanded by peer $j$ at node $i\in\mathcal{C}\cup\mathcal{V}\cup\mathcal{P}$. In addition, $\omega_{i,j}$ shows the available bandwidth between node $i$ and peer $j$ accordingly. When a peer $j$ requests a representation $r^*=(x^*,y^*)$, it is possible to receive it with the exact requested quality $x^*$ and resolution $y^*$ from all nodes, with a lower resolution $x^*,y<y^*$ (to reconstruct it by the PSR function) or with a higher bitrate level $x>x^*,y^*$ (to rebuild it by the PTR function) from adjacent peers. 

In fact, if $r^*$ is available in the cache of node $i$, $r^*$ can be directly delivered to $j$. Otherwise, since both edge and peer nodes are able to run the TR function, the demanded representation can be reconstructed from a higher-quality representation. As another option, the requested representation can also be built by the PSR function from a lower resolution. Therefore, for each peer's request, we define $\mathcal{R}_j=\{(x^{*}, y^{*}_{min}),...,(x^*, y^{*}-1),(x^*,y^*),(x^{*}+1,y^*),...,(x^{*}_{max}, y^{*})\}$ as the set of possible representations (pairs of (quality, resolution)), for serving requested representation $r^{*}=(x^*,y^*)$. Note that $y^{*}_{min}$ and $x^{*}_{max}$ indicate the minimum resolution and maximum quality for the requested segment, respectively. 

In the following, we propose an MILP optimization model that decides an optimal action for each request issued by each peer, \eg $j\in\mathcal{P}$, based on the given information in such a way that the total latency and network cost are minimized. To this end, the following aspects are mandated to be considered for each request to force the model to: \textit{(i) }ensure that each request is served  through only one action; \textit{(ii) }select the best action in terms of latency (\ie fetching, fetching/TR, or fetching/SR) and network costs (\ie bandwidth and computation) among all possible actions; \textit{(iii) }not violate the given available resource constraints by the chosen action; \textit{(iv) }not violate rules defined by stakeholders. To address these aspects, we define \textit{five} groups of constraints, described as follows.

\textbf{\textit{(i) }Action Detector Constraint.}
This constraint selects a suitable action from the proposed action tree (Fig.~\ref{ALIVE:actionTree}) in response to peer $j$'s request. Let $t\in\mathcal{T}$ define the TR status, where $t=0$ indicates that the requested representation is transferred without transcoding, while $t=1$, $t=2$, or $t=3$ denote that the requested representation is transcoded from a higher quality representation by the VTS's ETR, by the local peer's PTR, or by an adjacent peer's PTR, respectively. Moreover, $s\in\mathcal{S}$ illustrates the SR status, where $s=1$ means the demanded representation is constructed from a lower resolution representation by the local peer's PSR, otherwise $t=0$. Therefore, Eq.~(\ref{ALIVE:eq:1}) determines proper values of the binary decision-making variables $D^{r,t,s}_{i,j}$ and forces the model to opt for only one action for each request issued by peers, where $D^{r,t,s}_{i,j}=1$ indicates node $i$ (\ie CDN, origin, VTS, or peer) replies the representation $r$ requested by peer $j$ through TR status $t$ and SR status $s$ (see Table~\ref{tab:ALIVE:notation} for notations):
%%%%%%%%%%Const1%%%%%%%%%%%%
\begin{flalign} 
\label{ALIVE:eq:1}
&\sum_{i\in\mathcal{C}\cup\mathcal{V}\cup\mathcal{P}}\sum_{t\in\mathcal{T}}\sum_{s\in\mathcal{S}}\sum_{r\in\mathcal{R}_j} D^{r,t,s}_{i,j}~.~ \alpha^{r}_{i,j}
=1,&&\forall j\in\mathcal{P}
\end{flalign}
where $\alpha^{r}_{i,j}$ is used to assure the representation $r$ requested by peer $j$ is available in node $i$. 
%%%%%%%%%%%%%%%%%%%%%%%%%%%%

\textbf{\textit{(ii) }Latency Calculator Constraints.}
This group of constraints calculates the serving latency required for answering peers' requests. The first equation, \ie Eq.~(\ref{ALIVE:eq:2}), determines the transmitting latency $\mathcal{L}^{r}_{i,j}$, for transmitting representation $r$ from each node $i$ to peer $j$:
%%%%%%%%%%Const2%%%%%%%%%%%%
\begin{flalign} 
\label{ALIVE:eq:2}
&\sum_{t\in\mathcal{T}} \sum_{s\in\mathcal{S}} D^{r,t,s}_{i,j}~.~\delta^{r}_j\leq \mathcal{L}^{r}_{i,j}~.~\omega_{i,j} ,&&\\&\forall j\in\mathcal{P}, i\in\{\mathcal{C}\cup\mathcal{V}\cup\mathcal{P}\}\setminus{j}, r\in\mathcal{R}_j&\nonumber
\end{flalign}
%%%%%%%%%%%%%%%%%%%%%%%%%%%%
where $\delta^{r}_j$ indicates the size (in bytes) of representation $r$ requested by the peer $j\in\mathcal{P}$. In addition, Eq.~(\ref{ALIVE:eq:3}) measures the required computational latency $\tau_{i,j}^{r}$, \ie TR time at node $i\in\mathcal{V}\cup\mathcal{P}$ or SR time at node $i=j$, in case of serving the representation $r\in\mathcal{R}_j$ requested by $j\in\mathcal{P}$, from a higher quality or a lower resolution representation, respectively.
%%%%%%%%%%Const3%%%%%%%%%%%%
\begin{flalign} 
\label{ALIVE:eq:3}
&\sum_{t\in\mathcal{T}}\sum_{s\in\mathcal{S}}\sum_{r\in \mathcal{R}_j} D^{r,t,s}_{i,j}~.~\mu^{r}_{i,j}\leq \tau_{i,j}^{r},&&\forall j\in\mathcal{P}, i\in\mathcal{P}\cup\mathcal{V}
\end{flalign}
%%%%%%%%%%%%%%%%%%%%%%%%%%%%
Hence, the total serving latency, \ie $\Psi$, is the sum of transmitting latency ($\mathcal{L}^{r}_{i,j}$ in Eq.~(\ref{ALIVE:eq:2})) plus computational latency ($\tau^{r}_{i,j}$ in Eq.~(\ref{ALIVE:eq:3})) and can be expressed as follows:
%%%%%%%%%%Const4%%%%%%%%%%%%
\begin{flalign} 
\label{ALIVE:eq:4}
&\sum_{i\in\mathcal{C}\cup\mathcal{V}\cup\mathcal{P}}\sum_{j\in \mathcal{P}}\sum_{r\in \mathcal{R}_j}\mathcal{L}^{r}_{i,j}+\tau_{i,j}^{r}\leq \Psi&&
\end{flalign}
%%%%%%%%%%%%%%%%%%%%%%%%%%%%

\textbf{\textit{(iii) }Policy Constraints.}
These constraints apply essential policies to the model to prevent choosing incorrect actions. Eq.~(\ref{ALIVE:eq:5}) forces the model to fetch the exact representation $r^*$ from the origin or CDN servers when one of them is chosen to serve peer $j$. 
%%%%%%%%%%Const5%%%%%%%%%%%%
\begin{flalign}  
\label{ALIVE:eq:5}
&\sum_{i\in\mathcal{C}}\sum_{r\in\mathcal{R}_j} D^{r,t=0,s=0}_{i,j}~.~r = r^* && \forall j\in\mathcal{P}
\end{flalign}
%%%%%%%%%%%%%%%%%%%%%%%%%%%%
Note that $i=0$ in Eq.~(\ref{ALIVE:eq:5}) indicates that the origin server is selected to serve the requested representation $r$. Moreover, since the SR module can only be run on the local peer to serve the requested representation $r$, Eq.~(\ref{ALIVE:eq:6}) prevents other nodes ($i\not=j$) from using their computational resources for running SR:
%%%%%%%%%%Const6%%%%%%%%%%%%
\begin{flalign}  
\label{ALIVE:eq:6}
&\sum_{i\in \{\mathcal{C}\cup\mathcal{V}\cup\mathcal{P}\}\setminus{j}}\sum_{r\in\mathcal{R}_j}\sum_{t\in\mathcal{T} }D^{r,t,s=1}_{i,j}~.~r = 0, && \forall j\in\mathcal{P}
\end{flalign}
%%%%%%%%%%%%%%%%%%%%%%%%%%%%
Furthermore, Eq. (\ref{ALIVE:eq:7}) controls seeders not to fetch the requested representation $r$ from a leecher:
%%%%%%%%%%Const7%%%%%%%%%%%%
\begin{flalign}  
\label{ALIVE:eq:7}
&\sum_{i\in\mathcal{P}_2}\sum_{t\in\mathcal{T}\setminus{1}}\sum_{s\in\mathcal{S}}\sum_{r\in\mathcal{R}_j} D^{r,t,s}_{i,j}~.~r = 0, && \forall j\in\mathcal{P}_1
\end{flalign}
%%%%%%%%%%%%%%%%%%%%%%%%%%%%
The final constraint of this group prevents peers from serving the requested representation $r$ by employing both PTR and PSR simultaneously:
%%%%%%%%%%Const8%%%%%%%%%%%%
\begin{flalign}  
\label{ALIVE:eq:8}
&\sum_{r\in\mathcal{R}_j} D^{r,t=2,s=1}_{i,j}~.~r = 0, && \forall i,j\in\mathcal{P}, \text{where}~i=j  
\end{flalign}
%%%%%%%%%%%%%%%%%%%%%%%%%%%%

\textbf{\textit{(iv)} Resource Constraints.}
This group of constraints guarantees that the required resources for constructing and/or transmitting the requested segments will not violate the available resources. Therefore, Eq.~(\ref{ALIVE:eq:9}) ensures that the required bandwidth for transmitting representation $r$ on the link between nodes $i$ and $j$ will respect the available bandwidth: 
%%%%%%%%%%Const9%%%%%%%%%%%%
\begin{flalign}
\label{ALIVE:eq:9} 
&\sum_{t\in\mathcal{T}}\sum_{s\in\mathcal{S}}\sum_{r\in \mathcal{R}_j} D^{r,t,s}_{i,j}~.~\lambda^{r}_j\leq \omega_{i,j},&&\\& \forall j\in\mathcal{P},i\in\{\mathcal{C}\cup\mathcal{V}\cup\mathcal{P}\}\setminus{j}\nonumber
\end{flalign}
%%%%%%%%%%%%%%%%%%%%%%%%%%%%
where $\lambda^r_j$ is the bitrate associated with representation $r$. Furthermore, Eq.~(\ref{ALIVE:ALIVE:eq:10}) restricts the maximum processing capacity required for the ETR function to the available computational resource on the VTS server $i$ (denoted by $\Omega_i$):
%%%%%%%%%%Const10%%%%%%%%
\begin{flalign} 
\label{ALIVE:ALIVE:eq:10}
&\sum_{j\in\mathcal{P}}\sum_{r\in \mathcal{R}_j}D^{r,t=1,s=0}_{i,j}~.~\theta^{r}_{i,j}\leq\Omega_i,&&\forall i\in\mathcal{V}
\end{flalign}
%%%%%%%%%%%%%%%%%%%%%%%%%
where $\theta^r_{i,j}$ is the required computational resources (\ie CPU time in seconds) for running the ETR function to achieve demanded representation $r$.
Similarly, Eq.~(\ref{ALIVE:eq:11}) limits the maximum processing capacity of 
peer $i$ for serving adjacent peers' requests through PTR and serving its own requests through PTR or PSR functions to the available computational resource (\ie $\Omega_i$):
%%%%%%%%%%Const11%%%%%%%%
\begin{flalign}
\label{ALIVE:eq:11}
& \sum_{r\in\mathcal{R}_i}D^{r,t=2,s=0}_{i,i}~.~\theta^{r}_{i,j}+ \sum_{r\in\mathcal{R}_i}D^{r,t=0,s=1}_{i,i}~.~\theta^{r}_{i,j} +
\sum_{j\in\mathcal{P}\setminus{i}}\sum_{r\in\mathcal{R}_j}D^{r,t=3,s=0}_{i,j}~.~\theta^{r}_{i,j}\leq\Omega_i,&&\hspace{-.5cm}\forall i\in\mathcal{P}
\end{flalign}
%%%%%%%%%%%%%%%%%%%%%%%%%%
In addition, Eq.~(\ref{ALIVE:eq:12}) restricts the maximum power resources of peer $i$ required for running the PTR or PSR functions to the available power resources (shown by $\phi_{i}$):  
%%%%%%%%%%%Const12%%%%%%%%
\begin{flalign} 
\label{ALIVE:eq:12}
& \sum_{r\in\mathcal{R}_i}D^{r,t=2,s=0}_{i,i}~.~\eta^{r}_{i,j} + \sum_{r\in\mathcal{R}_i}D^{r,t=0,s=1}_{i,i}~.~\eta^{r}_{i,j} +
\hspace{-.3cm}\sum_{j\in\mathcal{P}\setminus{i}}\sum_{r\in\mathcal{R}_j}D^{r,t=3,s=0}_{i,j}~.~\eta^{r}_{i,j}\leq\phi_i,&&\hspace{-1cm}\forall i\in\mathcal{P}
\end{flalign}
%%%%%%%%%%%%%%%%%%%%%%%%%%
where $\eta^{r}_{i,j}$ is the required power (in milliampere-hour) for building the representation $r$ requested by peer $j$ on peer $i$ through PTR or PSR functions.

\textbf{\textit{(v)} Cost Calculator Constraints.}
The final group of constraints formulates the total network costs, including bandwidth and computational costs. For this purpose, Eq.~(\ref{ALIVE:eq:13}) calculates the network bandwidth cost (\ie $\Lambda$ in dollars) for transmitting segments on the link between server $i\in\mathcal{C}\cup\mathcal{V}$ and peer $j\in\mathcal{P}$ as follows:
%%%%%%%%%%%Const13%%%%%%%%
\begin{flalign} 
\label{ALIVE:eq:13}
&\Delta_{bw}~.~\sum_{t\in\mathcal{T}}\sum_{s\in\mathcal{S}}\sum_{r\in \mathcal{R}_j} D^{r,t,s}_{i,j}~.~\lambda^{r}_j \leq \Lambda
,&&\forall j\in\mathcal{P},i\in\mathcal{C}\cup\mathcal{V}
\end{flalign}
%%%%%%%%%%%%%%%%%%%%%%%%%
where $\Delta_{bw}$ refers to the bandwidth cost per bit per second. Similarly, Eq.~(\ref{ALIVE:eq:14}) measures the total required computational cost $\Pi$ required for the ETR function at the VTS $v\in\mathcal{V}$:
%%%%%%%%%%%Const14%%%%%%%%
\begin{flalign}
\label{ALIVE:eq:14}
&\Delta_{co}~.~\sum_{j\in\mathcal{P}}\sum_{r\in \mathcal{R}_j}D^{r,t=1,s=0}_{i,j} \leq \Pi
,&\forall i\in\mathcal{V}
\end{flalign}
%%%%%%%%%%%%%%%%%%%%%%%%%%
where $\Delta_{co}$ indicates computational resource cost per CPU core per second. Since bandwidth and computational costs are essential parts of nowadays' networks, we thus represent the total network costs (\ie $\xi$) as follows:
%%%%%%%%%%%Const15%%%%%%%%
\begin{flalign}
\label{ALIVE:eq:15}
&\Pi+\Lambda \leq \xi&&
\end{flalign}

\textbf{Optimization Problem.} 
As described in Section~\ref{chap:Background:HAS}, a HAS client always operates an ABR algorithm to estimate the network's bandwidth. Indeed, it measures the time between requesting a segment's representation to download and receiving its last packet. Thus, minimizing the latency in the optimization model affects the HAS clients' performance and improves users' QoE directly. On the other hand, the optimization model must be able to be adjusted by the network operator based on their business models. Therefore, making a trade-off between latency and network costs in the objective function makes the model adaptable to the operators' desired policies. For this purpose, the MILP model~(\ref{ALIVE:eq:16}) considers multiple objectives to minimize the clients' total latency as well as the network costs:
%%%%%%%%%%%objective%%%%%%%%
\begin{flalign}
\textit{Minimize}&\hspace{.3cm} \beta ~.~\frac{\Psi}{\Psi^*}+ (1-\beta)~.~ \frac{\xi}{\xi^*} 
\label{ALIVE:eq:16}\\
  s.t.&\hspace{.5cm}\text{constraints}\hspace{.5cm}\text{Eq.}(\ref{ALIVE:eq:1})-\text{Eq.}(\ref{ALIVE:eq:15})&&\nonumber\\
  vars.&\hspace{.5cm} \mathcal{T}^{r}_{i,j},\tau^{r}_{i,j},\Psi, \xi, \Pi,\Lambda \geq 0, D^{r,t,s}_{i,j}\in\{0,1\}\nonumber 
\end{flalign}
%%%%%%%%%%%%%%%%%%%%%%%%%%%%

By minimizing the objective function~(\ref{ALIVE:eq:16}), an optimal action will be selected for each request issued by peer $j\in \mathcal{P}$. The weighting coefficient ($\beta$) enables operators to adjust desirable preferences for latency and network costs based on their specific policies. Since $\Psi$ and $\xi$ have different dimensions (\ie seconds and dollars, respectively), a normalization method is required to form the objective function. Hence, we normalize each factor by dividing it by its associated maximum value (\ie $\Psi^*$ and $\xi^*$, respectively) calculated over all possible actions. To measure $\Psi^*$, we run the MILP model with $\beta=0$ to achieve the minimum network costs, in other words, the maximum serving latency. Similarly, we set $\beta=1$ to measure $\xi^*$. The calculated $\Psi^*$ and $\xi^*$ values are then reused to run the model for different sets of $\beta$. 
%%%%%%%%%%%%%%%%%%%%%%%%%%%%%%%%%%%%%%%%%

\textbf{Theorem I}: The proposed MILP model (\ref{ALIVE:eq:16}) is an NP-hard problem.\\
\textbf{Proof.} Utilizing the introduced binary and non-binary variables drives the optimization model~(\ref{ALIVE:eq:16}) to be an MILP model. Employing methods introduced by \cite{conforti2014integer}, the proposed MILP model can be reduced to the \textit{Multiple Knapsack Problem} (MKP), which is a well-known NP-hard problem~\cite{lewis1983michael}, in polynomial time.
%%%%%%%%%%%%%%%%%%%%%%%%%%%%%%%%%%%%%%%%%%%%%%%%%%%%%%%%%%%%%%%%%%%%%%%%%%%%%%%%%%%%%%%%%%%%%%%%%%%%%%%%%%%%%%%%%%%%%%%%%%%%%%%%%%%%%%%%%%%%%%%%%%%%%%%%%%%%%%%%%%%%%%%%%%%

\subsection{ALIVE Heuristic Approach}
\label{subsec:ALIVE:Heuristic} 

The introduced optimization model~(\ref{ALIVE:eq:16}) suffers from high time complexity, which makes it impractical for large-scale scenarios. To cope with this issue, we propose a lightweight and near-optimal \textit{Greedy-Based Algorithm} (GBA), which is run by the DMM module instead of the MILP model~(\ref{ALIVE:eq:16}). Before explaining the details of GBA, let us describe the new modules, workflow, and time slot structure used by each VTS server to form the proposed heuristic strategy in the following sub-section. 
%%%%%%%%%%%%%%%%%%%%%%%
\subsubsection{Heuristic Workflow}
\begin{figure}[t]
	\centering
	\includegraphics[width=1\textwidth]{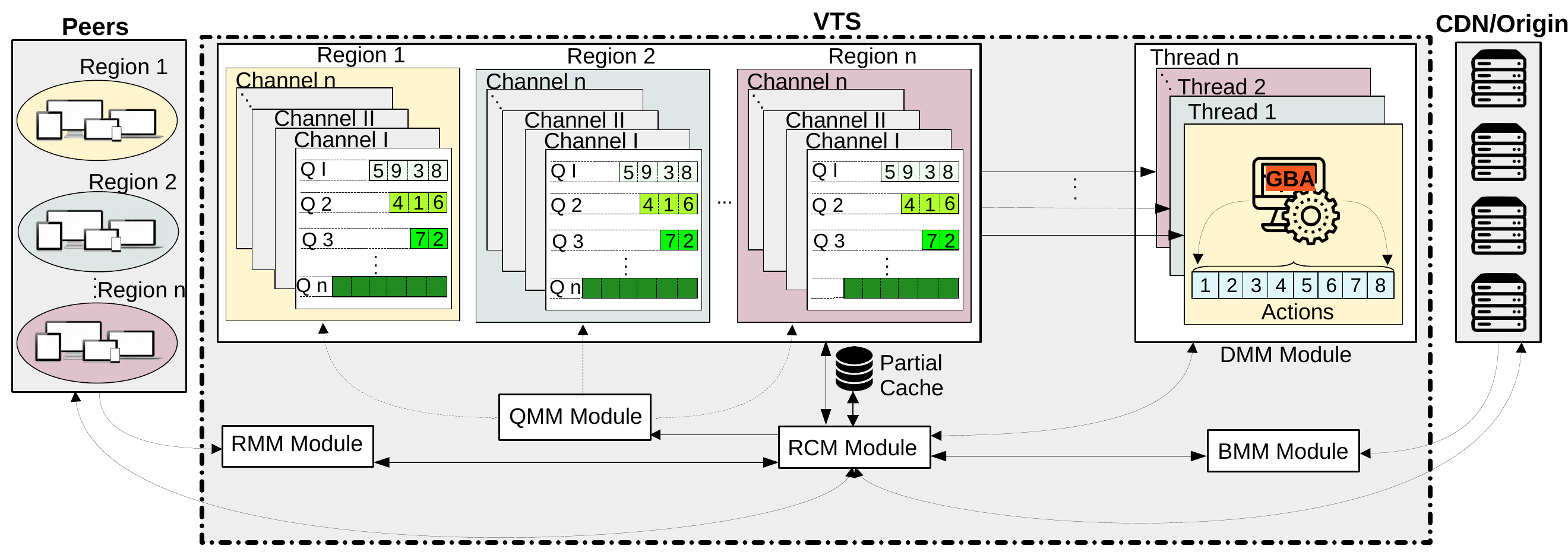}
	\caption{ \small Proposed ALIVE heuristic workflow.}
 \vspace{.5cm}
	\label{heu-fig}
\end{figure}

As shown in the heuristic workflow (Fig.~\ref{heu-fig}), in addition to the modules introduced in Section~\ref{sec:Design:Architecture}, each VTS server hosts multiple \textit{queues of clients' requests} (\ie representations), one per peer region, live video channel, and each channel's representation (bitrate, resolution). Moreover, each VTS is augmented with a \textit{Request Controller Module} (RCM) and a \textit{Queuing Manager Module} (QMM). A time slot structure, including two intervals, \textit{(i)} \textit{Collecting Interval} (CI) and \textit{(ii)} \textit{Serving Interval} (SI), is also utilized by each VTS. 
%%%%%%%%%%%%%%%%%%%%%%%%%%%%%%%%%%%%%%%%%%%%%%%%%
%%%%%%%%%%%%%%%%%%%%%%%%%%%%%%%%%%%%%%%%%%%%%%%%%
In the CI interval, the RMM and BMM modules deliver obtained CMCD and CMSD messages, measured available resources (\ie bandwidth, power, computation, joining/leaving times), and queues' information to the RCM module. The pseudo-code of the RCM module is shown in Alg.~\ref{RCM_alg}, where it receives HTTP clients' requests, the RMM and BMM modules' information, and an $on\_the\_fly$ list (holding requests currently being served) as input parameters. It calls the \textit{ExtractInfo} function to extract regions (based on IP addresses), demanded channels, and representations from the incoming HTTP requests and save the extracted information in the $Ext\_info$ list (line 2). It then aggregates the incoming HTTP requests and the extracted information (\ie region/channel/bitrate) via the \textit{Agg} function and forwards the outputs stored in $Agg\_reqs$ to the QMM module (lines 3--4). 

Since more than one queue can proceed and might violate all/several resource constraints (\eg the bandwidth, computation, or power limitations), they are assessed in priority order where the queue with a higher number of requests is served first. Hence, the \textit{Sort} function is used to store the region's queues in the $SortedQueue$ list (line 6). It then starts to serve requests $req$ in the highest priority queue by comparing it to the requests in the $on\_the\_fly$ list. The \textit{HoldReq} function is called when $req$ is in the $on\_the\_fly$ list (\ie $req$ has been issued by another peer and still is in progress) to hold the request to prevent sending identical requests if a response to the request is in flight from the CDN/origin server (lines 8--10). 
%%%%%%%%%%%%%%%%%%%%%%%RCM%%%%%%%%%%%%%%%%%%%%%%%
\begin{center}
	\begin{algorithm}[!t]		
            \small
            \caption{\small ALIVE RCM module.}\label{RCM_alg}
		\begin{algorithmic}[1]
            \State \textbf{Input} $requests$, $RMM\_info$, $BMM\_info$, $on\_the\_fly$
            \State $Ext\_info\leftarrow$ ExtractInfo($requests$)
            \State $Agg\_reqs$$\leftarrow$Agg($requests,Ext\_info$)
            \State $queues\leftarrow$ QMM($AggReqs$)
            \For{each $reg$ in $Regions$}
                \State $Sorted\_queues_{reg}\leftarrow$ Sort($queues_{reg}, Ext\_info_{reg}$)
                 \State *//One DMM() thread takes care of each region $reg$ 
                  \For{each $req$ in $Sorted\_queue_{reg}$}
                     \If{req $\in$ $on\_the\_fly$}
                        \State HoldReq($req$)
                      \Else
                       \State $on\_the\_fly$.add($req$)
                        \State $opt\_node, opt\_action\leftarrow$ DMM($req$,$RMM\_info$, $BMM\_info$,$\beta$)
                        \State ServeRequest($req, opt\_node, opt\_action$)
                        \State Update()   
                       \EndIf
                \EndFor 
            \EndFor
        \end{algorithmic}
      \end{algorithm}          
\end{center}                    
%%%%%%%%%%%%%%%%%%%%%%%%RCM%%%%%%%%%%%%%%%%%%%%%%%
\raggedbottom
When $req$ is not found in the $on\_the\_fly$ list, that means $req$ is a completely new one, and the $on\_the\_fly$ list must be updated by $req$ to be processed (line 12). Considering the $req$, received $BMM\_info$, $RMM\_info$, and coefficient value $\beta$, the RCM calls the \textit{DMM} module to determine the optimal node and action based on the objective function in Eq.~(\ref{ALIVE:eq:16}) and adjusted weighting coefficient (\ie $\beta$) for the request $req$. 
The results are stored in $opt\_ node$ and $opt\_ action$ (line 13). Moreover, the \textit{ServeRequest} function is called to communicate with the peers, CDN, or origin server concerning the decisions made by the DMM, and to serve the peer's request $req$ (line 14). Finally, it updates the \textit{available resources}, stores popular segments fetched from a CDN or the origin server into the \textit{partial cache}, and updates the $on\_the\_fly$ list via the \textit{Update} function (line 15). 
%%%%%%%%%%%%%%%%%%%%%%%GBA%%%%%%%%%%%%%%%%%%%%%%%
\begin{center}
	\begin{algorithm}[!t]
		\small
            \caption{\small ALIVE GBA algorithm.}\label{GBA}
		\begin{algorithmic}[1]
            \State \textbf{Input} $req$, $RMM\_info$, $BMM\_info$, $\beta$
            \State \textbf{Output} $opt\_node, opt\_action$
            \State$opt\_node\leftarrow\{\varnothing\}$, $opt\_action\leftarrow \{\varnothing\}$, $obj\leftarrow \beta$ 
            \State$act^{*}\leftarrow \{\varnothing\}$, $obj^{*}\leftarrow 0$ 
            \State$Fsb\_nodes\leftarrow$ FsbNodes($req$, $RMM\_info$, $BMM\_info$)
            \For{each $j\in Fsb\_nodes$}
                \State $Fsb\_actions\leftarrow$ FsbActions(j, $req$)
                \State $act^{*}, obj^{*}$ $\leftarrow$ CostFunction($j, req, Fsb\_actions,$\\ $RMM\_info_{j}, BMM\_info_{j}, \beta$)
                \If {$obj > obj^{*}$}
                    \State $opt\_node\leftarrow j, opt\_action\leftarrow act^{*}, obj\leftarrow obj^{*}$
                \EndIf          
            \EndFor 
            \State \textbf{Return} $opt\_node, opt\_action$
        \end{algorithmic}
      \end{algorithm}          
\end{center}                    
%%%%%%%%%%%%%%%%%%%%%%%%GBA%%%%%%%%%%%%%%%%%%%%%%%

Considering the information provided in the CI interval (\ie available information on resources and queues of clients' requests (\ie representations) provided by the RCM and QMM modules), the DMM in the SI interval must run multiple threads of the GBA algorithm (one thread per peer region) to answer the questions mentioned in Section~\ref{sec:ALIVE:Design:PROBLEMSTATEMENT}. This time slot structure will be repeated until the live streaming session ends and all queues are served. The details of the proposed heuristic algorithm are described in the following subsection. 
\subsubsection{Greedy-Based Algorithm (GBA)}
\label{sec:Heuristic:Greedy}

Alg.~\ref{GBA} describes the pseudo-code of our proposed \textit{greedy-based algorithm} (GBA) invoked by the \textit{DMM} module to solve the optimization problem~(\ref{ALIVE:eq:16}).  It receives a request $req$, $RMM\_info$, $BMM\_info$, and coefficient $\beta$ as inputs and returns the determined $opt\_action$ and $opt\_node$ for serving the $req$ as outputs. 

The optimal node $opt\_node$, optimal action $opt\_action$, and associated objective value $obj$, plus two auxiliary variables $act^{*}$ and $opt^{*}$ are initiated (lines 3-4). Moreover, the \textit{FsbNodes} function is called to determine all feasible nodes (\ie peers, VTS, CDNs, or origin) that hold the requested representation or convertible representations (\ie higher or lower ones for running TR and SR, respectively), and to prepare a list of all feasible nodes in $Fsb\_nodes$ (line 5). 

In the next step, first, all feasible actions for each feasible node are determined and stored in the $Fsb\_actions$ list by invoking the \textit{FsbActions} function (line 7). It then employs the \textit{CostFunction} (Alg.~\ref{CostFunction}) to find the action with the minimum objective value among all determined feasible actions of a specific node. 

The outputs of the \textit{CostFunction} are stored in the auxiliary variables $opt^*$ and $act^*$ (line 8--9). Next, it checks whether a new minimum objective value has been determined (lines 10--11). Finally, at the end of the for loop, the best node and action with a minimum value of $obj$ is selected and returned as $opt\_node$ and $opt\_action$ (line 14).
%%%%%%%%%%%%%%%%%%%%%%%CostFunction%%%%%%%%%%%%%%%%%%%%%%%
\begin{center}
	\begin{algorithm}[!t]
            \small
            \caption{\small ALIVE CostFunction. }\label{CostFunction}
		\begin{algorithmic}[1]
            \State \textbf{Input} $j, req, Fsb\_actions ,RMM\_info, BMM\_info, \beta$
            \State \textbf{Output} $act, obj$
            \State $act\leftarrow \{\varnothing\}$, $obj\leftarrow \beta$, $obj^{*}\leftarrow 0$
            \State$max\_delay, max\_cost \leftarrow$ CalcMax($j, req, RMM\_info, BMM\_info$)
            \If {($j\in\mathcal{P}$)}
                \For{each $a\in Fsb\_actions$}
                    \State $delay\leftarrow CalcDly(j, a, RMM\_info, BMM\_info)$
                    \State $obj^{*}\leftarrow\frac{\beta_1~.~ delay}{max\_delay}$
                    \If {($obj > obj^{*}$)}
                        \State $act\leftarrow a, obj\leftarrow obj^{*}$
                    \EndIf
                \EndFor
            \Else
                \For{each $a\in Fsb\_actions$}
                    \State $delay\leftarrow CalcDly(j, a, RMM\_info, BMM\_info)$
                    \State$cost\leftarrow CalcCst(i, j, a, RMM\_info, BMM\_info)$
                    \State $obj^{*}\leftarrow\frac{\beta ~.~ delay}{max\_delay}$ + $\frac{(1-\beta)~.~ cost}{max\_cost}$
                    \If {($obj > obj^{*}$)}
                        $act\leftarrow a, obj\leftarrow obj^{*}$
                    \EndIf
                \EndFor
            \EndIf
            \State \textbf{Return} $act, obj$
        \end{algorithmic}
      \end{algorithm}          
\end{center}                    
%%%%%%%%%%%%%%%%%%%%%%%CostFunction%%%%%%%%%%%%%%%%%%%%%%%%%%

The \textit{CostFunction} (Alg.~\ref{CostFunction}) calculates the objective function~(\ref{ALIVE:eq:16}) for each node and its associated actions stored in $Fsb\_actions$, respectively. In the first step, it initiates the primary variables, \ie objective $obj$, action $act$, and auxiliary variable $obj^*$ (line 3). It then calls \textit{CalcMax} function to calculate the maximum values of the delay and cost based on the strategy discussed in Section~\ref{sec:ALIVE:Design:MILP} and stores them in $max\_delay$ and $max\_cost$, respectively (line 4). Afterward, it checks the node $j$ to find if it is a neighboring peer or a node of another type (VTS, CDN, or origin). Suppose it will be a peer; thus, Alg.~\ref{CostFunction} in a \textit{for} loop measures the minimum objective value via calling the \textit{CalcDly} function in each iteration. Finally, the determined objective and its associated action are stored in $act$ and $obj$, respectively (lines 5--12). Otherwise, the same strategy is utilized to calculate the minimum objective value via calling the \textit{CalcDly} and \textit{CalcCst} functions (lines 13--21). 
Assume $\rho$, $\kappa$, and $\gamma$ indicate the number of peer regions, number of live channels, and number of representations per channel. In the worst case, the time complexity of the proposed heuristic method would be $O(\rho ~.~ \kappa ~.~ \gamma)$ in each time slot. 

%%%%%%%%%%%%%%%%%%%%%%%%%%%%%%%%%%%%%%%%%%%%%%%%%%%%%%%%%%%%%%%%%%%%%%%%%%%%%%%%%%%%%%%%%%%%%%%%%%%%%%%%%%%%%%%%%%%%%%%%%%%%%%%%%%%%%%%%%%%%%%%%%%%%%%%%%%%%%%%%%%%%%%%%%%%
\subsection{ALIVE Performance Evaluation}
\label{sec:ALIVE:Performance Evaluation}
This section describes the employed SR models, explains the evaluation setup, metrics, and methods, and assesses the performance of \texttt{ALIVE} in four scenarios.
%%%%%%%%%%%%%%%%%%%%%%%%%%%%%%%%%%%%%%%%%%%%%%%%%%%%%%%%%%%%%%%%%%%%%%%%%%%%%%%%%%%%%%%%%%%%%%%%%%%%%%%%%%%%%%%%%%%%%%%%%%%%%%%%%%%%%%%%%%%%%%%%%%%%%%%%%%%%%%%%%%%%%%%%%%%
\subsubsection{Super-Resolution Models}
\label{sec:SRTR}
Super-resolution (SR) is a computational task for restoring high-frequency signals lost in the low-resolution image to obtain the high-resolution counterpart. The goal of such a process is minimizing visual degradation as much as possible while obtaining a high-resolution image~\cite{SRSurvey}. Although traditional SR methods like \textit{bilinear} or \textit{bicubic} interpolate pixel values directly, they always suffer from visual artifacts. In recent years, attention has shifted towards machine learning-based image SR~\cite{SRSurvey}. These models (\eg \textit{SRCNN}~\cite{SRCNN}) can learn the mapping from low-resolution image patches to high-resolution patches directly and outperform traditional SR approaches. Some learning-based SR methods focus on better visual quality, while others concentrate on making practical models by alleviating their complexity.

\begin{figure}[t]
    \centering
    \includegraphics[width=0.6\linewidth]{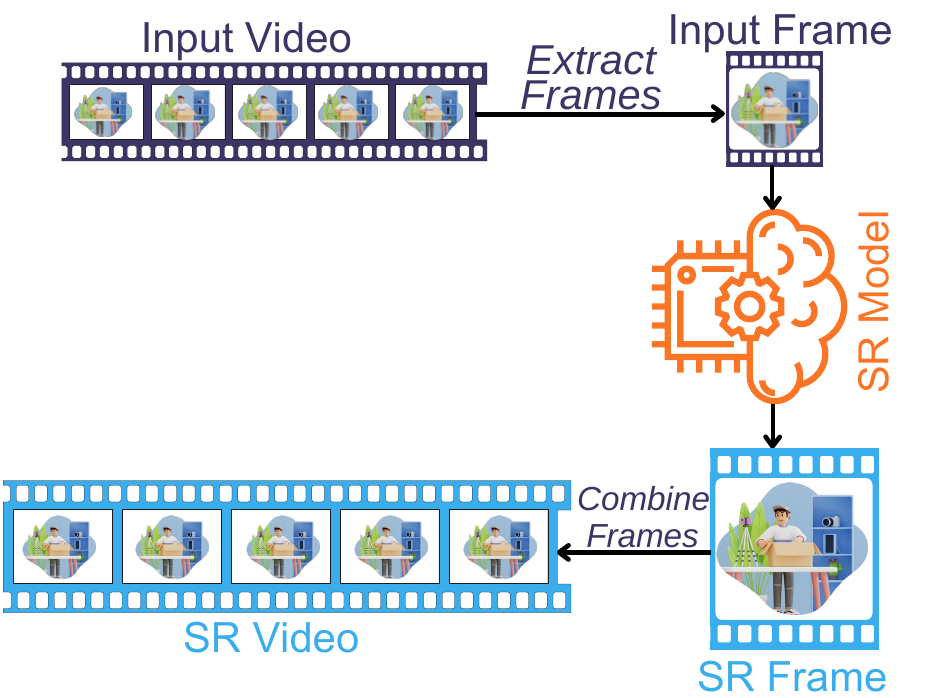}
    \caption{ Illustration of video super-resolution pipeline.}
    \vspace{.5cm}
    \label{fig:srvideo}
\end{figure}

\begin{figure}[!t]
    \centering
    \includegraphics[width=.8\linewidth]{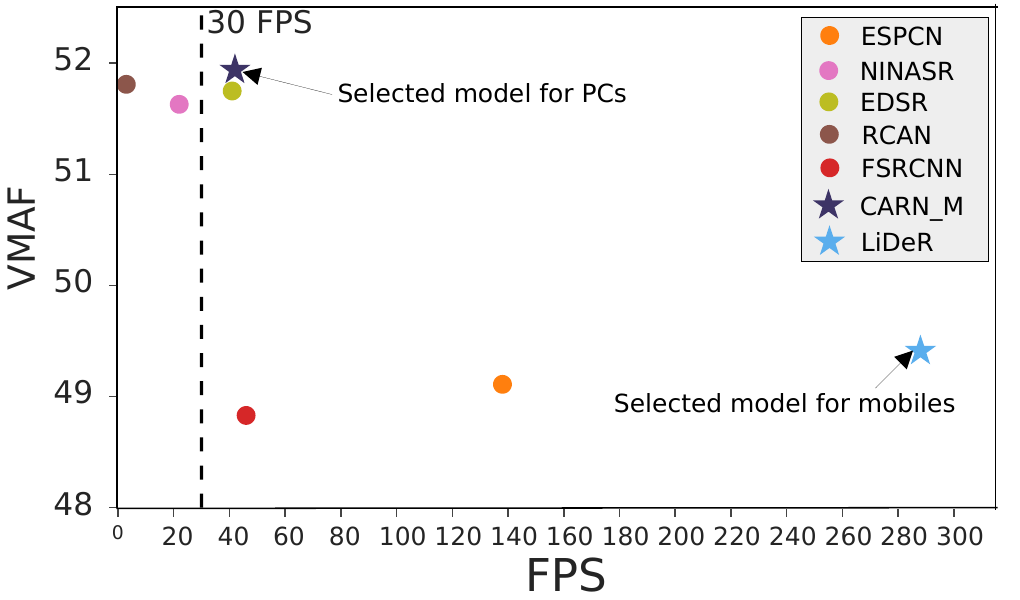}
    \caption{\small Performance of different video SR models in terms of VMAF and FPS. Results are obtained with input resolution $640\times360$ and target resolution $1280\times720$. $30$ FPS line is shown to indicate real-time execution.}
    \vspace{.5cm}
    \label{fig:srmodels}
\end{figure}

Inspired by the image SR, applying SR techniques for video sequences has also been popular recently. As illustrated in Fig.~\ref{fig:srvideo}, the video frames must be extracted for the video SR process. Then the SR model can be applied to each frame or a small group of frames individually. Finally, the frames upscaled by SR must be combined to produce the reconstructed video.
%%%%%%%%%%%%%Fig4%%%%%%%%%%%%%%%%%
%%%%%%%%%%%%%Fig4%%%%%%%%%%%%%%%%%
We run the most recent and popular video SR models, namely, \textit{ESPCN}~\cite{ESPCN}, \textit{LiDeR}~\cite{LiDeR}, \textit{FSRCNN}~\cite{FSRCNN}, \textit{EDSR}~\cite{EDSR}, \textit{CARN} and \textit{CARN-M}~\cite{CARN}, \textit{NINASR}~\cite{NINASR}, and \textit{RCAN}~\cite{RCAN} to choose our PSR models with maximum quality (measured by \textit{Video Multi-method Assessment Fusion} (VMAF)~\cite{VMAF}) and minimum imposed delay (measured by frame per seconds (FPS)). The VMAF-FPS graph for the PSR models is depicted in Fig.~\ref{fig:srmodels}. As shown in Fig.~\ref{fig:srmodels}, CARN-M provides a high VMAF score while still managing to run in real-time using customer-grade hardware. Moreover, LiDeR is a model that can run in real-time with a higher VMAF score compared to FSRCNN and ESPCN. Therefore, we use the following video PSR models in all scenarios: 
\begin{enumerate}[noitemsep]
\item \textbf{CARN for PC-type peers}: CARN~\cite{CARN} is proposed to use cascading residual connections in the neural network. CARN's neural network is designed based on the ResNet~\cite{ResNet} structure. Cascading connections are added on top of the residual connections to enable information propagation between layers. These changes help CARN to process information using fewer layers. CARN achieves competitive SR performance with state-of-the-art networks while running much faster. Moreover, a lightweight version of CARN, called \textit{CARN-M}, using a more efficient residual block structure, is designed to be applied in practical scenarios. We use it in our system on client PCs.
\item \textbf{LiDeR for mobile peers}: Due to the improvements in smartphone hardware, particularly supporting GPUs, complex deep neural networks (DNNs) models can be run on mobile devices. However, mobile battery consumption is still a constraint compared to standard desktop PCs. LiDeR~\cite{LiDeR} is designed considering the constraints of mobile devices based on a lightweight dense residual neural network to operate on decoded video sequences in real-time without hardware-specific optimization.
\end{enumerate}
%%%%%%%%%%%%%Fig5%%%%%%%%%%%%%%%%%
%%%%%%%%%%%%%Fig5%%%%%%%%%%%%%%%%%
%%%%%%%%%%%%%%%%%%%%%%%%%%%%%%%%%%%%%%%%%%%%%%%%%%%%%%%%%%%%%%%%%%%%%%%%%%%%%%%%%%%%%%%%%%%%%%%%%%%%%%%%%%%%%%%%%%%%%%%%%%%%%%%%%%%%%%%%%%%%%%%%%%%%%%%%%%%%%%%%%%%%%%%%%%%
\subsubsection{Evaluation Setup}  
\label{sec:ALIVE:setup}
\begin{figure}[!t]
	\centering
	\includegraphics[width=.7\columnwidth]{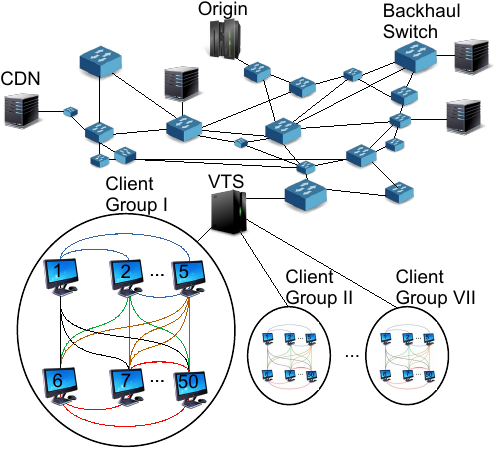}
	\caption{ ALIVE evaluation testbed.}
 \vspace{.5cm}
	\label{ALIVE:testbed-topo}
\end{figure}

We use a real backbone network topology called \textit{InternetMCI}~\cite{zoo} to assess the effectiveness of the \texttt{ALIVE} framework in a realistic large-scale setting. We instantiate a cloud-based large-scale testbed on the CloudLab~\cite{ricci2014introducing} environment. As illustrated in Fig~\ref{ALIVE:testbed-topo}, our testbed utilizes 375 elements, each of which runs Ubuntu 20.04 LTS inside Xen virtual machines. These components are: \textit{(i)} Seven groups of 50 open-source \textit{AStream}~\cite{juluri2015sara} DASH players (total 350 peers) working in a headless mode. As shown in Fig~\ref{ALIVE:testbed-topo}, each group includes five seeders and 45 leechers in a tree-mesh structure. \textit{SQUAD}~\cite{wang2016squad} and \textit{BOLA}~\cite{spiteri2016bola} are used in DASH clients as hybrid and buffer-based ABR algorithms, respectively. \textit{(ii)} Four CDN servers, containing a total cache size of 40\% of the video dataset, and an origin server, hosting all video sequences. The server machines run an Apache HTTP server and MongoDB. \textit{(iii)} 19 OpenFlow (OF) backbone switches, making the backbone topology with 45 backbone layer-2 links. \textit{(iv)} A VTS server with a partial cache size of only 5\% of the video sequences. Note that \texttt{ALIVE} is independent of the caching policy and is compatible with various caching strategies. 
%%%%%%%%%%%%%Fig6%%%%%%%%%%%%%%%%%
%%%%%%%%%%%%%Fig6%%%%%%%%%%%%%%%%%
For the sake of simplicity, we set the Least Recently Used (LRU) in all CDN and partial caches as the cache replacement policy. Furthermore, since peers can store a limited amount of content, we assume each peer can cache five segments of the video sequences at most. Python-based HTTP servers and the Python PuLP library with the CPLEX solver library are employed to implement the VTS and the proposed MILP model, respectively. Moreover, all modules of the VTS server, including heuristic algorithms, are implemented in Python to serve clients' requests for five live channels (\ie CH I--CH V). Each live channel plays a unique video~\cite{lederer2012dynamic} with 300 seconds duration and two-second segments in bitrate ladder (representation set) \{(0.089,320p), (0.262,480p), (0.791,720p), (2.4,1080p), (4.2,1080p)\} [Mbps, content resolution]. For simplicity, all scenarios assume that the channel probability is known in advance and sorted in descending order. The channel access probability is generated by a Zipf distribution~\cite{cherkasova2004analysis} with the skew parameter $\alpha=0.7$, \ie the probability of an incoming request for the $i^{th}$ channel in each peer group is given as $prob(i)=\frac{1/i^{\alpha}}{\sum_{j=1}^{K}1/j^{\alpha}}$, where $K=5$.

We use Docker image \textit{jrottenberg/ffmpeg}~\cite{ffmpeg} to measure the segment transcoding time on the VTS. Moreover, we execute the transcoding function via \textit{FFmpegKit}~\cite{ffmpeg-mobile} on an iPhone 11 (Apple A13 Bionic, iOS 15.3), a Xiaomi Mi11 (Snapdragon 888, Android 11), and a PC (Apple M1, MacOS 12.0.1) to measure the transcoding time on the heterogeneous P2P network. Power consumption is measured by device tools, \eg \textit{Android Energy Profiler} and \textit{Android Battery Manager}. Moreover, we utilize the \textit{CodeCarbon}~\cite{codecarbon} project to measure power consumption for the client PCs.
In the literature, the bandwidth value from an edge server to CDN servers is assumed higher than to the origin server~\cite{yi2017lavea, al2019multi}. Therefore, the Linux \textit{Wondershaper} tool~\cite{wondershaper} is employed to set 50 and 100 Mbps as a bottleneck bandwidth in different paths from the CDN and origin servers to the VTS, respectively. 

To emulate the mobile network conditions, we assume 250 peers start the experiments initially, and then every three seconds a new peer joins the sessions. Furthermore, we use a real \textit{4G network trace}~\cite{raca2018beyond} collected on bus rides for links between peers to edge servers in all experiments. The average bandwidth of this trace is approximately 3780 kbps with a standard deviation of 3190 kbps. Note that the VTS directs the first peer to the best CDN server (in terms of lowest latency), while other participating peers can be connected on both CDN and P2P links. In hybrid P2P-CDN systems like \texttt{ALIVE}, most costs are related to backhaul bandwidth and edge computational costs. Therefore, other service costs are negligible compared to the aforementioned costs. The computational and bandwidth costs (\ie $\Delta_{co}$ and $\Delta_{bw}$) are set to $0.029\$$ per CPU per hour and $0.12\$$ per GB, respectively~\cite{aws-calc}. The weighting parameter $\beta$ and the monitoring interval are set to 0.5 and one second in all experiments, respectively. 
%%%%%%%%%%%%%%%%%%%%%%%%%%%%%%%%%%%%%%%%%%%%%%%%%%%%%%%%%%%%%%%%%%%%%%%%%%%%%%%%%%%%%%%%%%%%%%%%%%%%%%%%%%%%%%%%%%%%%%%%%%%%%%%%%%%%%%%%%%%%%%%%%%%%%%%%%%%%%%%%%%%%%%%%%%%
\subsubsection{Evaluation Methods, Scenarios, and Metrics}
\label{sec:ALIVE:Methods}
\textit{\textbf{Methods for Comparison: }}In the \texttt{ALIVE} performance evaluation, we first investigate the SR models' behavior. Next, we analyze the performance of the proposed optimization model and heuristic algorithms. Finally, we refer to our proposed heuristic methods and present practical testbed results to evaluate \texttt{ALIVE}'s effectiveness compared to the following \textit{baseline} schemes.
%%%%%%
\begin{enumerate}[noitemsep]
\item \textbf{Non Hybrid (NOH)}: The NOH system is regular CDN-based streaming with no P2P support. 
\item \textbf{Simple Edge-enabled Hybrid (SEH)}: The SEH system employs a simple VTS server without caching and transcoding capabilities. In this system, peers only can be served via one of the actions 1, 7, or 8 (Fig.~\ref{ALIVE:actionTree}).
\item \textbf{Non Transcoding-enabled Hybrid (NTH)}: Like in most works, there is no transcoding capability in this approach. In an NTH-based system, peers only can be served through one of the actions 1, 4, 7, or 8 (Fig.~\ref{ALIVE:actionTree}). 
\item \textbf{Edge Caching/Transcoding Hybrid (ECT)}: ECT does not include transcoding at the peer side, and requests can be served via all actions except actions 2 and 3. 
\item \textbf{Non SR-enabled Hybrid (NSH)}: In the NSH-based system, there is no SR feature on the peer side and requests can be served via all actions except action 2. 
\end{enumerate}
%%%%%%
Note that we use our testbed with similar settings and topology in all systems for fair and robust comparisons. Moreover, all aforementioned baselines execute a simplified version of the proposed GBA algorithm to answer the questions mentioned in Section~\ref{sec:ALIVE:Design:PROBLEMSTATEMENT}. 
\rf{We use the method discussed in Section~\ref{sec:SFC:Performance Evaluation} to calculate the results reported in the next section.}

\textit{\textbf{Description of Scenarios: }}We compare the performance of the \texttt{ALIVE} methods with the aforementioned baseline approaches in four scenarios. 
\begin{enumerate}[noitemsep]
\item \textbf{Scenario I}: This scenario studies the impact of changing SR models on the \textit{perceptual quality} metrics. Moreover, it investigates the \textit{latency}, \textit{quality distortion}, and \textit{energy overheads} of running computational-intensive tasks, \ie transcoding and super-resolution, on peers.
\item \textbf{Scenario II}: In this scenario, we provide a performance comparison between the proposed optimization model and the heuristic algorithm by conducting several experiments. These experiments study the impact of changing key parameters like the \textit{bitrate ladder size}, the number of \textit{peers' requests}, and \textit{live channels} on the \textit{optimization model and heuristic algorithms performance} metrics.
\item \textbf{Scenario III}: In this scenario, we run a series of experiments on the designed testbed to assess the performance of the explained systems in terms of common \textit{QoE-related} parameters compared to other approaches. 
\item \textbf{Scenario IV}: As a final scenario, the behavior of different frameworks in terms of \textit{network utilization/cost} metrics will be analyzed on the testbed.
\end{enumerate}

%%%%%%%%%%%%
\textit{\textbf{Evaluation Metrics: }}Perceptual quality, algorithm performance, QoE-related, network utilization/cost metrics are defined as follows:
\begin{enumerate}[noitemsep]
\item \textbf{Perceptual quality metrics}:
\begin{itemize}[noitemsep]
\item \textbf{VMAF:} Video Multi-method Assessment Fusion~\cite{VMAF} is a learning-based objective quality metric (with a range from 0 to 100) that predicts the perceived video quality.
\end{itemize}
\item \textbf{Algorithm performance metrics}:
\begin{itemize}[noitemsep]
\item \textbf{ETV}: Execution Time Values (in seconds) for the different \texttt{ALIVE} schemes.
\item \textbf{NOV}: Normalized Objective Values (\ie of the objective function Eq.~\ref{ALIVE:eq:16}) for the \texttt{ALIVE} schemes (in \%).
\end{itemize}
\item \textbf{Common QoE metrics}:
\begin{itemize}[noitemsep]
\item \textbf{ASB}: Average Segment Bitrate (in Mbps) of all downloaded segments.
\item \textbf{AQS}: Average Quality Switches, \ie the number of segments whose bitrate levels change compared to the previous ones.
\item \textbf{ASD}: Average Stall Duration, \ie the average of total video freeze times (in seconds) in all clients.
\item \textbf{ANS}: Average Number of Stalls, \ie the average number of rebuffering events.
\item \textbf{ASL}: Average Serving Latency (in seconds) for serving all clients, including transmission latency plus computational latency. 
\item \textbf{APQ}: Average Perceived QoE (with a range from 0 to 5), calculated by ITU-T Rec. P.1203 mode 0~\cite{p1203}.
\end{itemize}
\item \textbf{Network utilization/cost metrics}:
\begin{itemize}[noitemsep]
\item \textbf{CHR}: Cache Hit Ratio (in \%), defined as the fraction of segments fetched from the CDN, VTS, or peers.
\item \textbf{ETR}: Edge Transcoding Ratio (in \%), \ie the fraction of segments transcoded at the VTS.
\item \textbf{PTSR}: Peer SR and TR Ratio (in \%), \ie the fraction of segments constructed by TR or SR at the peers.
\item \textbf{BTL}: Backhaul Traffic Load (in Gbps), \ie the volume of segments downloaded from the origin server.
\item \textbf{EEC}: Edge Energy Consumption (in kWh) for running ETR.  
\item \textbf{NCV}: Network Cost Values (in \$), including computational and bandwidth costs.
\end{itemize}
\end{enumerate}
%%%%%%
We discuss details of the scenarios and \texttt{ALIVE} performance evaluation in the next section, where each experiment is executed 20 times to ensure accuracy, and the average values and standard deviations are reported in the experimental results. 
%%%%%%%%%%%%%%%%%%%%%%%%%%%%%%%%%%%%%%%%%%%%%%%%%%%%%%%%%%%%%%%%%%%%%%%%%%%%%%%%%%%%%%%%%%%%%%%%%%%%%%%%%%%%%%%%%%%%%%%%%%%%%%%%%%%%%%%%%%%%%%%%%%%%%%%%%%%%%%%%%%%%%%%%%%%
\subsubsection{Evaluation Results}
\label{sec:ALIVE:Evaluation Results}
%%%%%%%%%%%%%%%%%%%%%%%%%%%%%%%%%%%%%%%%%%%%%%%%%%%%%%%%%%%%%%%%%%%%%%%%%%%%%%%%%%%%%%%%%%%%%%%%%%%%%%%%%%%%%%%%%%%%%%%%%%%%%%%%%%%%%%%%%%%%%%%%%%%%%%%%%%%%%%%%%%%%%%%%%%%%%%%%%%%%%%%%%%%%%%%%%%%%%%%%%%%%%%%%%%%%%%
\textbf{Scenario I: }Running transcoding or super-resolution on peers (\ie PTR or PSR) must be sufficiently fast, not significantly imposing an extra delay to the \texttt{ALIVE} system, and not considerably consuming peers' battery; otherwise, the clients' requests may use other actions that make the network and edge server congested. Hence, in the first scenario, we conduct experiments to investigate the latency, quality distortion,  and energy overheads of running transcoding and super-resolution tasks on peers. 

\begin{figure}[t]
    \centering
    \includegraphics[width=.9\linewidth]{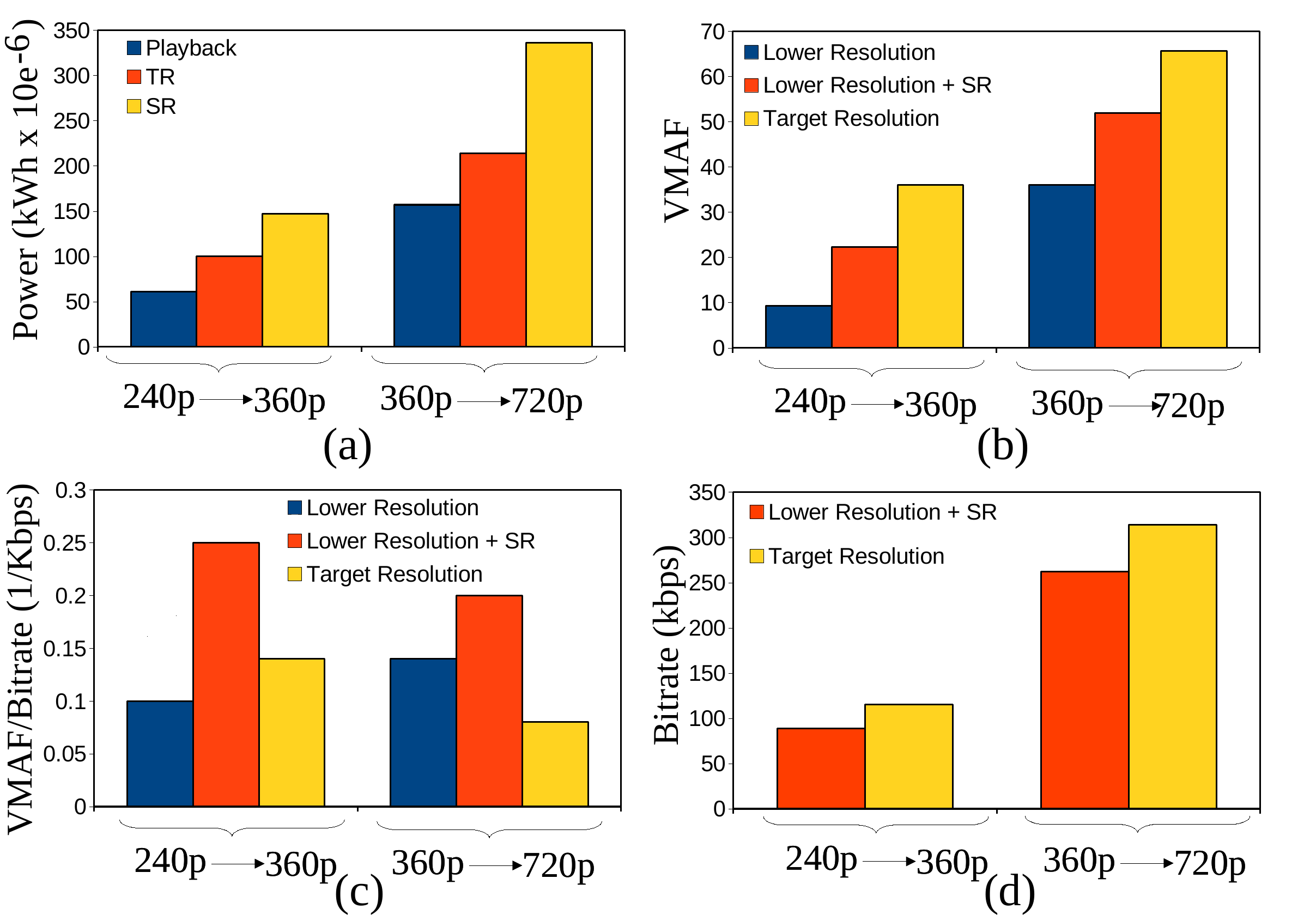}
    \caption{Performance comparison of applying peer's SR on a lower representation in terms of (a) power consumption, (b) VMAF score, (c) VMAF/Bitrate, and (d) bandwidth saving.}
    \vspace{.5cm}
    \label{fig:SR-Results}
\end{figure}

The first experiment of this scenario measures the client's power consumption when the client utilizes the PSR function to jump to a higher representation level in the bitrate ladder from the lower received representation~(Fig~\ref{fig:SR-Results}(a)).
It is noteworthy that \texttt{ALIVE} only operates the SR on the targeted client as the last step of playing out the video. This is because the output of the SR model is a decoded representation, and then \texttt{ALIVE} directly displays it. Therefore, the SR process is accomplished without inducing any extra delay, except the SR processing delay, for preparing the requested representation. However, applying SR at the edge or other network layers requires a re-encoding phase before transmitting it to the client, consequently imposing an extra encoding delay on the system. The results of this experiment are obtained for two upscaling operations on the client PC peers by the discussed CARN-M SR model. The used representations in this experiments are (89~kbps - 240p) $\rightarrow$ (262~kbps - 360p) and (262~kbps - 360p) $\rightarrow$ (791~kbps - 720p). The power consumption of peers, when they \textit{(i)} play the low representation video, \textit{(ii)} transcode the video from a higher representation, or \textit{(iii)} apply SR from a low resolution, is plotted in Fig.~\ref{fig:SR-Results}(a). As obvious from the results, the SR application consumes much more battery than playing out the lower representation directly (about 52\%), while this rise is about 38\% when transcoding the requested representation from the highest representation (\ie (4.2~Mbps - 1080p) $\rightarrow$ (262~kbps - 360p) or (791~kbps - 720p)). 
%%%%%%%%%%%%%Fig7%%%%%%%%%%%%%%%%%
%%%%%%%%%%%%%Fig7%%%%%%%%%%%%%%%%%

We run further experiments to calculate the power consumption (\ie power values are given as kWh$\times10^{-3}$ and percentage of battery usage) for a five-minute video (\ie 150 segments) when different actions are executed simultaneously on one peer (\eg play video I, transcode video II, upscaled video III, simultaneously). As reported in Table~\ref{tab:powerconsumption}, the SR operation continues to use the most power, followed by the transcoding (TR) function, while battery consumption of playing back a video (\ie PLY) is not comparable with them. Moreover, a combination of PLY, TR, and SR tasks does not put a significant burden on the peers' batteries (\ie 4.04\% and 7.74\%) compared to the energy used to play, transcode, or upscale a video individually. Note that the reported values for each segment with two seconds durations are negligible (\ie $\frac{measured~Power~or~Battery}{segment~numbers}$).
%%%%%Power consumption table%%%%%
\begin{table}[!t]
\caption{Power consumption and battery usage of different approaches for a 5-min. video on peers.}
\resizebox{0.97\linewidth}{!}{%
\centering
\label{tab:powerconsumption}
\scriptsize
        \begin{tabular}{l|cc|cc}\toprule
        \multirow{2}{*}{\textbf{Method}} &\multicolumn{2}{c|}{\textbf{$240p\rightarrow360p$}} &\multicolumn{2}{c}{\textbf{$360p\rightarrow720p$}} \\\cmidrule{2-5}
        &\textbf{Power (kWh$\times10^{-3}$)} &\textbf{Battery} &\textbf{Power (kWh$\times10^{-3}$)} &\textbf{Battery} \\\midrule
        PLY &119 &0.39\% &152 &0.49\% \\
        TR &130 &0.42\% &352 &1.14\% \\
        SR &955 &3.09\% &1,862 &6.03\% \\
        TR + PLY &237 &0.77\% &479 &1.55\% \\
        SR + PLY &1,020 &3.30\% &1,915 &6.20\% \\
        SR + TR &1,043 &3.71\% &2,036 &7.25\% \\
        SR + TR + PLY &1,084 &4.04\% &2,079 &7.74\% \\
        \bottomrule
        \end{tabular}}
\vspace{.5cm}
\end{table}
%%%%%%Power consumption table%%%%%%

The second experiment examines the quality of the video upscaled by SR in terms of the VMAF metric. As illustrated in Fig.~\ref{fig:SR-Results}(b), by applying the peer's SR, we significantly increase the VMAF score of the displayed video compared to the lower representation (by at least 41\%). Although the requested target representation still outperforms the VMAF score obtained by the SR (Fig.~\ref{fig:SR-Results}(b)), applying SR provides the best VMAF score per kbps of data delivered to the client (Fig.~\ref{fig:SR-Results}(c)). We repeat the first and second experiments for different representations in the bitrate ladder to find a sweet spot in the trade-off between quality improvement and power consumption. As depicted in Fig's.~\ref{fig:SR-Results} (a) and (b), we discover that the sweet spot is to go up one step in the bitrate ladder, as increasing resolution too much would result in significant distortion. Therefore, upscaling the targeted representation from the closest one would result in a pleasant user's QoE (in terms of the VMAF score) and less power consumption. 

In the third experiment, we investigate a solution to discover the exact bandwidth saving that can be achieved by applying SR. Imagine we use the $Rep_1 = $~(89 kbps - 240p) representation to build $Rep_2 = $~(262 kbps - 360p) by SR. We first apply the SR on $Rep_1$ to increase the resolution to $Rep_2$. We then calculate the VMAF score of $Rep_2$. In the next step, we encode the input source video with a requested resolution (\ie 360p) in different bitrates (starting from the lower bitrate, \ie 89 kbps, and gradually increasing it) to find the bitrate in which the VMAF score matches $Rep_2$. We have applied the same approach to all representations in the bitrate ladder and found that applying SR can save around $24.53\%$ bandwidth on average while maintaining the same visual quality (Fig.~\ref{fig:SR-Results}(d)).
%%%%% Transcoding Time & VMAF Table%%%%%
\begin{table}[t]
\centering
\caption{Transcoding time and VMAF scores for 3-min. video on different peer types.}
\label{tab:transcodingTimeVMAF}
% \scriptsize
    \begin{tabular}{cc|cc|cc}\toprule
    \multirow{2}{*}{\textbf{Input BR}} &\multirow{2}{*}{\textbf{Target BR}} &\multicolumn{2}{c|}{\textbf{Client PC}} &\multicolumn{2}{c}{\textbf{Client Mobile}} \\\cmidrule{3-6}
    & &\textbf{Time (s)} &\textbf{VMAF} &\textbf{Time (s)} &\textbf{VMAF} \\\midrule
    4219k &89k &4.31 &15.38 &15.98 &13.75 \\
    4219k &262k &5.33 &44.61 &18.32 &42.13 \\
    4219k &791k &11.74 &76.21 &39.28 &73.14 \\
    4219k &2484k &20.44 &93.33 &74.91 &91.53 \\\midrule
    2484k &89k &3.80 &14.35 &16.55 &13.01 \\
    2484k &262k &4.83 &42.27 &18.82 &40.02 \\
    2484k &791k &11.36 &71.56 &39.76 &69.06 \\\midrule
    791k &89k &2.05 &12.21 &10.43 &11.24 \\
    791k &262k &3.35 &36.33 &14.81 &34.76 \\\midrule
    262k &89k &1.28 &11.01 &5.85 &10.32 \\
    \bottomrule
    \end{tabular}
    \vspace{.5cm}
\end{table}
%%%%% Transcoding Time & VMAF Table%%%%%

In the final experiment of this scenario, we measure the peers' transcoding times and VMAF scores for a 3-min video. We note that the VMAF score differences in mobile and PC peers are due to using different transcoding presets, \ie \textit{veryfast} and \textit{fast}, respectively. It is worth mentioning that transcoding needs decoding a video into raw frames and then re-encoding those frames into new frames. Thus, transcoding time at the peers is equal to the encoding time due to leveraging the video processing that is already being done to capture or view video. Therefore, peer transcoding by the PTR function can already shorten one decoding time. As shown in Table~\ref{tab:transcodingTimeVMAF}, running transcoding for the whole video takes 1.28--20.44 seconds (0.014--0.22 seconds per segment) and 5.85--74.91 seconds (0.065--0.8 seconds per segment) on PC and mobile peers, respectively. Moreover, the results in the VMAF column indicate that using the closest input bitrate to the target bitrate achieves better VMAF scores. 
%%%%%%%%%%%%%%%%%%%%%%%%%%%%%%%%%%%%%%%%%%%%%%%%%%%%%%%%%%%%%%%%%%%%%%%%%%%%%%%%%%%%%%%%%%%%%%%%%%%%%%%%%%%%%%%%%%%%%%%%%%%%%%%%%%%%%%%%%%%%%%%%%%%%%%%%%%%%%%%%%%%%%%%%%%%

\begin{figure}[!t]
    \centering
    \includegraphics[width=.9\linewidth]{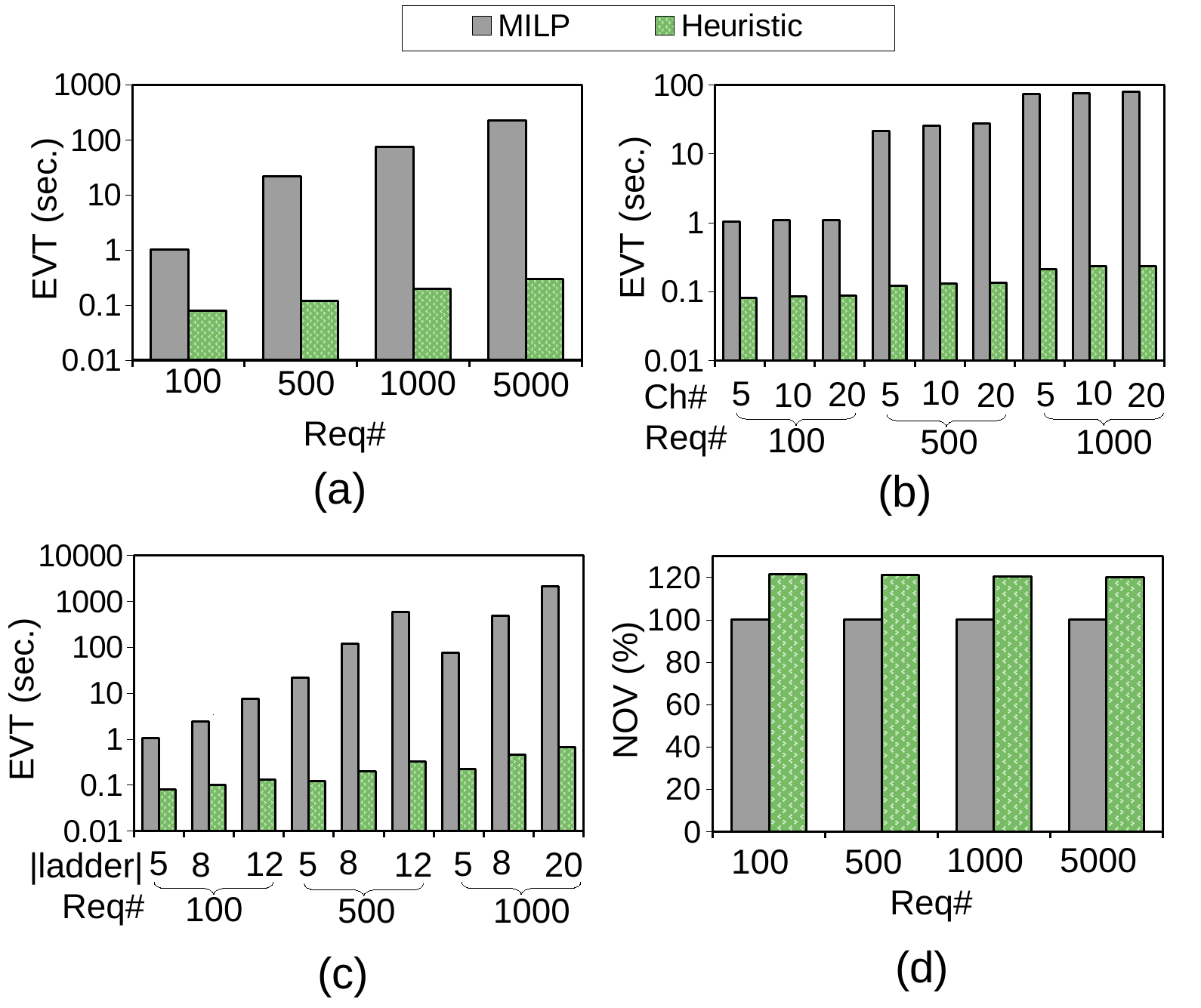}
    \caption{\small Impact of (a) increasing the number of arriving requests ($Req\#$), (b) increasing the number of live channels ($Ch\#$), and (c) changing the bitrate ladder, on the performance of the MILP model and the proposed heuristic method in terms of ETV and (d) NOV.}
    \vspace{.5cm}
    \label{fig:alg-perf}
\end{figure}

\textbf{Scenario II: }
In the second scenario, we compare the performance of the proposed MILP optimization model with the GBA-based heuristic method in terms of the algorithm performance metrics (Fig.~\ref{fig:alg-perf}). 

In the first experiment of this scenario, we measure the ETV metric by \textit{(i)} increasing the number of peers' requests ($Req\#$), \textit{(ii)} increasing the number of live channels ($Ch\#$), and \textit{(iii)} increasing the number of representations in the bitrate ladder ($|ladder|$) to assess \texttt{ALIVE}'s GBA scalability and practicality. Fig.~\ref{fig:alg-perf}(a) highlights the ETV comparison of the \texttt{ALIVE} GBA scheme with the MILP model scheme when the VTS must handle different numbers of simultaneous requests (\ie $Req\#$~= 100, 500, 1000, and 5000) received from peers. It should be noted that all experimental settings, including the number of other components, are the same as discussed in Section~\ref{sec:setup} for this experiment. The results show that, by increasing $Req\#$, the ETV value of the MILP model increases exponentially, while the proposed lightweight GBA achieves much less ETV. 
%%%%%%%%%%%%%Fig8%%%%%%%%%%%%%%%%%
%%%%%%%%%%%%%Fig8%%%%%%%%%%%%%%%%%

In the second and fourth experiments, we investigate the impact of increasing the number of live channels (\ie $Ch\#$) and the bitrate ladder on the values of ETV, respectively. As depicted in Fig.~\ref{fig:alg-perf}(b), increasing the number of live channels (\ie $Ch\#$~=~5, 10, 20) does not have a significant effect on the ETV values and slightly increases it in both the MILP and heuristic methods. In the next experiment, we examine the behavior of the proposed schemes when using a different number of representations in the bitrate ladders (\ie $|ladder|$~=~5, 8, 12). 
As plotted in Fig.~\ref{fig:alg-perf}(c), replacing a bitrate ladder with a ladder with more representations increases the ETV values noticeably. In fact, having a bitrate ladder with more representations boosts the problem search space, consequently causing schemes to spend much more time deciding on the best nodes and actions. It is evident in all experiments that the proposed GBA-based heuristic scheme outperforms the MILP model significantly in terms of the ETV, showing minimal growth with changing the above-mentioned parameters (see Fig's.~\ref{fig:alg-perf}(a--c)).

In the last experiment of this scenario, we measure the normalized objective function value (\ie NOV) metric by varying the number of requests ($Req\#$). As displayed in Fig.~\ref{fig:alg-perf}(d), the MILP model surpasses the GBA method regarding the NOV value for all $Req\#$. We note that having more live channels and/or replacing the bitrate ladder result in similar trends for the NOV. As described earlier, the MILP model is executed for all peers' requests, while the GBA method is run based on regions' queues in a time-slotted fashion. The results in this experiment confirm that the GBA-based heuristic solution is a sub-optimal solution, but as shown in Fig.~\ref{fig:alg-perf}(d), they are close to the MILP model. 
%%%%%%%%%%%%%%%%%%%%%%%%%%%%%%%%%%%%%%%%%%%%%%%%%%%%%%%%%%%%%%%%%%%%%%%%%%%%%%%%%%%%%%%%%%%%%%%%%%%%%%%%%%%%%%%%%%%%%%%%%%%%%%%%%%%%%%%%%%%%%%%%%%%%%%%%%%%%%%%%%%%%%%%%%%%

\textbf{Scenario III: }
This scenario investigates the performance of the proposed \texttt{ALIVE} GBA method on the introduced testbed and compares the common QoE metrics with baseline methods. This scenario does not use the \texttt{ALIVE} MILP method due to its high time complexity and impracticality in large-scale environments. 

\begin{figure}[!t]
    \centering
    \includegraphics[width=.9\linewidth]{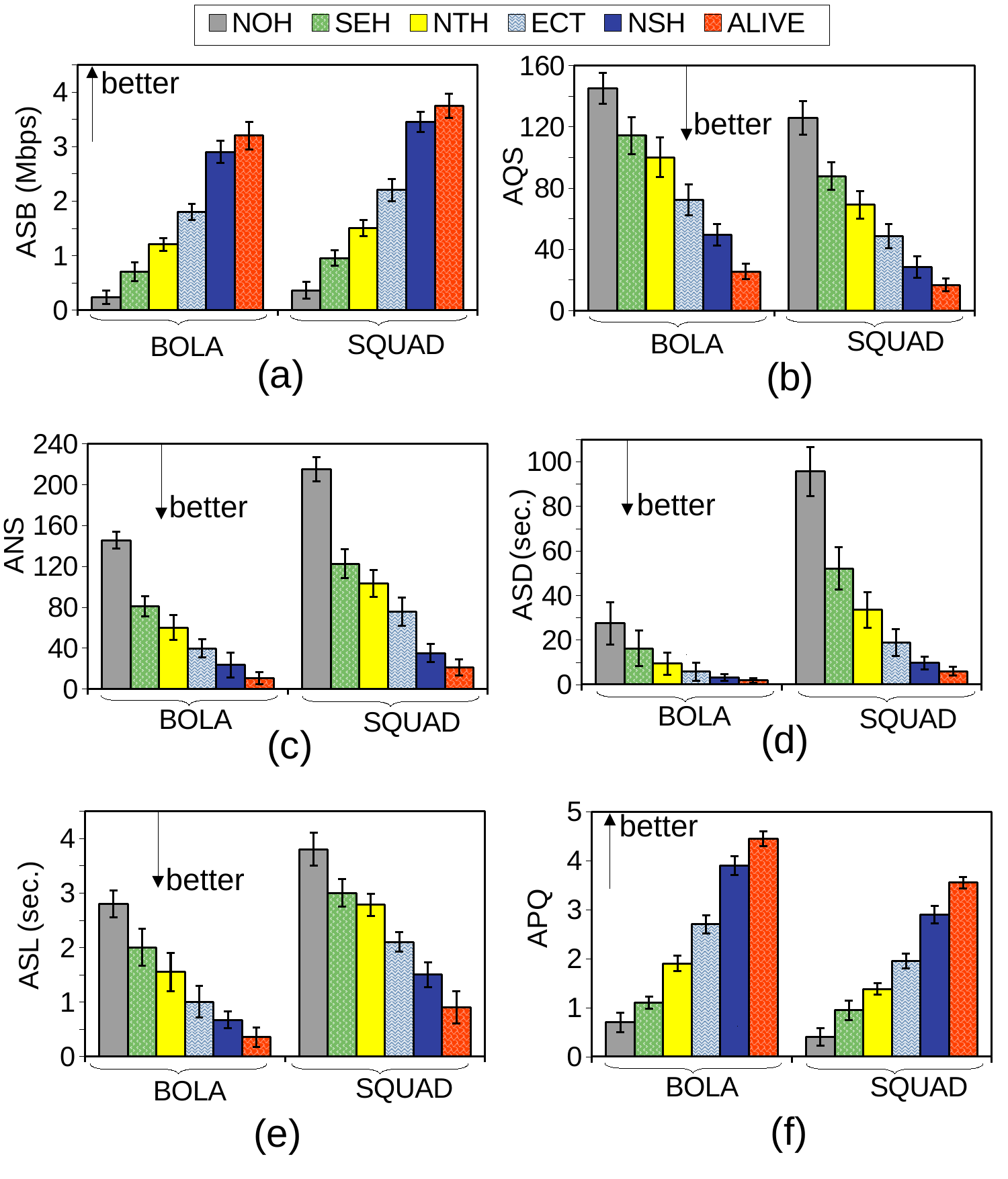}
    \caption{\small Performance of the ALIVE system compared with other baseline methods in terms of (a) ASB, (b) AQS, (c) ANS, (d) ASD, (e) ASL, and (e) APQ for 350 clients running BOLA and SQUAD ABR algorithms.}
    \vspace{.5cm}
    \label{fig:QoE-res}
\end{figure}

As illustrated in Fig.~\ref{fig:QoE-res}(a), \texttt{ALIVE} serves HAS clients with higher ASB in both BOLA and SQUAD-based systems. This is because \texttt{ALIVE} uses all computational, caching, and bandwidth resources in a hybrid P2P-CDN network. Similar to ASB, Fig.~\ref{fig:QoE-res}(b) demonstrates that \texttt{ALIVE} decreases the AQS for both ABR algorithms by at least 34.5\%. This is again due to using all provided resources in the \texttt{ALIVE} system for fetching, running TR, or operating SR compared to other methods.
%%%%%%%%%%%%%Fig9%%%%%%%%%%%%%%%%%
%%%%%%%%%%%%%Fig9%%%%%%%%%%%%%%%%%

The effectiveness of different systems in terms of ANS and ASD, as essential factors of users' QoE, are shown in Fig's.~\ref{fig:QoE-res}(c) and (d), respectively. It is worth noting that buffer-based ABR algorithms (\eg BOLA) are focused on the buffer status for adaptation decisions; thus, they work better than SQUAD-based players in terms of the ANS and ASD metrics. Utilizing the GBA algorithm as a decision-maker for selecting the best node and best action with minimum delay, enables the \texttt{ALIVE} system to serve HAS requests with fewer ANS and ASD values. To be precise, \texttt{ALIVE} decreases ANS by at least 40\% and shortens ASD for BOLA by at least 43\% compared to other systems. Results for the SQUAD-based systems follow the same trend, where ANS and ASD values obtained by the \texttt{ALIVE} system outperform baseline approaches by at least 42\% and 45\%, respectively. Moreover, we measure the ASL, which is considered an essential part of the end-to-end delay. As illustrated in Fig.~\ref{fig:QoE-res}(e), the \texttt{ALIVE} system demonstrates superior performance for both ABR algorithms in terms of the ASL metric. In fact, downloading requested segments via the most appropriate actions with minimum serving time (computation plus transmission) and from the best node with high available resources (\ie bandwidth, computation) significantly shortens both ASD and ASL values, where this improvement for ASL is at least 40\% compared to other methods. It is noteworthy that since SQUAD-based players download higher-quality segments and their ASD values are longer than those of BOLA-based players (Fig.\ref{fig:QoE-res}(a) and (d)), their ASL values are higher than those of BOLA-based players.
%It is noteworthy that due to more high-quality segments with longer ASD values downloaded by SQUAD-based players (Fig.\ref{fig:QoE-res}(a)), their ASL values are higher than those of BOLA-based players. 

Although enhancing each QoE parameter individually (\eg ASB, AQS, ANS, ASD, and ASL) can generally improve the users' satisfaction, having a comprehensive standard model is essential to analyze the system behavior in terms of users' QoE improvement. Therefore, we use a standard QoE model~\cite{p1203} to evaluate \texttt{ALIVE}'s performance in terms of the APQ metric. As plotted in Fig.~\ref{fig:QoE-res}(f), the value of the APQ metric is between zero and five, where zero is the worst-perceived quality and five is the best-perceived quality. It is worth mentioning that stall duration (measured by the ASD metric)
significantly affects APQ compared to other common QoE parameters. Hence, BOLA-based players with better ASD values achieve better APQ compared to SQUAD-based players. As shown in Fig.~\ref{fig:QoE-res}(f), the \texttt{ALIVE} system presents superior performance for both ABR algorithms. In fact, \texttt{ALIVE} enhances APQ by at least 22\% compared to other baseline techniques. Improving the ASB, AQS, ANS, and particularly shortening the ASD values, is the main reason for outperforming the NSH system as the second-best scheme in terms of the APQ metric.
%%%%%%%%%%%%%%%%%%%%%%%%%%%%%%%%%%%%%%%%%%%%%%%%%%%%%%%%%%%%%%%%%%%%%%%%%%%%%%%%%%%%%%%%%%%%%%%%%%%%%%%%%%%%%%%%%%%%%%%%%%%%%%%%%%%%%%%%%%%%%%%%%%%%%%%%%%%%%%%%%%%%%%%%%%%

\textbf{Scenario IV: }In the fourth scenario, we analyze the performance of the proposed \texttt{ALIVE} GBA scheme on the testbed regarding the network utilization/cost utilization metrics and compare the obtained results with baseline methods. Similar to scenario III, we do not utilize the \texttt{ALIVE} MILP model in this scenario.

\begin{figure}[!t]
    \centering
    \includegraphics[width=.9\linewidth]{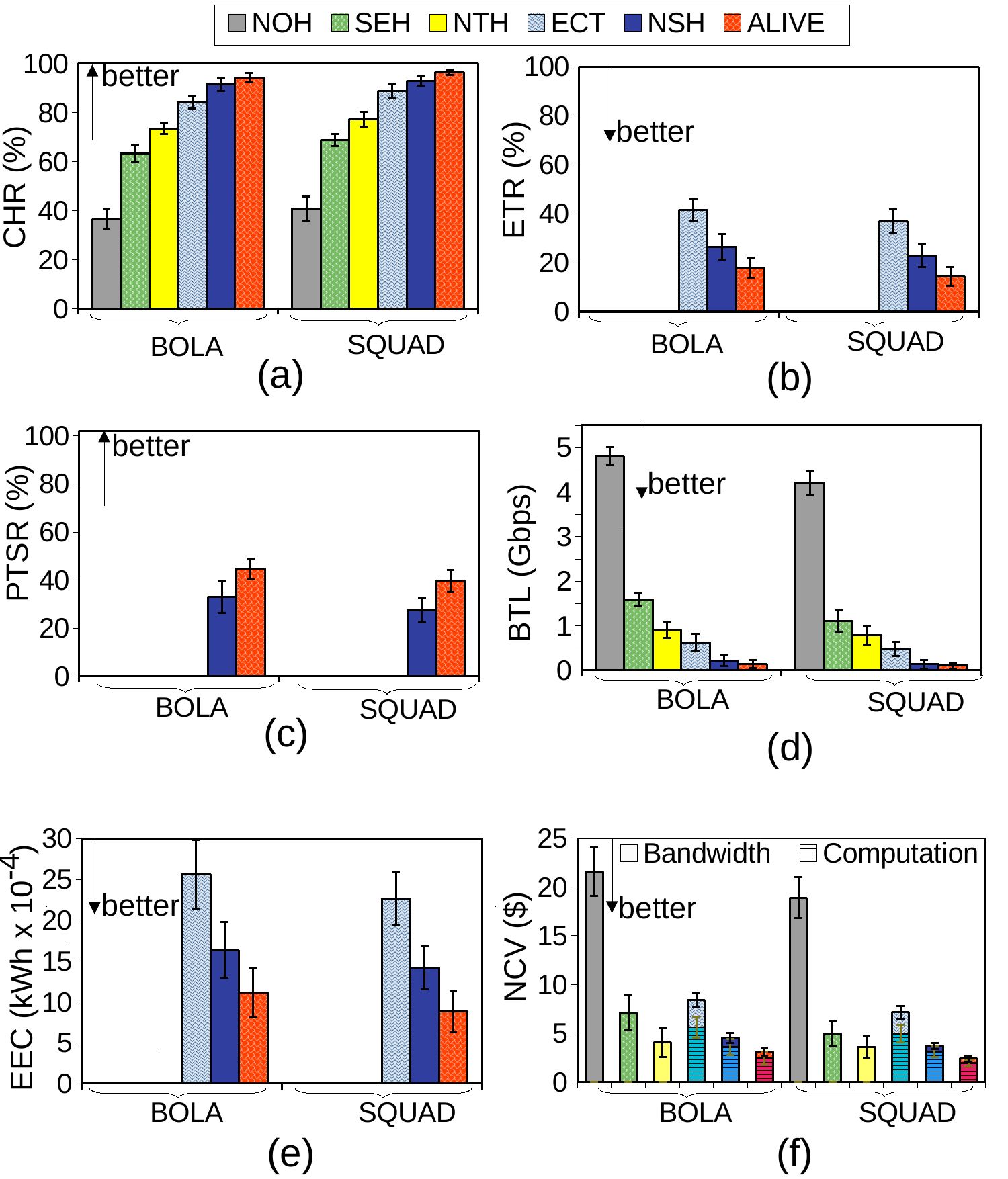}
    \caption{\small Performance of the \texttt{ALIVE} system compared with other baseline methods in terms of (a) CHR, (b) ETR, (c) PTSR, (d) BTL, (e) EEC, and (e) NCV for 350 clients running BOLA and SQUAD ABR algorithms.}
    \vspace{.5cm}
    \label{fig:net-res}
\end{figure}

The performance analysis of the \texttt{ALIVE} scheme concerning the CHR and ETR metrics are depicted in Fig's.~\ref{fig:net-res}(a) and (b), respectively. It should be noted that different frameworks encounter cache miss events under various conditions, which are listed as follows:
\begin{itemize}[noitemsep]
\item \textbf{Condition I: }The requested representations are not held in CDN servers (\eg for NOH). 
\item \textbf{Condition II: }The requested representations do not exist in CDN servers or peers (\eg for SEH).
\item \textbf{Condition III: }The requested representations are not in CDN servers, VTS, or neighboring peers (\eg for NTH).
\item \textbf{Condition IV: }Available bandwidth is insufficient to fetch the requested representation from the selected server (\ie CDN, or VTS) or peer (\eg for NOH, SEH, or NTH).
\item \textbf{Condition V: }The requested or higher representations do not exist in CDN servers; the VTS server does not hold the requested or higher representations; the VTS's resources are insufficient for serving requests via the ETR function. In addition, the requested representations are not cached in any adjacent peer. (\eg for ECT)  
\item \textbf{Condition VI: }Peers do not have the candidate representations (\ie higher or lower) for serving requests through the PTR or PSR functions; peers' resources (\ie computational or bandwidth) are inadequate for serving requests by the PTR or PSR operations (\eg for NSH and \texttt{ALIVE}).
\end{itemize}

The CHR metric plotted in Fig.~\ref{fig:net-res}(a) shows that \texttt{ALIVE} outperforms the NOH, SEH, NTH, and ECT approaches by at least 12\% due to its ability to fetch requested, higher, or lower representations in a hybrid system and then to reconstruct the requested representations via PRT, PSR, or ETR functions. Moreover, \texttt{ALIVE} slightly improves over the NSH system regarding the CHR slightly (by 4\%) due to having an extra action 3 for fetching the lower representation from adjacent peers. As shown in Fig.~\ref{fig:net-res}(b), \texttt{ALIVE} improves the ETR metric of the ECT system by 53\% due to using idle computational resources of the P2P network for running the PTR and PSR functions. Note that since NOH, SEH, and NTH are not transcoding-enabled systems, the ETR metrics does not apply. Moreover, since \texttt{ALIVE} employs the PSR function besides PTR, it improves over the NSH system by 32\% in terms of ETR.
%%%%%%%%%%%%%Fig10%%%%%%%%%%%%%%%%%
%%%%%%%%%%%%%Fig10%%%%%%%%%%%%%%%%%

The PTSR metric in Fig.~\ref{fig:net-res}(c) indicates that \texttt{ALIVE} increases the usage of available computational resources of the P2P network (by 36\%) compared to the NSH framework, which is the only baseline approach that employs peers' computations for running PTR. This is because \texttt{ALIVE} operates super-resolution by the PSR function besides direct fetching and transcoding for serving requests. However, improving the cache hit ratio (\ie CHR) and constructing requested representations at the edge or P2P network (\ie by ETR, PTR, or PSR) directly affects the BTL metric since the system downloads fewer segments from the origin server; \texttt{ALIVE} thus outperforms other systems and significantly reduces fetching representations from the origin server, by at least 24\%, compared to other frameworks (Fig.~\ref{fig:net-res}(d)).

The next experiment analyzes the EEC metric to assess edge server energy consumption for producing requested representations in different systems. As is evident from Fig.~\ref{fig:net-res}(e), \texttt{ALIVE} has better EEC results than the NSH and ECT methods (the only transcoding-enabled baseline systems) and reduces energy consumption by at least 31\%. Indeed, serving requested representations through \textit{(i)} the representations cached in all layers of the network (CDN, P2P, and edge), \textit{(ii)} the representations transcoded by transcoding-enabled peers, or \textit{(iii)} the resolutions up-scaled from the lower representations with smaller sizes at the peer, allows \texttt{ALIVE} to transcode fewer segments at the edge, consequently consuming less energy. In the final experiment, we measure the incurred costs to providers, based on network bandwidth and computational costs, in different systems. As plotted in Fig.~\ref{fig:net-res}(f), the \texttt{ALIVE} scheme achieves the best performance among all the methods and reduces network costs by at least 34\%. Employing all resources provided by peers improves the CHR and ETR metrics and decreases BTL as a network cost factor. 
%%%%%%%%%%%%%%%%%%%%%%%%%%%%%%%%%%%%%%%%%%%%%%%%%%%%%%%%%%%%%%%%%%%%%%%%%%%%%%%%%%%%%
\section{Summary}\label{chap:Hybrid-P2PCDN:Conclusion}
In this chapter, we utilized the benefits provided by the CDN and P2P delivery networks to introduce a hybrid P2P-CDN NAVS system for HAS clients in live streaming applications, named \texttt{RICHTER}. We leveraged the NFV and edge computing paradigms to introduce a new in-network component with a broader and central view of the network called Virtualized Tracker Server (VTS), placed at the edge of hybrid P2P-CDN networks as follows. \texttt{RICHTER} utilized all peers' potential idle computational, caching, and bandwidth resources to serve HAS clients by transcoding/fetching from peers, by transcoding/fetching from VTS, or by fetching from CDN/origin. Besides considering resource limitations, we designed an \textit{Action Tree}, including all possible actions for serving clients' requests employed by VTSs at the edge of a P2P-CDN network. Taking into account information provided by peers (\ie cache occupancy, available bandwidth values, and available computation), by CDNs (\ie cache server occupancy, available bandwidth values), and by video players (\ie requested video quality levels), \texttt{RICHTER}'s VTSs run an MILP optimization model to determine suitable nodes for serving requests. Due to the NP-completeness of the proposed MILP model, we extended the VTSs by designing an OL-based approach that uses an unsupervised Self-Organizing Map (SOM) technique for action selection decisions. To test the practical deployment of the proposed solution, we deployed \texttt{RICHTER} and analyzed its performance through experiments conducted in a large-scale testbed including 350 clients and compared its results with selected baseline approaches. The experimental results demonstrated that \texttt{RICHTER} achieved higher users' QoE (by at least 59\%), lower latency (by at least 39\%), and better network utilization (by at least 70\%) compared to other methods.

We also proposed a latency- and cost-aware hybrid P2P-CDN system for live streaming clients, called \texttt{ALIVE}. In addition to using video transcoding, \texttt{ALIVE} enables video super-resolution over the P2P network to reconstruct the requested representations from the available lower quality/bitrate representation. Considering this new feature, \texttt{ALIVE} redesigned \texttt{RICHTER}'s \textit{Action Tree} to cover all feasible resources (\ie storage, computation, and bandwidth) provided by peers, edge, and CDN servers for serving peer requests with acceptable latency and quality. Moreover, we formalized the problem as a multi-objective MILP optimization model, including both clients' latencies and delivery costs. To alleviate the optimization model's high time complexity, we proposed a lightweight greedy-based heuristic algorithm as well. Experimental results of comprehensive scenarios over a large-scale cloud-based testbed displayed that \texttt{ALIVE} enhanced the users' QoE (by at least 22\%), decreased delivery cost (by at least 34\%), shortened clients' serving latency (by at least 40\%), improved edge server energy consumption (by at least 31\%), and reduced backhaul bandwidth usage (by at least 24\%) compared to baseline approaches.

%% file: Chapters/Chapter7/7-1-Intro.tex
% \singlespacing
\chapter[Conclusions and Future Work]{Conclusions and Future Work}\label{chap:Conclusion}
\doublespacing
\vspace{-1cm}
In the following, we highlight the contributions of the thesis and several directions for future work.

%% file: Chapters/Chapter7/7-2-Summary.tex
\section{Conclusions}\label{chap:Conclusion:conc}
Over the past decade, VoD and live video content started to dominate Internet traffic. This trend has continued since the number of powerful end devices capable of requesting streams with higher qualities and resolutions has increased. However, this has also led to rising network costs and resource requirements; thus, today's delivery of OTT services is challenging for Internet and content providers. Therefore, new approaches with more intense network contributions, either by improving the performance of the existing delivery networks (\ie CDNs) or by involving the user devices as active contributors in the content delivery process, are needed to satisfy clients with acceptable quality and latency values. This thesis aimed to design, develop, deploy, and evaluate \rf{fully transparent} Network-Assisted Video Streaming (NAVS) frameworks for HTTP Adaptive Video Streaming (HAS) clients by utilizing new networking paradigms.
We presented \textit{seven} contributions that addressed the common design goals of NAVS frameworks for HAS clients. The design and evaluation of our proposed mechanisms were guided by the challenges and research questions (\textit{RQs}) identified in Chapter~\ref{chap:Introduction}. In the following, we summarize the contributions of the designed frameworks based on the \textit{RQs} addressed. Note that the italic keywords denote the primary constituent of our contributions for addressing \textit{RQs}.
 
\begin{enumerate}[label=\textit{RQ}\arabic*,noitemsep] 
\item\textbf{How can SDNs/CDNs provide assistance for HAS clients in order to improve media delivery services?}\\
\noindent To answer this \textit{RQ}, we introduced five \rf{NAVS frameworks \rf{(see Fig.~\ref{Thesis-publication} to find their differences and relations)}}, each equipping MEC servers with multimedia VNFs (\eg Virtual Proxy Function, Virtual Transcoding Function, Virtual Cache Function) and enabling them to communicate and collaborate with the SDN controller, consequently serving HAS clients' requests with better QoE and latency. The principal contributions of the proposed frameworks are as follows:
\begin{itemize}[noitemsep]
\item\textbf{Design of new server/segment selection policies:} In Chapter~\ref{chap:EdgeSDN}, we introduced the \texttt{ES-HAS} (Section~\ref{chap:EdgeSDN:ES-HAS}) and \texttt{CSDN} (Section~\ref{chap:EdgeSDN:CSDN}) frameworks that utilized edge servers with the \textit{Virtual Reverse Proxy} function to act as a gateway between HAS clients and the network and gather their information with support from the SDN controller. These edge servers also were enabled to \textit{aggregate} clients' requests and send only one request in order to prevent sending identical requests to other servers issued by multiple clients. We introduced a \textit{server/segment selection policy} method, which enables \texttt{ES-HAS} to serve HAS requests in three ways: \textit{(i)} by the originally requested quality level from a cache server with the shortest fetch time, \textit{(ii)} by overwriting the original request through a \textit{replacement quality} (only better quality levels with minimum deviation) from a cache server, or \textit{(iii)} by the originally requested quality level from the origin server.
\texttt{CSDN} expanded the \texttt{ES-HAS} system by incorporating the \textit{Virtual Transcoding Function} into the edge server, plus extending the server/segment policy to decide whether to transcode to the requested quality from a higher quality level available at the edge in case that quality was not present in the cache servers. 
%%%%%%%
\item \textbf{Design of VNFs that can be chained under an SDN controller's coordination to establish different types of video streaming services:} In Chapter~\ref{chap:SFCEnabled}, we introduced \texttt{SARENA} (Section~\ref{chap:SFCEnabled:SARENA}), which employed three multimedia VNFs: VPF, VTF, and VCF. We designed \textit{chains of these VNFs} to serve various multimedia service requests, \eg VoD and live video streaming, \textit{(i)} by using VCF to fetch them from the edge, \textit{(ii)} by using VTF to transcode them from higher quality levels fetched from a CDN server, \textit{(iii)} by using VTF to transcode them from higher quality levels fetched from the edge (\ie by VCF), \textit{(iv)} fetching from a CDN, or \textit{(v)} fetching from the origin. Moreover, we introduced \textit{request scheduler} and \textit{edge configurator} modules on the SDN controller to apply appropriate scheduling decisions for various multimedia requests and scale edge server resources based on service requirements dynamically, respectively. In addition, a \textit{lightweight heuristic} algorithm that can be run on the edge servers was proposed to mitigate the time complexity of the proposed requests scheduling policy. 
%%%%%%%
\item \textbf{Collaboration between edge servers under an SDN controller's coordination:} In Chapter~\ref{chap:CollaborativeEdge}, we presented two collaborative edge-assisted frameworks for HAS clients, namely \texttt{LEADER} (Section~\ref{chap:CollaborativeEdge:LEADER}) and \texttt{ARARAT} (Section~\ref{chap:CollaborativeEdge:ARARAT}). These frameworks utilized the capabilities of the SDN controller to establish collaboration between each Local Edge Server (LES) that receives HAS requests and its Neighboring Edge Servers (NESs). These frameworks introduced a set of actions called \textit{Action Tree} to serve client requests with acceptable latency and quality, considering all available resources (\ie storage, computation, and bandwidth provided by edge, CDN, and origin servers). Fetch the requested quality from the LES or an NES edge server that holds that and has the highest available bandwidth or transcode the requested quality in the LES or an NES that has the highest available computational resources are examples of the actions defined in the Action Tree. Furthermore, the SDN controller in both systems employed a \textit{bandwidth allocation} module that assists edge servers in achieving appropriate bandwidth resources to serve their requests from the most suitable servers in terms of time. \texttt{ARARAT} extended \texttt{LEADER}'s Action Tree, considered \textit{network cost} in the action decision-making process, and proposed extra \textit{coarse-grained} and \textit{fine-grained} heuristic algorithms to solve the problem. 
%The evaluation results demonstrated that \texttt{LEADER} and \texttt{ARARAT} enhance users' QoE by at least 22\%, decrease the streaming cost, including bandwidth and computational costs, by at least 47\%, and enhance network utilization by at least 13\% compared to state-of-the-art approaches.
\end{itemize}
%%%%RQ2%%%%%
\item \textbf{How can resources (\ie computation, storage, bandwidth) provided by the HAS clients be utilized to enhance media delivery services?}\\
\noindent In order to answer this \textit{RQ}, in Chapter~\ref{chap:Hybrid-P2PCDN}, we utilized the benefits provided by the CDN and P2P delivery networks to introduce two hybrid P2P-CDN NAVS systems for HAS clients in live streaming applications, which are latency-sensitive services. These systems are called \texttt{RICHTER}~(Section~\ref{chap:Hybrid-P2PCDN:RICHTER}) and \texttt{ARARAT}~(Section~\ref{chap:Hybrid-P2PCDN:ALIVE}). We leveraged the NFV and edge computing paradigms to introduce a new in-network component with a broader view of the network, namely \textit{Virtualized Tracker Server} (VTS), placed at the edge of these hybrid P2P-CDN systems. These frameworks employed idle computational resources and available bandwidth of HAS clients (\ie peers) to offer \textit{distributed video processing services}, such as \textit{video transcoding} and \textit{video super-resolution}. We also considered all feasible resources (\ie storage, computation, and bandwidth), provided by peers, edge, and CDN servers and proposed \textit{Action Trees}, including all possible actions to serve requests with the most appropriate action. Using the P2P network and transmitting the requested quality representation directly from the most suitable adjacent peer, transcoding the requested quality representation from a higher quality at the peer, upscaling to the requested resolution by the super-resolution function from a lower resolution received from the adjacent peer in the P2P network, or transcoding to the requested quality representation from a higher quality at the VTS and then transmitting it from the VTS to the peer are example actions of these Action Trees. Moreover, we proposed \textit{heuristic methods} (based on \textit{online learning} or \textit{lightweight algorithms}) that are designed to play decision-maker roles in large-scale practical scenarios. 
%The evaluation results showed that \texttt{RICHTER} and \texttt{ALIVE} improve users' QoE by at least 22\%, decrease streaming service provider costs by at least 34\%, shorten clients' serving latency by at least 39\%, reduce edge server energy consumption by at least 31\%, and reduce backhaul bandwidth usage by at least 24\% compared to the baseline approaches.
%%%%RQ3%%%%%
\item \textbf{What is the utility of the proposed assistance and collaboration service?}\\
\noindent To answer this \textit{RQ}, a broad spectrum of evaluation metrics was used as the principal utilities of our proposed NAVS frameworks (Chapters~\ref{chap:EdgeSDN}--\ref{chap:Hybrid-P2PCDN}). These metrics can be categorized as follows:
\begin{itemize}[noitemsep]
\item \textbf{\rf{User QoE metrics (application QoS)}}: quality bitrate, number of quality switches, number of stalls, stalling duration, serving times, VMAF~\cite{VMAF} values, standardized perceived quality~\cite{p1203}.
\item \textbf{Network utilization metrics}: cache hit ratio, transcoding ratio at the edge, transcoding ratio at the P2P network, super-resolution ratio at the P2P network, computational cost, backhaul bandwidth cost, number of communicated messages to/from the SDN controller.
\item \textbf{Algorithm performance metrics}: execution times and objective function values.
\end{itemize}
%%%%RQ4%%%%%
\item \textbf{How can the utility of the proposed NAVS frameworks be thoroughly evaluated, both theoretically and practically?}\\
\noindent Conducting performance evaluations on large-scale testbeds instead of toy simulators or numerical assessments offers several advantages, such as a more realistic representation of real-world scenarios and improved identification of potential performance bottlenecks.
Therefore, in order to address this RQ, we designed \textit{large-scale CloudLab-based}~\cite{ricci2014introducing} testbeds to run realistic network topologies for all proposed frameworks (Chapters~\ref{chap:EdgeSDN}--\ref{chap:Hybrid-P2PCDN}). These testbeds include hundreds of elements, each of which runs Linux-based operating systems inside Xen virtual machines. These components include tens/hundreds of DASH players that run buffer-based and hybrid-based ABR algorithms, CDN/origin servers that contain bitrate ladders of real video datasets, the SDN controller and edge servers, which enable to run virtualized functions, and programmable OpenFlow backbone switches. Moreover, realistic network traces, tools, and assumptions were used for designing such testbeds. Additionally, our proposed frameworks were evaluated against state-of-the-art and baseline approaches using realistic network traces, tools, and assumptions. The testbed results demonstrated that our proposed frameworks outperformed their competitors in terms of QoE, network utilization, and algorithm performance metrics in most scenarios.
%In order to address this \textit{RQ}, all proposed frameworks were implemented over large-scale CloudLab-based~\cite{ricci2014introducing} testbeds, including realistic tools, network topologies and video datasets, and compared with state-of-the-art and baseline approaches in several scenarios. 
%%%%%
\end{enumerate}

%% file: Chapters/Chapter7/7-3-outlook.tex
\section{Future Directions and Challenges}\label{chap:Conclusion:Future}
As discussed throughout this Ph.D. thesis, several techniques have been utilized in designing NAVS systems to improve the QoE metrics of adaptive streaming services. However, new challenges have emerged in the video streaming field that require further attention from the network and service management perspectives. 
Therefore, this section identifies some of these challenges and discusses potential opportunities associated with them.
%%%%%
\subsection{Designing NAVS Systems Considering Traffic Encryption} 
Privacy and security have become two primary demands for Internet users in recent years. Therefore, Internet stakeholders and application providers typically utilize encryption methods, from communication protocols to application levels, to guarantee users' privacy and security. This trend involves video streaming applications as well~\cite{nigam2021systematic}. While traffic encryption can provide private and secure services, it poses severe difficulties for designing NAVS solutions. This is because most of the NAVS systems require analyzing users' exact requests to enhance adaptive video streams, which may contradict users' privacy. Therefore, content encryption brings extra challenges for developing NAVS solutions, such as limited content visibility and extra latency for encrypted videos, unless the service provider and the network provider overlap or agree to collaborate. Moreover, determining, calculating, and estimating the QoE metric necessary to provide network-based solutions would also be challenging.
%QoE metrics would need to be estimated and analyzed to allow network-based solutions to improve HAS streams. 
%Consequently, the primary RQ is ``whether an in-network component can infer QoE factors of encrypted HAS traffic.''

Many solutions have been suggested to address this challenge in specific scenarios. Most utilize machine learning techniques to deduce key QoE metrics from observable traffic patterns and statistics. For instance, the authors in \cite{orsolic2017machine,shen2020deepqoe,orsolic2020framework} presented QoE classification techniques for YouTube encrypted traffic. However, in the future, more research is necessary to classify encrypted traffic and evaluate QoE models on in-network components with the highest possible accuracy and the lowest possible latency, particularly in the context of live video scenarios. Furthermore, QoE analyses in the presence of Digital Rights Management (DRM)~\cite{ma2017digital}, which content providers use to restrict unauthorized distribution, must be performed by in-network components to investigate whether the service works as intended and not poses any degradation in the users' QoE. 
%%%%%
\subsection{Designing NAVS Systems Utilizing Emerging Protocols} 
QUIC (Quick UDP Internet Connection) is a new transport protocol that was first developed by Google~\cite{cui2017innovating} and is now standardized~\cite{rfc9000} to address the limitations of the traditional TCP protocol. Several works showed that QUIC, based on the UDP protocol, can be combined with networking paradigms to benefit networks and applications like HAS-based streaming, especially when network congestion, loss, and RTT are high~\cite{marx2020same,palmer2018quic,kumar2021quicsdn,hou2022qfaas}. This is because QUIC can: \textit{(i)} be implemented in the user space rather than the kernel and can thus be deployed and updated more easily than TCP; \textit{(ii)} bring 0-RTT connection establishment when client and server have already communicated in the past, which helps reducing latency; \textit{(iii)} provide true multiplexing of HTTP/2 streams at the transport level, as opposed to standard TCP that still introduces head-of-line blocking when the packets of a certain stream are lost. However, devising NAVS solutions over QUIC or supporting multiple transport protocols (both TCP and QUIC) has only been marginally investigated and is still an open challenge~\cite{CP-Steering,ContextAware}. 
%Hence, the main RQ is ``how can we augment the current NAVS solutions with multi-protocol stacks, including QUIC and TCP, to satisfy all users with different network connections (\eg lossy or stable) with acceptable QoE and latency?'' Considering this RQ, we initially discussed two solutions~\cite{CP-Steering,ContextAware} that leverage QUIC's features to improve users' QoE and latency.

As another promising technology, Web Real-Time Communication (WebRTC)~\cite{roach2016webrtc}, including a set of protocols and application programming interfaces (APIs), has been introduced to allow audio and video streaming directly between browsers without needing third-party plugins or software. WebRTC can be used for a variety of use cases, including applications with low-latency and high-quality requirements, \eg live streaming. 
Nonetheless, designing NAVS solutions that combine HAS-based techniques with WebRTC to improve the performance and quality of real-time video streaming over the Internet has only been minimally explored and is still an open challenge.
% Therefore, the main RQ for investigation is ``how can HAS-based NAVS techniques be integrated with WebRTC to improve the performance and quality of real-time video streaming over the Internet''. Moreover, designing a multi-protocol NAVS, including both HAS and WebRTC protocol, is still a challenging question.
%%%%%
\subsection{Designing NAVS Immersive Video Streaming}
With the popularity and increasing use of immersive streaming devices, Virtual Reality (VR) and Augmented Reality (AR) content such as point clouds and 360\degree~video have become a rich and interactive way to consume media. In such types of streaming, high-resolution content, significant data transfer speed, and reduced motion-to-glass latency are needed to obtain high-quality and immersive QoE. 
However, VR streaming currently suffers from inadequate QoE due to the extensive bandwidth necessary for immersive content. One technique for bandwidth reduction and for QoE enhancement of VR streaming is based on \textit{viewport-dependent} schemes. The viewport refers to the portion of the content that the user is currently viewing. The user's head movements control the viewport, allowing them to look around the scene in different directions. The viewport is typically rendered in real-time to provide a seamless and immersive experience for the user~\cite{van2023tutorial}. By dynamically adjusting the content resolution/bitrate based on the user's viewport, the streaming service can reduce the amount of data that needs to be transmitted, thereby reducing bandwidth consumption and improving the overall quality of the stream. For instance, when the user is looking straight ahead and not using peripheral vision, the streaming service can reduce the content resolution visible in the peripheral vision. In fact, it is less likely that the user will notice a lower resolution in those areas.
% By adjusting the viewport based on the user's movements, the streaming platform can reduce the amount of data that needs to be transmitted, thereby minimizing bandwidth requirements and improving the overall quality of the stream~\cite{yaqoob2020survey}. 

% MPEG DASH has standardized a new viewport-dependent VR solution called tile-based streaming~\cite{niamut2016mpeg}. 
In recent years, video streaming protocols like MPEG DASH have started to use a new viewport-dependent VR solution called tile-based streaming. In tiled-based streaming, a video is divided into small rectangular areas, so-called ``tiles'', each of which can be independently streamed based on the user's field of view. As the user moves their head and changes their viewport, the streaming platform loads and renders only the necessary tiles rather than streaming the entire video~\cite{van2023tutorial}. However, problems like the use of in-network servers, \eg edge servers, which employ machine learning techniques to predict user behavior and network conditions to prefetch and prioritize content, or how to prioritize the delivery of viewport tiles using networking protocols like HTTP2/3, 
need further investigations and are still open challenges in designing NAVS systems for tile-based immersive video streaming. 
% the following items are challenging questions that must be taken into account in designing NAVS systems for tile-based immersive video streaming to provide users with seamless, high-quality, and personalized experiences:
% \begin{enumerate}[noitemsep]
% \item How can in-network servers, \eg edge servers, using machine learning techniques be utilized to predict user behavior and network conditions to prefetch and prioritize content?
% \item How to prioritize the delivery of viewport tiles using networking protocols like HTTP2/3? 
% % \item What type of user and network information would be beneficial for predicting the location where a user is likely to view content in the upcoming time?
% \end{enumerate}
%%%%%
\subsection{Designing Sustainable NAVS HAS System} 
In next-generation networks, particularly in 6G, sustainability will become a fundamental aspect, aiming to create eco-friendly networks, also known as green networks~\cite{green-ps,yrjola2020sustainability}. Sustainable NAVS multimedia systems refer to the use of network and streaming technologies that improve the environmental impact while still providing high-quality and low-latency services to users. This can involve improving the use of network resources, reducing energy consumption, and decreasing the carbon footprint associated with video streaming. 
% Hossfeld~\etal~have discussed the trade-off between the QoE of video streaming services and $CO_2$ emissions and investigated the answers to the following questions in~\cite{hossfeld2023greener}: \textit{(i)} What is more important to emphasize in the present and future: green user behavior or green networking technology? \textit{(ii)} What are the implications of solution approaches on networking and communications technology? 

However, problems such as \textit{(i)} leveraging emerging computing technologies to improve the sustainability and efficiency of video streaming systems, \textit{(ii)} using in-network components to assist video encoders in reducing video bitrates and energy consumption, \eg by constructing dynamic bitrate ladders or by deploying lightweight video quality assessment methods~\cite{menon2022etps,TQPM}, while maintaining high quality and user experience, or \textit{(iii)} enhancing network resource utilizations in multi-CDN environments~\cite{WhichCDN}, such as bandwidth and server capacity, and reducing energy consumption and carbon emissions, have rarely been investigated yet, and are still open challenges in designing sustainable NAVS systems~\cite{timmerergreen}.
% the following challenging items have rarely been investigated yet, and further investigation of them can be a step forward in designing sustainable NAVS systems:
% \begin{enumerate}[noitemsep]
% \item How to leverage emerging computing technologies to improve the sustainability and efficiency of video streaming systems?
% \item How can in-network components assist video encoders in reducing video bitrates and energy consumption, \eg by constructing dynamic bitrate ladders or by deploying lightweight video quality assessment methods~\cite{menon2022etps,TQPM}, while maintaining high quality and user experience?
% \item How can network resources be enhanced in multi-CDN environments~\cite{WhichCDN}, such as bandwidth and server capacity, and energy consumption and carbon emissions be reduced?
% % \item How can we shift ABR algorithms from HAS clients into an in-network component to apply green adaptation policies?
% \end{enumerate}
%%%%%